\documentclass[review]{elsarticle}

\usepackage[colorlinks,citecolor=blue,linktoc=all,linkcolor=cyan]{hyperref}
\usepackage{graphicx}

\usepackage[T1]{fontenc}
\usepackage{mathrsfs}             
\usepackage{slashed}              
\usepackage{amsmath}
\usepackage{amssymb}
\usepackage{amsbsy}
\usepackage{amsfonts}

\usepackage[dvipsnames]{xcolor}
\usepackage{upgreek}
\usepackage{comment}
\usepackage{geometry}
\usepackage{microtype}
\usepackage{enumitem}
\usepackage{longtable}
\usepackage{booktabs}
\usepackage{caption}
\usepackage{multirow}
\usepackage{array}
\usepackage{setspace}
\usepackage{fancyhdr}
\usepackage{xspace}
\usepackage{colortbl}
\usepackage{sidecap}
\usepackage{float}
\usepackage{tablefootnote}
\usepackage{tabularx}
\usepackage{makecell}
\usepackage{booktabs}
\usepackage{placeins}

\usepackage[most]{tcolorbox}
\usepackage{enumitem}
\setlist[itemize]{
  itemsep=1pt,
  parsep=0pt,
  topsep=4pt,
  partopsep=0pt,
  leftmargin=1.5em
}
\newtcolorbox{advchallengebox}[1][]{
  colback=gray!5,
  colframe=gray,
  boxrule=0.5pt,
  arc=0pt,
  left=8pt,
  right=8pt,
  top=6pt,
  bottom=6pt,
  before skip=10pt,
  after skip=10pt,
  breakable,
  fonttitle=\bfseries,
  title=#1
}

\newcommand{\boxsection}[1]{%
  \vspace{2pt}%
  \noindent\textbf{#1}\par\vspace{2pt}%
}

\newcommand{\twocoladvchallenge}[3][]{
\begin{advchallengebox}[#1]
\begin{minipage}[t]{0.48\linewidth}
\boxsection{Advantages}
#2
\end{minipage}
\hfill
\begin{minipage}[t]{0.48\linewidth}
\boxsection{Challenges}
#3
\end{minipage}
\end{advchallengebox}
}

\newcommand{\twocolChallAch}[3][]{
\begin{advchallengebox}[#1]
\begin{minipage}[t]{0.48\linewidth}
\boxsection{Challenges}
#2
\end{minipage}
\hfill
\begin{minipage}[t]{0.48\linewidth}
\boxsection{Achievements}
#3
\end{minipage}
\end{advchallengebox}
}

\numberwithin{equation}{section}
\numberwithin{table}{section}
\numberwithin{figure}{section}

\journal{Progress in Particle and Nuclear Physics}

\usepackage{titlesec}
\usepackage{sectsty}
\titleformat{\section}{\normalfont\Large\bfseries}{\thesection}{1em}{}
\titleformat{\subsection}{\normalfont\large\bfseries}{\thesubsection}{1em}{}
\titleformat{\subsubsection}{\normalfont\normalsize\bfseries}{\thesubsubsection}{1em}{}

\begin{document}
	
	\begin{frontmatter}
		
		\title{Hunting dark matter: the present and future of direct particle detection}
		
        \author[ITP,IAP]{Felix~Kahlhoefer}
        \ead{kahlhoefer@kit.edu}
		\author[KIP]{Belina~von Krosigk\corref{mycorrespondingauthor}}
		\cortext[mycorrespondingauthor]{Corresponding author}
		\ead{bkrosigk@kip.uni-heidelberg.de}
		\author[MPIK]{Teresa~Marrodán~Undagoitia}
        \ead{marrodan@mpi-hd.mpg.de}
		\author[xxx]{Valentina~Novati}
        \ead{valentina.novati@lpsc.in2p3.fr}
		\author[ZU]{María~Luisa~Sarsa}
        \ead{mlsarsa@unizar.es}

		\address[ITP]{Institute for Theoretical Particle Physics, Karlsruhe Institute of Technology, 76128 Karlsruhe, Germany}            \address[IAP]{Institute for Astroparticle Physics, Karlsruhe Institute of Technology, 76128 Karlsruhe, Germany}
		\address[KIP]{Kirchhoff-Institute for Physics, Heidelberg University, 69120 Heidelberg, Germany}
            \address[MPIK]{Max-Planck-Institut für Kernphysik, 69117 Heidelberg, Germany}
            \address[xxx]{Univ. Grenoble Alpes, CNRS, Grenoble INP, LPSC-IN2P3, 38000 Grenoble, France}
            \address[ZU]{Centro de Astropart\'{\i}culas y F\'{\i}sica de Altas Energ\'{\i}as (CAPA), Universidad de Zaragoza, 50009 Zaragoza, Spain}
        
	\begin{abstract}

  The nature of dark matter remains one of the major open questions in particle physics and cosmology. 
  Direct detection experiments aim to identify dark matter particles through their interactions with ordinary matter, but despite substantial improvements in experimental sensitivity, no conclusive detection has been achieved. 
  In this review, we provide a comprehensive overview of the present status and future prospects of direct particle detection, focusing on searches for dark matter particles from the Galactic halo. 
  We introduce the theoretical framework for interpreting experimental results, including nuclear and electron scattering, collective excitations, particle absorption, and time- and direction-dependent signatures. 
  We review the detection technologies employed in these searches, highlighting their operating principles, advantages, and limitations, as well as the calibration techniques and background mitigation strategies required to achieve high sensitivity. Particular emphasis is placed on experimental results reported since 2020, covering nuclear and electron recoil searches, annual modulation measurements, and directional detection. We also discuss experiments under construction and proposed future detectors, ranging from large-scale noble-liquid experiments to low-threshold technologies targeting sub-GeV dark matter. 
  Finally, we examine the challenges that will shape the next generation of experiments, including the need for increasingly low energy thresholds, improved background understanding and discrimination, and the growing importance of neutrino-induced backgrounds. 
  The diversity and complementarity of detector technologies will remain essential for exploring the dark matter parameter space and establishing the nature of any potential signal.
		\end{abstract}
		
		
	\end{frontmatter}	
	
	\thispagestyle{empty}
	\tableofcontents
	
	\newpage


\section{Introduction}
\label{sec:introduction}

When considering all observations of the Universe alongside established theories, a remarkably coherent scenario emerges. General relativity, combined with the cosmological principle, leads to the standard cosmological model. This model consistently accounts for a wide range of observables with high quantitative precision, although its success relies on the introduction of two unknown components: dark matter, introduced to account for the dynamics of galaxies and clusters and the evolution of primordial density perturbations, and dark energy, which drives the accelerated expansion of the Universe. 

Dark matter accounts for 27\% of the Universe's mass-energy budget~(see \cite{Aghanim:2018eyx} and Sec.~25 of \cite{ParticleDataGroup:2026mpi}); however, no known particles or aggregates possess the properties required to align with observational evidence across galactic and cosmological scales. 
Although all these observations provide evidence for dark matter through its gravitational effects, they also allow us to place constraints on its possible non-gravitational interactions, both self-interactions and couplings to the other constituents of the Universe. For instance, the apparently collisionless behavior of dark matter in galaxy cluster collisions, such as the Bullet Cluster, places upper bounds on its self-interaction strength~\cite{Clowe:2006eq}.
Furthermore, primordial nucleosynthesis (see Sec.~24 of~\cite{ParticleDataGroup:2026mpi}) as well as measurements of acoustic oscillations in the cosmic microwave background, as well as the growth of large-scale structures in the Universe strongly disfavor dark matter is exclusively composed of ordinary baryonic or leptonic matter~(see Sec.~27 of \cite{ParticleDataGroup:2026mpi}). 
Understanding the nature of dark matter remains one of the most elusive and challenging objectives in modern cosmology and particle physics, driving an extensive program of experimental efforts towards its detection~(see Sec.~27 of \cite{ParticleDataGroup:2026mpi}). 

Dark matter may represent a placeholder for our limited understanding of the physics laws governing the Universe. A theoretical framework incorporating modified gravity or modified dynamics~\cite{BOEHM_2024116503} could provide an alternative interpretation of the observations, potentially rendering the concept of dark matter unnecessary. However, 
the particle nature of dark matter remains a leading hypothesis, motivated by a broad range of theoretical and phenomenological considerations. While the need to extend the Standard Model of particle physics provides a compelling motivation to explore new particle candidates, non-particle candidates, such as primordial black holes, remain viable alternatives~\cite{Green:2020jor}. Many theories beyond the Standard Model provide particle candidates that could explain dark matter~\cite{Feng:2010gw,Zurek:2024vma}, and complementary experimental strategies have been developed. The first approach involves the direct detection of energy depositions from dark matter particles in specialized detectors~\cite{Billard:2022uyg}, whereas indirect detection aims to observe the products resulting from the annihilation or decay of dark matter in regions where it accumulates~\cite{Conrad:2017pda}. 
A complementary experimental approach consists of searching for new physics that can constrain the properties of potential dark matter candidates, primarily through accelerator experiments, where either the candidates themselves or new mediators connecting the Standard Model with the dark sector could be discovered~\cite{Boveia:2018yeb}. 
All of these approaches rely on some coupling between dark matter---or more generally, a dark sector---and Standard Model particles. Since the nature and strength of these interactions are highly model-dependent, the interpretation of experimental results, whether positive or negative, remains challenging.

This review focuses on the direct detection of particle dark matter, but excludes axion and axion-like particles searches, which mostly utilize distinct and specific detection methods via haloscopes or helioscopes focusing on wave-like dark matter. 
The interested reader may consult other reviews for further information (see \cite{DiLuzio:2020wdo,Irastorza:2018dyq}, and Sec.~89 of 
\cite{ParticleDataGroup:2026mpi}).

The first experimental efforts to directly detect dark matter candidates can be traced back to the 1980s~\cite{Ahlen:1987mn}. Experimental sensitivities have improved by orders of magnitude since then, both by increasing the target mass and strongly reducing the backgrounds. Furthermore, the development of better performing detectors in terms of energy threshold have enabled the extension of the accessible parameter space into the low dark-matter mass region. However, non-standard interactions may produce signals outside the energy ranges typically explored or with spectral features that differ from usual expectations. 

Despite significant experimental efforts, no dark matter signal has been conclusively identified. DAMA/LIBRA has reported an annual modulation over more than two decades that is compatible with a dark matter interpretation~\cite{Bernabei:2020mon,Bernabei:2026fdk}; however, this observation has not been established as a dark matter detection, as it remains incompatible with the results of numerous other sensitive experiments under most commonly considered scenarios for the coupling between dark matter and Standard Model particles~\cite{Billard:2022uyg}.

This review is organized as follows:  
Sec.~\ref{sec:theory} categorizes the experimental signatures associated to dark matter interactions with standard model particles and revises the dependence of the expected interaction rates in a convenient detector on particle models and relevant astrophysical parameters. It provides the tools 
for interpreting and comparing experimental results using different targets and signatures, which require assuming specific coupling and halo models; 
Sec.~\ref{sec:tech} presents the different technologies for particle detection applied in direct dark matter searches, highlighting the advantages offered by each as well as the challenges they face for this application; the signal generation mechanisms within the different detection schemes are also introduced; 
Sec.~\ref{sec:bck} 
discusses the calibration strategies required for the proper characterization of the potential signatures of dark matter, and it also reviews the primary backgrounds limiting the sensitivity of direct detection experiments and current state-of-the-art mitigation strategies;
Sec.~\ref{sec:experiments} summarizes the experiments taking data or recently decommissioned 
as well as those currently under construction, commissioning, or in the planning stage, providing an overview of the current state of the field and its prospects for the next two decades. To improve readability, this review covers only the results presented since 2020. 

The present review aims to give a comprehensive overview of the different technologies and the experimental landscape for direct dark matter particle detection. Other recent reviews of direct detection with a different or more narrow focus can be found in Refs.~\cite{Essig:2022dfa,Misiaszek:2023sxe,Baudis:2025yva}. For reviews of direct detection with a stronger focus on theoretical concepts, we refer to Refs.~\cite{Cerdeno:2010jj,Lin:2019uvt}. For much broader reviews of dark matter evidences, theories and searches we recommend Refs.~\cite{Bertone:2016nfn,Bauer:2017qwy,Balazs:2024uyj,Cirelli:2024ssz}.
\section{Theory}
\label{sec:theory}

\subsection{The differential event rate}

At the most basic level, the expected event rate in a direct detection experiments is given by the product of the flux of dark matter particles $j_\chi = n_\chi \cdot v = \rho_\chi \cdot v / m_\chi$, the total number of target atoms $N_T = m_T / m_N$ and the interaction cross section $\sigma$. In these formulas, $n_\chi$ and  $\rho_\chi$ denote the dark matter number density and mass density, respectively, $v$ and $m_\chi$ the dark matter velocity and mass, and $m_T$ and $m_N$ denote the mass of the entire target and of each target nucleus. The dark matter density is expected to be time-independent and the same for all experiments, and is often denoted by $\rho_0$.\footnote{The reason that we express the number density in terms of the mass density and the dark matter mass is that the mass density directly affects the orbits of stars and gas in the Milky Way and is hence the quantity that can be constrained by astrophysical observations.} By convention, we divide the expected event rate by the total mass of the target to obtain the normalised rate $R$ that depends only on the target material and not on the specific experiment.

In practice, we are often interested not in the total event rate, which is proportional to the total cross section $\sigma$, but in the differential event rate with respect to some observable quantity, such as the deposited energy in the detector $E$, which depends on a suitable differential cross section such as $\mathrm{d}\sigma/\mathrm{d}E$. Moreover, since not all dark matter particles have the same velocity $v$, we need to multiply with the (normalised) velocity distribution $f(\mathbf{v},t)$ and integrate over $v$. These considerations yield
\begin{equation}
\frac{\mathrm{d}R}{\mathrm{d}E} = \frac{\rho_0}{m_\chi m_N}\, \int \mathrm{d}^3v \; v \, f(\mathbf{v},t)\; \frac{\mathrm{d}\sigma}{\mathrm{d}E} \; .
\label{eq:dRdER_integral}
\end{equation}

The differential cross section depends on both $v$ and $E$ and vanishes if the kinetic energy of the initial state is too small compared to the energy deposition $E$. For example, in the case of elastic nuclear scattering, the nuclear recoil energy $E_R$ is given by
\begin{equation}
\label{eq:SI-recoil-energy}
E_R = \frac{\mu^2 v^2 (1 - \cos\theta)}{m_N} \; ,
\end{equation}
where $\mu = m_N m_\chi / (m_N + m_\chi)$ is the reduced mass and $\theta$ is the scattering angle in the centre-of-mass frame. The recoil energy is maximal for $\theta = \pi$ and satisfies $E_R < 2 \mu^2 v^2 / m_N$. For a given nuclear recoil energy, we therefore have to restrict the range of the velocity integration to $v > v_\text{min} = \sqrt{m_N E_R / (2 \mu^2)}$. For the case of inelastic scattering on nuclei or scattering on electrons, the minimal velocity depends separately on the momentum transfer $q$ and the energy transfer $\Delta E$:
\begin{equation}
\label{eq:vmin}
 v_\text{min} = \left|\frac{\Delta E}{q} + \frac{q}{2 m_\chi} \right| \; ,
\end{equation}
which recovers the formula for elastic nuclear scattering for $\Delta E = q^2 / (2 m_N)$.

We note that in realistic experiments the deposited energy cannot be directly observed. Instead, the recoiling nucleus is slowed down by the detector medium. This leads to excitation and ionization, resulting in the emission of heat, photons, and phonons. Depending on the detector used, one or more of these signals are detected. The recoil energy must therefore be reconstructed based on the generated and detected quanta. A detector response function $p(O, E)$ quantifies the probability to measure the observable $O$ for a fundamental energy deposition $E$. The actual signal in the detector, i.e.\ the differential event rate with respect to the observable $O$ is then obtained by convolving $\mathrm{d}R/\mathrm{d}E$ with $p(O,E)$. In simple cases, $p(O,E)$ is just a normal distribution with mean given by $E$ and standard deviation given by the detector resolution $\sigma_E$. In general, a precise determination of the detector response function is one of the key challenges for direct detection experiments.

Two different types of input are required to calculate the differential event rate: the astrophysical inputs $\rho_0$ and $f(\mathbf{v})$, and the unknown particle physics properties $m_\chi$ and $\mathrm{d}\sigma/\mathrm{d}E$ (which together also determine $v_\text{min}$). Let us discuss each of these in turn.

\subsection{Astrophysical inputs}
\label{ssec:astrophysical}

The astrophysical inputs $\rho_0$ and $f(\mathbf{v})$ are an important source of uncertainty for the interpretation of direct detection experiments~\cite{McCabe:2010zh,Green:2011bv,Green:2017odb}. While a change in the assumed local dark matter density $\rho_0$ simply implies an overall rescaling of the differential event rate, its value must be known in case of a discovery to determine the total cross section and hence measure the fundamental particle physics properties of dark matter. We refer to Ref.~\cite{Read:2014qva} for a detailed review and simply note here that there exist different approaches for determining $\rho_0$, which agree on the fact that $\rho_0 > 0$, i.e.\ that there is clear evidence for a non-zero dark matter density in the solar neighbourhood. However, the different approaches make different assumptions regarding the symmetry and isotropy of the dark matter halo and lead to inferred values for $\rho_0$ that disagree with each other by more than the claimed uncertainties. A reasonable estimate for the uncertainty in $\rho_0$ is therefore obtained by considering the spread of results obtained using different methods. This procedure yields roughly $0.3 \, \mathrm{GeV/cm^{3}} \leq \rho_0 \leq 0.6 \, \mathrm{GeV/cm^{3}}$. For the purpose of comparing results from different collaborations, many experiments adopt the value $\rho_0 = 0.3 \, \mathrm{GeV/cm^{3}}$, see also the recommendations in Ref.~\cite{Baxter:2021pqo}. Constraints on the scattering cross section obtained for this reference value can be trivially rescaled to different values of $\rho_0$ (or to sub-dominant dark matter components).

The dark matter velocity distribution $f(\mathbf{v})$ enters the calculation in a more complicated way. Assuming an isotropic scattering cross section, such that $\mathrm{d}{\sigma}/{\mathrm{d}E_R} \propto v^{-2}$, direct detection experiments are sensitive to the integral
\begin{equation}
 \eta(v_\text{min}) = \int_{v_\text{min}} \frac{f(\mathbf{v})}{v} \mathrm{d}^3 v \; .
 \label{eq:eta}
\end{equation}
Since $v_\text{min}$ depends on the mass of the target atoms and the energy range under consideration, changes in $f(\mathbf{v})$ affect different experiments in different ways. Moreover, different types of interactions may have a different velocity dependence, and some experiments may even have an anisotropic detector response that depends on the direction of the velocity. A comparison between different direct detection experiments therefore usually requires assumptions on both the velocity dependence of the cross section and on $f(\mathbf{v})$. We note in passing that there exist a number of halo-independent methods~\cite{Fox:2010bz,Frandsen:2011gi,Gondolo:2012rs,Ferrer:2015bta,Feldstein:2014gza,Gelmini:2017aqe,Kahlhoefer:2018knc,Bernreuther:2023aqe} that perform a comparison of different experiments without relying on such assumptions.

The most common assumption is that $f(\mathbf{v})$ is isotropic in the rest frame of the Milky Way, such that an anisotropy only arises from the peculiar velocity of the Sun $\mathbf{v}_\text{S}$ and the (time-dependent) velocity of the Earth relative to the Sun $\mathbf{v}_\text{E}(t)$:
\begin{equation}
 f(\mathbf{v}) = f_G(|\mathbf{v} - \mathbf{v}_\text{S} - \mathbf{v}_\text{E}(t)|) \; .
\end{equation}
The velocity distribution in the Galactic rest frame depends on the assumed density profile $\rho(r)$ of the dark matter halo. In the simplest case of an isothermal halo with $\rho(r) \propto r^{-2}$, one expects a Maxwell-Boltzmann distribution with velocity dispersion $v_0$, which can be observationally determined from the rotational velocity of the Galactic plane. An isothermal halo would however have infinite mass and therefore infinite escape velocity, while we infer a finite escape velocity $v_\text{esc}$ from observations of high-velocity objects in the solar neighbourhood. To account for this fact, one should consider a truncated Maxwell-Boltzmann distribution:
\begin{equation}
 f_G(v) = \begin{cases}
\frac{1}{N} \exp(-v^2/v_0^2) \, , & v < v_\text{esc} ,\\
0 & v > v_\text{esc} 
\end{cases}\; ,
\end{equation}
where $N$ is a suitable normalisation constant that ensures $\int f(v) \mathrm{d}^3 v = 1$. This distribution is commonly referred to as the Standard Halo Model (SHM). While $v_0$ has been measured rather precisely in recent years~\cite{Abuter:2021yys}, uncertainties in $v_\text{esc}$ remain sizable. The currently recommended values for these parameters are $v_0 = 238 \, \mathrm{km/s}$ (replacing the long-standing recommendation $v_0 = 220 \, \mathrm{km/s}$) and $v_\text{esc} = 544 \, \mathrm{km/s}$, see Ref.~\cite{Baxter:2021pqo}, which also gives recommendations for $\mathbf{v}_\text{E}$ and $\mathbf{v}_\text{S}$. 

In the SHM and for $v_\text{min} \ll v_\text{esc}$, the velocity integral scales approximately as $\eta(v_{\rm min}) \sim \exp(- v_{\rm min}^2/v_0^2)$, which implies
for elastic and isotropic nuclear scattering that
\begin{equation}
\frac{dR}{dE_R} \sim \exp\left(-\frac{E_R}{E_0}\right),
\end{equation}
with the characteristic energy $E_0 = 2 \mu_N^2 v_0^2/m_N$. Hence, the nuclear recoil energy spectrum is expected to fall exponentially for sufficiently low recoil energies. At higher recoil energies, an even stronger suppression is expected due to the Galactic escape velocity and nuclear form factors (see below).

\begin{figure}
    \centering
    \includegraphics[width=0.7\linewidth]{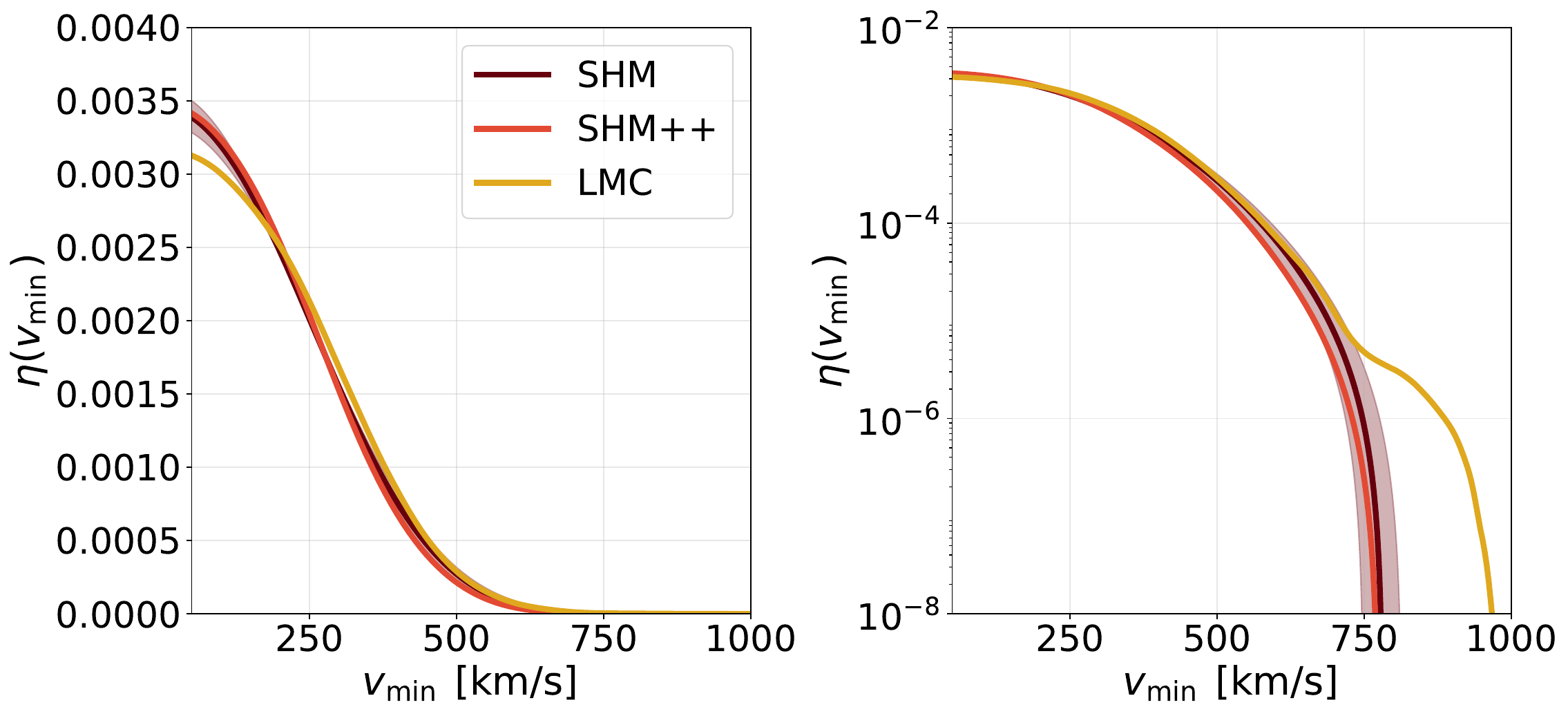}
    \caption{The velocity integral $\eta(v_\text{min})$ defined in eq.~\eqref{eq:eta} for the Standard Halo Model (SHM) and two alternative halo models with anisotropic contributions from non-virialised dark matter substructures (SHM++, see Refs.~\cite{Evans:2018bqy,OHare:2019qxc}) or from the Large Magellanic Cloud (LMC, see Refs.~\cite{Smith-Orlik:2023kyl,Reynoso-Cordova:2024xqz}). For the SHM we show a band to indicate the effect of varying the parameters $v_0$ and $v_\text{esc}$ within the $1\sigma$ uncertainties recommended in Ref.~\cite{Baxter:2021pqo}. The two panels differ only in the scaling of the y-axis.}
    \label{fig:halo_models}
\end{figure}

Over the years, many alternative forms for $f(\mathbf{v})$ have been proposed, inspired by numerical simulations of structure formation, analytical models of galactic dynamics or observations of stellar velocities from Gaia, see figure~\ref{fig:halo_models}. While some of these modifications simply change the functional form of $f(v)$, others introduce anisotropic contributions, arising for example from non-virialised dark matter substructures~\cite{Evans:2018bqy,OHare:2019qxc} or from the proximity of the Large Magellanic Cloud~\cite{Smith-Orlik:2023kyl,Reynoso-Cordova:2024xqz}. A particularly interesting idea is that in principle also dark matter particles with $v > v_\text{esc}$ can contribute to $f(v)$ as they travel through the Milky Way. Such particles (often called boosted dark matter) can arise for example in the decays of heavier states, through the scattering of slower dark matter particles off highly energetic cosmic rays~\cite{Bringmann:2018cvk,Dent:2019krz,Xia:2021vbz} or via solar reflection~\cite{An:2017ojc,An:2021qdl}. While the contribution to $\rho_0$ of such particles is expected to be tiny, they may nevertheless give relevant signals in direct detection experiments due to their comparably large kinetic energy.

\begin{figure}
    \centering
\includegraphics[width=0.75\linewidth]{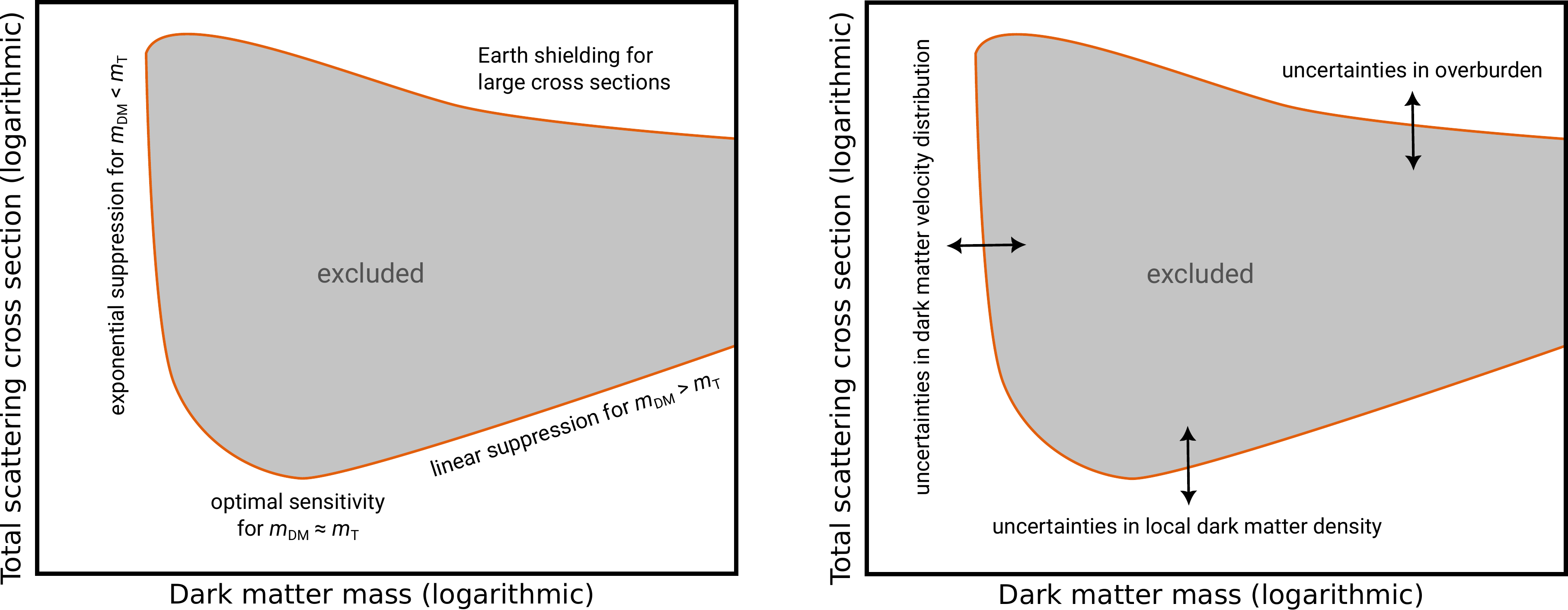}
    \caption{Illustration of the typical shape of an exclusion from a direct detection experiment. In the right panel, we show the dominant (theoretical) uncertainties that affect the exclusion.}
    \label{fig:exclusion}
\end{figure}

Indeed, energy and momentum conservation play a fundamental role in understanding the sensitivity of direct detection experiments. If there are no dark matter particles with $v > v_\text{max}$ in the laboratory frame (where in the SHM $v_\text{max} = v_\text{esc} + |\mathbf{v}_\text{S} + \mathbf{v}_\text{E}|$), the integrated velocity distribution $\eta(v_\text{min})$ vanishes for $v_\text{min} > v_\text{max}$. In other words, there will be some maximal possible energy deposition $E_\text{max}$, which for elastic nuclear scattering is given by $E_\text{max} = 2 \mu^2 v_\text{max}^2 / m_N$. Clearly, $E_\text{max}$ decreases with decreasing dark matter mass (and hence reduced mass $\mu$). Once $E_\text{max}$ falls below the energy threshold of an experiment, dark matter scattering becomes unobservable. As a result, direct detection experiments lose sensitivity for small dark matter masses. For large dark matter masses (i.e. $m_\chi \gg m_N$), on the other hand, the reduced mass and hence the scattering kinematics become independent of the dark matter mass. In this limit, the only dependence on the dark matter mass is through the overall factor $\rho_0 / m_\chi$, which leads to a linear loss in sensitivity. For elastic nuclear scattering, the optimal sensitivity is typically achieved for $m_\chi \approx m_N$. This is illustrated in figure~\ref{fig:exclusion}, which also shows the impact of astrophysical uncertainties on the exclusion limits.

\subsubsection{Flux dependence on time and direction}
\label{ssec:modulation}

The motion of the Sun through the Milky Way gives rise to two additional effects that can be used to identify a dark matter signal and reject possible backgrounds, namely a dependence of the incoming dark matter flux (and hence of the differential event rate) on time and direction.

The motion of the Earth around the Sun implies an annual variation of the velocity of the Earth in the Galactic rest frame. In summer, $\mathbf{v}_\text{E} \cdot \mathbf{v}_\text{S} > 0$, i.e.\ the velocities are partially aligned, while the opposite is the case in winter. This means that in the laboratory frame, dark matter particles are expected to have slightly higher velocities in summer than in winter. Since $|\mathbf{v}_\text{E}| / |\mathbf{v}_\text{S}| \approx 10\%$, this seasonal variation is a rather small effect, such that the event rate can approximately be expanded as
\begin{equation}
\label{eq:anmod}
 R(t) = R_0 \left[1 + A \cos\left(2 \pi \frac{t - t_0}{1 \, \text{year}}\right)\right] \;, 
\end{equation}
where $t_0$ is approximately in early June, corresponding to the maximal rate for $A>0$.\footnote{
A positive modulation amplitude is generally expected at sufficiently high recoil energies, where a larger flux of particles can contribute in summer than in winter. At low recoil energies, on the other hand, the modulation amplitude changes sign, $A < 0$, which is sometimes referred to as anti-modulation. From the experimental point of view, the existence of this change of sign implies that cancellations are possible if wide energy regions are considered in an analysis of the modulation.} While the modulation amplitude $A$ is expected to be of the order of $10\%$ for heavy dark matter, it can be significantly enhanced for light dark matter, for which only the tail of the velocity distribution contributes to scattering and small changes in the velocity can have a large effect. Given a sufficiently stable detector, the time dependence of the dark matter signal can potentially be observed and used to distinguish signal from background.

Furthermore, the motion of the Sun implies a preferred direction of incoming dark matter particles (the so-called WIMP wind), which in the rest frame of the detector changes over the course of the day due to the rotation of the Earth. This effect can be exploited in two ways. The first is to employ targets for which the detector response depends on the direction of the momentum transfer $\mathbf{q}$, such as electron excitations in anisotropic crystals. 
In such materials, dark matter particles coming from certain directions have a larger chance of scattering and/or may deposit more energy, which in turn leads to a dependence of the differential event rate on the velocity direction of the incoming dark matter particle~\cite{Griffin:2018bjn,Geilhufe:2019ndy,Boyd:2022tcn,Stratman:2024sng}. The change in direction of the WIMP wind then leads to a diurnal modulation of the event rate.

The second possibility is to directly reconstruct the direction of the nuclear recoil, which is correlated (but not identical) with the direction of the incoming dark matter particle. Doing so requires special types of experiments, employing for example low-density targets that allow for the reconstruction of tracks from nuclear recoils. 
Denoting the direction of the nuclear recoil by the unit vector $\hat{\mathbf{r}}$ and the surrounding infinitesimal solid angle element by $\mathrm{d}\Omega$, one can show that
\begin{equation}
\frac{d^2\sigma}{dE_R\, d\Omega}
=
\frac{d\sigma}{dE_R}\,
\frac{1}{2\pi}\,
v\,
\delta(\mathbf{v}\cdot\hat{\mathbf{r}} - v_{\min}),
\end{equation}
and hence $\mathrm{d}^2 R/\mathrm{d}E_R\, \mathrm{d}\Omega$ is proportional to the Radon transform~\cite{Vahsen:2020pzb}
\begin{equation}
\hat{f}(v_{\min},\hat{\mathbf{r}},t)
=
\int
\delta(\mathbf{v}\cdot\hat{\mathbf{r}} - v_{\min})
\, f(\mathbf{v},t)\,
d^3\mathbf{v}.
\end{equation}
If the velocity distribution $f(\mathbf{v})$ is isotropic in the Galactic rest frame, the left-hand side of this integral only depends on $\hat{\mathbf{r}}$ through the angle $\theta$ between the direction of the WIMP wind and the direction of the nuclear recoil. The generic prediction is that the rate of events scattering in the forward direction (i.e.\ in the direction of the WIMP wind) exceeds the rate of backward scattering events. A detector capable of measuring this forward-backward asymmetry would provide a powerful tool for confirming the detection of dark matter particles, see also Refs.~\cite{Mayet:2016zxu,Vahsen:2021gnb} for reviews on the topic.

We note that diurnal modulations may also be of interest for experiments without directional sensitivity if the flux of dark matter particles itself varies over the course of the day. This situation arises in certain models of strongly-interacting~\cite{Kavanagh:2016pyr,Bertou:2025adb} or inelastic~\cite{Emken:2021vmf} dark matter, where the dark matter particles may scatter in the atmosphere or the Earth on the way to the detector and the resulting flux depends on whether the WIMP wind comes from above or below the horizon (see also section~\ref{sec:shielding}). A similar but smaller effect may also arise from gravitational focusing~\cite{Kouvaris:2015xga}.

\subsection{Particle physics inputs}

Let us now turn to the differential cross section, which captures the underlying particle physics and determines the shape and magnitude of the expected signals. At the most basic level, we distinguish between the case of dark matter-nucleus and dark matter-electron interactions, although a more detailed description also requires the treatment of collective excitations, such as band structures in semiconductors or phonons in crystals (see below).

\subsubsection{Nuclear scattering}
\label{subsubsec_Nucl_Scat}

Elastic dark matter-nucleus scattering arises from the more fundamental interaction between dark matter and protons or neutrons (which in turn can be calculated from the underlying interaction between dark matter and quarks or gluons~\cite{Bishara:2017pfq}). The momentum transfer in elastic dark matter-nucleus scattering is typically too small to resolve the structure of the nucleus, such that the contribution from the individual nucleons can be combined using a nuclear form factor, which captures the gradual loss of coherence for large momentum transfer~\cite{Lewin:1995rx,Hoferichter:2018acd}. While the nuclear form factors depend on the specific target material of the detector, the dark matter-nucleon scattering cross section is the same for all experiments, although it might be different for scattering on protons and neutrons. Experimental exclusion limits and sensitivities are therefore commonly quoted in terms of the total dark matter-nucleon scattering cross section $\sigma_n$ (which implicitly assumes equal couplings to protons and neutrons) or the dark matter-proton scattering cross section $\sigma_p$ (which, depending on the context, assumes either equal coupling to protons and neutrons or no coupling to neutrons).\footnote{Of course also the opposite case, in which dark matter couples dominantly to neutrons, is possible. Dark matter models with different couplings to protons and neutrons are often referred to as isospin-violating dark matter~\cite{Feng:2011vu}.}

\begin{figure}
    \centering
\includegraphics[width=\linewidth]{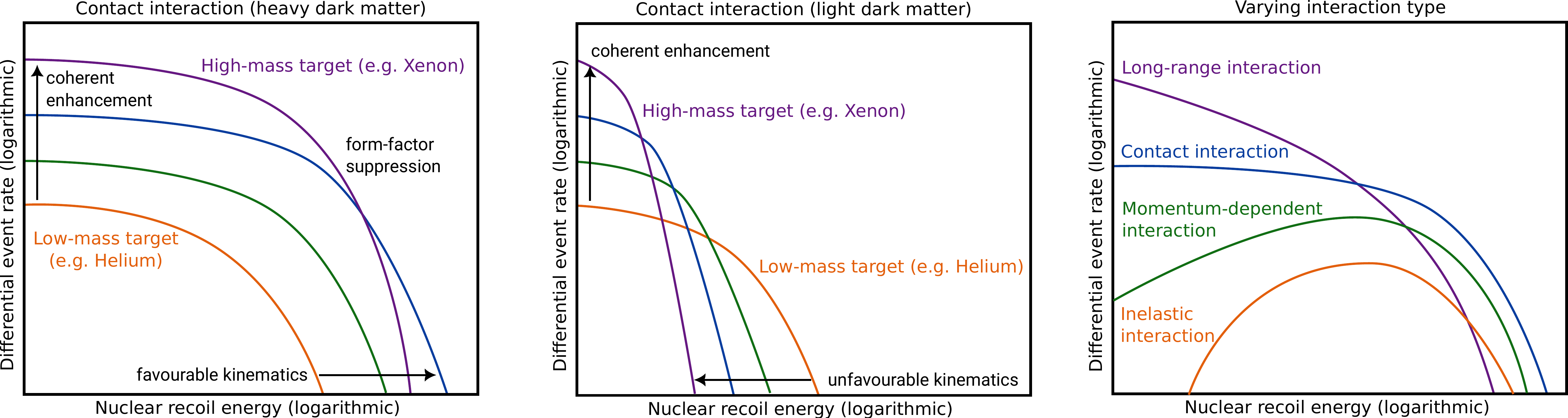}
    \caption{Illustration of the typical shape of the differential event rate. The first two panels focus on the case of (spin- and momentum-independent) elastic scattering for different target elements and correspond respectively to the case of heavy and light dark matter. In the third panel we compare different types of interactions.}
    \label{fig:rates}
\end{figure}

The dark matter-nucleon interactions can be classified using a non-relativistic effective field theory~\cite{Fitzpatrick:2012ix,Anand:2013yka}, which consists of roughly a dozen different effective operators (with the precise number depending on the dark matter spin) that may depend on four different vectors: the dark matter spin $\mathbf{s}_\chi$, the nucleon spin $\mathbf{s}_n$, the momentum transfer $\mathbf{q}/m_n$ and the relative velocity $\mathbf{v}$. The simplest operator, called $\mathcal{O}_1$ depends on none of these vectors and gives rise to so-called spin-independent scattering. In this case the contribution from all nucleons can be added coherently, such that the scattering cross section for a nucleus with charge number $Z$ and mass number $A$ is given by
\begin{equation}
\label{equ:diff_xsec_SI}
 \frac{\mathrm{d}\sigma^{\rm SI}}{\mathrm{d}E_R} = 
\frac{m_N}{2\mu_p^2 v^2} \sigma_p^{\rm SI} \left(Z + (A-Z) \frac{f_n}{f_p}\right)^2 F^2(q) \; ,
\end{equation}
where $\mu_p$ denotes the dark matter-proton reduced mass and $f_{p,n}$ are the couplings of dark matter to protons and neutrons, respectively. The nuclear form factor $F(q)$ parametrises the loss of coherence for large momentum transfer and is normalised such that $F(0) = 1$. For $f_p = f_n$ the scattering cross section is enhanced by a factor $A^2$ for heavy nuclei, see the first two panels of figure~\ref{fig:rates}. These panels also show how targets with higher atomic masses (such as xenon) benefit from favourable kinematics for heavy dark matter ($m_\chi \gtrsim m_N$), while targets with lower atomic masses (such as helium) have clear advantages for light dark matter ($m_\chi \lesssim m_N$).

A second operator of particular interest, called $\mathcal{O}_4$, involves the product $\mathbf{s}_\chi \cdot \mathbf{s}_n$, giving rise to so-called spin-dependent scattering. In this case, the dark matter-nucleus scattering cross section is non-zero only for nuclei with non-zero total spin. Assuming that dark matter couples only to the proton spin (but not to the neutron spin), the total scattering cross section can be written as
\begin{equation}
\frac{d\sigma^{\rm SD}}{dE_R}
=
\frac{2m_N}{3\mu_p^2 v^2}
\frac{J+1}{J}
\langle S_p\rangle^2
\,\sigma_p^{\rm SD}
\,F_{p}^2(q) \; ,
\end{equation}
where $\langle S_p \rangle$ denotes the proton contribution to the total nuclear spin $J$. We emphasize that $\langle S_p \rangle$ can be non-zero even for nuclei in which all protons are paired, such as $^{129}$Xe.

An analogous expression for dark matter coupled only to the neutron spin is obtained by replacing $p \to n$ in the expression above. While it is common practice in the direct detection community to consider spin-dependent couplings to protons and neutrons separately, realistic dark matter models typically predict both couplings to be present at the same time. In this case, interference between the two contributions must be considered, complicated by the fact that (in contrast to the spin-independent case) the spin-dependent form factors may differ substantially for protons and neutrons~\cite{Klos:2013rwa}.

All other non-relativistic operators involve the momentum transfer and/or the relative velocity, both of which are small (in natural units) in the non-relativistic limit. These operators therefore only play a role if the dominant spin-independent and spin-dependent operators are absent~\cite{GAMBIT:2021rlp}. Crucially, these interactions correspond to differential event rates that are not monotonically falling and could therefore be distinguished from standard spin-independent or spin-dependent scattering, see the third panel of figure~\ref{fig:rates}.

For elastic scattering, the incoming and outgoing particles are the same, such that the recoil energy and the momentum transfer are directly related. If, on the other hand, the scattering particles have additional internal degrees of freedom, it becomes possible to absorb some of the energy of the collision into excitations. The resulting changes in the scattering kinematics may substantially change the shape of the expected differential event rates and may even enhance the fraction of the dark matter kinetic energy that can be deposited in the detector.
Let us first consider the case that the dark matter particle has two different states that are separated by the mass splitting $\delta$~\cite{Tucker-Smith:2001myb}. For a transition from the ground state to the excited state, we can calculate $v_\text{min}$ from eq.~\eqref{eq:vmin} with $\Delta E = \delta + E_R$ and $q = \sqrt{2 m_N E_R}$, which yields
\begin{equation}
    v_\text{min} = \frac{1}{\sqrt{2 m_N E_R}}\left(\frac{m_N E_R}{\mu}+\delta\right) \; .
\end{equation}
Notably, $v_\text{min}$ diverges for $E_R \to 0$, which -- given the upper bound $v_\text{max}$ on dark matter velocities in the Galactic halo -- has two implications. First, scattering into the excited state becomes impossible if $\delta > \tfrac{1}{2} \mu v_\text{max}^2$. Second, if up-scattering is kinematically allowed, there is still a lower bound on $E_R$, such that the differential event rate no longer drops monotonically as in the elastic case, but instead exhibits a peak at finite recoil energy, see the third panel of figure~\ref{fig:rates}.

In certain dark matter models (such as Pseudo-Dirac dark matter~\cite{Izaguirre:2015zva}), the couplings of dark matter particles to ordinary matter are fully off-diagonal, meaning that each interaction must involve a transition from the ground state to the excited state (or vice versa). These so-called inelastic dark matter models are interesting for direct detection experiments because of the unusual signatures they predict. While for elastic scattering it is usually beneficial to lower the energy threshold and consider light target nuclei, inelastic dark matter typically favours high energies and heavier targets. For completeness, let us mention that it is also possible to have $\delta < 0$, i.e.\ a down-scattering from the excited state to the ground state, often referred to as exothermic dark matter~\cite{Graham:2010ca}. Again, the differential event rate has a non-trivial shape, since the deposited energy in the detector is bounded from below by $|\delta|$.

A different type of inelastic scattering occurs if the target atom is excited in the scattering process. Such excitations are possible for both the nucleus (see e.g.\ Ref.~\cite{McCabe:2015eia}) or the electron shell. In fact, electronic excitations are possible even if the dark matter particle couples only to nuclei. Intuitively, this is because the scattering process happens on such a short timescale that the nucleus is displaced from the centre of the electron cloud~\cite{Dolan:2017xbu}. As the electrons catch up with the recoiling nucleus, they may transition into an excited state. The most extreme case, in which one or more electrons are ejected from the atom, is known as the Migdal effect~\cite{Ibe:2017yqa}. While several experiments are currently attempting to measure this effect~\cite{MIGDAL:2022yip,Xu:2023wev}, it was only recently observed directly for the first time~\cite{Yi:2026fmf}. Finally, we note that, instead of emitting an electron, the nucleus may also emit a photon in a process analogous to Bremsstrahlung~\cite{Kouvaris:2016afs}.

\subsubsection{Electron scattering}
\label{sssec:theory_ER_scattering}

Of course it is also conceivable that dark matter particles scatter directly off electrons~\cite{Essig:2011nj,Bloch:2020uzh}. In this case the scattering kinematics are complicated by the fact that the electrons in a target are in energy eigenstates but not in momentum eigenstates, such that we need to take into account their quantum-mechanical wavefunctions. For the case that scattering leads to the ionization of a bound electron, we can denote possible initial-state wavefunctions by $\psi_j$ (where $j$ may for example label different energy shells) and final-state wavefunctions by plane wave solutions $\psi_{\mathbf{k'}}$. The scattering probability then depends on the transition amplitude
\begin{equation}
    f_{j\to\mathbf{k}'}(\mathbf{q}) = \int \mathrm{d}^3 r \psi_j (\mathbf{r}) \psi_{\mathbf{k}'}^\ast(\mathbf{r}) e^{i \mathbf{q} \cdot \mathbf{r}}
\end{equation}
with the momentum transfer $\mathbf{q}$.
While the initial state has the binding energy $E_b < 0$, the final-state electron has positive energy $E_R = k'^2/(2 m_e)$, such that the total amount of energy transferred is $\Delta E = E_R - E_b$. This energy must be equal to the energy lost by the scattering dark matter particle, 
\begin{equation}
    \omega = {\mathbf{v}\cdot\mathbf{q}}- \frac{\mathbf{q}^2}{2 m_\chi}
\end{equation}
with the dark matter velocity $\mathbf{v}$.
The scattering cross section for initial state $j$ is then given by
\begin{equation}
    (\sigma v)_j = \int \frac{\mathrm{d}^3 p'}{(2\pi)^3} \frac{\mathrm{d}^3 k'}{(2\pi)^3} (2\pi) \delta(\Delta E - \omega) \frac{|\mathcal{M}_\text{free}|^2}{16 m_\chi^2 m_e^2} |f_{j\to\mathbf{k}'}(\mathbf{q})|^2
\end{equation}
where $\mathbf{p}'$ is the final dark matter momentum, $\mathcal{M}_\text{free}$ denotes the matrix element for the scattering off a free electron and $m_e$ is the electron mass. By expanding the final-state wavefunctions into spherical harmonics, this expression can be rewritten as
\begin{equation}
    (\sigma v)_j = \int {\mathrm{d}^3q} \,\mathrm{d} \ln E_R \delta(\Delta E_e - \omega) \frac{|\mathcal{M}_\text{free}|^2}{(16 \pi)^2 m_\chi^2 m_e^2} |f_j^\text{ion}(k',q)|^2 \; 
\end{equation}
with the ionisation form factor $|f_j^\text{ion}(k',q)|^2$. 
If the squared matrix element is simply a constant (which is commonly the case for scattering via a heavy mediator), it can be directly related to the total dark matter-electron scattering cross section
\begin{equation}
\label{eq:DMelectronCS}
    \sigma_e = \frac{\mu_e^2 |\mathcal{M}_\text{free}|^2}{16 \pi m_e^2 m_\chi^2} \; ,
\end{equation}
where $\mu_e$ is the reduced mass of the dark matter-electron system. However, in the context of dark matter-electron scattering, one is often interested in the case where the matrix element depends on the momentum transfer $q$, for example because scattering is mediated by a light exchange particle.\footnote{Recent studies have explored a more general effective field theory of dark matter-electron interactions~\cite{Krnjaic:2024bdd}, analogous to the case of dark matter-nucleon interactions.} In such a case, the squared matrix element can be written as the product of a constant term and a so-called dark matter form factor:
\begin{equation}
|\mathcal{M}_\text{free}(q)|^2 = \overline{|\mathcal{M}_\text{free}|^2} \, F_\text{DM}(q)^2 \; .
\end{equation}
By convention, the form factor is normalised such that $F_\text{DM}(q_\text{ref}) = 1$ for the reference momentum $q_\text{ref} = \alpha m_e$, such that $\overline{|\mathcal{M}_\text{free}|^2} = |\mathcal{M}_\text{free}(q_\text{ref})|^2$. For example, a massless mediator would give $F_\text{DM}(q) = (\alpha m_e)^2/q^2$. The differential scattering cross section is then conventionally written as
\begin{equation}
    \frac{\mathrm{d}(\sigma v)_j}{\mathrm{d} \ln E_R} = \frac{\overline{\sigma}_e}{16 \pi \mu_e^2} \int \mathrm{d}^3q \delta(\Delta E_e - \omega) F_\text{DM}^2(q) |f_j^\text{ion}(k',q)|^2 \; ,
\end{equation}
where $\overline{\sigma}_e$ is defined in terms of $\overline{|\mathcal{M}_\text{free}|^2}$ in analogy to eq.~\eqref{eq:DMelectronCS}.

The differential event rate then follows from an integration over dark matter velocities as in the case of nuclear scattering, see eq.~\eqref{eq:dRdER_integral}. In the present case, the integral over the angle between $\mathbf{v}$ and $\mathbf{q}$ can be carried out to eliminate the delta function and obtain
\begin{equation}
\label{eq:dRdER_electron}
\frac{\mathrm{d}R}{\mathrm{d} \ln E_R} = \frac{\rho_0}{m_\chi}\frac{\overline{\sigma}_e}{8 \mu_e^2} \int q \mathrm{d}q \, F_\text{DM}^2(q) |f_j^\text{ion}(k',q)|^2 \eta(v_\text{min})\; ,
\end{equation}
with the velocity integral $\eta(v_\text{min})$ introduced above and
\begin{equation}
    v_\text{min} = \frac{\Delta E}{q}+\frac{q}{2 m_\chi} \; .
\end{equation}

In the discussion above we have assumed that dark matter-electron scattering leads to ionisation, i.e.\ a free electron. An alternative case of great experimental interest are semiconductor targets, in which a dark matter particle may excite an electron from the valence band to the conduction band. The resulting expression for the scattering cross section is identical to the one for ionisation, except that the ionization form factor $|f_j^\text{ion}(k',q)|^2$ is replaced by a transition form factor $|f^\text{crystal}(q,\Delta E)|^2$. Calculating this form factor is generally challenging and requires advanced numerical methods~\cite{Knapen:2021bwg,Trickle:2022fwt}. However, it has been understood in recent years that the transition form factor can be related to experimentally measurable quantities, specifically to the dielectric function~\cite{Hochberg:2021pkt,Knapen:2021run,Hochberg:2025rjs}, which substantially reduces the theory uncertainties.

\begin{figure}
    \centering
\includegraphics[width=0.45\linewidth]{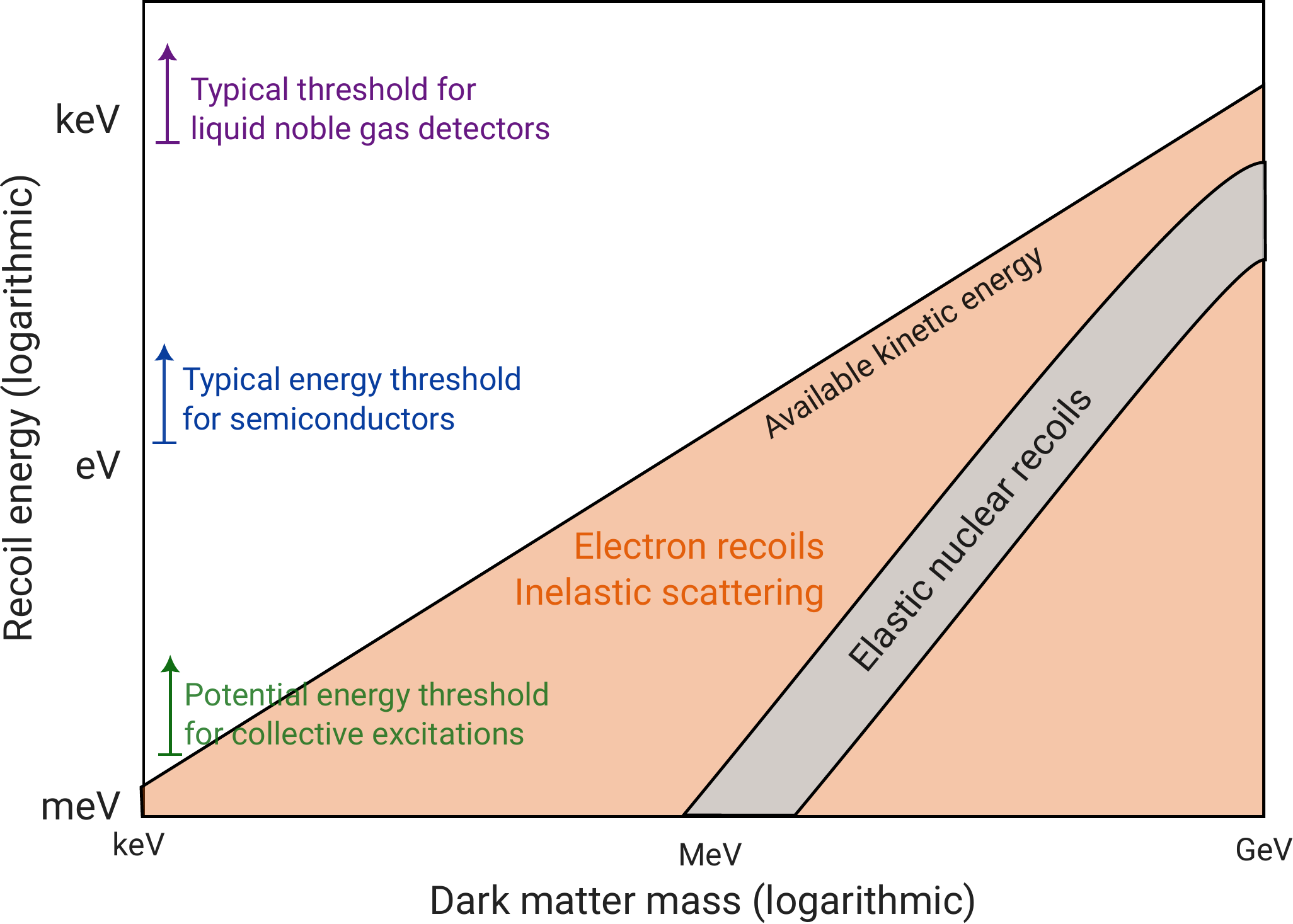}
    \caption{Illustration of the scattering kinematics of light dark matter for elastic nuclear recoils (gray band) and general inelastic or electron recoils, for which the momentum and energy transfer are independent (orange region). Figure adapted from Ref.~\cite{Essig:2022dfa}.}
    \label{fig:kinematics}
\end{figure}

The fact that for dark matter-electron scattering there is no one-to-one correspondence between the momentum transfer $q$ and the electronic recoil energy $E_R$ makes the scattering kinematics more favourable than in the case of nuclear scattering, because the entire kinetic energy of the incoming dark matter particle can in principle be transferred to the bound electron, see figure~\ref{fig:kinematics}. Assuming that sensitivity to a single ionization electron can be achieved, observation of dark matter scattering becomes possible as soon as $m_\chi v^2 / 2 > |E_b|$, which implies $m_\chi > 2 |E_b|/v_\text{max}$. For semiconductors, the band gap $E_b$ is of the order of a few electronvolt, such that electron scattering can be sensitive to dark matter masses as small as 1 MeV -- well below the typical reach of nuclear scattering.

\subsubsection{Collective excitations}

To extend the sensitivity of direct detection experiments to even smaller masses, one needs to consider collective excitations in the target material of the detector that require even less energy to excite than the electronic excitations in semiconductors. Such excitations exist in crystals for example in the form of phonons, i.e.\ collective oscillations of the atoms in the lattice~\cite{Trickle:2019nya,Kahn:2021ttr,Campbell-Deem:2022fqm}. Other types of excitations that have been considered include collective excitations of electrons called plasmons~\cite{Kozaczuk:2020uzb,Kurinsky:2020dpb}, spin waves called magnons~\cite{Trickle:2019ovy}, rotational modes called rotons~\cite{Caputo:2019cyg,Acanfora:2019con}, as well as multi-phonon processes~\cite{Knapen:2016cue}.

Schematically, the total event rate can be written in terms of the interaction rate $\Gamma(\mathbf{v})$ as
\begin{equation}
    R = \frac{\rho_0}{m_\chi \, \rho_T} \int \mathrm{d}^3 v f(\mathbf{v}) \Gamma(\mathbf{v}) \; , 
\end{equation}
where $\rho_T$ is the target's mass density. The interaction rate, in turn, can be written in a similar form as for dark matter-electron scattering:
\begin{equation}
    \Gamma(\mathbf{v}) = \frac{\pi \overline{\sigma}}{\mu^2} \int \frac{\mathrm{d}^3 q}{(2\pi)^3} F_\text{DM}^2(q) S(\mathbf{q}, \omega)
\end{equation}
where $F_\text{DM}(q)$ is the dark matter form factor (see above) and $\overline{\sigma}$ is an appropriate reference cross section. The dynamic structure factor $S(\mathbf{q},\omega)$, finally, captures the material response to a dark matter scattering event. It depends on the effective scattering potential and plays an analogous role as the transition form factor for electron scattering. For isotropic materials, the angular integration can be readily carried out, leading to an expression that is in analogy with eq.~\eqref{eq:dRdER_electron}.

A particularly promising opportunity turn out to be excitations of acoustic and optical phonons. Acoustic phonons have a wavelength much larger than the lattice spacing $a$. They can have an arbitrarily low energy and an approximately linear dispersion relation $\omega_\text{acoustic} = c_s k$ with the the sound speed $c_s \sim 10^{-5}$. Optical phonons, on the other hand, correspond to a relative motion of different nuclei within the same elementary cell. As a result, these phonons are gapped, i.e.\ they require a minimum energy transfer of the order of $\omega_\text{optical} \sim c_s / a$ to be excited. Typical values are $\omega_\text{optical} \sim (10\text{--}100) \, \mathrm{meV}$. For acoustic phonons, the typical energy deposition for a dark matter particle with velocity $v \sim 10^{-3}$ follows from the kinematic matching of the momentum transfer $q \sim m_\chi v$ with the phonon dispersion relation, giving by $\omega \sim c_s m_\chi v$. Optical phonons, on the other hand, can be excited for a wide range of momentum transfer, such that potentially the entire kinetic energy of order $m_\chi v^2$ can be deposited. This makes optical phonons particularly attractive for the search for dark matter particles in the mass range between keV and MeV~\cite{Knapen:2017ekk}.

\subsubsection{Bosonic absorption and dark Compton scattering}

So far, we have focused exclusively on the case of dark matter scattering, in which the dark matter-particles leave the detector after interacting with the target material. In models of bosonic dark matter, however, it becomes possible for a dark matter particle to be fully absorbed, such that its entire rest mass is deposited as energy in the detector~\cite{Pospelov:2008jk,Hochberg:2016sqx}. The most commonly considered example is an axion-like particle $a$ with mass in the eV to keV range, which directly couples to electrons with coupling strength $g_{ae}$~\cite{Irastorza:2018dyq}. Absorption then proceeds in analogy to the photo-electric effect and is therefore often called the axio-electric effect. Analogous processes are possible for scalar~\cite{Mitridate:2021ctr} and vector~\cite{An:2014twa} dark matter particles. For even lighter dark matter particles, one can consider the coupling to collective excitations, which may be produced through the absorption of dark matter particles~\cite{Mitridate:2023izi}.

For dark matter particles bound to the Milky Way halo, the kinetic energy is completely negligible compared to the rest mass, such that the energy deposition is essentially mono-energetic. The total event rate is then simply given by~\cite{Bloch:2016sjj}
\begin{equation}
    R = \frac{\rho_0}{m_\chi} \sigma_\text{abs} v \; .
\end{equation}
The absorption cross section $\sigma_\text{abs}$ is proportional to $g_{ae}^2 / v$, such that the velocity factor cancels and the total rate simply becomes proportional to the coupling squared. If the dark matter rest mass is large compared to the typical electronic binding energy, the latter can be neglected and the resulting signal is a mono-energetic peak in the electron recoil energy. This is a very clean signal that can be searched for with great sensitivity even in the presence of backgrounds.

Dark matter models that allow for absorption also allow for so-called dark Compton scattering~\cite{Hochberg:2021zrf}, in which the electron that absorbs the dark matter particle emits a photon: $\chi + e^- \to \gamma + e^-$. In this case, the photon may carry most of the deposited energy and proceed to ionise other atoms. While the total deposited energy is the same as in the case of direct absorption, the detailed kinematics are different, which may lead to larger rates and/or more favourable signals in certain cases. A similar process is the absorption of fermionic dark matter, in which a fermionic dark matter particle is converted into a neutrino upon scattering off an electron~\cite{Dror:2019onn,Dror:2020czw,Ge:2022ius}.

We note that bosonic particles with couplings to electrons are generally unstable against loop-induced decays into pair of photons~\cite{Ferreira:2022egk}. Nevertheless, it is possible to suppress the decay width enough to satisfy astrophysical constraints from searches for x-ray lines (see e.g.~Ref.~\cite{Perez:2016tcq}). More critical, in particular for sub-keV particles, are bounds arising from the production of such particles in the interior of stars, which may constitute a sizable source of cooling and change stellar evolution~\cite{Raffelt:1996wa}. On the other hand, this effect can be exploited to directly search for bosonic particles produced in the Sun, which would have a typical kinetic energy of a few keV and induce a broader distribution of electronic recoil energies~\cite{Arisaka:2012pb}. A combined analysis of all these considerations was performed in Ref.~\cite{Athron:2020maw}.

\subsubsection{Earth and atmospheric shielding}
\label{sec:shielding}

In the discussion above, we have used the input from particle physics only to describe the interactions between dark matter particles and the target material within the detector. The implicit assumption has been that dark matter interactions are so rare that the probability for any dark matter particle to scatter more than once is completely negligible. This assumption however no longer holds when considering dark matter candidates with very large scattering cross sections, which are sometimes referred to as strongly-interacting dark matter.\footnote{Note that the term "strongly-interacting" is potentially confusing, since it is used both to denote dark matter with strong self-interactions and dark matter with strong coupling to visible matter. We use it to refer to the latter case here.} The result of such interactions is a modification of the dark matter flux $n_\chi v$ that arrives at the experiment. In the most extreme case, where the mean free path $\lambda = (n  \sigma)^{-1}$ is small compared to the overburden $L$ of the experiment, dark matter particles may be stopped -- or at least decelerated to completely negligible velocities -- before they reach the detector, effectively making the experiment blind for such models~\cite{Kavanagh:2017cru,Emken:2018run}.

Any direct detection experiment therefore places not only a lower bound on the scattering cross section of dark matter, but also an upper bound, beyond which the experiment is no longer sensitive. Extending the sensitivity to even larger couplings requires placing the experiment on the surface of the Earth at the cost of substantially increased backgrounds~\cite{Emken:2019tni}. Since for even larger couplings the atmosphere itself contributes to the shielding of dark matter experiments, it may be necessary to consider balloon- or satellite-based experiments, see Ref.~\cite{Du:2024afd}. It is worth noting, however, that such large couplings between dark and visible matter are subject to astrophysical~\cite{Mack:2007xj} and cosmological constraints, for example due to their effects on structure formation~\cite{Xu:2018efh,Boddy:2018wzy,Nadler:2019zrb}.

\subsection{Statistical methods}

Let us briefly review the main statistical methods used in the direct detection community to derive constraints on the particle physics parameters of a given dark matter model based on experimental data, referring to Ref.~\cite{Conrad:2014nna} for a more in-depth discussion. The choice of method depends on whether there exists an accurate background model that can be used to predict the number of expected events and their distribution and whether the experimental information makes it possible to distinguish signal from background based on the distribution of observed events.

The simplest possibility is the Poisson method (sometimes also called "cut-and-count method"), which relies on the definition of a single signal region. This signal region usually depends on the expected distribution of background events and is optimised\footnote{This optimisation should of course be performed based on calibration data and control regions, not on the dark matter search data itself. In practice, methods like blinding or salting are used to avoid biases.} to enhance the purity of the signal. For a model predicting $s$ signal events, the probability to observe $n$ events is given by the Poisson probability~\cite{ParticleDataGroup:2024cfk}
\begin{equation}
    p(n; s) = \frac{s^n \, e^{-s}}{n!} \; .
\end{equation}
Correspondingly, the cumulative probability to observe at most $n$ events is given by
\begin{equation}
    P(n; s) = \sum_{i=0}^n p(i;s) \; .
\end{equation}
A signal hypothesis can be excluded at confidence level $c$ if $P(n; s) < (1 - c)$. For example, for $n = 0$ any signal hypothesis with $s \geq 2.3$ (3.0) can be excluded at 90\% (95\%) confidence level.
If the expected number of background events $b$ is known, one can replace $s \to s + b$ in the equations above. In such a case it becomes possible to exclude the background-only hypothesis if $n \gg b$, such that the probability $p(n; b)$ becomes very small. Alternatively, one can still use the criterion based on $P(n; s+b)$ from above to place a conservative upper bound on the signal strength $s$. A potential difficulty with this approach is that an overestimation of the background (or a downward fluctuation in the data) can lead to overly strong exclusion limits. This issue can be addressed for example using the Feldman-Cousins method~\cite{Feldman:1997qc}.

The Poisson method can be easily extended to include several signal regions, for example different bins in an observable $O$, by multiplying the individual Poisson probabilities. In fact, in the limit of an infinite number of bins, this approach leads to the unbinned (or extended) likelihood method~\cite{Barlow:1990vc}, which introduces a likelihood function
\begin{equation}
    \mathcal{L} = p(n; s+b) \prod_{i=1}^n \text{pdf}(O_i) \; .
\end{equation}
Here $n$ denotes the total number of observed events with corresponding observables $O_i$ and $\text{pdf}(O)$ denotes the probability distribution of the observable $O$, which depends on the signal and background model (both their shape and relative normalisation). The background model usually contains a number of free parameters describing various sources of uncertainties that may be constrained by additional calibration measurements. Maximising the likelihood function with respect to these nuisance parameters yields the profile likelihood~\cite{Rolke:2004mj}, which in turn can be used to construct various test statistics, such as the profile likelihood ratio (PLR)~\cite{Cowan:2010js}. The PLR can then be used both for limit setting (i.e.\ hypothesis tests) and parameter inference (in the case of an excess in the data). As for the Poisson method, there is the risk of overly strong exclusions in the case of downward fluctuations, which can be addressed using so-called power constrained limits~\cite{Cowan:2011an}, see also Ref.~\cite{Baxter:2021pqo}.

If the expected distribution of background events is not known, it is still possible to exploit a difference in the distribution of observed events and expected signal events using the so-called maximum gap and optimum interval methods~\cite{Yellin:2002xd}. These methods consider all possible signal regions and select the most sensitive one while applying an appropriate statistical penalty to ensure correct coverage of the resulting exclusion limit. By construction, this method never rejects the background hypothesis, i.e.\ it cannot be used to find and characterise an excess in the data, which would be necessary to discover dark matter. This feature makes these methods a useful tool for deriving conservative exclusion limits even when the observed data is not well described by the background model.  

\subsection{Complementarity with other dark matter searches}

\begin{figure}
    \centering
\includegraphics[width=0.45\linewidth]{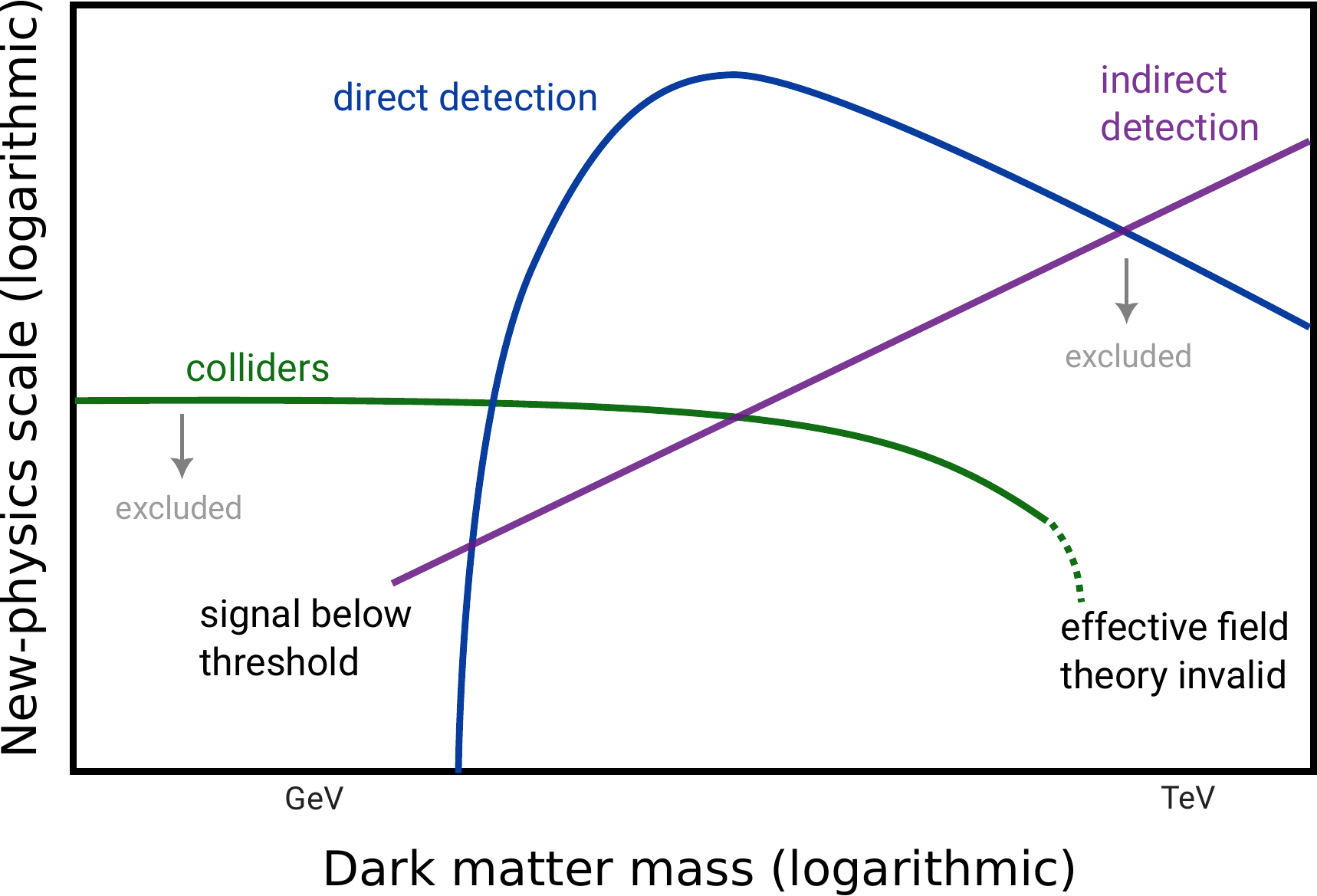}
    \caption{Illustration of the complementarity between different dark matter search strategies, assuming that the interactions of dark matter can be parametrised in terms of the new-physics scale in an effective field theory.}
    \label{fig:complementarity}
\end{figure}

To conclude this chapter, let us briefly discuss the complementarity between direct detection experiments and other probes of dark matter. We emphasize that direct detection experiments probe the dark matter \emph{scattering} cross section in the \emph{non-relativistic} limit. A comparison with other types of dark matter searches that probe the annihilation cross section -- called indirect detection -- or the production cross section, in particular at colliders, is therefore necessarily model-dependent. Even within a given model, it is non-trivial to connect the vastly different energy scales relevant for the different processes, which requires solving renormalisation group equations~\cite{DEramo:2016gos,Bishara:2018vix}. In models with a heavy mediator and unsuppressed scattering and annihilation in the non-relativistic limit, direct detection experiments typically give the strongest bounds on the dark matter couplings in the mass range where they are most sensitive~\cite{Belyaev:2018pqr}, while indirect detection experiments and collider experiments may be more constraining for light dark matter~\cite{Bondarenko:2019vrb}.

A more model-independent comparison is possible with constraints from neutrino telescopes such as IceCube~\cite{IceCube:2021xzo}, which probe the capture of dark matter particles in the Sun followed by their accumulation and eventual annihilation. Under the assumption of equilibrium, the strength of this signal depends only on the capture rate, i.e.\ the scattering cross section, but it of course relies on the ability of dark matter particles to annihilate into neutrinos. The resulting constraints can be competitive with direct detection experiments for the case of spin-dependent interactions with protons.

Finally, for very large scattering cross sections, astrophysical and cosmological constraints become relevant and provide complementary constraints in parameter regions where direct detection experiments are shielded from dark matter by the Earth and the atmosphere (see section~\ref{sec:shielding}).  
\section{Technology}
\label{sec:tech}

The main detector technologies used in direct dark matter searches are summarized below. Owing to their different detection principles and performance characteristics, each technology is particularly well suited to probing specific dark matter interaction channels and mass ranges. For example, cryogenic calorimeters and CCDs, with their extremely low energy thresholds, are especially sensitive to low-mass dark matter candidates, whereas noble-liquid detectors, thanks to their large target masses and excellent background rejection capabilities, provide the highest sensitivity to heavier dark matter particles.
Similarly, the choice of target material contributes to the detector's sensitivity to different dark matter mass ranges: lighter target nuclei are better suited for detecting low-mass dark matter, while heavier nuclei provide enhanced sensitivity to more massive dark matter particles.
Last but not least, different target or detector properties give access to different dark matter interactions, including spin-independent and spin-dependent dark matter--nuclear scattering as well as dark matter--electron scattering, or signal characteristics such as the direction of the incoming particle.

This section is organized by particle detection approach. For each, the operating principle and the signal generation mechanism are discussed. Schematic illustration of the components and working principle accompany the descriptions to facilitate understanding, and typical signal shapes are provided for each technique. The specific challenges faced and advantages offered by each experimental approach are highlighted in the boxes at the end of the corresponding subsection. The advantages and challenges boxes are not exhaustive and highlight only the two most prominent points in each category. A comprehensive list of experiments using each technique is included, along with references to instrumental papers describing the detectors.  

\subsection{Cryogenic crystal calorimeters} 
\label{ssec:tech_bolom}

\begin{figure}[ht]
	\centering
    \includegraphics[width=0.65\columnwidth]{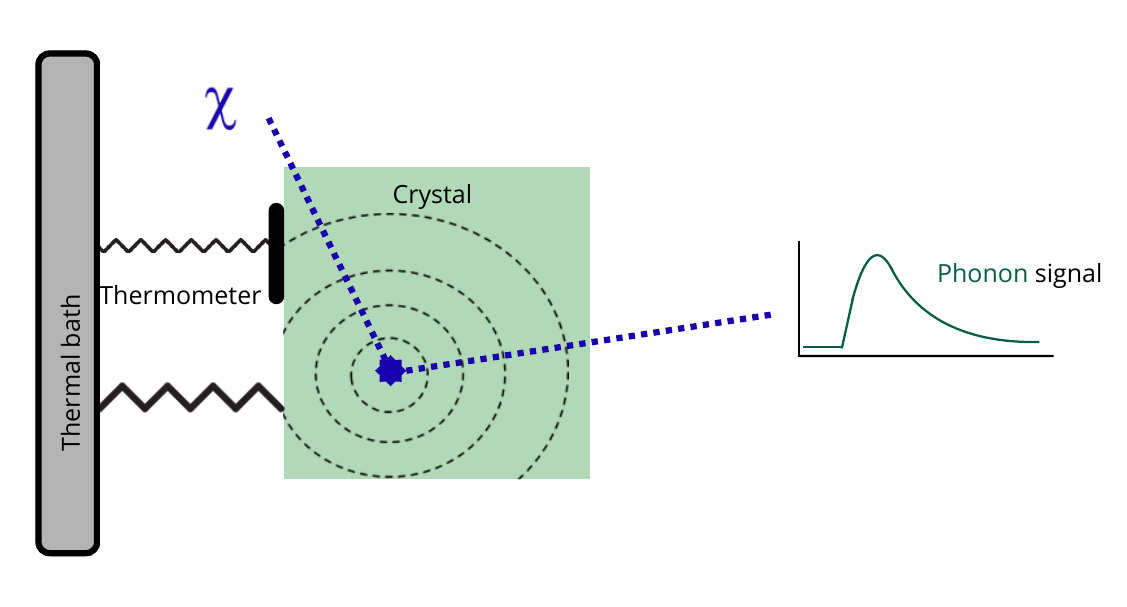} 	
	\caption{Illustration of a cryogenic calorimeter. The interaction occurs in the absorber crystal (shown in green), generating a phonon signal (dashed lines) measured by a sensor on the absorber surface (thermometer). Both the absorber and sensor are linked to a thermal bath in the mK temperature range via a conductance (represented by the zig-zag lines). The inset on the right shows the typical signal shape of a cryogenic calorimeter in arbitrary units. }
	\label{fig:cryo_calorimeter_sketch}
	\end{figure}

Cryogenic calorimeters are phonon-mediated detectors operated at cryogenic temperatures (order of 10\,mK).
These detectors are composed of a target equipped with phonon-sensors used to read out the signals. 
Dark matter particles may interact with the target, transferring energy to either the nucleus or the electrons.
The sensitivity to the resulting phonons enables these detectors to achieve sub-eV energy resolutions and, consequently, exceptionally low energy and thus dark matter mass thresholds.
This section focuses on solid target detectors whereas cryogenic calorimeters with a liquid target will be discussed in Sec.~\ref{ssec:tech_superfluid}.
Figure~\ref{fig:cryo_calorimeter_sketch} shows a schematic illustration of a cryogenic calorimeter and highlights the basic parts referred to in the text: the target material, the phonon sensor and the thermal links to the bath.

The choice of the solid target material is highly flexible as long as it has a mono-crystalline structure to allow efficient phonon propagation.
The target flexibility and the high energy resolution make these detectors suitable for different physics applications, including dark matter, neutrinoless double beta decay~\cite{Giuliani:2026syi}, and precision measurements (e.g. ionization yield measurements, decays half lives).

Several phonon sensor technologies, also referred as thermometers, are commonly used in these experiments:
\begin{itemize}
    \item \textit{transition-edge sensors} (TES) use a superconducting transition to map a resistance change as a function of a temperature variation. TESs are characterized by high sensitivities over a small temperature range. 
    The typical resistance variations in the order of m$\Omega$ require a read-out based
    on superconducting quantum interference devices (SQUIDs).
    The working point is tuned to operate these sensors in their superconducting transitions. 
    \item \textit{neutron-transmutation doped} (NTD) thermistors translate temperature changes into resistance changes. By doping a semiconductor crystal (germanium or silicon) with neutrons, the thermistor is made to function near the metal–insulator transition. NTDs exhibit a good linearity over several orders of magnitude in energy.
    \item \textit{magnetic microcalorimeters} (MMC) detect changes in temperature by monitoring resulting changes in the magnetization of a paramagnetic sensor. 
    The magnetization variations are measured with a SQUID-based readout. The response of MMCs is highly linear over a large dynamic range.
    \item \textit{kinetic inductance detector} (KID) is a superconducting inductor–capacitor (LC) resonator whose resonance frequency ($1 / \sqrt{LC}$) becomes highly stable when operated at low temperatures. Phonons resulting from particle interactions break Cooper pairs in the superconductor, which increases the inductance and consequently shifts the resonance frequency. Typically, KIDs have a linear response below tens of keV.
\end{itemize}

Different sensor technologies will be prevalently sensitive either to athermal or thermal phonons. 
NTDs are mainly sensitive to thermal phonons, KIDs are sensitive to athermal phonons, TESs and MMCs can be either sensitive to athermal~\cite{Irwin:1995zzb} or thermal phonons~\cite{Pyle:2015pya, Krantz:2023vpe} depending on the sensor design. 
Cryogenic calorimeters can achieve energy resolutions ranging from tens of eV down to the sub-eV level, depending on the sensor technology employed. This capability allows cryogenic calorimeters to reach eV-scale energy thresholds.

One last essential part of a cryogenic calorimeter is its connection to the thermal bath: a thermal link is used to discharge the heat once it has been measured by the sensor. The principal link can be established either through the absorber or the sensor, depending on whether the sensor is sensitive to athermal or thermal phonons.
The thermal conductance to the thermal bath needs to be tuned to ensure that the energy has been readout by the thermal sensor before being discharged to the thermal bath.

\begin{figure}[h]
	\centering
    \includegraphics[width=0.45\columnwidth]{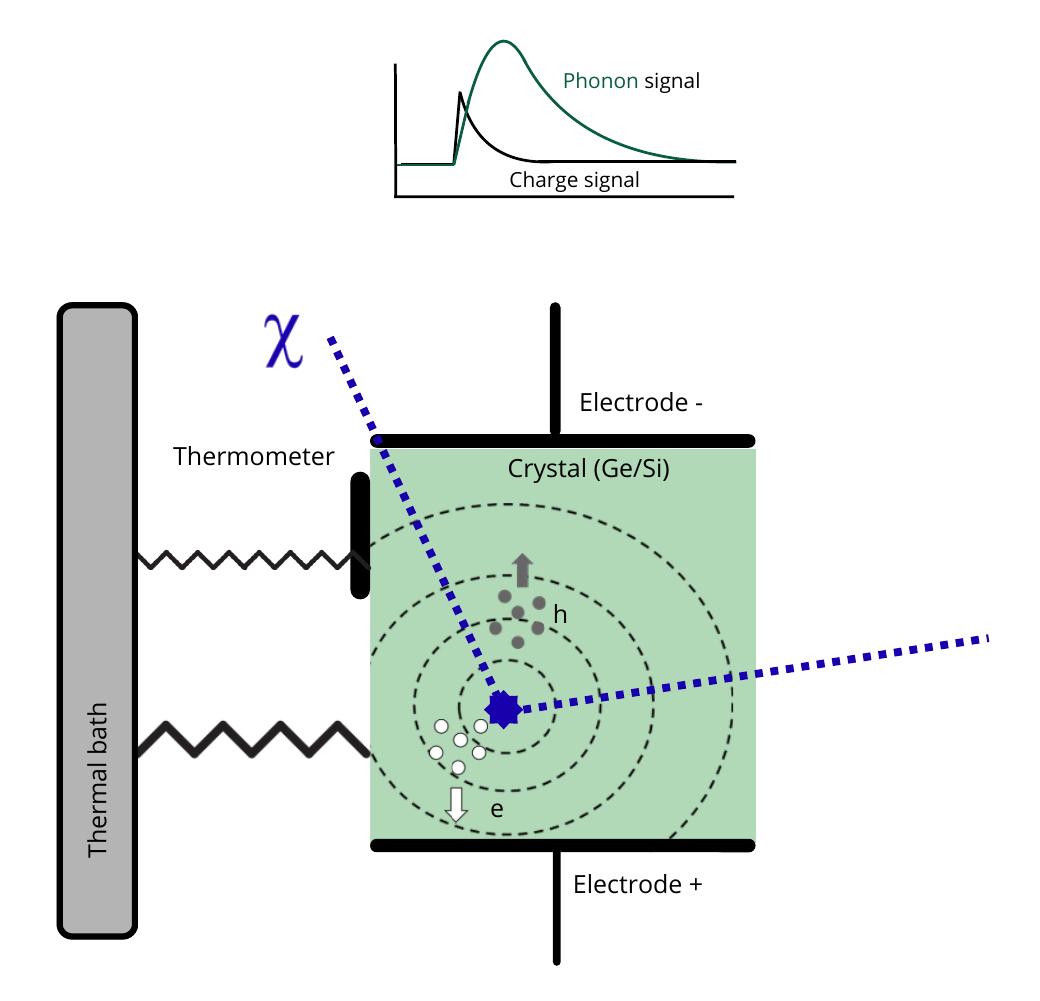} 
    \includegraphics[width=0.45\columnwidth]{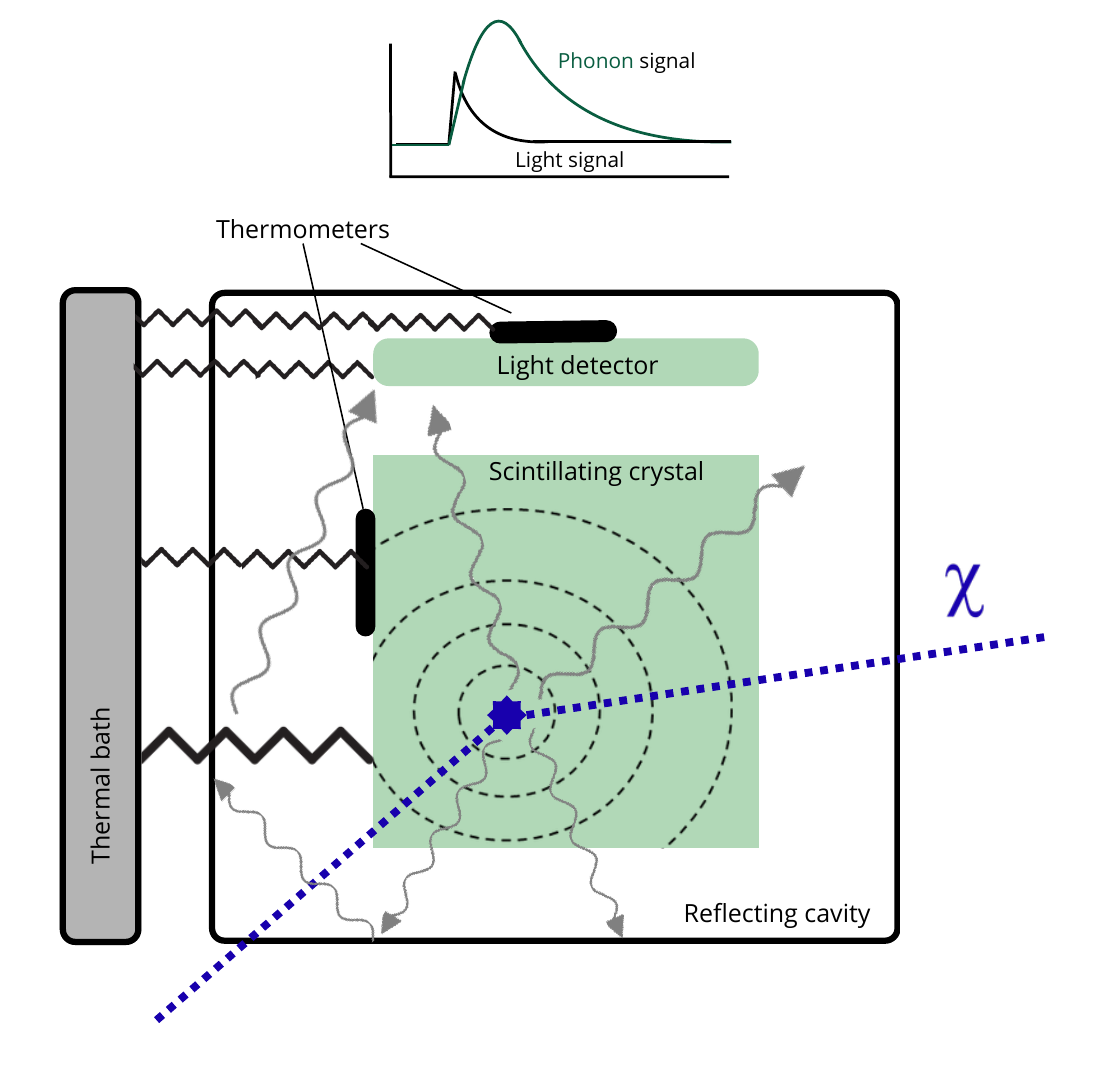} 
	\caption{Illustration of an ionization (left) and a scintillating (right) calorimeter. Both utilize the cryogenic calorimeter architecture shown in Fig.~\ref{fig:cryo_calorimeter_sketch} to extract the phonon signal. In the ionization calorimeter, electrodes deposited on the semiconductor crystal surface enable the drift and collection of the primary ionization charges. In the  scintillating calorimeter, scintillation photons (gray wavy arrows) escape the absorber and are detected by a light detector, typically another cryogenic calorimeter, placed within a reflecting cavity. The top inset displays the typical simultaneous phonon and charge (left) and phonon and light (right) signals, which depend on the interacting particle, enabling background discrimination. 
    }
	\label{fig:cryo_calorimeter_sketch2}
	\end{figure}

The addition of a second readout channel---measuring either the light or charge signal---to the phonon readout allows cryogenic calorimeters to identify the type of particle interaction (electron recoils versus nuclear recoils).
This feature is common also to other detector technologies (see Sec.~\ref{ssec:tech_noble} and \ref{ssec:tech_superfluid}). 
A simplified illustration of a cryogenic calorimeter with a double readout is shown in Fig.~\ref{fig:cryo_calorimeter_sketch2}, for charge (light) signal on the left (right).
The \textbf{light readout} requires that the target is a scintillating crystal and the addition of a secondary detector to collect the emitted light. The secondary detector is usually also a cryogenic calorimeter as they are best suited to operate at temperatures around 10~mK while preserving a low background level. Reflective foil is often used inside the detector cavity to enhance light collection.
CRESST~\cite{Angloher:2025fzw}, TESSERACT~\cite{Billard:2024zvc} and COSINUS~\cite{Angloher:2025shf} are cryogenic calorimeter-based experiments that use scintillating crystals coupled with light detectors.
The \textbf{charge readout} requires that the target is a semiconductor crystal capable of producing electron–hole pairs when energy is deposited, which can then be drifted and collected for measurement.
An electric field is applied across the target via aluminium electrodes which can also be used to readout the charge signal.
A simultaneous readout of phonon and charges allows for the discrimination between electron-recoil and nuclear recoil events, thanks to the lower ionization yield of nuclear interactions.
The discrimination power between electron- and nuclear-recoils is enhanced when the detectors are operated in a \textit{low-voltage mode} corresponding typically to voltages below 10\,V.
When the detectors are operated in \textit{high-voltage mode} (typically at bias voltages ranging from tens to hundreds of volts), the charge readout is no longer informative and particle identification is no longer possible.
However, this range of voltages is still interesting thanks to the Neganov-Trofimov-Luke (NTL) effect~\cite{Neganov:1985khw, Luke:1988xcw} which strongly amplifies charge signals and thus also boosts signals from below the threshold into the sensitive region improving the sensitivity to lower-energy dark matter events. 
When the charges are drifted toward the electrodes, they interact with the crystalline lattice producing additional NTL phonons.
The total energy measured by the phonon channel ($E_{\rm ph}$) becomes the sum of the initial deposited recoil energy ($E_{\rm R}$) and the contribution of the NTL effect ($E_{\rm NTL}$):
\begin{equation}\label{eq:ntl}
    E_{\rm ph} = E_{\rm R} + E_{\rm NTL} = E_{\rm R} \left( 1 + Q \cdot e \dfrac{V}{\epsilon}\right),
\end{equation}
where $V$ is the voltage applied across the target, $\epsilon$ is the average energy required to create one electron-hole pair and $e$ is the electron charge. $Q$ is a quenching factor which takes into account the different ionization yield of electron- and nuclear-recoil interactions (see Sec.~\ref{sec:bck} for more details).
SuperCDMS~\cite{SuperCDMS:2016wui}, TESSERACT and EDELWEISS~\cite{EDELWEISS:2017lvq} are or were cryogenic calorimeter-based experiments that will use or used a double readout of charge and phonons.

Cryogenic calorimeters can identify the time at which an energy deposition occurred in the target with high precision, but typically they are not capable of reconstructing the position of the event in the target.
However, some detector technologies were developed to distinguish at least between surface and bulk events. 
For example, this can be achieved for detectors reading out charges with (interleaved) electrodes. This technology was developed both by SuperCDMS~\cite{Brink:2006rs} and EDELWEISS~\cite{Broniatowski:2009qu} collaborations.

Although the mass of an individual detector module (e.g. order of 1\,g to 1\,kg) is constrained primarely by heat capacity and/or the production of large radiopure crystals, the total target mass can be increased by deploying additional detector modules.
While the scalability of cryogenic detectors was demonstrated by CUORE~\cite{CUORE:2021mvw}, a neutrinoless double beta decay experiment currently operating 988 detectors, its implementation is more challenging in the case of a low-threshold dark matter experiment as there can be a larger spread in detector performance.

An additional challenge of cryogenic calorimeters is an instrumentation background, referred to as low-energy  excess~\cite{Baxter:2025odk}, that causes a steep increase in the rate close to threshold. 
The cause of this background is not known to date and might have different origins (stress on the interfaces between sensors/electrodes/supports and the absorber, stress or damages in the absorber crystal, etc.).
Cryogenic calorimeter experiments generally observe this excess of low-energy events that currently limits their sensitivity. A broad effort within the community~\cite{Fuss:2022fxe} is focused on understanding the origin of this background in order to mitigate its impact and, if possible, eliminate it (see Sec.~\ref{sssec:instrumental_bck}).

\vspace{4mm}
\twocoladvchallenge[Cryogenic crystal calorimeters]
{
\begin{itemize}
\item Excellent ER/NR discrimination 
\item  Tens of eV down to sub-eV energy threshold
\end{itemize}
}
{
\begin{itemize}
\item Low energy excess can limit the science reach 
\item No monolithic scaling
\end{itemize}
}
\vspace{4mm}

\subsection{Semiconductors}
\label{ssec:tech_solid}

Semiconductor-based detectors exploit the generation of electron–hole pairs within the target material, using the resulting charge signals to determine the amount of energy deposited.
While a variety of semiconductor crystals exist, silicon and germanium are generally preferred because of the high radiopurity that can be achieved and the large crystal boule sizes that can be produced, thanks to the development in the semiconductor industry.
The fundamental signal quanta are electron-hole pairs for semiconductor detectors.
In germanium and silicon, the mean energy required to create an electron–hole pair is about 3\,eV~\cite{Knoll:2010xta} and $(3.6-3.8)$\,eV~\cite{Pehl:1968, Ramanathan:2020fwm}, respectively, and their band-gap energies 0.74\,eV~\cite{Varshni:1967} for germanium and 1.17\,eV~\cite{Bludau:1974} for silicon. Consequently, the band gap determines the minimum detectable energy.
Thanks to the low energies required to create an electron-hole pair, these detectors can achieve excellent energy resolutions.
The following subsections will focus on two semiconductor-based detector technologies: charge-coupled devices and high purity germanium detectors which are both notably used in the direct dark matter search.

\subsubsection{Charge-coupled devices}
\label{sssec:tech_ccd}

\begin{figure}[ht]
	\centering
	\includegraphics[width=0.75\columnwidth]{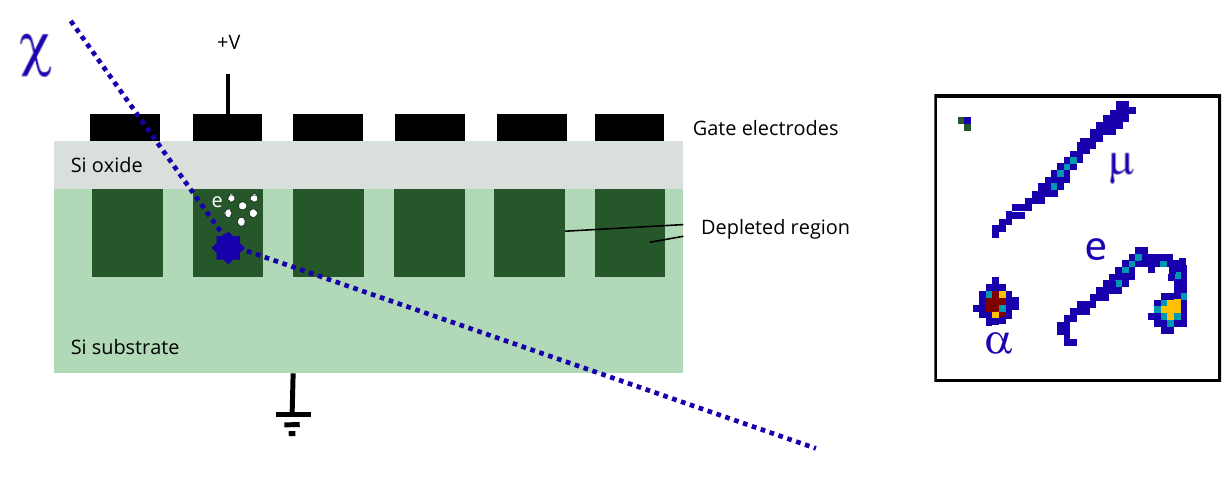} \\
    
	\caption{Illustration of a charged-coupled device. The sensitive volume is the Si-depleted region (shown in dark green), where the energy released by an interacting particle during the exposure period is converted into electron-hole pairs, which are efficiently collected and stored until readout. The distribution of the number of electrons accumulated per pixel is shown in the inset on the right for typical event topologies associated with different interacting particles, enabling background discrimination. A dark matter interaction would look like the cluster in the upper-left corner.}
	\label{fig:CCD_sketch}
	\end{figure}

First developed in the early 1970s~\cite{Smith:2010}, charge-coupled devices (CCD) are microcircuits designed to record visible-light images and they have found a scientific application in radiation detection and imaging.
CCDs are made from a silicon wafer, which is hundreds of micrometers thick, with a thin layer of silicon dioxide deposited on top to electrically isolate the gate electrodes from the semiconductor. When a voltage is applied to the gate electrodes, an electric field is created in the silicon, forming a depleted region that acts as a potential well for collecting and storing electrons generated by photon or particle interactions.
Figure~\ref{fig:CCD_sketch} shows a schematic illustration of a CCD including a typical image highlighting the different event topologies that can be distinguished using this technology.

Following a particle interaction, a number of electron-hole pairs proportional to the deposited energy is generated in the semiconductor. The electrons are transported by a weak electric field to potential wells (pixels), where they are collected and stored during the exposure.
When the exposure time has concluded, the charges are physically moved row by row through the chip toward the serial readout register.
Each image requires an exposure time of several hours, therefore the timing information of each particle event is not available.
However, this technology provides both position and track information of the interacting particle allowing for particle type identification and for fiducialization.
The x-y position of an event is determined by the pixel(s) collecting the charge, while the z-position can be inferred from the lateral extent of the charge cluster, since charge deposited farther from the collection plane undergoes greater diffusion during its drift toward the pixels.

CCDs are typically operated at temperatures between 130 and 140\,K to reduce dark currents, which are thermal excitations of electron–hole pairs that can mimic a dark matter signal. 
In a conventional CCD the charge is converted to voltage once and is then destroyed. This readout technique achieved a root mean square (RMS) noise resolution of around 2 electrons~\cite{DAMIC:2016lrs} for dark matter applications.
The introduction of skipper CCD readout enabled a reduction in the RMS readout noise to below 0.1 electrons~\cite{Tiffenberg:2017aac}. 
In skipper CCDs, the charge can be measured multiple ($N$) times without destroying it. 
This is achieved by repeatedly transferring the charge packet between the the readout node and a storage region, allowing successive measurements of the same charge.
The average of the $N$ independent measurements thus permits to lower the noise by a factor of 1/$\sqrt{N}$.
This technique is perfect for rare event searches with a low event rate, as the readout time will scale linearly with $N$ and the area of the CCD.
The reduction of the RMS noise from around 2 electrons to $<0.1$ electrons makes CCD sensitive to single electrons and achieve energy thresholds around ($2-3$)\,eV~\cite{Essig:2015cda} depending on the dark current. 
CCD technology is currently used by SENSEI~\cite{SENSEI:2020dpa}, DAMIC~\cite{DAMIC:2020cut,DAMIC-M:2024ooa} and DMSQUARE~\cite{Avalos:2021fxm} and will also be adopted by OBSCURA~\cite{Oscura:2022vmi}, which unifies and builds upon the efforts of both collaborations.

\vspace{4mm}
\twocoladvchallenge[Charge-coupled devices]
{
\begin{itemize}
    \item A few eV energy threshold 
    \item Spatial resolution enables particle identification
\end{itemize}
}
{
\begin{itemize}
    \item Only charge-producing interactions are detectable
\item No monolithic scaling 
\end{itemize}
}
\vspace{4mm}

\subsubsection{High purity germanium detectors}
\label{sssec:tech_hpge}

\begin{figure}[htb]
	\centering
	\includegraphics[width=0.7\columnwidth]{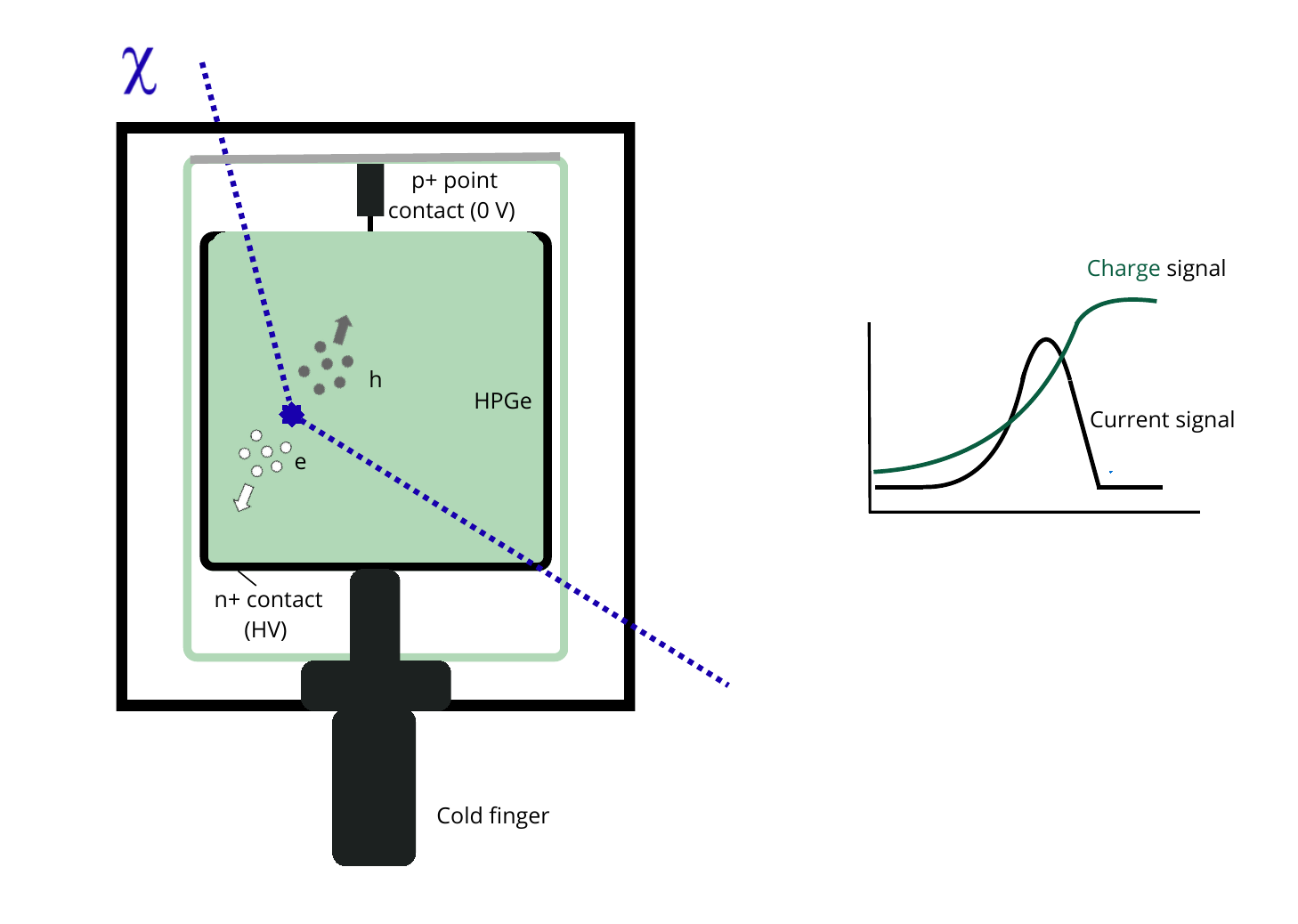} \\
    
	\caption{Illustration of a high-purity germanium  detector with a point-contact geometry. The germanium crystal (shown in green) is housed in a vacuum vessel and cooled to 77\,K via a cold finger. Energy deposited by an interacting particle is converted into electron-hole pairs, which are drifted and collected at the electrodes by the electric field. The inset on the right displays the typical signal shape. The time evolution of the charge signal --and its derivative, the current pulse-- can be used to identify multiple site energy depositions from point-like interactions.}
	\label{fig:HPGe_sketch}
	\end{figure}

Germanium crystals can be purified with the zone refining technique~\cite{Pfann:1966} and Czochralski crystal growth~\cite{Sumathi:2023}. 
The combination of these two procedures achieves electrically active impurity levels as low as (10$^{9}$ -- 10$^{10}$)\,/cm$^3$~\cite{Abrosimov:2019qeb}. 
After purification, controlled doping with acceptor or donor species can be used to tailor the electrical properties of the high-purity germanium (HPGe) crystal, resulting in p-type or n-type material, respectively.
HPGe detectors for dark matter searches employ a point-contact geometry, as shown in Fig.~\ref{fig:HPGe_sketch}, opposed to planar and coaxial geometries. 
In the point-contact geometry, one of the electrodes is reduced to a very small contact, resulting in an extremely low detector capacitance.
The readout of the charge signals is performed at the point-like electrode.
The point-contact geometry is favored by dark matter experiments thanks to its lower capacitance ($\sim 1$\,pF~\cite{Wei:2021drf}) which permits to achieve lower noise levels and consequently lower thresholds of the order of 100 eV$_\mathrm{ee}$.
Reducing the size of the point-like electrode ($\sim$1\,mm in diameter) leads to a lower capacitance.
While the electric field of planar and coaxial detectors is designed to be uniform and provide a consistent response through all the detector volume, the electric field of the point-contact geometry is not uniform. The electric field of the point-contact detectors is weak in most of the volume and extremely strong close to the point-like electrode.
The position of an event within the crystal volume influences the detector response, allowing surface and bulk events to be discriminated based on their pulse shapes~\cite{Yang:2016crf}.
The mass of a single HPGe target can be of the order of $(0.5-1)$\,kg without degrading the detector performance.
HPGe detectors are operated at 77\,K to suppress thermally induced leakage current arising from germanium’s small band gap.
Point-contact detectors are currently employed by the CDEX~\cite{CDEX:2023vvc} experiment, but they have a long history dating from early 1990s. Since germanium detectors measure only ionization, they cannot discriminate between electronic recoils (ERs) and nuclear recoils (NRs), making background suppression more challenging in searches for NRs.

\vspace{4mm}
\twocoladvchallenge[High purity germanium detectors]
{
\begin{itemize}
\item Discrimination between surface and bulk events
\item Relatively simple technology
\end{itemize}
}
{
\begin{itemize}
\item Only charge-producing interactions are detectable
\item Background suppression in NR searches
\end{itemize}
}
\vspace{4mm}

\subsection{Scintillation crystals}
\label{ssec:tech_scint}

\begin{figure}[htb]
	\centering
        \includegraphics[width=0.8\columnwidth]{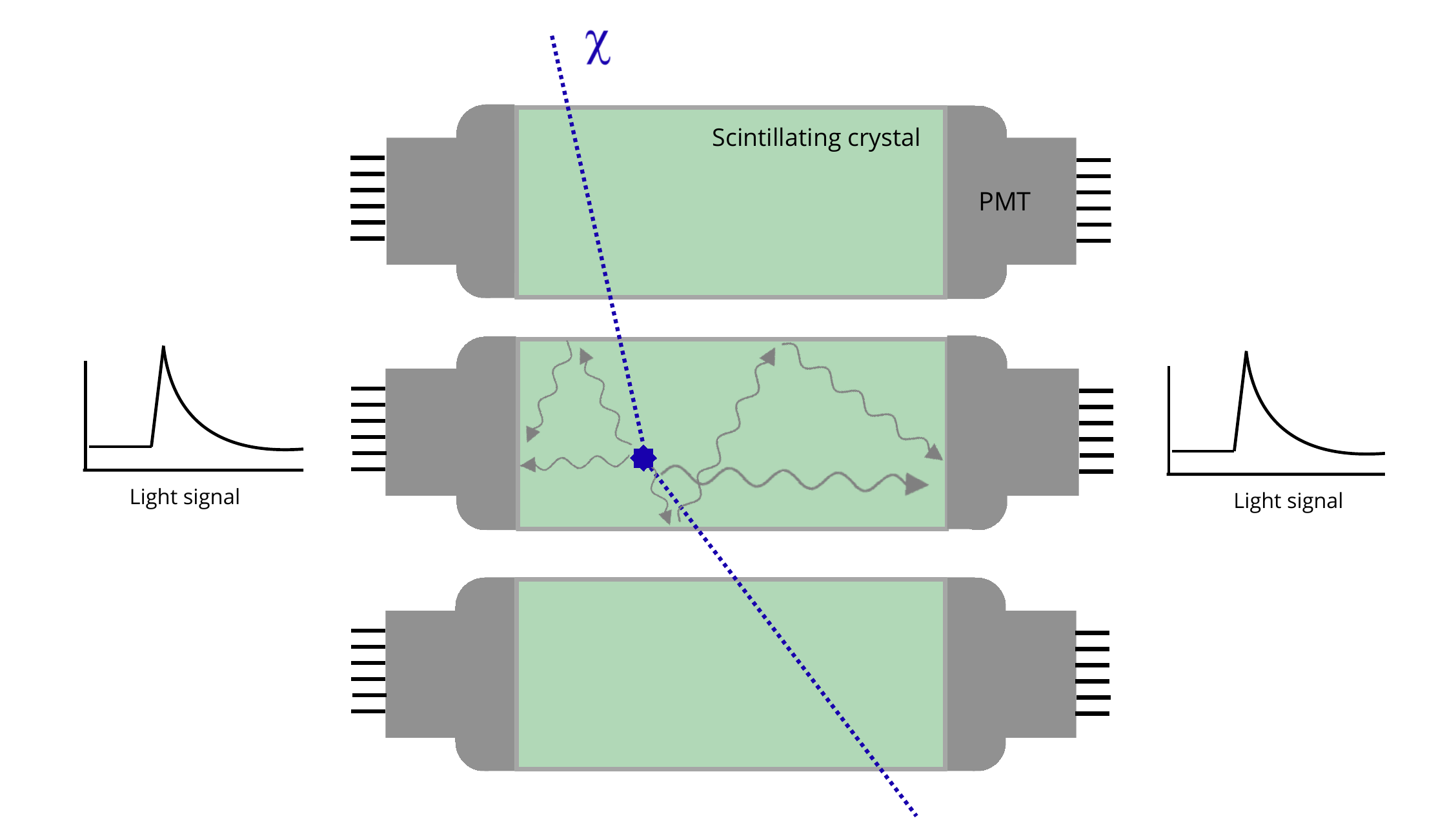}
	\caption{Illustration of a scintillation crystal detector array. The energy released by a particle interaction results in the emission of scintillation photons (gray wavy arrows), typically in the visible-UV range, that propagate toward light sensors. The left and right insets display typical signal shapes from the photosensors (exemplified here by photomultiplier tubes). The signal shape depends on the scintillation time constant of the chosen target. To prevent dark events from triggering the system, two or more photosensors per module are read out in coincidence mode. }
	\label{fig:scintillationCrystal}
	\end{figure}

When charged particles interact in a medium, light emission from excited states is one of the available channels for energy conversion. Although the percentage of energy converted into light is generally quite low, scintillation readout is advantageous due to the ease of light propagation over large distances in transparent media which allows for large size targets. A wide variety of materials produce scintillation under particle excitation, including liquid noble gases (see Sec.~\ref{ssec:tech_noble}), organic liquids or plastics, and inorganic scintillating crystals. This section will focus on scintillating crystals with only-scintillation readout which in most cases operate under controlled room-temperature conditions. These detectors consist of a scintillating crystal coupled to one or more light sensors (see Fig.~\ref{fig:scintillationCrystal}), typically arranged in detector arrays. 

The field-deployable nature of these detectors enabled applications in contexts such as nuclear emergency response, in-situ material assay, environmental and aerial mapping, geological and mineral surveys and space exploration.
The crystal size can be large, up to $\sim (10-15)$\,kg, and in addition several modules can be combined in a common set-up, allowing for large-mass experiments. The high density of most inorganic scintillating crystals, ranging from about $(3-8)$\,g/cm$^3$, compared with those of other detection targets allows compact high-mass detection systems. Moreover, stable operation in very long time periods can be achieved with a remarkably high duty cycle of up to about 95\%. 

The successful development of scintillation detection technology is linked to the improvement of the available light sensors. The development of photomultiplier tubes (PMTs) with high quantum efficiency in a broad wavelength range was a key asset. From the first scintillation detections by eye using ZnS screens in 1903~\cite{Crookes1903}, scintillation detection has undergone remarkable advances, culminating in today's state-of-the-art PMTs. Despite their remarkable performance, PMTs are being replaced in many applications by silicon photomultipliers (SiPMs),
an emerging light-detection technology that offers several advantages, including higher radiopurity, compactness, and low-voltage operation. However, SiPMs also have some limitations, such as high dark-count rates at room temperature and relatively low photon detection efficiency in the ultraviolet (UV) range.

There is a wide variety of scintillating crystals, incorporating interesting combinations of target nuclei with high and/or low masses, and with and without nuclear spin, for instance. 
This allows tuning sensitivities for different applications and it is particularly interesting in the application to searches for dark matter or the neutrinoless double beta decay. 
Scintillation follows the release of energy by charged particles in the material: electron-hole pairs are produced (both free and in the form of excitons), which relax their energy through some light-emitting process. 
The crystal is transparent to these scintillation mechanisms because the sub-gap photons emitted cannot produce new electron-holes pairs. 
Inorganic scintillation crystals can be classified into doped and pure crystals. 
In the case of doped inorganic crystals, light is emitted by energy levels in the forbidden gap (color centers) linked to the dopant atoms which decay radiatively. 
The scintillation photon energies are found in the visible or near UV wavelength range, at sub-gap energies. 
These crystals show, in general, high light yields and scintillation decay time constants dependent on the dopant. 
In the case of pure inorganic crystals, the light is emitted in the radiative recombination of self-trapped excitons, which can occur at any point of the crystal lattice. 

The light yield generally depends strongly on the interacting particle. 
The scintillation "quenching factors" measure the relative efficiency for the production of scintillation with respect to an electron releasing the same energy, and they lack a successful theoretical modeling, requiring dedicated measurements in order to correctly calibrate the energy scale for any interacting particle.  
In addition, the scintillation time constants also depend on the interacting particle, enabling particle identification for most scintillating targets at high energy. 
However, this method has significant limitations at low energies and is typically unsuitable for event-by-event discrimination in the region of interest for dark matter searches. 

For organic materials (which can be solid or liquid), the mechanism for producing scintillation is different, being based on molecular excited states that decay radiatively. 
They are characterized by a fast scintillation component (in the ns range), low to moderate light yield, and good particle discrimination capabilities. 
Their comparatively low light yield, particularly relative to inorganic scintillating crystals, limits the achievable energy threshold and generally makes them less suitable as targets for dark matter searches. 

In general, scintillation is a subdominant path for the energy conversion, implying that the average energy required to produce a scintillation photon is high, and the number of photons per unit of energy released is low. 
This results in a poorer energy resolution in comparison with other detection techniques, especially those described in Sec.~\ref{ssec:tech_bolom} and \ref{ssec:tech_superfluid}.

One inorganic crystal to highlight in particular is thallium-doped sodium iodide (NaI(Tl)), which is an excellent scintillator and which has been applied in direct searches for dark matter for more than thirty years.
For a recent review, see Ref.~\cite{Martinez:2026thh}.
NaI(Tl) is a high-performing scintillating material: large-size crystals can be grown and the amount of light produced by deposited energy unit is among the highest, at the level of 40000\,photons/MeV, i.e. the production of a scintillation photon requires on average 25\,eV. 
Light collection at the level of $(15-22)$\,photoelectrons/keV in large crystals has been demonstrated, implying that threshold energies below 1\,keV$_\text{ee}$ are at hand with limited energy resolution above (20\,--\,25)\% at 1\,keV (statistical). 
The wavelength of the light produced, peaking around 415\,nm, perfectly matches the peak sensitivity of
bialkali PMTs, easing the fabrication of scintillation detectors. 

In the application to direct dark matter searches, NaI has other relevant advantages,
like a 100\% nonzero spin isotopic content for both $^{127}$I and $^{23}$Na, which makes this material a remarkable target for spin-dependent interacting DM particle candidates, while the combination of heavy and light nuclei allows for good kinematic matching in a wide range of DM particle masses. 
In addition, the long-standing DAMA/LIBRA result on annual modulation (see Sec.~\ref{sec:experiments}) has kept interest in this target material alive for decades. 

Thallium-doped cesium iodide (CsI(Tl)) crystals are also an interesting target, applied both in direct dark matter searches and the detection of coherent neutrino-nucleus scattering. 

Dark matter search experiments using inorganic scintillating crystals with only-scintillation readout are DAMA/NaI~\cite{Bernabei1999DAMA_NaI}, DAMA/LIBRA~\cite{Bernabei2008DAMA_LIBRA}, KIMS~\cite{Lee2006KIMS}, ANAIS-112~\cite{Amare2019ANAIS112}, COSINE-100~\cite{Adhikari:2017esn}, and COSINE-100U~\cite{Lee:2024wzd}, while experiments under construction or planning stage are SABRE~\cite{Antonello2019SABRE}, COSINE-200~\cite{COSINE:2020egt} and PICOLON~\cite{Fushimi:2021mez} (see Sec.~\ref{sec:experiments}).
Recently, the replacement of PMTs by SiPMs has been proposed to reduce the achievable energy thresholds with inorganic scintillators (doped or pure) and increase the radiopurity of the setups, because PMTs are currently a high contribution to the background budget. However, to overcome the limitation of the high dark current of SiPMs, crystal operation at temperatures around 100\,K is required. Experiments following this strategy in R\&D phase are ASTAROTH~\cite{Martinenghi:2025ckb} and ANAIS+~\cite{Amaré:2026XQ}. 

\vspace{4mm}
\twocoladvchallenge[Scintillation crystals]
{
\begin{itemize}
\item High light yield 
\item Operation at room temperature
\end{itemize}
}
{
\begin{itemize}
\item Crystal radiopurity
\item Understanding NR-Quenching Factors 
\end{itemize}
}
\vspace{4mm}

\subsection{Bubble chambers}
\label{ssec:tech_bubble}

\begin{figure}[h]
	\centering
    \includegraphics[width=0.9\columnwidth]{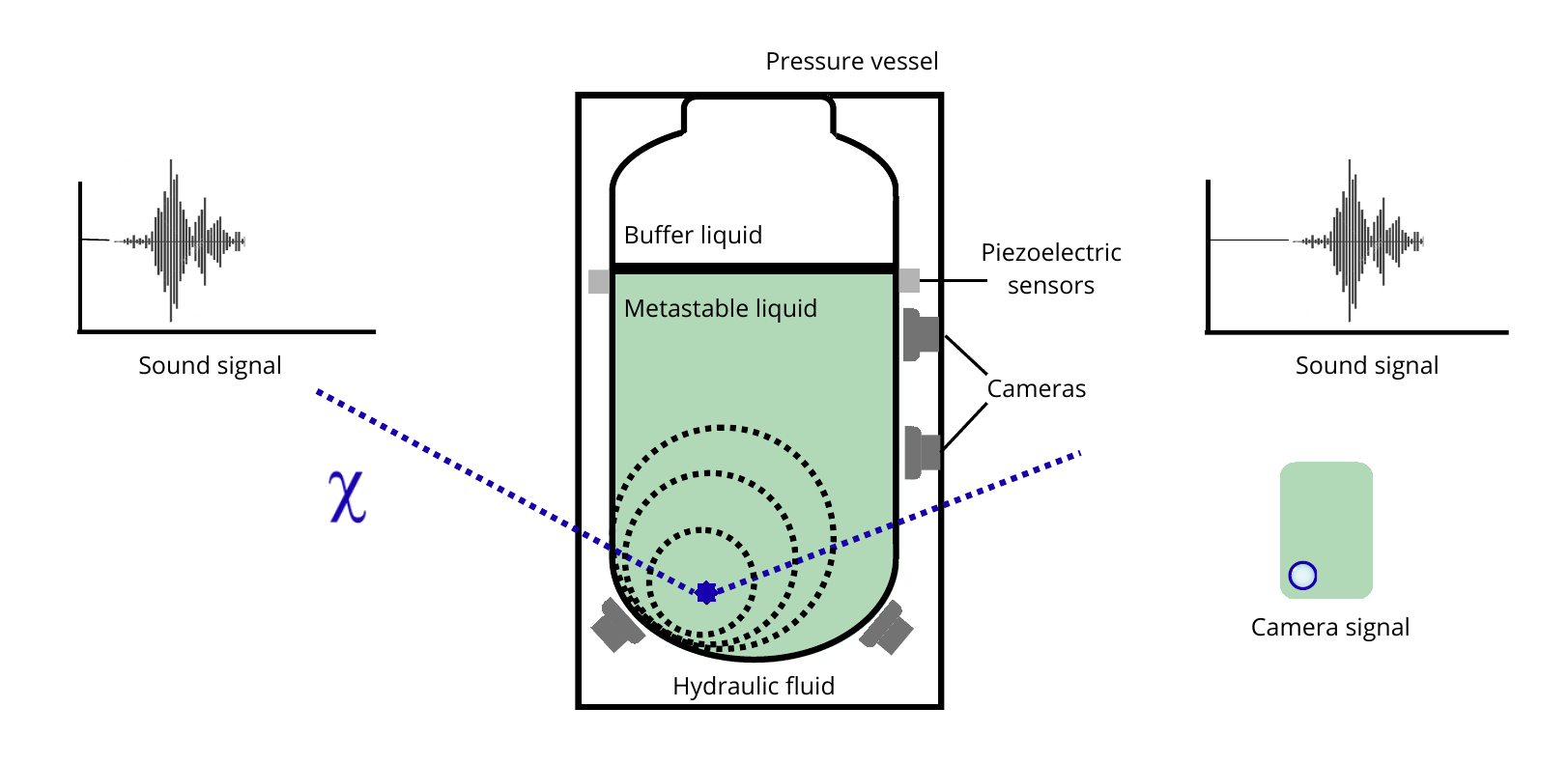}
	\caption{Illustration of a bubble chamber. The sensitive volume consists of a metastable liquid (shown in green) housed in a tight vessel, under controlled temperature and pressure. The energy released by a highly-ionizing particle interacting in the liquid can nucleate a gas bubble. This nucleation generates a sound wave (dotted lines) propagating through the liquid. The left and right insets display representative sound signal waveforms registered by piezoelectric sensors and the right-bottom inset, a bubble as viewed by the cameras under appropriate lighting.}
	\label{fig:bubble_chamber}
	\end{figure}

Superheated liquids are maintained above their boiling point at a given pressure without immediately vaporizing. This metastable state persists in the absence of suitable nucleation sites, allowing even a small energy deposition to trigger rapid bubble formation. This phenomenon forms the basis of particle detection in bubble chambers, first invented in 1952~\cite{Glaser1952}, and is exploited in dark matter detection experiments. In these detectors, energy deposited by a particle interacting with a nucleus in the superheated liquid initiates the formation of a microscopic bubble~\cite{Seitz}.
These bubbles are then recorded, allowing to identify particle interactions (see Fig.~\ref{fig:bubble_chamber}). 

The bubbles are typically recorded by high-speed cameras, which provide images of the chamber from multiple angles. 
This allows the bubbles to be reconstructed in three dimensions with a resolution of approximately 1\,mm. 
As a consequence, an innermost volume can be defined for analysis, featuring a lower background. 
In addition to visual imaging, the bubble formation process also produces acoustic signals: the rapid growth of bubbles generates small pressure waves that can be detected by sensitive microphones or acoustic sensors attached to the chamber. 

By adjusting the temperature and pressure of the superheated fluid, the detector can be calibrated to primarily respond to high energy depositions  d$E$/d$x$, i.e. those caused by nuclear recoils or alpha particles, while remaining insensitive to energy depositions caused by electrons. 
This is a clear advantage of this technology, since electronic recoils are one of the dominant background sources in most experiments. As a result, however, no searches are generally conducted using electron recoils. 
Interestingly, the sound-based signal of a nuclear recoils differs from the one of alpha particles, allowing to also reject this background~\cite{Behnke:2010xt}.

Many modern bubble chamber detectors typically employ target masses ranging from sub-kilogram scales to tens of kilograms, with larger detectors under development. They use liquid compounds containing fluorine, such as C$_3$F$_8$, C$_4$F$_{10}$, CF$_3$I and C$_2$CIF$_5$. The nucleus of fluorine ($^{19}$F) has an unpaired proton, giving it a particularly large expectation value for proton spin content and enhancing its sensitivity to spin-dependent interactions with protons (see Sec.~\ref{sec:experiments}).
Building on the pioneering experiments COUPP~\cite{Behnke2012COUPP}, PICASSO~\cite{PICASSO:2012ngj}, and SIMPLE~\cite{SIMPLE:2014pun}, the PICO collaboration was established and subsequently operated the PICO-2L, PICO-60L, and PICO-40L detectors. PICO~\cite{PICO:2019vsc} have achieved nuclear-recoil thresholds of a few keV, with PICO-60 operating at thresholds as low as 2.45\,keV in its dark matter search.  A larger-scale experiment, PICO-500L~\cite{Giroux:2021vpy}, is also planned.
For further information, see Sec.~\ref{sec:experiments}.

{\bf Scintillating bubble chambers} extend the conventional bubble-chamber technique by combining bubble nucleation in a superheated liquid with the detection of scintillation light~\cite{Alfonso-Pita:2022akn,Alfonso-Pita:2023frp}. This approach retains the intrinsic insensitivity of bubble chambers to electron-recoil backgrounds, while providing an additional observable that can be used for background discrimination and energy reconstruction.  In noble liquids, the absence of molecular degrees of freedom limits the pathways through which electronic recoils can transfer energy into the localized heating required for bubble nucleation. The Scintillating Bubble Chamber (SBC) Collaboration is developing this technology using liquid argon to search for low-mass, GeV-scale dark matter through nuclear recoils with energy thresholds in the sub-keV range. The argon is doped with xenon to shift to longer wavelengths, improving light collection, while SiPMs detect the scintillation signal. The first physics-scale detector, SBC-LAr10, contains approximately 10\,kg of liquid argon and is being used at Fermilab to characterize the detector response and calibrate its sensitivity to low-energy nuclear recoils. A second, low-background detector, SBC-SNOLAB, is intended for underground dark matter searches at SNOLAB. 

A related approach to dark matter detection is employed by the Indian Dark Matter Search Experiment (InDEx), which uses {\bf superheated droplet detectors}. In this technique, droplets of a superheated liquid are suspended in a gel matrix, and energy deposited by a particle interaction can trigger bubble nucleation within an individual droplet. As in conventional bubble chambers, the nucleation threshold can be controlled through the thermodynamic operating conditions. InDEx employs tetrafluoroethane (C$_2$H$_2$F$_4$), which contains hydrogen and may therefore provide sensitivity to low-mass dark matter through nuclear recoils on protons. Although not yet experimentally demonstrated, the droplet-based approach may also offer the potential to achieve lower nuclear-recoil energy thresholds than those of conventional, more mature bubble-chamber detectors. The second InDEx run operated with an active target mass of approximately 70\,g and reached a recoil threshold of about 2\,keV~\cite{Das:2025coa}.

\vspace{4mm}
\twocoladvchallenge[Bubble chambers]
{
\begin{itemize}
\item Sensitivity to spin dependent interactions
\item Background suppression
\end{itemize}
}
{
\begin{itemize}
\item Limited size: recompression of liquid challenging 
\item No electronic recoil searches available
\end{itemize}
}
\vspace{4mm}

\subsection{Liquid noble-gas scintillators}
\label{ssec:tech_noble}

\begin{figure}[h]
	\centering
    \includegraphics[width=0.45\columnwidth]{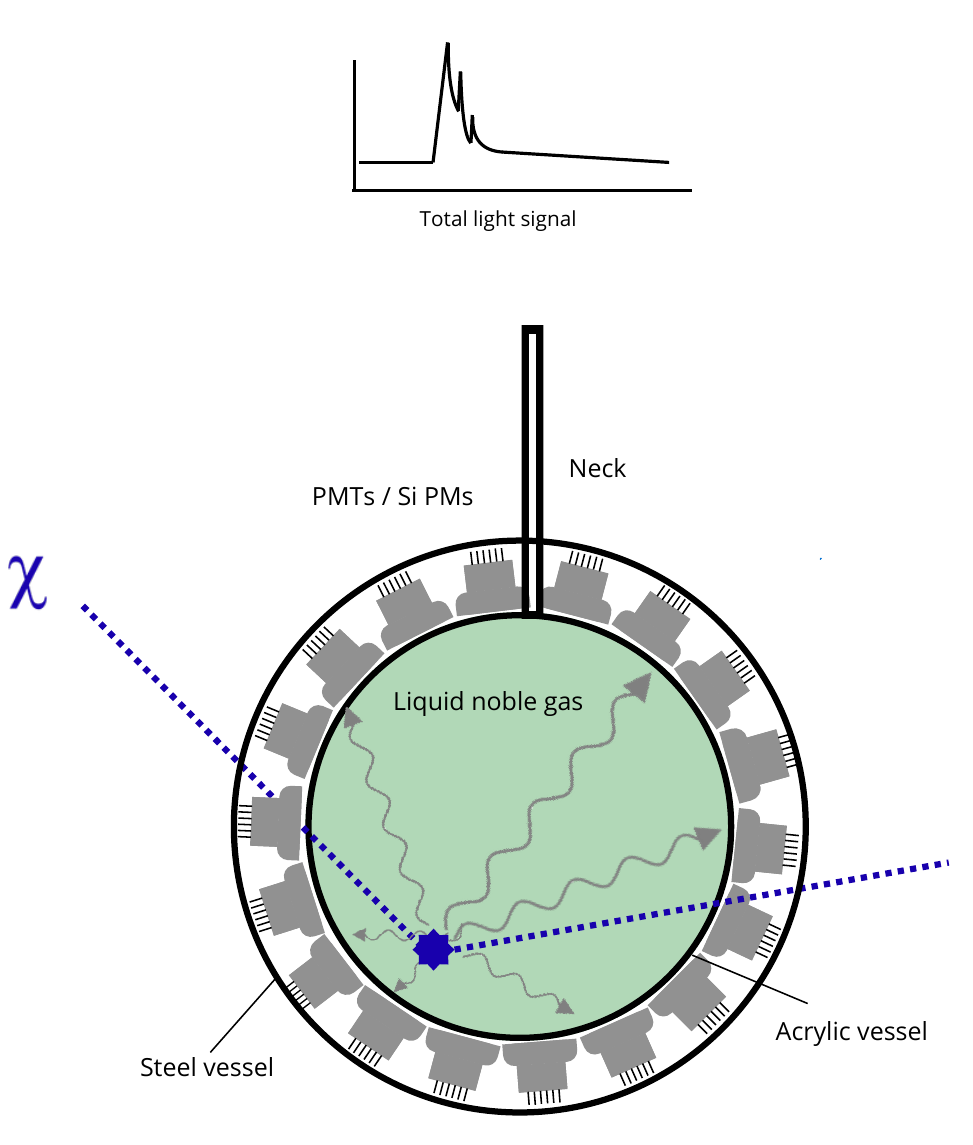}
    \includegraphics[width=0.45\columnwidth]{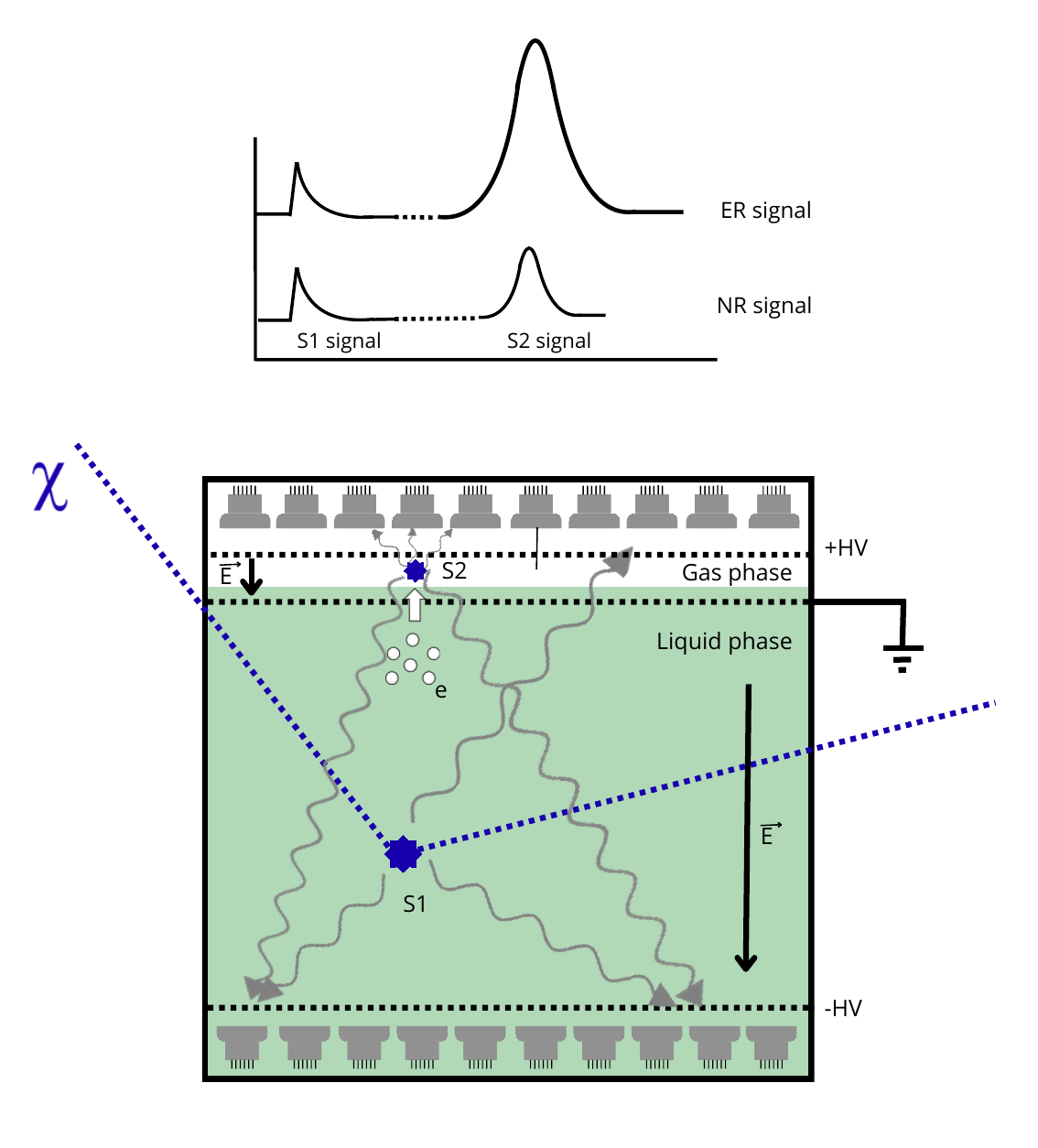}
	\caption{Illustration of single-phase (left) and dual-phase (right) detectors using liquid noble-gas scintillators (shown in green), typically argon or xenon. In both concepts, scintillation light (gray wavy arrows) is detected by photosensors, such as PMTs or SiPMs. In the dual-phase detector, this prompt light constitutes the S1 signal, which is complemented by the S2 signal generated via electroluminiscence as electrons drift into the upper gas phase under an applied electric field. The top insets display typical signals for both detector concepts. In the single-phase (left), individual PMT signals allow the reconstruction of the interaction position and in the case of using argon, the scintillation pulse shape allows particle discrimination. In the dual-phase (right) the ratio between S1 and S2 signals allows for particle discrimination, while the time interval between them and the pattern of light sensor hits, the reconstruction of the interaction point. }
	\label{fig:noble_gas_concepts}
	\end{figure}

Liquid noble gas detectors employ mostly argon or xenon as target.
Although neon has also been proposed~\cite{Lippincott:2011zr}, no experiments using this element have yet been operated. There are two types of detectors (see Fig.~\ref{fig:noble_gas_concepts}): single-phase (liquid) and dual-phase time projection chambers (TPCs), in which the main target is liquid with a thin gas layer on top.  

When a charged particle deposits energy in the noble liquid medium, it becomes excited, creating excimers (i.e. excited dimers) that de-excite by emitting ultraviolet scintillation light (usually called S1).
The energy deposition also ionises the medium, creating free electrons.  In the absence of an applied electric field, the electrons recombine with the ions, producing additional excimers and contributing to the S1 scintillation signal. 
\\

\noindent{}\textbf{Single-phase detectors} are typically spherical and collect the light signal using photosensors placed on the target surface. This configuration enables large light collection efficiency, which supports a low energy threshold, typically at the $\sim$\,keV level.
In this type of detector, position reconstruction is based on the timing of photons hitting the photosensors. Although the position resolution is a bit lower than that of TPCs, at the $(0.1-1)$\,cm level~\cite{DEAP:2025igy,Abe:2013tc}, it is sufficient for determining the analysis volume (fiducialization). Particle discrimination is possible using the shape of the S1 pulse. The dimer de-excitation occurs from two excited states, singlet and triplet, which differ on their lifetime. The emitted scintillation light has a characteristic wavelength of approximately 128\,nm for argon and 175\,nm for xenon. While the decay times in argon are quite different, at 6\,ns and 1.6\,$\upmu$s~\cite{Boulay:2006mb}, those in xenon are about  3\,ns and 30\,ns~\cite{Cichon:2022gbu}, respectively.

Experiments using the single-phase technology are DEAP~\cite{DEAP:2019yzn} which is a 3.6\,t liquid argon detector operating at the SNOLAB laboratory, and XMASS~\cite{XMASS:2022tkr} which was a $\sim 1$\,t liquid xenon experiment at the Kamioka Observatory.\\

\noindent{}\textbf{Dual-phase detectors} are typically of cylindrical shape, with electrodes positioned at the top and bottom of the target volume to establish an electric field that drives the ionization electrons towards the liquid surface, where they can be collected.

Consequently the electrons drift through the liquid up to the surface. A thin layer of the noble gas is typically maintained at the top, where a strong electric field in the amplification region accelerates the electrons, producing additional scintillation photons through proportional scintillation (electroluminescence~\cite{Lansiart:1976}). This signal is delayed by the drift time of the electrons and is usually referred to as S2. The position of every event can be reconstructed with mm precision by combining the photon hit pattern in the top photosensor array ($xy$ position) and the drift time, which gives the $z$ position. This allows in turn to select an innermost volume for analysis, which is cleanest in terms of radioactive backgrounds. The ability to select the innermost volume, combined with good background attenuation due to the high density of the noble liquids (argon 1.4\,g/cm$^3$ and xenon 3\,g/cm$^3$), enables these detectors to effectively suppress external radiation from background sources.  

This technology has the advantage of enabling multi-tonne detector masses for dark matter searches, thanks to the relatively straightforward scalability of the target volume.
One major advantage of recording both the scintillation and the ionization signals is that their ratio can be used to distinguish between different particles or particle interactions. The number of photons and electrons produced in an energy deposition differs for electrons, alpha particles and nuclear recoils, and it is used to suppress backgrounds. 
Both scintillation light and proportional scintillation in TPCs are typically detected using photomultipliers (PMTs). Although many R\&D activities are being carried out towards using silicon-based photosensors (SiPMs, see for instance~\cite{DarkSide-20k:2024usz,Baudis:2018pdv}), only photomultipliers have been employed until now. This is due to their very low dark count rate (0.3\,Hz/cm$^2$ at liquid xenon temperature), which allows signals as small as three photons to be selected without a large increase in accidental coincidences. For future large liquid argon detectors, SiPMs are planned~\cite{DarkSide-20k:2024usz}, since conventional PMTs are not sufficiently stable. 
In contrast, PMTs are still being considered for future liquid xenon detectors to maintain a low energy threshold, thanks to their low dark count rates.

\vspace{4mm}
\twocoladvchallenge[Liquid noble-gas scintillators]
{
\begin{itemize}
\item Relatively easy scaling up $\to$ tonne-scale masses
\item Fiducialization to suppress backgrounds
\end{itemize}
}
{
\begin{itemize}
\item Internal and accidental backgrounds
\item Lowering the energy threshold below $\sim$\,keV
\end{itemize}
}
\vspace{4mm}

Dual-phase TPC technology has grown rapidly in recent years. For example, early xenon detectors around 2005 had a mass of around 10\,kg, whereas experiments nowadays are running with a mass of around 10\,t. Similarly, liquid argon detectors have grown from an initial size of around 50\,kg to the current detector under construction, which will contain 20\,t of argon. While the large increase in size enables a rapid gain in sensitivity for rare event searches, it also presents challenges. In order to reach the desired sensitivity, the experiments need to continue reducing backgrounds. For argon detectors, the cosmogenically produced $^{39}$Ar poses a significant challenge. Upcoming experiments will employ underground argon to suppress its background~\cite{Xu:2015xx}.  For xenon-based detectors, there are no short-lived xenon isotopes, but radon emanating from the detector materials is the dominant current background. Nevertheless, successive generations of xenon detectors have achieved substantial reductions in background levels, demonstrating the effectiveness of improved detector design and background mitigation strategies. Current efforts to further reduce radon backgrounds combine material selection, radon distillation, hermetic time projection chambers (TPCs), and radon barriers~\cite{XLZD:2024nsu}. Furthermore, as the detectors become larger, background noise from random coincidences start to play a major role. This background originates from randomly paired isolated light (S1) and charge (S2) pulses, which also happen to pass all selection criteria by chance. While some possible origins of the isolated pulses are being discussed, no full explanation exists yet, and this is a very active field of research
~\cite{Kopec:2023uii}. Possible sources include photosensor dark-count rates, energy depositions in regions where there is no electric field, and electrons being captured for a certain time in either impurities of the liquid or the liquid-gas surface, see section~\ref{sssec:instrumental_bck}.

Liquid noble detectors, particularly those based on liquid xenon TPCs, have led the sensitivity to dark matter with masses above a few GeV during the last years. Below this mass, the nuclear recoil has such a low energy that not enough quanta (photons and electrons) are produced to be recorded. This energy is typically around 3\,keV nuclear recoil energy for xenon~\cite{LZ:2024zvo,PandaX:2024qfu,XENON:2025vwd}. For argon, the energy threshold is rather determined by the ability to separate nuclear recoils from the $^{39}$Ar background and in the latest experiment it was at about 50\,keV~\cite{DarkSide:2018ppu}. 
Various analysis strategies are being developed to overcome this limitation, ranging from using only the S2 charge signal, which is usually larger than S1~\cite{DarkSide:2018bpj,XENON:2026qow}, to reducing the requirement on the S1 number of photons~\cite{XENON:2024ijk,LUX:2019npm,LUX:2020yym} or relying on the Migdal and Bremstrahlung effects (see Sec.~\ref{subsubsec_Nucl_Scat}).

Following the WARP experiment~\cite{Cocco:2011zz}, the most recent experiment to produce dark matter results using liquid argon TPC technology is DarkSide-50~\cite{DarkSide:2018bpj}. DarkSide-20k~\cite{DarkSide-20k:2017zyg} is currently under construction. This experiment is a collaboration between experiments worldwide and aims to build a 20\,t detector. 
The first liquid xenon TPCs were XENON10~\cite{Aprile_2011} and ZEPLIN-I~\cite{Ivaniouchenkov:2000}. These were followed by several larger detectors: ZEPLINII-III, XENON100, LUX and XENON1T. Currently, there are three competing multi-tonne experiments using this technology: PandaX~\cite{PandaX-4T:2021bab}, LZ~\cite{Akerib:2020lzi} and XENONnT~\cite{XENON:2024}. 
These experiments provide the best current sensitivities for WIMP masses above $\sim 3$\,GeV, with LZ setting the most stringent exclusion limit. Most current results and status of the experiments is described in Sec.~\ref{sec:experiments}.

\subsection{Superfluid noble-gas detectors}
\label{ssec:tech_superfluid}

Current superfluid noble gas detectors employ superfluid helium as the target medium and can be broadly categorized into two experimental approaches: a cryogenic calorimeter concept and a resonator-based bolometric concept.
A central motivation common to both approaches is the low mass of the target nucleus ($^4$He or $^3$He), which for a fixed dark matter mass leads to comparatively larger nuclear recoil energies (cf. Eq.~\ref{eq:SI-recoil-energy}), thereby improving sensitivity to sub-GeV dark matter. 
These experiments are therefore, by design, optimized for light dark matter searches. 
At higher masses (above a few GeV), however, they cannot compete with technologies employing heavier target nuclei (most notably liquid noble-gas scintillators discussed in Sec.~\ref{ssec:tech_noble}) which benefit from the coherent enhancement of the dark matter–nucleus scattering cross section scaling approximately as $A^2$ (cf. Eq.~\ref{equ:diff_xsec_SI}), which are thus intrinsically more sensitive at large momentum transfer.

The calorimetric approach effectively combines elements of the cryogenic calorimeter technology discussed in Sec.~\ref{ssec:tech_bolom} with the light-emission characteristics typically associated with TPCs, as described in Sec.~\ref{ssec:tech_noble}. 
In contrast, the resonator approach represents a novel detection paradigm that departs from conventional dark matter–nucleon scattering searches, instead exploiting the unique quantum and collective excitations of the superfluid medium as the primary signal channel.\\

\begin{figure}[h]
\centering
\includegraphics[width=0.45\columnwidth]{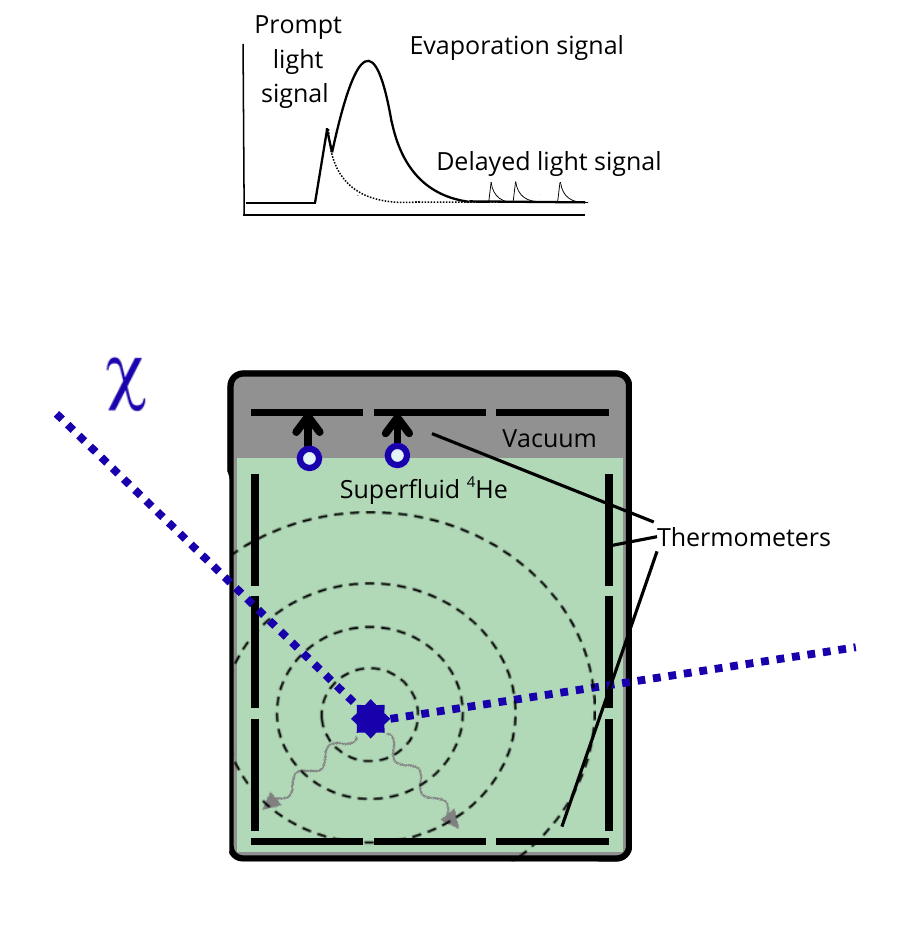}
\includegraphics[width=0.45\columnwidth]{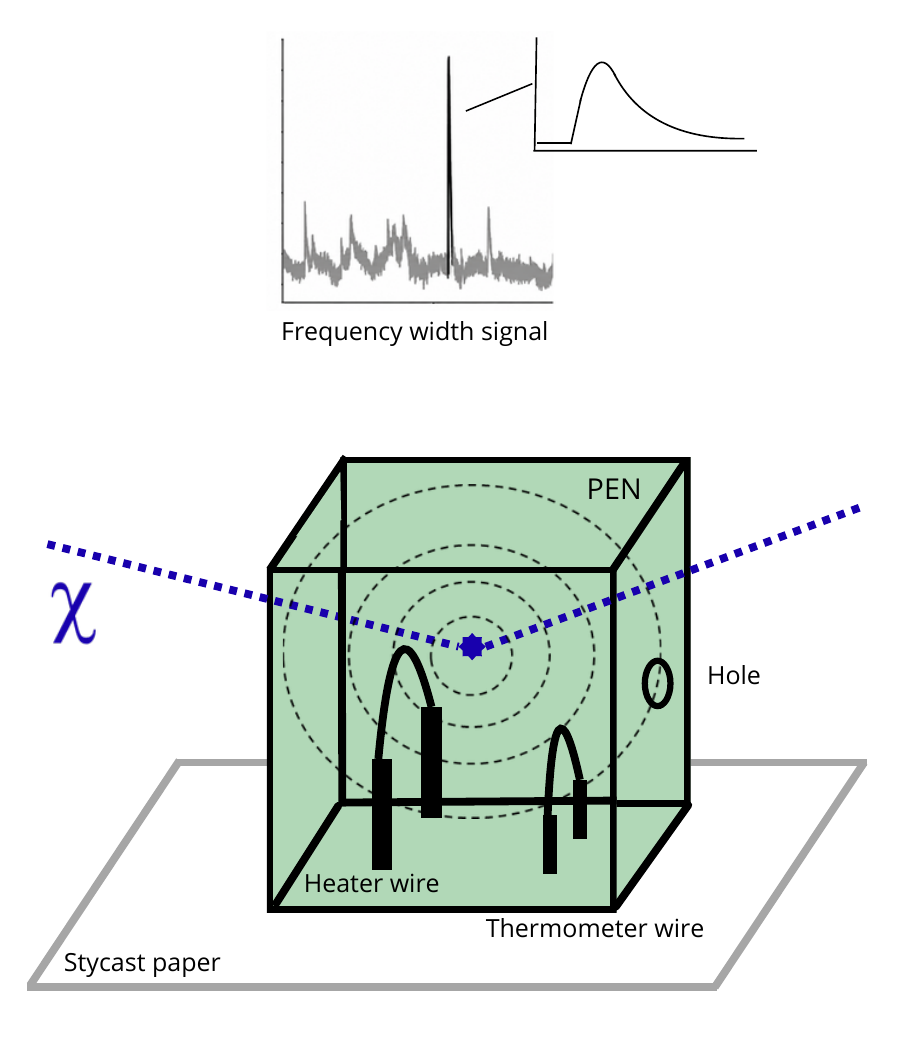}
\caption{Illustration of the superfluid noble gas detector concepts. Left: The calorimetric approach uses superfluid $^4$He (shown in green) housed in a vacuum vessel, measuring the generated quanta with microcalorimeters (thermometers): Photons (gray wavy arrows) are absorbed directly, while phonons and rotons (dashed lines) are detected via quantum evaporation of helium atoms at the surface (white spheres), adsorbed onto the top thermometers. 
Right: The resonator concept exploits quantum and collective excitations of $^3$He superfluid medium (shown in green) excited by the interacting particles, which are measured using mechanical resonators. The scattering of quasiparticles in the resonator modifies the resonance frequency width, acting as a sensitive thermometer. The top insets display typical total waveforms, combining prompt and delayed light with evaporation signals for the calorimetric concept (left) and the frequency width signal for the resonant concept (right).
}
\label{fig:superfluid_helium_concepts}
\end{figure}

\noindent{}\textbf{The calorimetric approach} is depicted in Fig.~\ref{fig:superfluid_helium_concepts} (left) and consists at its core of a superfluid helium-4 target instrumented with calorimetric sensors both within the liquid target and above it in vacuum.
The currently proposed volumes are of the order of (1\,-\,10)\,L corresponding to about (0.15\,-\,1.5)\,kg, which can be easily scaled to larger volumes in future generation experiments.
The sensor technologies that are currently being considered are TESs and MMCs, which are capable of sub-eV energy resolutions, and thus ultra-low thresholds, when operated at about 10\,mK (see Sec.~\ref{ssec:tech_bolom}).
Such low operating temperatures are compatible with a liquid $^4$He target, since helium remains liquid down to 0\,K at ambient pressure.

The calorimetric approach pursues the key idea to exploit the unique excitation spectrum of superfluid helium-4, in particular the presence of long-lived phonons and rotons, collectively referred to as quasiparticles. 
Its basic principles were already developed and demonstrated in the 1990s within the solar neutrino project HERON~\cite{Lanou:1987}.
An energy deposition in the target produces a population of these quasiparticles, as well as excited and ionized helium atoms. 
The latter rapidly form excimer states (He$_2^*$), whose radiative decay gives rise to scintillation light predominantly in the ultraviolet (UV) range. 
Excimers in the singlet state decay within nanoseconds, yielding an essentially instantaneous UV signal, whereas triplet states are long-lived and give rise to delayed emission. 
Triplet excimers have a decay time of about 13\,s.
However, in current detector designs, triplet excimers drift to material surfaces before decaying where quenching processes 
are induced, triggering photon emission already within tens of milliseconds. 
Both prompt and delayed UV photons are collected by the calorimetric sensors located within the liquid and in the vacuum region above the helium surface, enabling efficient light readout over the full target volume.

The quasiparticle channel is accessed indirectly via the process of quantum evaporation at the liquid–vacuum interface. 
A fraction of phonons and rotons reaching the surface can transfer sufficient energy to eject individual helium atoms into the vacuum, effectively converting collective excitations into detectable free atoms, with one evaporated atom per incident quasiparticle above threshold. 
These helium atoms subsequently impinge on the sensor wafers positioned above the liquid, where they are adsorbed. 
The adsorption process releases binding energy that depends on the absorber material which is typically higher than the binding energy of helium to helium.
This mechanism thus naturally provides a signal amplification of a factor of, for example, about 10 for silicon and about 20 for sapphire~\cite{Zhang_2016}.
To maintain this signal amplification, though, the absorbers must remain free of helium and the helium-flow, which is intrinsic to superfluid helium, must be stopped before reaching the absorbers.

By combining the prompt UV signal, the delayed triplet signal, and the quasiparticle-induced evaporation signal, these detectors can reconstruct the deposited energy and potentially extract spatial information. 
Moreover, the partitioning of energy among scintillation and quasiparticle channels differs for nuclear and electronic recoils, enabling discrimination between signal-like and background-like interactions \cite{DELight:2024bgv}.
Superfluid \(^4\)He also provides an intrinsically pure target, as impurities freeze out at cryogenic temperatures and helium has no long-lived radioactive isotopes. 
At sufficiently low temperatures, the liquid forms a highly coherent quantum state with well-defined elementary excitations, providing an exceptionally clean medium for rare event searches.

Currently, two experiments are realizing the calorimetric superfluid noble-gas concept: DELight~\cite{vonKrosigk:2022vnf} and HeRALD~\cite{Hertel:2018aal} which is part of TESSERACT~\cite{Billard:2024zvc}. The sensor technologies they employ are MMCs and TESs, respectively, yielding anticipated energy thresholds of $\mathcal{O}(10)$\,eV$_{\rm nr}$ for the entire detector. 
While these low thresholds combined with the efficient energy transfer from light dark matter particles to the light nuclei are a great advantage of these detectors, the calibration near threshold poses a challenge for the lack of a suitable neutron source at these energies, a challenge common to most low-threshold direct-detection experiments.

\vspace{4mm}
\twocoladvchallenge[Superfluid helium calorimeters]
{
\begin{itemize}
\item Small nuclear mass $\to$ $\mathcal{O}(10)$\,eV$_{\rm nr}$ energy threshold
\item Intrinsically pure target
\end{itemize}
}
{
\begin{itemize}
\item Calibration near threshold
\item Stopping of superfluid helium film flow
\end{itemize}
}
\vspace{4mm}

\noindent{}\textbf{The resonator approach} uses superfluid helium-3 as a bolometric target and mechanical resonators to measure the quasiparticles produced by an energy deposition.
At temperatures well below the superfluid transition temperature, fermionic $^3$He atoms form Cooper pairs. 
An energy deposition in the liquid breaks Cooper pairs and creates quasiparticle excitations, thereby increasing the quasiparticle density and the effective temperature of the superfluid. 
The very small heat capacity and exponentially suppressed thermal quasiparticle population at sub-millikelvin temperatures make the medium particularly sensitive to small energy depositions.

The quasiparticles can be measured using a mechanical resonator immersed in the superfluid. 
When driven at resonance, scattering of quasiparticles from the moving resonator produces a drag force that changes the width and other properties of the mechanical resonance. 
The resonator thus acts as a sensitive thermometer for the superfluid and allows the deposited energy to be reconstructed. 
At sufficiently low temperatures, quasiparticles propagate ballistically over macroscopic distances, while Andreev reflection in the superfluid modifies their interaction with the moving resonator. 
The response can be measured using low-noise electrical readout, with SQUID-based readout providing particularly high sensitivity. 
In addition to quasiparticle production, particle interactions can produce scintillation light through excitation and ionization of the helium atoms, offering the possibility of a complementary photon-detection channel and thus potential discrimination between electronic and nuclear recoils.

The use of superfluid $^3$He as a target for dark matter detection was first proposed in 1988 and was subsequently investigated by the MACHe3 and ULTIMA projects \cite{Pickett1988, Mayet:2000aa, Winkelmann:2006pw}.
This detection concept is currently pursued by the QUEST-DMC experiment~\cite{QUEST-DMC:2023nug}. 
QUEST-DMC is developing a superfluid $^3$He-B bolometer with a target volume of about $1\,\mathrm{cm^3}$, instrumented with sub-micron vibrating wire resonators and operated at temperatures around $100\,\upmu$K. 
For an optimized detector using SQUID-based resonator readout, a nuclear-recoil threshold of about $0.5$\,eV is projected, providing sensitivity to dark matter masses down to approximately $25\,\mathrm{MeV}$. 
The non-zero nuclear spin of $^3$He makes the target particularly attractive for searches for spin-dependent dark matter interactions. 
The small target volume currently limits the achievable exposure, while scaling the concept to arrays of bolometers represents a possible route towards larger target masses.

\subsection{Gaseous detectors}
\label{ssec:tech_gas}

Gas detectors constitute a mature and versatile technology for direct dark matter searches, particularly in the low-mass regime. 
Owing to the possibility of operating with light target gases and achieving sizable intrinsic amplification, these detectors can attain very low ionization thresholds while maintaining good background discrimination capabilities.
Within this class, two conceptually distinct approaches exist: spherical proportional counters~\cite{Nikolopoulos:2025kws} and gas TPCs. 
As in other light-target technologies (cf. Sec.~\ref{ssec:tech_superfluid}), the use of light nuclei such as hydrogen-, helium-, carbon-, neon-, or argon-based gases enhances sensitivity to sub-GeV dark matter by increasing the recoil energy for a given dark matter mass (cf. Eq.~\ref{eq:SI-recoil-energy}), while limiting competitiveness at higher masses where heavy targets benefit from the coherent ($A^2$) enhancement of the scattering cross section.\\

\begin{figure}[h]
\centering
\hspace{10mm}
\includegraphics[width=0.34\columnwidth]{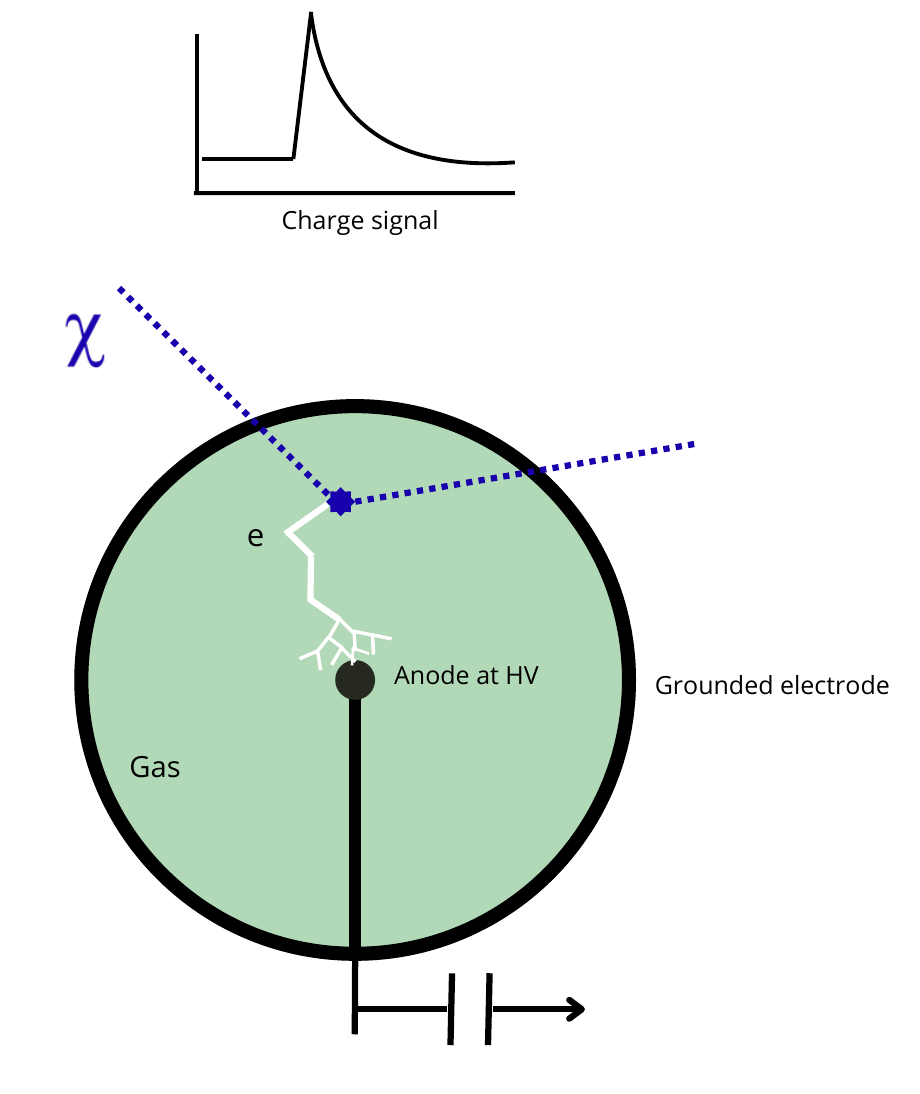}
\hspace{10mm}
\includegraphics[width=0.51\columnwidth]{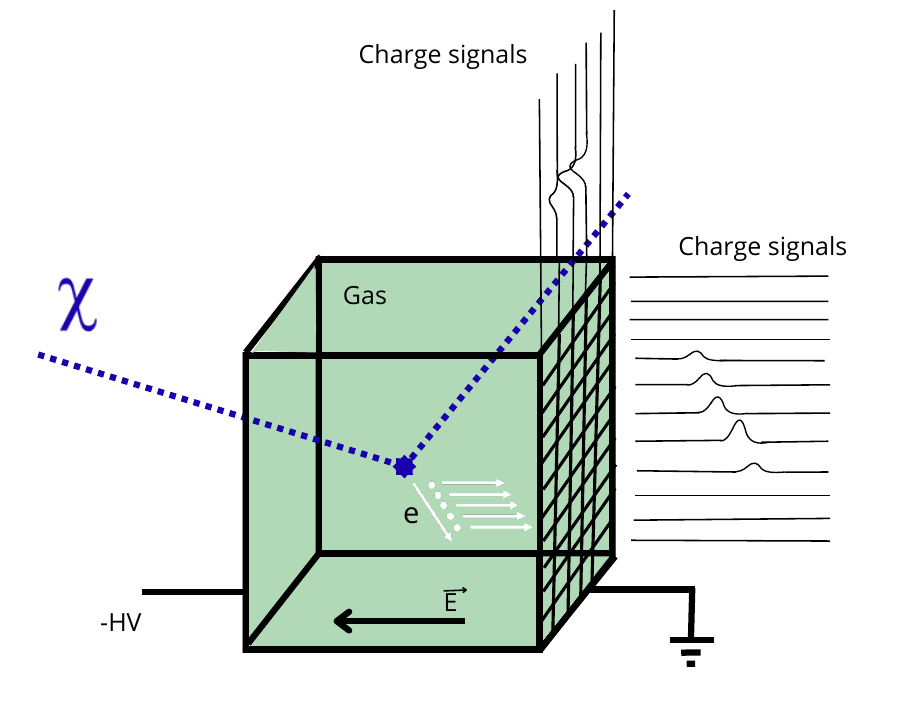}
\caption{Illustration of the spherical proportional counter and gas TPC concepts. Both use suitable gas mixtures (shown in green) as targets, operating in either low or high-pressure modes. Under an applied electric field, electrons (ions) produced after a particle interaction in the gas drift to the anode (cathode), generating a charge signal. In the spherical proportional counter (left), the field increases sharply near the anode, producing avalanche multiplication of the electron cloud, while in the gas TPC with a constant drift field, near the anode plane a higher electric field facilitates similar multiplication. The 2D readout in the latter provides, besides the time information, the ability to reconstruct the position of the interaction. The insets display typical signals registered by both detection concepts. }
\label{fig:gas_detector_concepts}
\end{figure}

\noindent{}\textbf{Spherical proportional counters} consist of a large metallic sphere filled with a suitable gas mixture and instrumented with a small central anode, as shown in Fig.~\ref{fig:gas_detector_concepts} (left). 
Ionization electrons produced in a particle interaction drift radially toward the center under an electric field that scales approximately as $1/r^2$.
Close to the anode, the electric field becomes sufficiently strong to initiate an avalanche multiplication process, providing substantial signal gain and enabling the detection of extremely small primary ionization charges, in principle down to single electrons.
At the same time, the intrinsically low capacitance of the small central sensor, typically below 1\,pF and largely independent of the detector size, results in very low electronic noise.
The combination of low capacitance and high gas amplification enables ionization thresholds of order $(10-100)$\,eV$_{\rm ee}$.

An important advantage of this technology is the flexibility in the choice of target gas. 
By exchanging or mixing gases such as helium-, neon-, or hydrogen-rich compounds, the effective target nucleus, and thus the interaction kinematics and coupling sensitivity, can be readily adapted without major modifications to the detector hardware. 
This versatility enables systematic studies of different interaction channels and facilitates optimization for specific dark matter mass ranges or coupling scenarios.
Although the readout is typically limited to a single or small number of channels, relevant spatial information can be inferred from the temporal development of the pulse, i.e. the pulse-shape, as the drift time depends on the radial position of the interaction. 
This allows for basic fiducialization and surface background rejection. 
The simplicity of the geometry, combined with the possibility of constructing large volumes with minimal instrumentation, offers advantages in terms of radiopurity and scalability, making spherical proportional counters a competitive approach for light dark matter searches.
At the same time, the intrinsically low density of gaseous targets limits the target mass that can be achieved within a given detector volume, making the accumulation of large exposures challenging even when scaling to larger detector dimensions.

Depending on the target gas, active masses currently range from about 0.1\,kg to several kilograms, while ionization thresholds at the level of approximately 100\,eV$_{\rm ee}$ have been achieved.
These characteristics provide sensitivity to dark matter masses well below 1\,GeV, with current results reaching masses below 200\,MeV for interactions with hydrogen.
Future developments target single-electron sensitivity, corresponding to ionization energies of only a few tens of eV, while larger detector volumes and increased operating pressures are being pursued to increase the achievable target mass and exposure.
The leading implementation of this approach is the NEWS-G experiment~\cite{NEWS-G:2022kon}, which has demonstrated the operation of spherical proportional counters with different light target gases and detector diameters up to about 1.4\,m.\\

\noindent{}\textbf{Gas TPCs} can be operated over a wide range of gas pressures, with higher-pressure operation primarily aiming at low energy thresholds and increased target mass, while lower pressures allow the reconstruction of nuclear recoil tracks and thus potentially provide directional information, as illustrated in Fig.~\ref{fig:gas_detector_concepts} (right). 
In both approaches, a particle interaction ionizes the target gas and the liberated electrons drift under a uniform electric field toward a segmented readout plane. 
There, amplification structures such as GEMs (Gas Electron Multipliers) or Micromegas (MICRO-MEsh GAseous Structures) provide gas gains typically of order $10^3$ or higher, enabling the detection of small primary ionization signals. 
The amplified charge can be recorded directly using highly segmented electrodes or, in optical TPCs, through the secondary light produced during the amplification process. 
The drift time provides information along the drift direction, while the segmented or optical readout reconstructs the transverse coordinates, enabling three-dimensional reconstruction of the interaction topology. 
This rich topological information provides an important handle for background rejection and fiducialization, although diffusion of the drifting charge increasingly degrades the spatial resolution for long drift distances. 
As for spherical proportional counters, an additional advantage is the flexibility in the choice of target gas, allowing the detector to be optimized for different dark matter masses and interaction types.

At higher pressures, typically up to about 10\,bar, the increased gas density provides larger target masses while retaining the low thresholds enabled by gas amplification. 
Light target gases such as Ne or Ar, often combined with small fractions of molecular gases to optimize charge transport and amplification, provide sensitivity to dark matter in the GeV and sub-GeV mass range. 
The short recoil tracks at these pressures generally preclude directional reconstruction at the energies relevant for dark matter searches, but the topology of events can still provide information for background rejection and fiducialization. 
Energy thresholds of a few hundred eV$_{\rm ee}$~\cite{Castel:2025yvh} or below are achievable, making high-pressure gas TPCs particularly suitable for searches for light dark matter. 
However, even at elevated pressures, the density of gaseous targets remains substantially lower than that of liquid or solid targets, limiting the target mass and achievable exposure for a given detector volume.

At lower pressures, the reduced gas density increases the range of recoiling nuclei. 
Nuclear recoils with energies of order tens of keV can produce tracks of millimeter to centimeter length, enabling reconstruction of their three-dimensional topology and, with sufficiently fine-grained readout, their orientation and potentially their sense, commonly referred to as head-tail discrimination. 
The key scientific motivation for such directional gas TPCs is the measurement of the recoil direction, which would provide a distinctive signature of the Galactic dark matter wind and thus a powerful handle against backgrounds. 
Directional sensitivity is not only effective in rejecting approximately isotropic backgrounds, but is particularly valuable in mitigating the background from coherent elastic neutrino--nucleus scattering (CEvNS), which otherwise constitutes an irreducible background for next-generation experiments. 
In non-directional detectors, separating a dark matter signal from the neutrino background requires a statistical analysis relying on differences in their energy and time distributions and on an accurate model of the neutrino fluxes. 
By contrast, the anisotropic recoil distribution correlated with the Earth's motion, as expected from Galactic dark matter, provides an additional observable that can statistically disentangle dark matter and neutrino-induced events. 
However, achieving sufficiently long tracks while preserving low energy thresholds requires relatively low gas densities, which strongly reduces the target mass per unit volume and poses a major challenge for scalability. 
Moreover, as the recoil energy decreases, the track length becomes shorter while diffusion and straggling increasingly obscure the initial recoil direction, making directional reconstruction progressively more difficult precisely in the low-energy regime relevant for light dark matter. 
Directional gas TPCs therefore trade target mass and, to some extent, energy threshold for the unique capability of measuring recoil directions, offering a complementary strategy within the broader landscape of direct detection technologies.

The higher-pressure approach is pursued by the TREX-DM experiment~\cite{Castel:2019ngt}, which operates a radiopure Micromegas-based TPC with light target gases at pressures of up to about 10\,bar. 
Directional gas TPCs are being developed by several experiments, including CYGNO~\cite{DAstolfo:2024cei}, NEWAGE~\cite{Ikeda:2021ckk}, and MIMAC~\cite{Beaufort:2023vlb}, which employ different gas mixtures and readout concepts to reconstruct nuclear recoil tracks. 
These efforts contribute to the wider CYGNUS program, which aims toward the development of larger-scale directional dark matter observatories~\cite{Vahsen:2020pzb}.

\vspace{4mm}
\twocoladvchallenge[Gaseous Detectors]
{
\begin{itemize}
\item Flexibility in the target mass
\item Flexibility in target gas
\end{itemize}
}
{
\begin{itemize}
\item Scaling up to large target masses
\item Understanding NR-Quenching Factors 
\end{itemize}
}
\vspace{4mm}

\subsection{Different approaches and emerging ideas}
\label{ssec:tech_others}

As far as no hint on the properties of the dark matter particle has been found, the experimental program must further diversify to explore as broadly as possible the available dark matter parameter space, encompassing different mass ranges and interaction channels with both electrons and nuclei (quarks). 
Although lowering the energy threshold has become the dominant strategy for extending the sensitivity to sub-GeV dark matter, complementary approaches based on novel target materials, new detection observables and technologies, and increased background discrimination power are opening discovery opportunities. 
Some of them are briefly described below. 

\begin{itemize}
    \item Extending the idea of cryogenic calorimeters to the detection of individual phonons, polar materials with several atoms per primitive cell, such as gallium arsenide or sapphire, have optical phonons that can be excited with an energy deposition of around 100\,meV~\cite{Knapen:2017ekk,Griffin:2018bjn,Coskuner:2021qxo,Baiocco:2025omh}.
    \item In order to push the semiconductor technology as far as possible, various works have explored materials with narrow band gap, such as Dirac materials~\cite{Hochberg:2017wce,Coskuner:2019odd,Geilhufe:2019ndy}, or 2-D materials~\cite{Hochberg:2016ntt}, such as graphene~\cite{Catena:2023qkj,Catena:2023awl} and carbon nanotubes~\cite{Cavoto:2019flp}.
    \item Even lower thresholds may be possible with superconductors, such as superconducting nanowires for single-photon detection~\cite{Hochberg:2019cyy,QROCODILE:2024nqm}, superconducting qubits~\cite{Hochberg:2026mdl}, transition edge sensors~\cite{Schwemmbauer:2025evp} or kinetic inductance detectors~\cite{Gao:2024irf}. 
    \item Future directions for scintillation crystals include low-threshold scintillators, such as organic crystals~\cite{Blanco:2019lrf, Blanco:2021hlm} or quantum dots~\cite{Blanco:2022cel}.
    \item An interesting new take on the idea of bubble chambers are single-molecule magnets, which are set up in a metastable magnetisation state that can be disturbed by dark matter \cite{Bunting:2017net,Alves:2026sgo}.
    \item A promising alternative strategy is to look for crystal defects~\cite{Rajendran:2017ynw} produced by the interactions of dark matter, such as color centers~\cite{Budnik:2017sbu,Araujo:2025rhr}. A similar strategy is adopted by so-called paleo-detectors~\cite{Baum:2018tfw,Baum:2023cct,Fung:2025cub,Theodosopoulos:2026ehn}, exploiting the tracks left by dark matter interactions in ancient minerals, which can be identified with powerful microscopy analysis. 
    \item Along similar lines, DNA-based detectors have been proposed to search for the defects (i.e.\ the breaking of DNA strands) caused by dark matter interactions~\cite{Drukier:2012hj,OHare:2021cgj}.
    \item Various groups have also performed a systematic exploration of a wide range of different materials using large databases and/or machine learning, to identify e.g.~molecular scintillators~\cite{Cook:2024cgm}, narrow-gap semiconductors and semimetals~\cite{Geilhufe:2018gry} or anisotropic materials~\cite{Griffin:2025wew} that are particularly well-suited for dark matter direct detection.
\end{itemize}

\section{Calibration and backgrounds} 
\label{sec:bck}

Direct detection experiments aim to observe the tiny energy deposits produced when a dark matter particle directly interacts with a nucleus or electron within a detector target. 
Depending on the interaction type, the signature is either a nuclear recoil or an electron recoil. 
For typical galactic dark matter velocities ($v \sim 10^{-3}c$) and WIMP masses in the range of $(10-1\,000)$\,GeV, the nuclear recoil energies fall in the keV-scale, a hallmark of direct detection experiments.
With the transition to masses well below the GeV scale, the observable recoil energies become notably lower which requires either low-threshold detectors or a lighter target material or, ideally, a combination of both.

Electron interactions generally probe lower dark matter masses or alternative coupling mechanisms. Also certain classes of dark matter candidates -- such as dark photons, axion-like particles, or models featuring inelastic scattering -- predict electron recoils as the dominant signature, demanding precise understanding of electronic response and background discrimination.
Given that the signature differs from that of a nuclear recoil, both in terms of the underlying interaction physics and the typical energy range, the corresponding calibration strategies also differ, often relying on low-energy electron sources or photoelectric calibration techniques.

\subsection{Detector calibration}

In direct dark matter detection experiments, calibration serves two primary purposes. First, the detector response must be calibrated to establish the energy scale, enabling the conversion of measured observables -- photons, phonons or charges -- into the true deposited recoil energy while accounting for spatial variations in detector response. Second, calibrations are required to characterize and optimize the discrimination between electronic recoils (ER) and nuclear recoils (NR), allowing to select the searched signature. Accurate calibration of both aspects is essential for reliable signal reconstruction and background rejection.

\vspace{4mm}
\twocolChallAch[Detector calibration]
{
\begin{itemize}
\item Calibrating lowest energies
\item Conversion from calibration data to the energy region of interest 
\end{itemize}
}
{
\begin{itemize}
\item Increasing knowledge on quenching factors 
\item Development of new sources
\end{itemize}
}
\vspace{4mm}

\subsubsection{Energy scale}
\label{subsubsection:E_scale}

Particle interactions inside the detectors are reconstructed through secondary observables such as scintillation photons, ionization electrons, or phonons. The relationship between these measured signals and the true deposited energy depends on detector-specific response functions, including quenching effects (non detectable signal losses), intrinsic charge and light yields, sensor efficiencies, and the transport of the signals -- for instance charge carriers or scintillation light -- through the detection system. Additional effects, such as charge collection efficiency, optical absorption and scattering, electric-field non-uniformities, and electronic response, can further modify the measured signal. Moreover, the detector response is generally not determined by deposited energy alone. It may depend on the position of the interaction within the active volume, the type of interacting particle, and non-linear or non-proportional effects in the production and collection of the measurable signals. As a result, identical energy deposits occurring at different locations or through different interaction channels may produce different detector responses that have to be properly accounted for in the reconstruction of the physical observables.

The partition of nuclear-recoil energy between electronic excitations and atomic motion in many common target materials is described using the Lindhard model, originally developed to describe the stopping of non-relativistic ions in matter at keV-scale energies, where nuclear and electronic stopping compete~\cite{Lindhard:1961, Lindhard:1963, Lindhard:1963b}.
The model predicts the fraction of the recoil energy transferred to the electronic system and is widely applied to nuclear recoils in solid-state targets, such as silicon and germanium, and liquid noble gases, such as argon and xenon.
At low recoil energies, particularly in the sub-keV regime, the applicability of the Lindhard description becomes increasingly uncertain and depends on the target material~\cite{Xu:2023rzm}.
Importantly, the Lindhard factor should not generally be identified with the experimentally observed quenching factor, which is the ratio of the detectable energy produced by a nuclear recoil to the energy that would be measured if the same energy were deposited by an electron recoil.

In scintillating compound crystals, such as NaI(Tl), CaWO$_4$, or Al$_2$O$_3$, the measured quenching factor may additionally depend on properties such as dopant concentration, impurities, crystal quality, and defect content~\cite{Angloher:2025fnl}.
Furthermore, the original Lindhard treatment is not directly applicable to compound targets in the same way as to elemental targets.
Experimental measurements are therefore required to establish the nuclear-recoil response of a particular detector material.
Measuring quenching factors is particularly relevant in the context of testing the observation by DAMA/LIBRA of a modulated signal (see Sec.~\ref{sec:experiments}).
Using NaI(Tl) detectors enables a direct comparison that is largely independent of assumptions about the underlying particle physics model and the dark matter halo, thereby removing significant sources of systematic uncertainty.
However, a dominant residual uncertainty remains from the limited knowledge of the quenching factor.
Precise measurements of this response are therefore essential for a robust interpretation of the signal and for establishing consistency or tension between different experimental results.  For liquid noble gases, these processes are commonly modeled using the Noble Element Simulation Technique (NEST)~\cite{Szydagis:2022ikv}, a simulation framework that describes scintillation and ionization yields as a function of deposited energy, particle type, and electric field.

Given all the reasons above, precise calibration is essential for all technologies to establish an accurate conversion between the observed detector response and the underlying deposited energy scale. This is typically achieved using radioactive calibration sources that provide reference signals at known energies and under controlled conditions.
Calibration of the electron recoil (ER) energy scale is well established across a wide range, from MeV down to sub-keV energies, using $\gamma$-ray, X-ray fluorescence, $\beta$-decaying sources and internal radioactive contaminants. Additional calibration techniques include single-photon calibration using laser-induced events and the identification of Compton steps~\cite{DAMIC-M:2022xtp,SuperCDMS:2025rgn} to characterize the detector response at low energies. Nevertheless, calibration at the lowest energies remains challenging.
Calibrating the nuclear recoil (NR) energy scale is even more challenging. Two approaches are used to measure the NR energy scale: the first involves performing neutron scattering experiments where the signal yield is related to a monoenergetic ER reference calibration. Neutrons  from a monoenergetic source scatter first in the medium of interest and then on a coincidence detector.  By selecting coincident neutron scatters at certain angle, the kinematics are fixed and the nuclear recoil energy is given. This is often done in separate laboratory setups~\cite{DarkSide-20k:2026xdv, Cintas:2024pdu, SuperCDMS:2023geu} but can also be done directly at the experiment~\cite{LUX:2016ezw}. Another option is to expose the actual detector to a neutron source, which would produce a continuous spectrum of nuclear recoils that can then be fitted to the corresponding Monte Carlo prediction. In this approach, however, the quenching factor needs to be incorporated directly in the Monte Carlo simulation, since it determines the mapping between the true nuclear recoil energy and the observed detector response. This becomes particularly challenging when the quenching factor is energy dependent, as its parameterization affects the simulated recoil spectrum.  Traditional radioactive sources such as americium–beryllium and californium are widely used because they provide reliable, continuous neutron emission. Alternative options include neutron generators, which produce neutrons on demand through fusion reactions and can be pulsed, as well as lower-energy yttrium–beryllium and antimony-beryllium photoneutron sources~\cite{XENON:2024kbh}, which generate neutrons with a softer energy spectrum that may be advantageous for calibration closer to the energy threshold. 

A limitation in calibrating large-volume detectors using external sources is that the resulting calibration primarily characterizes the outer layers of the detector, while providing limited sensitivity to the innermost regions.
In general, extending the calibration of the energy scale below 1\,keV remains a significant challenge across all detection channels, including both electron recoil (ER) and nuclear recoil (NR), as well as across different detector technologies. 
In this regime, the rapidly decreasing signal yield, combined with threshold effects and increased impact of noise and detector-specific systematics, makes precise measurements increasingly difficult. 
As a result, achieving a reliable and well-understood energy response at sub-keV energies is a common limitation that affects the sensitivity and interpretation of rare-event searches.
In cryogenic solid state calorimeters, one approach to low-energy NR calibration is provided by the CRAB (Calibration by Recoils for Accurate Bolometry) technique, which exploits radiative neutron capture inside the detector target~\cite{Wagner:2022ckd, Abele:2025vrj}. 
Following the capture of a thermal neutron, the compound nucleus de-excites through the emission of one or several $\gamma$ rays, causing the nucleus to recoil. 
For a single-$\gamma$ transition, the recoil energy is fixed by two-body kinematics, thereby providing a well-defined NR calibration line.
The method therefore provides a well-known nuclear-recoil reference directly in the sub-keV energy range relevant for low-mass dark matter and CE$\nu$NS searches, allowing the NR energy response and its relation to the electron-recoil calibration to be tested without relying on an extrapolation from higher recoil energies.
It was first demonstrated in CaWO$_4$, where neutron capture on $^{182}$W produces a prominent recoil line at about 112\,eV, and has since also been demonstrated with Al$_2$O$_3$ detectors~\cite{CRAB:2022rcm, CRESST:2023cxk, CRAB:2025ydy}. 
Suitable capture-induced recoil features are also predicted for other target materials, including Ge and Si~\cite{Abele:2025vrj}.

\subsubsection{Signal and background separation}

In detectors designed to search for WIMP-induced nuclear recoils, combining two independent signal channels provides a powerful strategy for distinguishing signal from background. By constructing ratios of these signals, experiments can take advantage of the fundamental difference in energy loss per unit length (d$E$/d$x$) between nuclear and electronic recoils. Nuclear recoils, produced for example by neutron or WIMP scattering, tend to generate denser ionization tracks and, depending on the detector material and operating principle, deposit a larger fraction of their energy into non-ionizing channels such as heat (phonons). In contrast, electronic recoils arising from gamma rays, beta decays, or other common background sources produce more diffuse tracks and typically yield higher ionization or scintillation signals for the same total deposited energy.

This difference in energy partitioning enables experiments to define discriminating observables based on measurable signal ratios. For instance, dual-phase noble liquid detectors use the ratio of ionization charge to prompt scintillation light, while cryogenic solid-state detectors exploit the ratio of ionization or scintillation to phonon energy. In single-phase scintillators, discrimination can be achieved through differences in the time structure, or pulse shape, of the scintillation signal. By using these approaches, experiments can effectively suppress dominant backgrounds while maintaining sensitivity to nuclear recoils.  However, the discrimination between nuclear and electronic recoils must be calibrated experimentally to establish the expected response of each recoil type and quantify the resulting separation.

\begin{figure}[h]
	\centering
	\includegraphics[width=0.35\columnwidth]{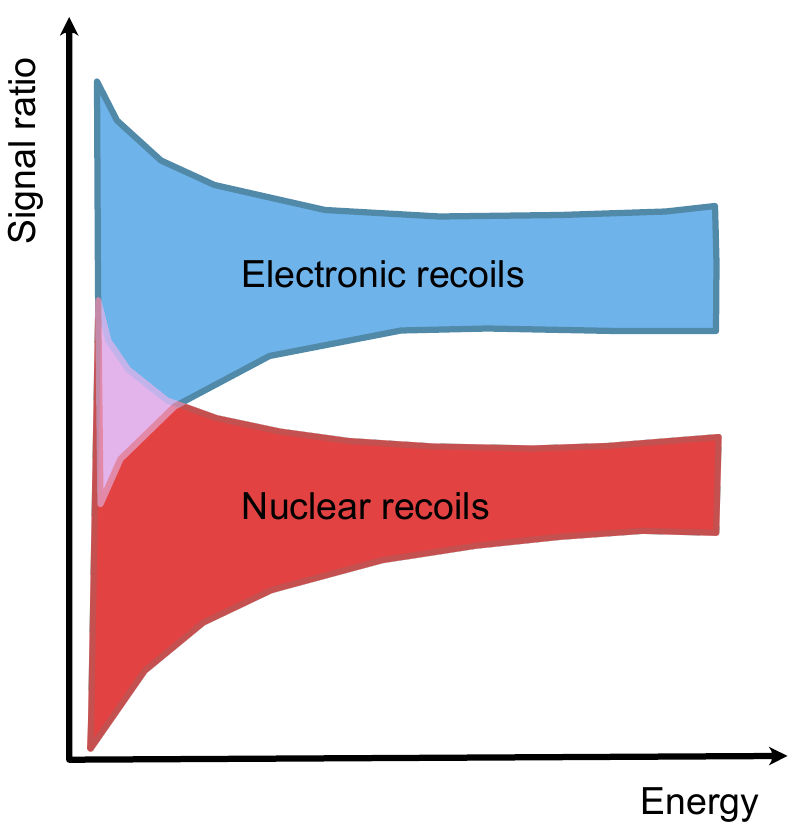}
	\caption{Schematic representation of electronic and nuclear recoil discrimination in a detector. In WIMP searches, nuclear recoils constitute the expected signal, while electronic recoils form a background component.}
	\label{fig:discrimination}
	\end{figure}

Radioactive neutron-emitting sources are commonly used to calibrate detector responses to nuclear recoils. Typically, these measurements result in a two-dimensional distribution of signal ratio versus deposited energy, allowing experiments to accurately define the region of parameter space where a potential dark matter signal is expected to appear and to quantify the efficiency and leakage of background events into this signal region. Figure~\ref{fig:discrimination} schematically shows the signal ratio for electronic and nuclear recoils.
Two distinct populations are visible, with ER events occupying a higher band and NR events forming a lower band due to their different relative signal yields. As the recoil energy decreases, the two bands broaden, leading to overlapping bands. This reduced separation is primarily caused by the lower number of detected signal quanta at low energies, which increases the impact of statistical fluctuations and detector noise. The resulting reduction in discrimination power at low energies may therefore affect the analysis threshold. The energy in the x-axis is reconstructed from one of the detectors signal, as explained in section~\ref{subsubsection:E_scale}.

\subsection{Experiment's backgrounds}

If interactions between dark matter and target nuclei were to be measured at all, they would be extremely rare. For this reason, great efforts are dedicated to minimizing and understanding background sources. This section summarizes the most important backgrounds in direct dark matter experiments and the most common strategies for suppressing them. Furthermore, the statistical treatment of signals and backgrounds is discussed.

\begin{figure}[h]
	\centering
	\includegraphics[width=0.5\columnwidth]{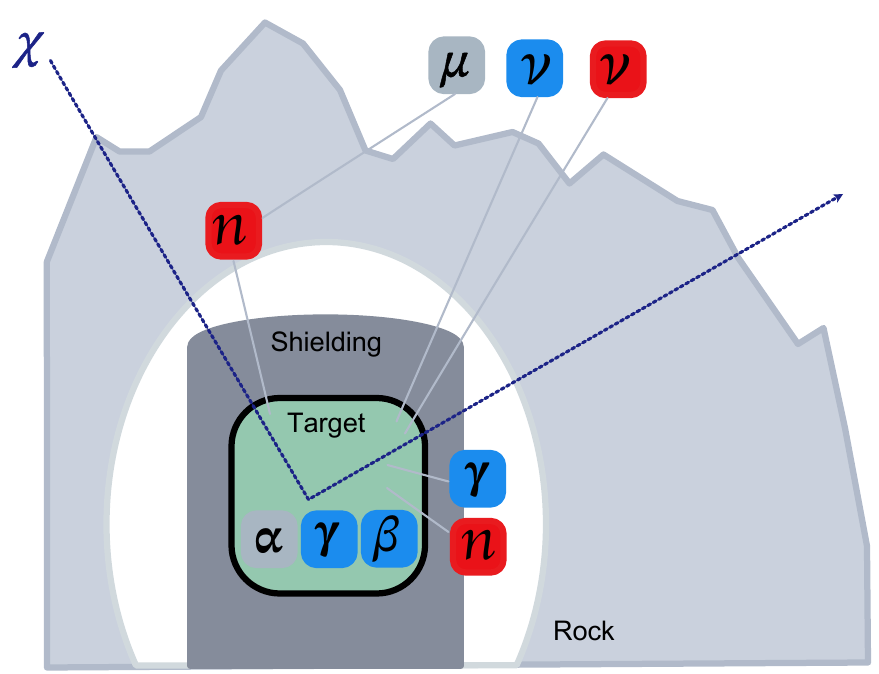}
	\caption{Schematic representation of background sources in a direct detection experiment.
    }
	\label{fig:bg}
	\end{figure}

Backgrounds are classified in this section as cosmogenic, external natural radioactivity (radiogenic), surface- and internal- backgrounds. Figure~\ref{fig:bg} shows these different sources of background schematically.
Backgrounds producing nuclear recoils — the signature of WIMP particles — are marked in red. Backgrounds producing electronic recoils are shown in blue. Muons and alpha particles are shown in gray.

\subsubsection{Cosmogenic backgrounds}
\label{sssec:cosmogenics}

Cosmogenic backgrounds arise from cosmic-ray interactions in the atmosphere and in the materials surrounding the detector~\cite{Formaggio:2004}. For direct dark matter detection experiments, muon-induced neutrons are a particularly important source of background, as they can produce nuclear recoils that mimic the expected signals from dark matter interactions. Cosmic rays, primarily protons, initiate particle showers in the upper atmosphere, producing secondary particles, including muons. These muons can penetrate deep underground, where they may generate neutrons through interactions with the surrounding rock, detector shielding, and detector materials. To mitigate this background, direct dark matter detectors are typically installed deep underground, for example, in mines or tunnels, where the overburden substantially reduces the cosmic-ray muon flux and, consequently, the production of muon-induced neutrons. 

Table~\ref{tab:UndergroundLabs} summarizes all underground laboratories hosting dark matter experiments along with some of their main characteristics.
\renewcommand{\arraystretch}{1.2}
\begin{table}[h]
    \centering
    \footnotesize
    \begin{tabular}{|c|c|c|c|c|c|}\hline
       {\bf Name} & {\bf Country} & {\bf Type} & {\bf Depth} & {\bf Muon flux} & {\bf Reference}\\
       & & & [km w.\ e.] & [$\times 10^{-7}$\,cm$^{-2}$\,s$^{-1}$]& \\ \hline
         CJPL  & Jinping, China & Mountain & 6.72 & $(2.0 \pm 0.4)\times 10^{-3}$ &   \cite{Wu:2013cno} \\
        SNOLAB & Sudbury, Canada & Mine & 6.01 & $(3.77 \pm 0.41)\times 10^{-3}$ &   \cite{Mei:2005gm} \\ 
        LSM & Modane, France & Mountain & 4.2 & $(5.5 \pm 0.1)\times 10^{-2}$ &   \cite{Mei:2005gm} \\
        SURF & Sanford, USA & Mine & 4.08 & $(5.09 \pm0.08) \times 10^{-2}$ &   \cite{LZ:2026kit} \\
        LNGS & Gran Sasso, Italy & Mountain & 3.1 & $(2.6 \pm 0.3)\times 10^{-1}$ &   \cite{Mei:2005gm} \\
        BUL & Boulby, UK & Mine & 2.81 & $(4.09 \pm 0.15)\times 10^{-1}$  & \cite{Mei:2005gm} \\
        SUPL & Stawell, Australia & Mine & $\sim 2.9^*$  & $(6.33 \pm 0.04)\times 10^{-1}$  & \cite{Fu:2026ubt}\\
        LSC & Canfranc, Spain & Mountain & $\sim 2.2^*$& $(5.26 \pm 0.21)$  & \cite{Trzaska:2019kuk} \\
        Kamioka & Hida, Japan & Mountain & 2.05 & $(1.58 \pm 0.21)$  & \cite{Mei:2005gm} \\
        Y2L & Yangyang, Corea & Mountain & $\sim 1.9^*$  & $(3.8\pm 0.12)$  & \cite{Adhikari:2017esn} \\
        Soudan & Soudan, USA & Mine & 1.95 & $(2.0 \pm 0.2)$ &  \cite{Mei:2005gm} \\
        JUSL & Jaduguda, India & Mine & 1.55 & $(2.79 \pm 0.13)$  & \cite{Sharan:2021tnj} \\
        NEXUS & Fermilab, USA & Flat &  0.2  & $(8.0\pm0.4)\times 10^{2}$ & \cite{Garrison:2014nla}
        \\
        \hline
    \end{tabular}
        \caption{Summary of underground laboratories, location and main characteristics. Values marked as $^*$ are calculated using the shielding density and the maximum height.}
    \label{tab:UndergroundLabs}
\end{table}

The overburden completely suppresses the hadronic component and reduces the muon flux by many orders of magnitude. It is worth noting that, for a given vertical rock overburden, laboratories located beneath mountains generally have less uniform shielding and consequently higher cosmic-ray muon fluxes than laboratories located below mine shafts, where the overburden is more uniform~\cite{JNE:2020bwn}. 
Furthermore, experiments often use active veto systems to record the few muons that reach the detector site. These can be plastic scintillators or large water tanks that use the Cherenkov effect. 

Since the underground location suppresses the muon rate, and since veto systems allow the remaining ones to be measured, this component constitutes a negligible background in most experiments. Muon-induced backgrounds are significantly more relevant. High-energy muons traveling through the cavity, shielding or support structures of the experiment can induce spallation reactions, resulting in the emission of multiple neutrons. These neutrons can have energies in the range of GeV, down to MeV and, in turn, travel through the experiment, reaching the innermost part of the target. The interaction of these neutrons with nuclei results in nuclear recoils, which is the same signature as WIMPs would leave in a detector.

It should be noted that both the target and detector materials can become cosmogenically activated while exposed at the surface, with the resulting radioactive isotopes potentially contributing to the background after the detector has been deployed and is in operation. In this review, such backgrounds are classified as either internal or external, depending on the location of the activated material within the detector.

\begin{figure}[h]
	\centering
	\includegraphics[width=0.39\columnwidth]{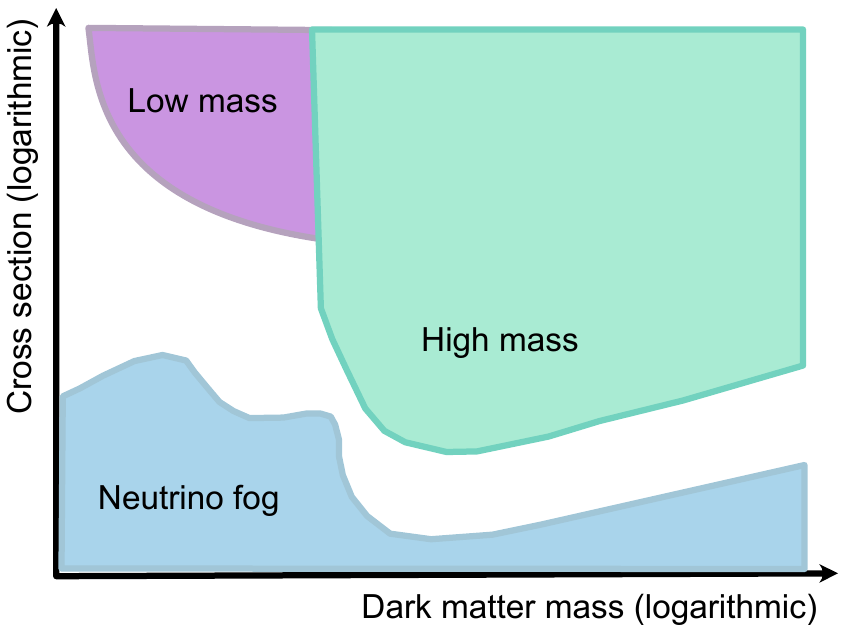}
	\caption{Scheme showing the typical shape of low mass and high mass dark matter results including the neutrino fog.}
	\label{fig:bg_fog}
	\end{figure}

Neutrinos also form part of the cosmogenic background and have to be taken into account in the experiment's background model. These include solar neutrinos, which are created in the interior of the Sun; atmospheric neutrinos, which are produced in particle showers; and the relic background of supernova neutrinos. Since neutrinos are weakly interacting particles, it is not possible to shield them. At the same time, detection of  neutrinos is interesting by itself offering opportunities to learn about neutrino properties or about astrophysical objects. In fact, two channels are available: neutrino-electron scattering, which produces an electron recoil, and the coherent scattering of neutrinos on nuclei (CE$\nu$Ns), which produces a nuclear recoil.  CE$\nu$Ns interactions from high energy solar $^8$B neutrinos have been measured in the last years~\cite{LZ:2025igz, XENON:2026smt, PandaX:2024muv}. 
In this channel, however, only the highest-energy component of the solar neutrino spectrum, namely the $^8$B neutrinos, is currently accessible. By contrast, the interaction rates of atmospheric and diffuse supernova neutrinos remain below the sensitivity of current-generation dark matter experiments. Figure~\ref{fig:bg_fog} shows a schematic representation of typical direct detection results for high masses (green) and low masses (purple) together with the so-called \emph{neutrino fog} (light blue), which represents the cross sections at which CE$\nu$Ns constitutes a relevant background. 
The neutrino fog is not a sharp boundary but a gradual, target-dependent region in which neutrino-induced signals become comparable to those expected from dark matter. Its characteristic shape arises because different neutrino sources dominate at different recoil energies and detector thresholds~\cite{OHare:2021utq}. At low dark matter masses, solar neutrinos produce nuclear recoils that closely resemble those from light dark matter, creating the bumps visible in Fig.~\ref{fig:bg_fog}. At higher masses, atmospheric and diffuse supernova neutrinos become the main background, leading to a flatter component above the bumps.

Recently first measurements of low energy $pp$-neutrinos via neutrino-electron scattering in liquid xenon has been shown~\cite{PandaX:2026qdp,XENON:2026rbz}, enabling the determination of the flux of this dominant component of the solar neutrino spectrum.

\subsubsection{External radiogenic backgrounds}

Natural radioactivity is also an important source of background radiation, including gamma rays and neutrons emitted from fission or $(\alpha, n)$ reactions. To shield the detector from external radiation originating from the surrounding experimental cavity and equipment, an appropriate shielding concept is required. Smaller experiments often use a combination of lead (with a high atomic number to stop gamma radiation), polyethylene (to stop neutrons) and copper (due to its high radio-purity) in the innermost layers~\cite{SuperCDMS:2024yiv,COSINUS:2021bdj}. Larger detectors often employ a water tank which fulfills all these requirements while tagging also remaining cosmic muons~\cite{Angloher:2024pbw,XENON1T:2014eqx}. 

Materials used in detector construction, including chambers, cryostats, holding structures, sensors, cables, and electronics, can contain trace amounts of radioactive isotopes. These materials are typically screened using high-purity germanium spectrometers, ICP-MS (Inductively Coupled Plasma Mass Spectrometry), NAA (neutron activation analysis), and alpha counters prior to the final detector's assembly to minimize activity levels to reach sub-ppt levels~\cite{Heusser:1995wd,Laubenstein:2020rbe,XENON:2021mrg}. The gamma emission from $^{40}$K, $^{60}$Co, $^{137}$Cs, uranium and thorium chains, among others, is quantified through screening measurements, which serve as inputs to Geant4 simulations used to optimize the detector and shielding designs. Often iterations with material or equipment providers take place to find the most radio-clean option. This material and equipment selection process often takes a few years before the experiment starts data taking. The public results of these material assays are compiled in the public \textit{Radiopurity.org}
database~\cite{radiopurity}, which serves as a valuable resource for the selection of ultra-low-background materials in rare-event search experiments.
Moreover, selected materials are ideally quickly transported underground to avoid cosmogenic activation of their isotopes~\cite{Amare:2017roa,DarkSide-20k:2023fbh}. This is of course most important for the dark matter target materials but can happen similarly in detector materials, for instance $^{60}$Co can be produced in copper if it is not brought underground sufficiently fast.

Regarding neutron emission, the trace concentrations of uranium and thorium in the materials are most important. Since several alpha decays occur in these chains, neutrons are produced as result of the $(\alpha, n)$ process in each material~\cite{Westerdale:2017kml,Cano-Ott:2024elo}. These neutrons have typically energies of a few MeV and if produced close to the target are one of the most dangerous sources of background for dark matter searches since they are also nuclear recoils. Some experiments have developed dedicated neutron detectors surrounding the target, with the aim of further suppressing this background.

\subsubsection{Surface and internal backgrounds}

Surface backgrounds constitute an additional background component to be considered. They can originate from radioactive contamination on detector or surrounding material surfaces, including the deposition of radon daughters. The corresponding signal might be degraded due to partial energy deposition in holder structures or dead detector layers, appearing in the region relevant for dark matter searches. Surface cleaning, pulse shape discrimination or removal of background events via analysis techniques are some of the methods used to mitigate these effects. For some detector technologies, selecting a fiducial volume is an effective way of strongly suppressing surface backgrounds.

Intrinsic backgrounds arise either from cosmogenic activation of the target materials or from radioactive impurities or noble gases present within the target.
$^{32}$Si is a cosmogenic isotope produced in the atmosphere by cosmic-ray spallation on argon. It enters the environment through precipitation and becomes incorporated into silica, quartz, and ultimately the silicon used for detector fabrication. For this reason, $^{32}$Si is distributed uniformly through the crystal and cannot be removed with fiducialization. Reducing this background for silicon calorimeters and  CCD detectors therefore requires careful selection of silicon with a low 
$^{32}$Si content. Its beta decay, followed by the decay of its daughter, produces low-energy electron recoils that can mimic the signals expected from dark matter interactions.

Tritium ($^{3}$H) is an important intrinsic background in direct dark matter searches, particularly for Ge detectors, but it can also contribute to the background in various detector targets~\cite{Amare:2017roa}. It undergoes $\beta$ decay with a low endpoint energy of 18.6\,keV and a half-life of 12.3 years. Tritium can be produced cosmogenically through interactions of cosmic-ray particles with detector materials during their production, transportation, or storage above ground. Consequently, minimizing cosmogenic activation through careful material handling and underground storage, together with accurate estimates of the residual tritium activity, is essential for achieving the low-background conditions. 

Potassium contamination is an important intrinsic background in NaI detectors, primarily due to the presence of radioactive $^{40}$K. Its removal during crystal growth and purification is particularly challenging because potassium and sodium have similar chemical properties, making their chemical separation difficult. Consequently, controlling potassium contamination requires careful selection and purification of the starting materials and optimized crystal-growth procedures to minimize the resulting radioactive background.

For liquid and gaseous detectors, radioactive trace contamination is often more relevant than surface background radiation, with radon being a relevant example. Radon is created in the primordial decay chains of uranium and thorium, which are present in trace amounts in all detector materials. Once emanated to the liquid/gas, $^{222}$Rn with a half live of 3.8\,days, can move around the detection medium where it decays. Specifically, the beta decays of the daughter nuclei constitute a dominant source of background radiation in many experiments, so great efforts are made to minimize it.  This is particularly important for underground laboratories, where the ambient $^{222}$Rn concentration is typically elevated, and common mitigation strategies include the use of de-radonized air or nitrogen flushing to suppress radon~\cite{Veselska_2024,Pelczar:2021}.

For liquid argon experiments, the cosmogenic activation of argon isotopes is an important source of background. In particular, neutron-induced reactions produce trace amounts of $^{39}$Ar, a radioactive isotope with a half-life of approximately 269 years. As a result, atmospheric argon contains $^{39}$Ar at an activity of about 1\,Bq/kg. To suppress the resulting background from its continuous $\beta$ decay, current experiments use argon extracted from underground sources, where the $^{39}$Ar concentration is strongly depleted~\cite{Acosta-Kane:2007you}.

\subsubsection{Time dependent backgrounds}
\label{ssec:sig_Mod}

Time-dependent backgrounds are particularly relevant in the case of searching for time-dependent specific signatures of the dark matter signal. These signatures are powerful tools for successfully identifying a dark matter signal among most of the background contributions. 
However, such time-dependent backgrounds, if not correctly accounted for, could result in fake dark matter detection and/or a reduced sensitivity of the experiments to a dark matter signal. 
Among the category of time-dependent specific signatures of the dark matter signal, we can consider both searches for the annual modulation of the DM interaction rate (peaking in early June) and searches for directional (forward-backward) asymmetries that follow a more complex time-dependent pattern (see Sec.~\ref{ssec:modulation}).
In addition, Earth-shielding effects would translate into specific daily modulations patterns in the detection rate, for dark matter particles interacting with relatively large cross-sections, either with nuclei or electrons, dependent on the laboratory position (longitude and latitude) and rock overburden, as well as on the dark matter halo model (see Sec.~\ref{ssec:modulation}). 
The two latter effects are much more difficult to mimic by any known background, while some background components and systematic effects are expected to show an annual modulation. 

The residual muon flux in underground locations (see cosmogenic background in Sec.~\ref{sssec:cosmogenics}) is known to modulate with 1-year period because of the seasonal atmospheric temperature changes, peaking for instance on early July at LNGS~\cite{LVD:2019zlh}. This means that muon-related events are a relevant source of annually modulated background contributions that should be different in northern and southern hemisphere locations.
Solar neutrinos, but also any other Sun-produced particles, are also expected to show an annual modulation in their interaction rate because of the change in the Earth-Sun distance, peaking, however, near the winter solstice, far from the expected phase of dark matter induced modulation.  

The most exhaustive analysis of possible backgrounds able to mimic a dark matter signal have been carried out within the context of the puzzling DAMA/NaI annual modulation observation for more than thirty years~\cite{Bernabei:2020mon} (see Sec.~\ref{ssec:exp_running}). The DAMA/LIBRA collaboration examined the effects from radon activity, temperature fluctuations, noise or electronic instabilities, calibration and efficiency drifts, and background variations due to neutrons or muons in the LNGS. The estimated impact on the count rate from all these sources was found to be at least two orders of magnitude smaller than the modulation amplitude measured by DAMA/LIBRA~\cite{Bernabei:2000ew}. 
However, beyond backgrounds able to mimic a dark matter signal, any slow-varying background could bias the result of a modulation analysis if not properly modeled. 
Furthermore, some procedures to extract a modulated component are susceptible to systematics in the presence of time-dependent backgrounds, particularly when yearly averaged backgrounds are subtracted to build residuals~\cite{Messina:2020pnt,Buttazzo:2020bto}.

\subsubsection{Instrumental backgrounds}
\label{sssec:instrumental_bck}

Instrumental backgrounds constitute a distinct class of backgrounds arising not from radioactive interactions but for example from detector effects, and electronics. Their characteristics are often experiment-specific and can limit the sensitivity to low-energy signals if not properly understood. 

As detector thresholds have reached the eV scale, several solid-state experiments have observed a steeply rising population of events toward low energies, commonly referred to as the \textit{low-energy excess} (LEE)~\cite{Fuss:2022fxe, Baxter:2025odk}. 
The LEE is not understood as a single universal background, but rather describes excess populations whose origin may differ between detector technologies and experiments. 
Proposed explanations include thermal radiation and photon-mediated processes such as Cherenkov radiation and luminescence, charge leakage and trapping effects, as well as the release of energy stored in mechanically or thermally induced stress. 
In cryogenic phonon detectors, the latter may arise from stress relaxation in the detector crystal, its mechanical supports, or deposited metal films and can produce athermal phonon bursts that closely resemble particle interactions. 
Several observations, including a decreasing event rate with time after cooldown and a dependence on detector mounting and mechanical stress, support stress-related explanations for at least part of the observed excesses. However, no common explanation has yet been established, and the LEE remains an important background for experiments approaching eV-scale thresholds.

In liquid noble gas detectors, an important example is accidental coincidences between uncorrelated primary scintillation (S1) and delayed electroluminescence (S2) signals, which can mimic the signature of a true low-energy interaction~\cite{DarkSide-50:2025umf,XENON:2021qze,LUX-ZEPLIN:2025uaf}. The isolated S1-like signals can originate from scintillation pulses produced in regions of the detector where the S2 signal is not efficiently collected or reconstructed, as well as from other sources of lone photons that can produce signals resembling a small S1. Conversely, S2-like signals may arise when the corresponding S1 signal is shadowed or otherwise not detected, falls below the analysis threshold, or when electrons are emitted with a delay following an earlier interaction. The combination of uncorrelated S1 and S2-like signals can therefore produce event topologies that resemble genuine low-energy interactions despite having no common physical origin. Such backgrounds are becoming increasingly relevant as detector sizes increase and are typically mitigated through analysis of the signal timing, position reconstruction, pulse characteristics, and data-quality selections.

\section{Experiments and results}
\label{sec:experiments}

All results shown in the figures of this section (Fig.~\ref{fig:exp_nearfuture_SI}-\ref{fig:exp_nearfuture_ER}) assume a local dark matter mass density of 0.3\,GeV/cm$^3$. 
Results that were reported for a density of 0.4\,GeV/cm$^3$ were rescaled accordingly. 
Other astrophysical input parameters cannot be unified by simple scaling and the reader is advised to keep potential differences in assumptions in mind when comparing the displayed results.

\subsection{Running or recently completed experiments with published results}
\label{ssec:exp_running}

This section reviews results from running or recently completed direct dark matter experiments published since 2020. 
Older results are only included if they are still world-leading at any part of the probed parameter space.
The results are organized according to their experimental signature into searches for nuclear recoils, electron recoils, modulation signals, and directional signatures. 
This classification is based on the observed signature rather than the underlying dark matter interaction or parameter space probed.
We restrict our discussion to searches based on well-established theoretical frameworks for which the relevant model assumptions are broadly accepted by the community. 
These include spin-independent and spin-dependent dark matter-nucleus scattering (see Sec.~\ref{subsubsec_Nucl_Scat}), inelastic processes involving bremsstrahlung or the Migdal effect (see Sec.~\ref{subsubsec_Nucl_Scat}), and dark matter-electron scattering (see Sec.~\ref{sssec:theory_ER_scattering}). 
Emerging scenarios, such as cosmic or solar boosted dark matter (see Sec.~\ref{ssec:astrophysical}), are beyond the scope of this review.
The figures do not extend to larger cross sections, as this region of parameter space is already strongly constrained by cosmological observations and affected by additional processes, such as the scattering of dark matter particles with cosmic rays.

\subsubsection{Nuclear recoil}
\label{sssec:exp_running_NR}

The following section focuses on dark matter searches in which the experimental signature is a nuclear recoil (NR).
Searches for dark matter-nucleus scattering through the Migdal effect or bremsstrahlung produce an electronic signature and are 
consequently discussed with the electronic recoil searches in Sec.~\ref{sssec:exp_running_ER}.
Table~\ref{tab:Running_NR} and the respective figures~\ref{fig:exp_running_NR_SI} to \ref{fig:exp_running_NR_SD} 
include only results using a given target material that have not been subsequently superseded by other results from the same experiment. The values reported in the tables are rounded for clarity. 
Where needed for context, further results 
are included in the accompanying text.
Furthermore, each figure includes only limits that probe the same interaction
scenario and can therefore be directly compared, while results obtained under different assumptions, such as light-mediator interactions, are not shown.\\

\renewcommand{\arraystretch}{1.2}
\begin{table}[H]
    \centering
    \caption{Summary of direct dark matter searches with nuclear-recoil signatures published since 2020 (or before if still world-leading), organized by detector technology. Provided are the leading results per experiment.
    Masses refer to the target or fiducial volume used for the respective analysis, where available. Exposures are the net exposures, if provided. Else the gross exposure is cited. If neither is provided, it is approximated (indicated with $\sim$ or $\lesssim$) based on the given mass and live time. Thresholds are quoted in the energy scale or experimental observable used by the collaboration, with approximate conversions to comparable energy scales where appropriate. Photoelectrons are abbreviated with p.e., extracted electrons with e$^-$, and detected photons with $\gamma$. Searches in which dark matter-nucleus scattering produces an electronic signature, in particular through the Migdal effect or bremsstrahlung, are instead summarized in Tab.~\ref{tab:Running_ER}.}
    \scriptsize
    \begin{tabular}{|c|c|c|c|c|c|c|}\hline
        \makecell{\bf Experiment\\\bf (Interaction)} & \makecell{\bf Target \\\bf(Mass)} & {\bf Location} &
        \makecell{\bf Data\\\bf Taking} &
        {\bf Exposure} & {\bf Threshold} & {\bf Reference(s)}\\\hline
        \hline
        \multicolumn{7}{|c|}{ Cryogenic crystal calorimeters } \\
        \hline
        COSINUS (SI) & NaI (3.67\,g) & LNGS & 2022 & 12\,g\,$\cdot$\,d & 4.0\,keV$_\mathrm{nr}$ &\cite{COSINUS:2023kqd}\\
        CRESST (SI) & CaWO$_4$ (23.6\,g) & LNGS & 2016-18 & 3\,600\,g\,$\cdot$\,d & 30\,eV$_{\rm nr}$ & \cite{CRESST:2019jnq} \\
        CRESST (SD) & LiAlO$_2$ (2.8\,g) & surface & 2019 & 2\,g\,$\cdot$\,d & 210\,eV$_{\rm nr}$ & \cite{CRESST:2020tlq,CRESST:2022jig} \\
        CRESST (SD) & LiAlO$_2$ (10.5\,g) & LNGS & 2021 & 1\,200\,g\,$\cdot$\,d & 84\,eV$_{\rm nr}$ & \cite{CRESST:2022dtl} \\
        CRESST (SI) & Si (0.35\,g) & LNGS & 2020-21 & 55\,g\,$\cdot$\,d & 10\,eV$_\mathrm{nr}$ & \cite{CRESST:2022lqw}\\
        CRESST (SI\,+\,SD) & Al$_2$O$_3$ (0.6\,g) & LNGS & 2020-21 & 140\,g\,$\cdot$\,d & 6.7\,eV$_{\rm nr}$ & \cite{CRESST:2024cpr}\\
        RES-NOVA (SI\,+\,SD) & PbWO$_4$ (13\,g) & LNGS & n.a. & 32\,g\,$\cdot$\,d & 2.5\,keV$_{\rm nr}$ & \cite{Alloni:2026vak}\\
        SuperCDMS (SI\,+\,SD) & Ge (600\,g) & Soudan & 2012-14 & 70\,000\,g\,$\cdot$\,d & 56\,eV$_\mathrm{ee}$ & \cite{SuperCDMS:2017nns}\\
        SuperCDMS-CPD (SI) & Si (10.6\,g) & surface & 2018 & 9.9\,g\,$\cdot$\,d & 16\,eV$_{\rm nr}$ & \cite{SuperCDMS:2020aus}\\
        SuperCDMS-HVeV (SI) & Si (0.93\,g) & surface & 2019 & 0.19\,g\,$\cdot$\,d & 9.2\,eV$_{\rm nr}$ & \cite{SuperCDMS:2022zmd}\\
        TESSERACT (SI) & Si (0.23\,g) & surface & n.a. & 0.12\,g\,$\cdot$\,d & 1.5\,eV$_{\rm nr}$ & \cite{TESSERACT:2025tfw}\\
        \hline
        \multicolumn{7}{|c|}{ Semiconductors } \\
        \hline
        CDEX-1B/-10 (SI\,+\,SD)& Ge (939\,g) & CJPL & 2014-17 & 740\,kg\,$\cdot$\,d/100\,kg\,$\cdot$\,d & 160\,eV$_{\rm ee}$ & \cite{CDEX:2021cll}\\
        CDEX-10 (SI) & Ge (939\,g) & CJPL & 2017-18 & 210\,kg\,$\cdot$\,d & 160\,eV$_\mathrm{ee}$ & \cite{CDEX:2024rfi}\\
        DAMIC (SI) & Si (42\,g) & SNOLAB & 2017-18 & 11\,kg\,$\cdot$\,d & 50\,eV$_\mathrm{ee}$ & \cite{DAMIC:2020cut}\\
        \hline
        \multicolumn{7}{|c|}{ Scintillation crystals } \\
        \hline
        COSINE-100 & NaI(Tl) (61.3\,kg) & Y2L & 2016-19 & $\lesssim 180$\,kg\,$\cdot$\,y & 0.7\,keV$_{\rm ee}$ & \cite{COSINE-100:2025kbw}\\
        \hline
        \multicolumn{7}{|c|}{ Bubble chambers } \\
        \hline
        InDEx (SI\,+\,SD) & C$_2$H$_2$F$_4$ (70.4\,g) & JUSL & n.a. & 7.2\,kg\,$\cdot$\,d & 2.0\,keV$_{\rm nr}$ &  \cite{Das:2025coa} \\
         PICO-60 (SI\,+\,SD) & C$_3$F$_8$ (52\,kg) & SNOLAB & 2016-17 & 1\,200\,kg\,$\cdot$\,d/1\,400\,kg\,$\cdot$\,d & 3.3\,keV$_{\rm nr}$/2.5\,keV$_{\rm nr}$ & \cite{PICO:2019vsc} \\
        \hline
        \multicolumn{7}{|c|}{ Liquid noble-gas scintillators } \\
        \hline
         DarkSide-50 (SI) & Ar (19.4\,kg) & LNGS & 2015-18 & 0.03\,t\,$\cdot$\,y & 4\,e$^-$ ($\sim 0.6$\,keV$_{\rm nr}$) & \cite{DarkSide-50:2022qzh, DarkSide-50:2025lns}\\
         DEAP-3600 (SI) & Ar (824\,kg) & SNOLAB & 2016-17 & 2.1\,t\,$\cdot$\,y & 95\,p.e. ($\sim 50$\,keV$_{\rm nr}$) & \cite{DEAP:2019yzn}\\
         LZ (SI\,+\,SD) & Xe (5.5\,t) & SURF & 2021-24 & 4.2\,t\,$\cdot$\,y & 3\,$\gamma$ \& 14.5\,e$^-$ (5.4\,keV$_{\rm nr}$)  & \cite{LZ:2024zvo} \\
         LZ (SI\,+\,SD) & Xe (5.1\,t) & SURF & 2023-25 & 5.7\,t\,$\cdot$\,y & 3.5\,e$^-$ (1\,keV$_{\rm nr}$)  & \cite{LZ:2025igz} \\
         PandaX-4T (SI) & Xe (2.4\,t/2.5\,t) & CJPL & 2020-22 & 1.5\,t\,$\cdot$\,y & 2\,p.e. \& 120\,p.e. ($\sim 3$\,keV$_{\rm nr}$)  & \cite{PandaX:2024qfu} \\
         PandaX-4T (SI\,+\,SD) & Xe (2.4\,t/2.5\,t) & CJPL & 2020-22 & (1.2/1.0)\,t\,$\cdot$\,y & 0.33\,keV$_{\rm nr}$  & \cite{PandaX:2025rrz, PandaX:2025rrz_erratum} \\
         XENONnT (SI\,+\,SD) & Xe ($\sim 4$\,t) & LNGS & 2021-23 & 3.1\,t\,$\cdot$\,y & $\sim 3.8$\,keV$_{\rm nr}$ & \cite{XENON:2025vwd} \\
         XENONnT (SI\,+\,SD) & Xe ($\sim 5$\,t) & LNGS & 2021-25 & 7.8\,t\,$\cdot$\,y & $(3-4)$\,e$^-$ (0.5\,keV$_{\rm nr}$) & \cite{XENON:2025vwd, XENON:2026qow} \\
         XMASS (SI) & Xe (97\,kg) & Kamioka & 2013-19 & $\lesssim 0.42$\,t\,$\cdot$\,y & (0.5 - 1.0)\,keV$_{\rm ee}$  & \cite{XMASS:2022tkr} \\
        \hline
        \multicolumn{7}{|c|}{ Gaseous detectors } \\
        \hline
        NEWS-G (SD) & CH$_4$ (114\,g) & LSM & 2019 & 0.60\,kg\,$\cdot$\,d & 2\,e$^-$ (130\,eV$_{\rm nr}$) & \cite{NEWS-G:2024jms} \\
        \hline
        \multicolumn{7}{|c|}{ Other technologies } \\
        \hline
        QROCODILE (SI) & W$_{45}$Si$_{55}$ (1.67\,ng) & surface & 2023 & $\lesssim 29$\,ng\,$\cdot$\,d & 0.11\,eV$_{\rm er}$ &  \cite{QROCODILE:2024nqm} \\\hline
    \end{tabular}
    \label{tab:Running_NR}
\end{table}


\subsubsection*{Cryogenic crystal calorimeters}

\begin{figure}[t]
    \centering
    \includegraphics[width=0.65\linewidth]{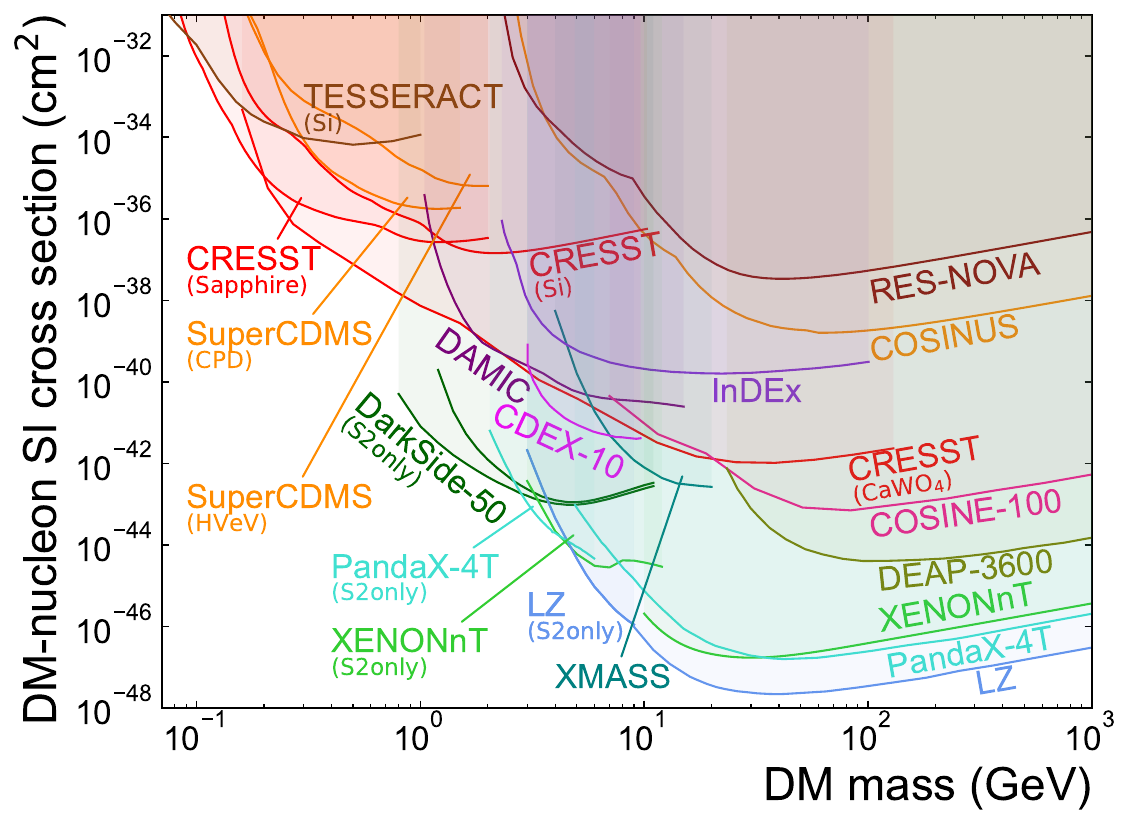}
    \caption{Exclusion region at 90\% C.L. for dark matter-nucleon spin-independent (SI) scattering. The left panel assumes elastic nuclear scattering. The experiments shown are: CDEX-10~\cite{CDEX:2021cll}, COSINE-100~\cite{COSINE-100:2025kbw}, COSINUS~\cite{COSINUS:2023kqd}, CRESST~\cite{CRESST:2019jnq,CRESST:2022lqw,CRESST:2024cpr}, DAMIC~\cite{DAMIC:2020cut}, \mbox{DEAP-3600}~\cite{DEAP:2019yzn}, DarkSide-50~\cite{DarkSide-50:2025lns},  
    InDEx~\cite{Das:2025coa},
    LZ~\cite{LZ:2024zvo, LZ:2025igz}, 
    PandaX-4T~\cite{PandaX:2024qfu, PandaX:2025rrz_erratum}, 
    RES-NOVA~\cite{Alloni:2026vak}, SuperCDMS~\cite{SuperCDMS:2020aus, SuperCDMS:2022zmd}, TESSERACT~\cite{TESSERACT:2025tfw}, XENONnT~\cite{XENON:2025vwd, XENON:2026qow} and XMASS~\cite{XMASS:2022tkr}.}
    \label{fig:exp_running_NR_SI}
\end{figure}

The \textbf{COSINUS} collaboration has presented their first underground dark matter search using NaI as a cryogenic scintillating calorimeter (see Sec.~\ref{ssec:tech_bolom}). 
For the first time in a NaI-based detector, the combined readout of phonons and scintillation light has been achieved, enabling event-by-event discrimination between electronic and nuclear recoils~\cite{COSINUS:2023kqd}.
The experiment is in operation at LNGS.
Using a 3.67\,g NaI target and an exposure of only 11.6\,g\,$\cdot$\,d, the analysis probes standard spin-independent elastic dark matter-nucleus scattering for dark matter masses above a few GeV. 
The analysis threshold is chosen at the total deposited energy at which the cut efficiency has reached about 50\% of its plateau value, yielding 4\,keV$_{\rm nr}$.
The strongest limit is approximately $(1-2)\times10^{-39}\,\mathrm{cm^2}$ at a dark matter mass of about $(60-70)$\,GeV.
Although the resulting exclusion limit (dark orange line in Fig.~\ref{fig:exp_running_NR_SI}) is not competitive with leading dark matter searches, the measurement demonstrates the strong benefit of event-by-event particle discrimination, improving the limit by up to two orders of magnitude compared with an analysis without discrimination. 
Most importantly, it demonstrates the operation of NaI as a cryogenic scintillating calorimeter, providing the basis for a future direct test of whether the DAMA/LIBRA modulation originates from nuclear recoils.\\

\textbf{CRESST}, operating at LNGS, has exploited the flexibility in target material offered by cryogenic calorimeters to probe both spin-independent and spin-dependent dark matter interactions at sub-GeV masses. 
In 2019, CRESST-III reported constraints on spin-independent elastic dark matter-nucleon scattering using a 23.6\,g CaWO$_4$ crystal and collecting an effective exposure of 3.64\,kg\,$\cdot$\,d~\cite{CRESST:2019jnq}.
These results are still leading to date in the mass range from 160 to 800\,MeV (red line in Fig.\ref{fig:exp_running_NR_SI}).
The energy threshold of 30.1\,eV$_{\rm nr}$ was defined by setting the accepted number of noise triggers to one per kg\,$\cdot$\,d of exposure.

\begin{figure}[t]
    \centering
    \includegraphics[width=0.49\linewidth]{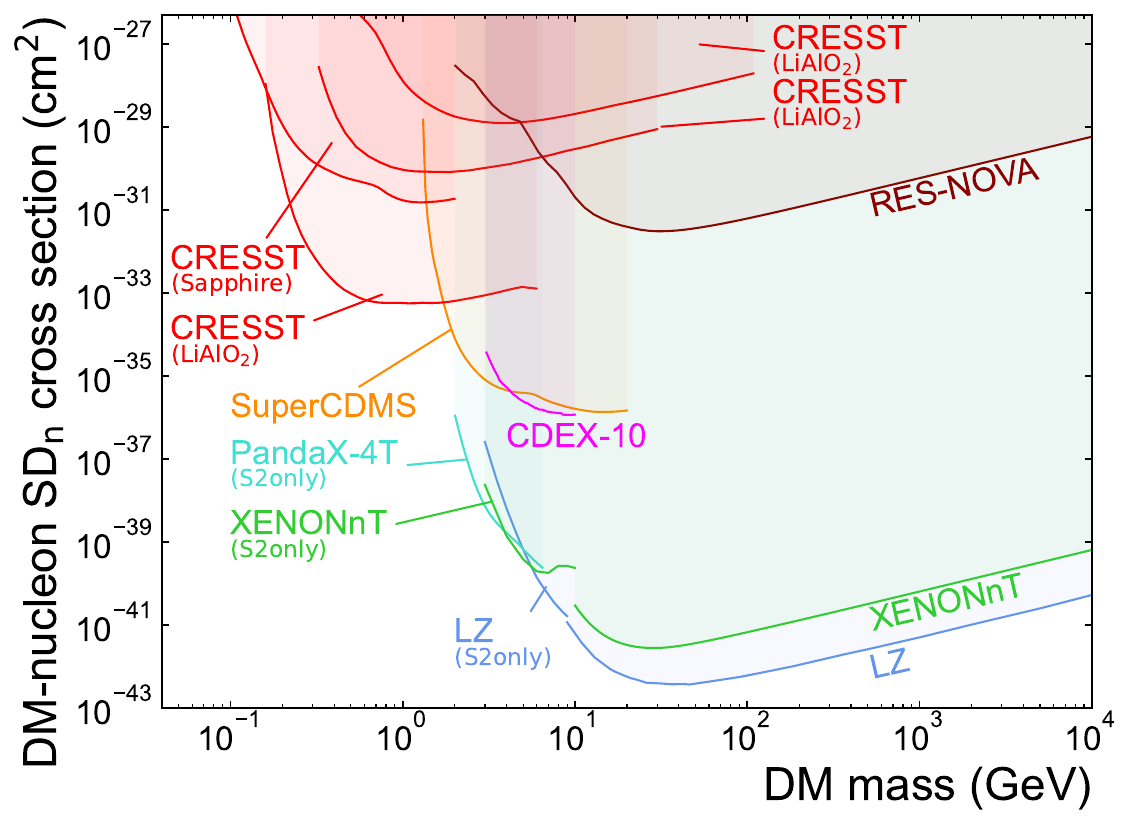}
    \hfill
    \centering
    \includegraphics[width=0.49\linewidth]{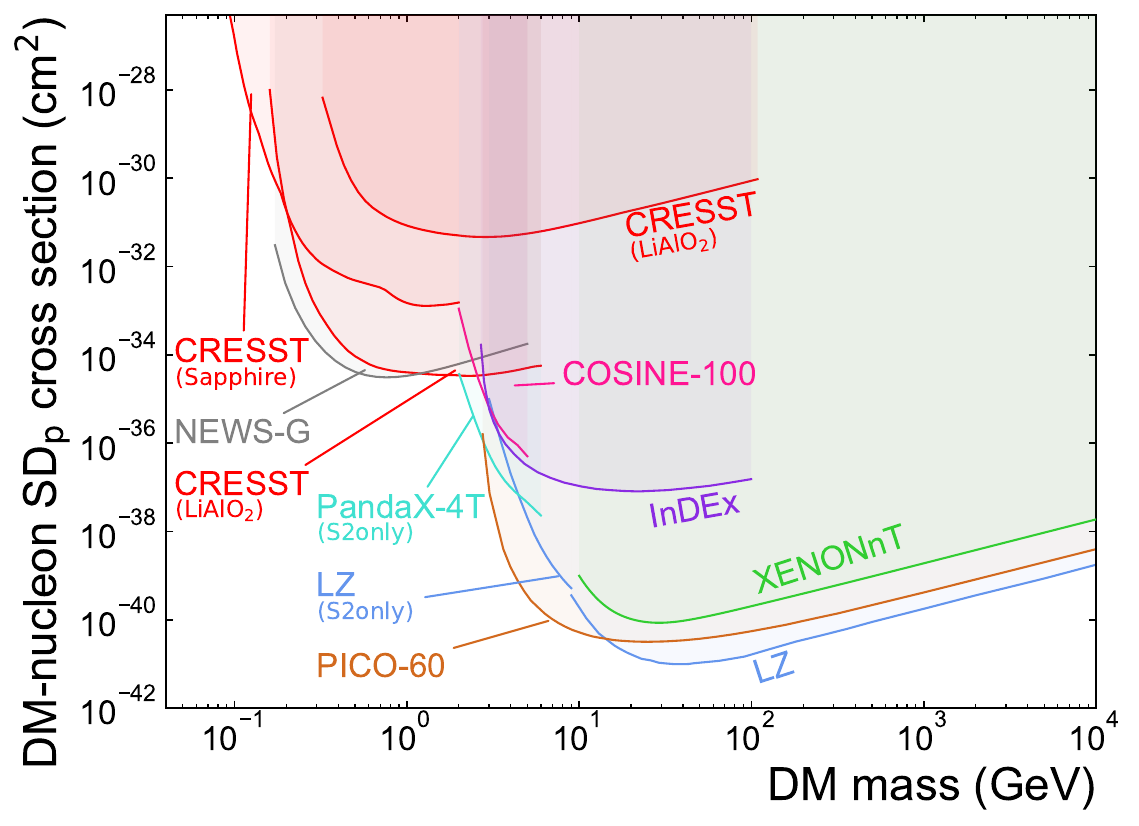}
    \caption{Exclusion region at 90\% C.L. for spin-dependent (SD) dark matter-nucleon scattering off neutrons (left) and protons (right). The experiments shown are: CDEX-10~\cite{CDEX:2021cll}, SuperCDMS (CDMSlite)~\cite{SuperCDMS:2017nns}, COSINE-100~\cite{COSINE-100:2025kbw}, CRESST-III~\cite{CRESST:2020tlq, CRESST:2022jig, CRESST:2022dtl, CRESST:2024cpr}, InDEx~\cite{Das:2025coa}, LZ~\cite{LZ:2024zvo, LZ:2025igz}, NEWS-G~\cite{NEWS-G:2024jms}, PandaX~\cite{PandaX:2024qfu, PandaX:2025rrz_erratum}, PICO-60~\cite{PICO:2019vsc}, RES-NOVA~\cite{Alloni:2026vak}, XENONnT~\cite{XENON:2025vwd, XENON:2026qow}.}
    \label{fig:exp_running_NR_SD}
\end{figure}

Using LiAlO$_2$ crystals containing the spin-sensitive isotopes $^6$Li, $^7$Li, and $^{27}$Al, CRESST subsequently established leading constraints on spin-dependent neutron and proton interactions (red lines in Fig.~\ref{fig:exp_running_NR_SD}). 
The first results were obtained with a 2.8\,g LiAlO$_2$ detector operated above-ground~\cite{CRESST:2020tlq,CRESST:2022jig}.
The detector achieved a threshold of 213.02\,eV on the X-ray-calibrated phonon energy scale, which was used as the nuclear-recoil analysis threshold. 
The original analysis was subsequently extended to include interactions with $^6$Li, significantly improving the constraints on spin-dependent neutron scattering without requiring additional data. 
A subsequent measurement at LNGS employed two 10.5\,g LiAlO$_2$ detector modules, which accumulated exposures of 1.161\,kg\,$\cdot$\,d and 1.184\,kg\,$\cdot$\,d,
respectively, before exposure-reducing cuts~\cite{CRESST:2022dtl}.
The respective thresholds of 83.60\,eV$_{\rm nr}$ and 94.09\,eV$_{\rm nr}$ were defined as the nuclear-recoil energy at which the trigger efficiency drops below 50\% of its plateau value.
The improved detector performance enabled leading constraints on spin-dependent neutron-only and proton-only interactions, reaching cross sections with the best performing module of about $4\times10^{-34}\,\mathrm{cm^2}$ at around 1.5\,GeV and $4\times10^{-35}\,\mathrm{cm^2}$ at around 2\,GeV, respectively. 
The constraints are currently leading between about 0.16 and 1.5\,GeV for neutron-only and between about 1.1 and 2\,GeV for proton-only interactions.

More recently, CRESST has pushed the sensitivity of spin-independent scattering towards even lower masses using very small Si and silicon-on-sapphire (SOS) calorimeters, with gross exposures of 55.06\,g\,$\cdot$\,d and 138\,g\,$\cdot$\,d and trigger thresholds of 10.0\,eV$_{\rm nr}$ and 6.7\,eV$_{\rm nr}$, respectively~\cite{CRESST:2022lqw,CRESST:2024cpr}. 
These thresholds were again defined as the recoil energy at which the accepted number of noise triggers is one per kg\,$\cdot$\,d of exposure.
The 0.6\,g SOS detector, with its Al$_2$O$_3$ target neglecting the thin silicon layer, extended the sensitivity to dark matter masses down to 73\,MeV and provides to date the leading constraints between about 90 and 200\,MeV, while reaching a cross section of approximately $1\times10^{-36}\,\mathrm{cm^2}$ at 500\,MeV. 
The SOS detector also probes spin-dependent dark matter neutron and proton scattering, exploiting the nuclear spins of $^{27}$Al and $^{17}$O.
The search provides currently leading constraints on proton scattering between about 95 and 170\,MeV
However, the collaboration cautions that the constraints below approximately 113\,MeV may be affected by dark matter scattering in the rock overburden.
Its baseline energy resolution of 1.0\,eV additionally enabled the first observation of single luminescence photons with a CRESST detector~\cite{CRESST:2024cpr}.\\

\textbf{RES-NOVA}, primarily designed for the detection of astrophysical neutrinos through CE$\nu$NS, has performed a dark matter search with a 13\,g PbWO$_4$ cryogenic calorimeter made from archaeological lead~\cite{Alloni:2026vak}. 
The use of archaeological Pb provides an intrinsically radiopure target material.
Using an exposure of 32.4\,g\,$\cdot$\,d, the measurement probes both spin-independent and spin-dependent neutron-only dark matter interactions (dark red line in Fig.~\ref{fig:exp_running_NR_SI} and Fig.~\ref{fig:exp_running_NR_SD}-left), reaching minimum cross sections of about $2\times10^{-38}\,\mathrm{cm^2}$ around $(25-30)$\,GeV and $3\times10^{-32}\,\mathrm{cm^2}$ around $(20-30)$\,GeV, respectively. 
The analysis threshold was set to 2.5\,keV on the gamma-calibrated phonon energy scale, corresponding to 2.5\,keV$_{\rm nr}$ under the assumption of an unquenched nuclear-recoil heat response.
The spin-independent search extends to masses down to about 2.2\,GeV, with the combination of heavy Pb and W nuclei and the much lighter O nuclei providing sensitivity over a broad mass range, illustrating the target-mass dependence discussed in Sec.~\ref{subsubsec_Nucl_Scat}. 
For the spin-dependent search, the sensitivity arises primarily from the non-zero nuclear spin of $^{207}$Pb and $^{17}$O and extends down to about 1\,GeV. 
Although the limits are not competitive with dedicated dark matter experiments, this measurement provides the first direct dark matter constraints using PbWO$_4$ and demonstrates that cryogenic detectors developed primarily for CE$\nu$NS measurements can simultaneously search for dark matter.\\

\textbf{SuperCDMS} has investigated both spin-independent and spin-dependent dark matter-nucleus scattering using cryogenic semiconductor crystals. 
In its earlier CDMSlite program at the Soudan Underground Laboratory, SuperCDMS operated germanium detectors at high bias voltages, exploiting the Neganov-Trofimov-Luke (NTL) effect (see Eq.~\ref{eq:ntl}) to amplify the signal from drifting charge carriers in the from of secondary phonons. 
In contrast to the traditional low-voltage operation of SuperCDMS detectors, this approach enabled an electron-equivalent energy threshold as low as 56\,eV$_{\rm ee}$, the lowest achieved by SuperCDMS at the time~\cite{SuperCDMS:2017nns}. 
This analysis threshold was defined as the energy at which the hardware trigger efficiency reached 50\%.
A search with the respective 600\,g germanium detector yielded constraints on spin-independent, spin-dependent proton, and spin-dependent neutron interactions based on net exposures of 6.25\,kg\,$\cdot$\,d and 70.10\,kg\,$\cdot$\,d in two runs.
The spin-dependent neutron constraints from the longer run remain world-leading between approximately 1 and 2\,GeV (orange line in Fig.~\ref{fig:exp_running_NR_SD}-left).

Within its subsequent low-mass dark matter R\&D program, SuperCDMS has performed spin-independent dark matter-nucleus scattering searches with two different gram-scale Si calorimeters both operated above ground~\cite{SuperCDMS:2020aus,SuperCDMS:2022zmd}. 
A 10.6\,g Cryogenic PhotoDetector (CPD) achieved a baseline energy resolution of 3.86\,eV and a trigger threshold of 16.3\,eV, set at 4.2 times the normally-distributed baseline noise. 
The threshold is set on an electron-recoil-calibrated phonon-energy scale. 
For the nuclear-recoil analysis, a heat-quenching factor of unity was assumed, such that the calibrated phonon energy was taken to equal the nuclear-recoil energy.
With a gross exposure of 9.9\,g\,$\cdot$\,d, the detector probed dark matter masses down to 93\,MeV, providing the leading constraints from cryogenic detectors between 93 and 140\,MeV at the time of publication~\cite{SuperCDMS:2020aus}. 
Its exclusion limit reaches a minimum cross section of about $1.8\times10^{-36}\,\mathrm{cm^2}$ at a mass of about 1.2\,GeV. 
A subsequent measurement with a 0.93\,g HVeV detector operated at zero bias voltage, without NTL amplification, achieved an even lower trigger threshold of 9.2\,eV on the calibrated phonon energy scale. 
The calibration was derived from laser-induced signals measured with NTL amplification and transferred to the zero-voltage data, with an associated energy-scale uncertainty of approximately 11\%. 
For the nuclear-recoil interpretation at zero bias, the calibrated phonon energy was assumed to equal the deposited nuclear-recoil energy.
The threshold was applied offline to continuously acquired data and chosen to yield a trigger rate of 20\,Hz. 
Despite a net exposure of only 0.185\,g\,$\cdot$\,d, the measurement extended the sensitivity to masses down to about 92\,MeV~\cite{SuperCDMS:2022zmd}. 
The HVeV measurement also investigated the origin of the LEE commonly observed in cryogenic low-threshold detectors (see Sec.~\ref{sssec:instrumental_bck}), finding evidence that luminescence from materials surrounding the detector may constitute one source of this background, which likely has multiple detector- and setup-dependent origins.\\

The \textbf{TESSERACT} collaboration has performed a spin-independent dark matter-\-nucleus scattering search with a 0.233\,g Si athermal phonon detector operated above ground~\cite{TESSERACT:2025tfw}. 
The detector achieved a baseline energy resolution of 361.5\,${\rm meV}$, the best demonstrated for an athermal phonon detector to date.
Using a laser-calibrated phonon-energy scale, the offline optimum-filter trigger threshold was set to four times the baseline noise, corresponding to 1.45\,eV, while the analysis threshold was set to 1.5\,eV. 
The nuclear-recoil interpretation of this energy scale within the dark matter analysis assumes that the deposited recoil energy is fully represented by the calibrated phonon response.
With an exposure of only 0.1165\,g\,$\cdot$\,d, the detector probed dark matter masses down to 44\,MeV, the lowest mass reached by a direct search for particle-like dark matter at the time of publication (dark brown line in Fig.~\ref{fig:exp_running_NR_SI}). 
The resulting limits provide the strongest constraints for masses between 44 and 87\,MeV, reaching a cross section of $4\times10^{-32}\,\mathrm{cm^2}$ at 87\,MeV ~\cite{TESSERACT:2025tfw}. 
The limit calculation accounts for dark matter attenuation in the atmosphere and the concrete overburden above the laboratory, as well as the possibility of signal pileup at high interaction rates. 
A key feature of the detector is its two-channel phonon readout, which allows interactions in the Si substrate to be distinguished from a class of low-energy backgrounds that couple predominantly to individual sensor channels. 
This strategy, which is also implemented by CRESST in its DoubleTES detector design~\cite{GRAPES-3:2024yym}, provides an important experimental handle to identify and reject sensor-related contributions to the LEE (see Sec.~\ref{sssec:instrumental_bck}).\\

\subsubsection*{Semiconductors}

\textbf{CDEX} has searched for dark matter-nucleus scattering using low-threshold p-type point-contact Ge detectors (see Sec.~\ref{sssec:tech_hpge}) at CJPL. 
The detectors measure the ionization produced by particle interactions, with energies expressed on an electron-equivalent scale.
Using CDEX-1B and CDEX-10 data with exposures of 737.1\,kg\,$\cdot$\,d and 102.8\,kg\,$\cdot$\,d, respectively, and a common analysis threshold of 160\,eV$_{\rm ee}$, CDEX derived constraints on spin-independent and spin-dependent neutron scattering (light pink line in Fig.~\ref{fig:exp_running_NR_SI} and \ref{fig:exp_running_NR_SD}-left, respectively), accounting for attenuation of standard-halo dark matter in the Earth~\cite{CDEX:2021cll}. 
The threshold corresponds to the lower boundary of the ionization-energy spectrum used in the analysis.
In this analysis, CDEX has also searched for dark matter interactions accompanied by the Migdal effect. 
These results are discussed and shown separately in Sec.~\ref{sssec:exp_running_ER}.
The aforementioned figures show only the elastic scattering search results which probe dark matter masses between 3 and 10\,GeV.
This search excludes a band in cross section, since sufficiently strongly interacting dark matter loses energy in the approximately 2\,400\,m rock overburden before reaching the detector. 
This shielding effect becomes particularly relevant for dark matter masses below about 1\,GeV. 

More recently, 205.4\,kg\,$\cdot$\,d of CDEX-10 data were reanalyzed to search for spin-independent nuclear recoils from dark matter interacting through a light mediator~\cite{CDEX:2024rfi}. 
The analysis used the same ionization-energy threshold of 160\,eV$_{\rm ee}$, corresponding to a combined detection and selection efficiency of approximately 4.5\% at threshold. 
For mediator masses of 1 and 10\,MeV, the analysis excludes previously unexplored parameter space at dark matter masses between approximately 2 and 5\,GeV, reaching cross sections at zero momentum transfer of $1.6\times10^{-36}\,\mathrm{cm^2}$ and $3.7\times10^{-39}\,\mathrm{cm^2}$, respectively, at a dark matter mass of 2\,GeV. \\

The \textbf{DAMIC} experiment searched for spin-independent dark matter-nucleus scattering using the ionization signal from nuclear recoils in Si CCDs (see Sec.~\ref{sssec:tech_ccd}) operated at SNOLAB~\cite{DAMIC:2020cut}. 
With a net exposure of 10.93\,kg\,$\cdot$\,d after data-quality selections, the analysis employed a threshold of 50\,eV$_{\rm ee}$.
This analysis threshold refers to the reconstructed ionization energy, calibrated using LED illumination with an electron-hole pair creation energy of 3.8\,eV. 
It was defined by requiring fewer than 0.1 expected noise-induced clusters in the final dataset, corresponding to a signal acceptance of approximately 10\% at threshold. 
The search probed dark matter masses down to about 1\,GeV and reached a minimum cross section of approximately $3\times10^{-41}\,\mathrm{cm^2}$ for masses between 7 and 10\,GeV (dark purple line in Fig.~\ref{fig:exp_running_NR_SI}). 
At the time of publication, these results provided the strongest constraints obtained with a silicon target for dark matter masses below 9\,GeV and directly tested the dark matter parameter space favored by the three candidate events previously observed with the silicon detectors of the CDMS\,II experiment~\cite{CDMS:2013juh}. 

Within the DAMIC data, an excess of events was observed below about 200\,eV$_{\rm ee}$, which could not be explained by the background model. 
A subsequent measurement using skipper CCDs confirmed the excess in the Si bulk with a significance exceeding $3\,\sigma$, while lowering the analysis threshold to 23\,eV$_{\rm ee}$~\cite{SENSEI:2023rcc}. 
Its origin remains unknown, and its characteristics appear to differ from those of other low-energy backgrounds observed in charge-sensitive detectors referred to as LEE (see Sec.~\ref{sssec:instrumental_bck})~\cite{Baxter:2025odk}.\\

\subsubsection*{Scintillation crystals}

\textbf{COSINE-100} has searched for spin-independent and spin-dependent proton-only dark matter-nucleus scattering using NaI(Tl) scintillation detectors (see Sec.~\ref{ssec:tech_scint}) at Y2L (dark pink line in Fig.~\ref{fig:exp_running_NR_SI} and \ref{fig:exp_running_NR_SD}-right, respectively). 
Using five NaI(Tl) crystals with a total target mass of 61.3\,kg and a live time of 2.86 years, the analysis employed a threshold of 0.7\,keV$_{\rm ee}$ ~\cite{COSINE-100:2025kbw}. 
This threshold refers to the electron-equivalent scintillation energy, calibrated using low-energy X-rays, and was set by requiring less than 1\% contamination from photomultiplier noise after event selection. 
The scintillation-signal acceptance is approximately 20\% at threshold. 
The conversion from nuclear-recoil energy to the measured scintillation energy uses energy-dependent quenching factors for Na and I. 
The spin-independent search covers dark matter masses from 7 to 1\,000\,GeV and reaches a minimum cross section of approximately $7\times10^{-44}\,\mathrm{cm^2}$ at around $(70-80)$\,GeV, improving the previous COSINE-100 result ~\cite{COSINE-100:2021xqn} by about an order of magnitude. 
The elastic spin-dependent proton scattering search extends to dark matter masses of about 5\,GeV and reaches a minimum cross section of approximately $4\times10^{-37}\,\mathrm{cm^2}$ at that mass. 
The analysis found no significant excess above the expected background and derived model-dependent constraints on the dark matter-nucleon scattering cross section.
These limits are not competitive with the leading dedicated nuclear-recoil searches over most of the parameter space.
Instead, experiments like COSINE-100, using the same NaI(Tl) target material as \mbox{DAMA/LIBRA}, are motivated by the fact that they can directly probe a dark matter interpretation of the observed DAMA modulation.\\

\subsubsection*{Bubble chambers}

The \textbf{InDEx} experiment searches for dark matter-nucleus scattering using superheated droplet detectors operated at JUSL. 
The detectors employ C$_2$H$_2$F$_4$ as the active target liquid and, similar to bubble chambers (see Sec.~\ref{ssec:tech_bubble}), detect the acoustic signal generated by the nucleation of a vapor bubble following a sufficiently energetic nuclear recoil. 
In its second science run, InDEx acquired an exposure of 7.2\,kg\,$\cdot$\,d~\cite{Das:2025coa}. 
By maintaining the detectors at an operating temperature of 35$^\circ$C, the thermodynamic threshold for bubble nucleation was reduced to 1.95\,keV. 
This threshold refers to the minimum energy required to nucleate a bubble and can be interpreted as a nuclear-recoil energy threshold, with an energy-dependent nucleation efficiency.
The probability of bubble nucleation depends on the nuclear recoil energy and the recoiling nucleus, and the corresponding efficiencies are included in the signal calculation. 
The resulting spin-independent limit reaches $1.55\times10^{-40}\,\mathrm{cm^2}$ at a dark matter mass of 20.4\,GeV, while the spin-dependent proton-only limit, derived from scattering on fluorine, reaches $7.97\times10^{-38}\,\mathrm{cm^2}$ at 21.0\,GeV (light purple line in Fig.~\ref{fig:exp_running_NR_SI} and \ref{fig:exp_running_NR_SD}-right, respectively). 
Although these constraints are substantially weaker than the leading limits from larger experiments, the measurement demonstrates the use of C$_2$H$_2$F$_4$ as a superheated target for direct dark matter searches and improves on the first InDEx run through both a lower threshold and a larger exposure~\cite{Kumar:2025ofs}.\\

\textbf{PICO} has searched for spin-independent and spin-dependent proton-only dark matter-nucleus scattering using C$_3$F$_8$ bubble chambers operated at SNOLAB. 
The final PICO-60 search used 52\,kg of C$_3$F$_8$ and combined two exposures of 1167\,kg\,$\cdot$\,d and 1404\,kg\,$\cdot$\,d, acquired at thermodynamic thresholds of 3.29\,keV and 2.45\,keV, respectively~\cite{PICO:2019vsc}. 
These thresholds correspond to the minimum energy required for bubble nucleation, as calculated from the temperature and pressure of the superheated liquid. 
They can be interpreted as nuclear-recoil energy thresholds, with energy-dependent nucleation efficiencies for fluorine and carbon determined from neutron calibrations. 
The detector combines optical reconstruction of bubble positions with acoustic measurements, allowing backgrounds from interactions near the detector walls and alpha decays to be rejected. 
Together with the strong intrinsic rejection of electronic recoils and the identification of multiple-scattering neutrons, this enabled the search to operate with an expected background of approximately one event in the 1404\,kg\,$\cdot$\,d exposure. 
The resulting spin-dependent proton-only limit improved the previous PICO limits for dark matter masses between 3 and 5\,GeV by approximately an order of magnitude and reaches a minimum cross section of $3.2\times10^{-41}\,\mathrm{cm^2}$ at a dark matter mass of 25\,GeV. 
The strong sensitivity to spin-dependent proton interactions originates primarily from the non-zero nuclear spin of $^{19}$F, while the low thermodynamic thresholds provide sensitivity to dark matter masses down to a few GeV. 
The spin-dependent result remains world-leading between approximately 3 and 15\,GeV (light brown line in Fig.~\ref{fig:exp_running_NR_SD}-right).\\

\subsubsection*{Liquid noble-gas scintillators}

\textbf{DarkSide-50} searched for spin-independent dark matter-nucleus scattering using a dual-phase liquid argon TPC (see Sec.~\ref{ssec:tech_noble}) at LNGS with an active target mass of 46.4\,kg. 
The low-mass analysis relied on the ionization signal as its energy observable, commonly referred to as a charge-only or S2-only search. 
After fiducialization, approximately 19.4\,kg of liquid argon remained, yielding an exposure of 12\,306\,kg\,$\cdot$\,d (about 0.034\,t\,$\cdot$\,y) after the spurious-electron veto~\cite{DarkSide-50:2022qzh}. 
The analysis threshold was set to four extracted electrons to suppress backgrounds from spurious ionization electrons, while the trigger efficiency was effectively 100\% in the analysis region. 
Based on the nuclear-recoil ionization-response model used in this analysis, the four-electron threshold corresponds to approximately 0.6\,keV$_{\rm nr}$. 
This conversion depends on the nuclear-recoil ionization yield and its fluctuations. 
The analysis initially provided the strongest constraints between 1.2 and 3.6\,GeV, reaching a cross section of $6\times10^{-43}\,\mathrm{cm^2}$ at 3\,GeV. 

These limits were subsequently improved through a refined model of the low-energy nuclear-recoil ionization response, combining calibration measurements from DarkSide-50, ARIS, SCENE, and ReD\footnote{ARIS (Argon Response to Ionization and Scintillation), SCENE (Scintillation Efficiency of Noble Elements), and ReD (Recoil Directionality) are dedicated liquid-argon calibration experiments studying the scintillation and ionization response to neutron-induced nuclear recoils.}~\cite{DarkSide-50:2025lns}. 
The additional ReD measurements favor a Lenz-Jensen description of atomic screening over the previously used ZBL\footnote{ZBL (Ziegler-Biersack-Littmark) denotes an empirical atomic screening potential used to model nuclear stopping, i.e. the energy transferred to atomic motion during collisions. The alternative Lenz-Jensen potential is based on the Thomas-Fermi description of atomic screening.} model, resulting in a higher predicted ionization yield at low nuclear-recoil energies. 
Consequently, low-energy nuclear recoils have a higher probability of producing signals above the unchanged four-electron analysis threshold, effectively extending the sensitivity towards lower recoil energies. 
Since the stochastic fluctuations associated with nuclear-recoil quenching are not yet experimentally established, limits are reported both without additional quenching fluctuations and assuming binomial fluctuations, with the latter further increasing the probability for low-energy recoils to produce signals above threshold. 
Using the same dataset, the updated limits provide the currently strongest constraints between 1.0 and about 2.5\,GeV without quenching fluctuations and between 0.8 and about 2.8\,GeV when binomial quenching fluctuations are included, reaching a cross section as low as about $5\times10^{-43}\,\mathrm{cm^2}$ and $2\times10^{-43}\,\mathrm{cm^2}$, respectively (dark green lines in Fig.~\ref{fig:exp_running_NR_SI}). 
The sensitivity at these low masses benefits from the lower nuclear mass of argon compared with xenon, which results in larger recoil energies for light dark matter particles (see Sec.~\ref{subsubsec_Nucl_Scat}).\\

\textbf{DEAP-3600} searched for spin-independent dark matter-nucleus scattering using a single-phase liquid argon scintillation detector at SNOLAB (see Sec.~\ref{ssec:tech_noble}). 
The detector contains approximately 3.3\,t of liquid argon, of which 824\,kg were retained as fiducial target mass after position-based event selections~\cite{DEAP:2019yzn}\footnote{The latest result from the DEAP-3600 Collaboration~\cite{DEAP:2026orr} has not yet been published in a peer-reviewed journal. We therefore quote the earlier published result and will update it if the newer analysis is published before the publication of this review.}. 
The analysis used a total exposure of 758\,t\,$\cdot$\,d ($\approx 2.08$\,t\,$\cdot$\,y), quoted before fiducialization, and obtained its sensitivity from the scintillation signal alone. 
The analysis threshold was set to 95 detected photoelectrons (p.e.), corresponding to approximately 15.6\,keV$_{\rm ee}$ on the electron-recoil-calibrated scintillation-energy scale. 
This lower boundary of the nuclear-recoil search region was chosen to achieve sufficient discrimination against electronic-recoil backgrounds using the different scintillation time profiles of nuclear and electronic recoils. 
The corresponding nuclear-recoil energy threshold is approximately 50\,keV$_{\rm nr}$, depending on the energy-dependent scintillation quenching factor derived from neutron calibration measurements. 
The analysis achieved strong electronic-recoil rejection through pulse-shape discrimination, with an average leakage probability of approximately $4\times10^{-9}$ at 90\% nuclear-recoil acceptance within the search region. 
No dark matter candidate events were observed, resulting in a minimum spin-independent cross-section limit of $3.9\times10^{-45}\,\mathrm{cm^2}$ at a dark matter mass of 100\,GeV (olive line in Fig.~\ref{fig:exp_running_NR_SI}). \\

The \textbf{LUX-ZEPLIN (LZ)} experiment searches for spin-independent and spin-dependent dark matter-nucleus scattering with a dual-phase xenon TPC at SURF, succeeding the earlier LUX experiment at the same site. 
LUX advanced low-threshold analyses of xenon TPC data, employing techniques to identify single-photon S1 signals and performing an S2-only analysis with a threshold of 3.5 detected electrons, thereby extending the sensitivity to nuclear recoils from low-mass dark matter down to about 2\,GeV~\cite{LUX:2019npm,LUX:2020yym}.
These constraints have since been superseded by the successor experiment LZ.

LZ, also operated at SURF, is using a TPC containing approximately 7\,t of liquid xenon as active mass. 
Its first dark matter search used a fiducial mass of 5.5\,t, providing leading constraints on spin-independent and spin-dependent neutron-only scattering, as well as constraints on spin-dependent proton-only scattering~\cite{LZ:2022lsv}. 
The subsequent nuclear-recoil analysis combined this dataset with additional measurements, yielding an exposure of 4.2\,t\,$\cdot$\,y with the same fiducial mass~\cite{LZ:2024zvo}. 
The analysis required both S1 and S2 signals, with a lower S1 boundary of three detected photons and an S2 threshold of 14.5 extracted electrons. 
These are signal-selection thresholds rather than directly measured nuclear-recoil energy thresholds. 
Using the calibrated scintillation and ionization response, including their fluctuations, the nuclear-recoil detection efficiency reaches 50\% at approximately 5.4\,keV$_{\rm nr}$.
The resulting spin-independent limit reaches a minimum cross section of $2.2\times10^{-48}\,\mathrm{cm^2}$ at 40\,GeV. 
The same analysis constrains spin-dependent neutron-only and proton-only interactions down to $3.7\times10^{-43}\,\mathrm{cm^2}$ at 46\,GeV and $9.8\times10^{-42}\,\mathrm{cm^2}$ at 40\,GeV, respectively. 
The stronger neutron sensitivity arises from the odd-neutron isotopes $^{129}$Xe and $^{131}$Xe. 
All three results are currently providing the leading constraints in the mass range from about 9\,GeV up to the TeV-scale (blue lines in Fig.~\ref{fig:exp_running_NR_SI} and \ref{fig:exp_running_NR_SD}).

A dedicated low-energy LZ analysis subsequently extended the elastic nuclear-recoil search to dark matter masses between 3 and 9\,GeV~\cite{LZ:2025igz}. 
Using a reduced fiducial mass of 5.09\,t, the analysis obtained a net exposure of 5.7\,t\,$\cdot$\,y. 
It retained the requirement of an S1 signal in at least three photomultipliers but restricted the analysis to S2 signals corresponding to $3.5-14.5$ extracted electrons. 
Together with an S1 range of $2-15$ detected photons, this defines a region of interest corresponding approximately to nuclear-recoil energies of $(1-6)$\,keV$_{\rm nr}$. 
The energy conversion and detection efficiency were determined using a nuclear-recoil response model constrained by dedicated neutron calibrations. 
The analysis provides leading spin-independent and spin-dependent neutron-only constraints down to about 5\,GeV (blue lines in Fig.~\ref{fig:exp_running_NR_SI} and \ref{fig:exp_running_NR_SD}). 
It also observes a nuclear-recoil signal consistent with solar-$^{8}$B CE$\nu$NS at a significance of 4.5\,$\sigma$, providing the strongest evidence for this process on xenon to date. 

Recently, LZ reported an unusual high-energy event consistent with a nuclear recoil of approximately 250\,keV, for which the expected background rate is very low~\cite{LZ:2026axp}. 
Possible interpretations include inelastic dark matter interactions and momentum-dependent operators, both of which can enhance the scattering rate at high recoil energies, see the right panel of Fig.~\ref{fig:rates}. 
While intriguing, the event requires further investigation.\\ 

\textbf{PandaX-4T} has obtained constraints on both spin-independent and spin-dependent dark matter-nucleus scattering, using its dual-phase liquid xenon TPC at CJPL.
The detector contains approximately 5.6\,t of liquid xenon, of which about 3.7\,t form the active target. 
A combined analysis of the commissioning run and first science run used fiducial masses of 2.38\,t and 2.48\,t, respectively, yielding a total exposure of 1.54\,t\,$\cdot$\,y~\cite{PandaX:2024qfu}. 
The analysis required both S1 and S2 signals, with lower selection boundaries of 2\,p.e. and 120\,p.e., respectively. 
Based on the calibrated scintillation and ionization response, these correspond to approximate nuclear-recoil energy thresholds of 4\,keV$_{\rm nr}$ and 3\,keV$_{\rm nr}$ for the two runs, respectively. 
The resulting spin-independent limit reaches a minimum cross section of $1.6\times10^{-47}\,\mathrm{cm^2}$ at a dark matter mass of 40\,GeV and provided the strongest constraints above 100\,GeV at the time of publication (cyan line in Fig.~\ref{fig:exp_running_NR_SI}).

A subsequent low-threshold analysis of the same two runs combined conventional S1-S2 paired events with ionization-only events~\cite{PandaX:2025rrz, PandaX:2025rrz_erratum}. 
The analysis used effective exposures of 1.20\,t\,$\cdot$\,y and 1.04\,t\,$\cdot$\,y for the paired and ionization-only samples, respectively. 
The latter required S2 signals corresponding to four to eight detected electrons, without a corresponding S1 signal. 
Using a nuclear-recoil response model based on NEST, as is common in this technology, and including uncertainties in the scintillation and ionization yields, the combined analysis achieved sensitivity to nuclear-recoil energies between approximately 0.33 and 3\,keV$_{\rm nr}$, where the detection efficiency exceeds 1\%. 
It provides currently leading constraints on spin-independent interactions between about 2.8 and 4.0\,GeV, spin-dependent neutron-only interactions between 2.0 and about 4\,GeV, and spin-dependent proton-only interactions between 2.0 and about 3\,GeV (cyan lines in Fig.~\ref{fig:exp_running_NR_SI} and \ref{fig:exp_running_NR_SD}). 
At a dark matter mass of 3\,GeV, the corresponding cross-section limits are $1.1\times10^{-43}\,\mathrm{cm^2}$, about $2\times10^{-38}\,\mathrm{cm^2}$, and about $6\times10^{-37}\,\mathrm{cm^2}$, respectively.\\

The \textbf{XENONnT} collaboration searches for spin-independent and spin-dependent elastic dark matter-nucleus scattering using a dual-phase liquid xenon TPC at LNGS, succeeding the earlier XENON1T experiment~\cite{XENON:2017lvq}.
The detector contains approximately 5.9\,t of active liquid xenon. 
Its first nuclear-recoil search, based on an exposure of 1.09\,t\,$\cdot$\,y, reached a minimum spin-independent cross section of $2.58\times10^{-47}\,\mathrm{cm^2}$ at 28\,GeV~\cite{XENON:2023cxc} and improved on the corresponding XENON1T limits~\cite{XENON:2018voc}. 
This result was subsequently superseded by a combined analysis of the first two science runs, using fiducial masses of approximately 4.2\,t and 4.0\,t, respectively, and a total exposure of 3.1\,t\,$\cdot$\,y~\cite{XENON:2025vwd}. 
The analysis required both S1 and S2 signals and achieved a nuclear-recoil detection efficiency above 10\% for energies exceeding approximately 3.8\,keV$_{\rm nr}$. 
The efficiency was determined using a detector-response model constrained by neutron calibrations, including the scintillation and ionization yields and their fluctuations. 
The resulting spin-independent limit reaches $1.7\times10^{-47}\,\mathrm{cm^2}$ at 30\,GeV (light green line in Fig.~\ref{fig:exp_running_NR_SI}).
The analysis also provides spin-dependent neutron-only and proton-only constraints with the stronger neutron sensitivity arising from the odd-neutron isotopes $^{129}$Xe and $^{131}$Xe (light green lines in Fig.~\ref{fig:exp_running_NR_SD}). 

A complementary ionization-only analysis combined three science runs, yielding an exposure of 7.83\,t\,$\cdot$\,y~\cite{XENON:2026qow}. 
The corresponding fiducial masses were 4.98\,t, 5.01\,t, and 4.87\,t, respectively. 
The analysis accepted S2 signals corresponding to approximately three extracted electrons in the first two runs and four extracted electrons in the third, without requiring a detectable S1 signal. 
These selection thresholds were chosen to suppress instrumental backgrounds from isolated electrons, while the conversion to nuclear-recoil energy relied on the calibrated ionization yield and its fluctuations. 
The analysis probes nuclear-recoil energies between 0.5 and 5.0\,keV$_{\rm nr}$ and dark matter masses between 3 and about 9\,GeV.
It excludes spin-independent cross sections above $6.0\times10^{-45}\,\mathrm{cm^2}$ at 5\,GeV and also improves constraints on spin-dependent neutron-only scattering (light green lines in Fig.~\ref{fig:exp_running_NR_SI} and \ref{fig:exp_running_NR_SD}) with both searches maintaining leading sensitivity around 4\,GeV.
The analysis additionally derives constraints on dark matter-nucleon scattering mediated by a light mediator, probing a different momentum dependence than the standard contact-interaction scenario.
The analysis demonstrates the low-mass sensitivity achievable with ionization-only signals in large liquid xenon detectors. 
XENONnT also provided, together with PandaX-4T, the first evidence for nuclear recoils from solar $^{8}$B neutrinos through CE$\nu$NS in a xenon dark matter detector, with a significance of 2.73\,$\sigma$~\cite{XENON:2024ijk,PandaX:2024muv}.\\


\textbf{XMASS-I} operated as a single-phase liquid xenon scintillation detector at the Kamioka Observatory. 
The detector contained 832\,kg of liquid xenon, of which 97\,kg were used as fiducial target mass for the spin-independent elastic dark matter-nucleon search. 
With a live time of 1590.9 days, this fiducial mass corresponds to a nominal exposure of approximately 0.42\,t\,$\cdot$\,y, which represents an upper limit on the effective exposure before accounting for event-selection efficiencies.
The final analysis used an electron-equivalent energy threshold of 1.0\,keV$_{\rm ee}$, with a reduced threshold of 0.5\,keV$_{\rm ee}$ available for part of the dataset~\cite{XMASS:2022tkr}. 
These thresholds refer to the scintillation signal calibrated using electronic recoils, while their interpretation in terms of nuclear-recoil energy depends on the scintillation quenching factor. 
Backgrounds originating near the photomultipliers required restrictive fiducial-volume selections and limited the experiment's sensitivity. 
The resulting spin-independent cross-section limit reaches a minimum of $1.4\times10^{-44}\,\mathrm{cm^2}$ at a dark matter mass of 60\,GeV (teal line in Fig.~\ref{fig:exp_running_NR_SI}).\\

\subsubsection*{Gaseous detectors}

\textbf{NEWS-G} searches for light dark matter using spherical proportional counters filled with gaseous targets (see Sec.~\ref{ssec:tech_gas}). 
The latest result uses the 135\,cm diameter detector S140, operated with 114\,g of methane at LSM~\cite{NEWS-G:2024jms}.
The hydrogen-rich target provides favorable kinematic matching to sub-GeV dark matter, while the unpaired proton gives sensitivity to spin-dependent proton-only interactions. 
The analysis used an exposure of 0.599\,kg\,$\cdot$\,d and selected events containing at least two resolved ionization electrons, rejecting single-electron events because of their high background rate. 
Individual electrons were identified from their temporally separated avalanche signals, with the detector response calibrated using a UV laser and $^{37}$Ar. 
The conversion to nuclear-recoil energy relied on measurements and a conservative extrapolation of the hydrogen ionization quenching factor in methane, resulting in an effective lower recoil-energy cutoff of approximately 127\,eV$_{\rm nr}$. 
The analysis probes new parameter space between 0.17 and 1.2\,GeV and reaches a minimum spin-dependent proton-only cross section of $3.09\times10^{-35}\,\mathrm{cm^2}$ at 0.76\,GeV (gray line in Fig.~\ref{fig:exp_running_NR_SD}-right). 
This demonstrates the sensitivity of gaseous detectors to spin-dependent interactions well below 1\,GeV.\\

\subsubsection*{Other technologies}

\textbf{QROCODILE} employs a superconducting WSi microwire single-photon detector that serves simultaneously as the dark matter target and sensor~\cite{QROCODILE:2024nqm}. 
With an active target mass of only 1.67\,ng, the detector achieves an energy threshold of 0.11\,eV. 
The threshold energy can reasonably be interpreted as electron recoil energy, although it strictly refers to the energy deposited by photons calibrated using infrared light, rather than to a directly measured electron-recoil energy.
The search derives constraints on spin-independent dark matter-nucleon scattering through nuclear recoils in the WSi target, considering both light- and heavy-mediator interactions. 
Although the sensor responds primarily to excitations of the electronic system, nuclear recoils can produce detectable signals through their coupling to phonons and quasiparticles. 
The interpretation therefore relies on a theoretical model of the nuclear-recoil response rather than a direct nuclear-recoil energy calibration. 
For a heavy mediator, the analysis probes previously unexplored parameter space below approximately 150\,MeV, reaching dark matter masses down to 15\,MeV and cross sections slightly below $10^{-30}\,\mathrm{cm^2}$ at 150\,MeV. 
For a light mediator, the sensitivity extends down to approximately 20\,MeV and overlaps with parameter space also investigated through the Migdal effect. 
The reported constraints conservatively do not include the potentially enhanced sensitivity from multiphonon production.
The collaboration estimates the cross sections at which atmospheric attenuation becomes relevant and uses these estimates to truncate the exclusion regions. 
However, a full attenuation-dependent calculation of the high-cross-section boundary is not provided. \\


\subsubsection{Electronic recoil}
\label{sssec:exp_running_ER}

Electronic-recoil signatures can arise from two fundamentally different classes of dark matter interactions. 
First, dark matter may scatter directly off bound electrons in the target, producing ionization and/or excitation signals whose interpretation is discussed in Sec.~\ref{sssec:theory_ER_scattering}. 
Second, dark matter may interact primarily with a nucleus but still produce an observable electronic recoil through secondary processes such as the Migdal effect or bremsstrahlung (see Sec.~\ref{subsubsec_Nucl_Scat}), thereby extending the sensitivity of nuclear-scattering searches to lower dark matter masses.
Table~\ref{tab:Running_ER} summarizes the results that employ electron recoils, the Migdal effect, or bremsstrahlung as detection channels. 
Their corresponding exclusion limits are shown in Fig.~\ref{fig:exp_running_ER} for electron recoils, assuming a light mediator with $F_{\rm DM}=(\alpha m_e/q)^2$ (left panel) and a heavy mediator with $F_{\rm DM}=1$ (right panel), and in Fig.~\ref{fig:exp_running_migdal} for searches exploiting the Migdal effect.
Also in this section, the table and figures include only results that are not fully superseded by a later result from the same experiment and each figure includes only limits that probe the same interaction scenario and can therefore be directly compared.\\

\renewcommand{\arraystretch}{1.2}
\begin{table}[H]
    \centering
    \caption{Summary of direct dark matter searches with electron-recoil signatures published since 2020 (or before if still world-leading), organized by detector technology.
    Provided are the leading results per experiment.
    Masses refer to the target or fiducial volume used for the respective analysis, where available. Exposures are the net exposures, if provided. Else the gross exposure is cited. If neither is provided, it is approximated (indicated with $\sim$ or $\lesssim$) based on the given mass and live time. Thresholds are quoted in the energy scale or experimental observable used by the collaboration, with approximate conversions to comparable energy scales where appropriate. Photoelectrons are abbreviated with p.e. and extracted electrons with e$^-$. The classification as an electron-recoil search refers to the experimentally observed signature rather than the underlying dark matter interaction and therefore includes direct dark matter-electron scattering as well as dark matter-nucleus scattering producing an electronic signature through the Migdal effect or bremsstrahlung. Searches in which dark matter-nucleus scattering produces a nuclear-recoil signature are instead summarized in Tab.~\ref{tab:Running_NR}.}
    \scriptsize
    \begin{tabular}{|c|c|c|c|c|c|c|}\hline
        \makecell{\bf Experiment\\\bf (Interaction)} & \makecell{\bf Target \\\bf(Mass)} & {\bf Location} &
        \makecell{\bf Data\\\bf Taking} &
        {\bf Exposure} & {\bf Threshold} & {\bf Reference(s)}\\\hline        \hline
        \multicolumn{7}{|c|}{ Cryogenic crystal calorimeters } \\
        \hline
        EDELWEISS (DM-e) & Ge (33.4\,g) & LSM & 2019-20 & 81\,g\,$\cdot$\,d & 1\,e-h pair ($\sim$3\,eV$_\mathrm{ee}$) & \cite{EDELWEISS:2020fxc}\\
        EDELWEISS (Mig.)& Ge (200\,g) & LSM & 2019-20 & $\lesssim 5.6$\,kg\,$\cdot$\,d & 30\,eV$_\mathrm{ee}$ & \cite{EDELWEISS:2022ktt}\\
        SuperCDMS-HVeV (DM-e) & Si (0.93\,g) & NEXUS & 2022 & 6.1\,g\,$\cdot$\,d & 1\,e-h pair ($\sim$3.8\,eV$_\mathrm{ee}$) &\cite{SuperCDMS:2025dha}\\
        SuperCDMS (Mig.+ Brems.)& Ge (600\,g) & Soudan & 2015 & 37\,kg\,$\cdot$\,d & 70\,eV$_\mathrm{ee}$ & \cite{SuperCDMS:2023sql}\\
        \hline
        \multicolumn{7}{|c|}{ Semiconductors } \\
        \hline
        CDEX-1B/-10 (Mig.)& Ge (939\,g) & CJPL & 2014-17 & 740\,kg\,$\cdot$\,d/100\,kg\,$\cdot$\,d & 160\,eV$_{\rm ee}$ & \cite{CDEX:2021cll}\\
        CDEX-10 (DM-e) & Ge (939\,g) & CJPL & 2017-18 & 210\,kg\,$\cdot$\,d & 160\,eV$_{\rm ee}$ & \cite{CDEX:2022kcd}\\
        DAMIC-M (DM-e) & Si (18\,g) & LSM & 2022 & 0.085\,kg\,$\cdot$\,d & 1\,e$^-$ ($\sim$1\,eV$_{\rm ee}$) & \cite{DAMIC-M:2023gxo}\\
        DAMIC-M (DM-e\,+\,Mig.) & Si (19.8\,g) & LSM & 2024-25 & 1.3\,kg\,$\cdot$\,d & 2\,e$^-$ ($\sim$5\,eV$_{\rm ee}$) & \cite{DAMIC-M:2025luv}\\              
        SENSEI (DM-e\,+\,Mig.) & Si (1.9\,g) & MINOS & 2019 & $(1.4-9.1)$\,g\,$\cdot$\,d & 1\,e$^-$ ($\sim$1\,eV$_{\rm ee}$) & \cite{SENSEI:2020dpa}\\
        SENSEI (DM-e\,+\,Mig.) & Si (13.1\,g) & SNOLAB & 2022-23 & 100\,g\,$\cdot$\,d & 2\,e$^-$ ($\sim$5\,eV$_{\rm ee}$) & \cite{ SENSEI:2023zdf}\\
        \hline
        \multicolumn{7}{|c|}{ Scintillation crystals } \\
        \hline
        COSINE-100 (Mig.)& NaI(Tl) (61.3\,kg) & Y2L & 2016-19 &  $\lesssim 180$\,kg\,$\cdot$\,y & 0.7\,keV$_{\rm ee}$ & \cite{COSINE-100:2025kbw}\\
        COSINE-100 (DM-e) & NaI(Tl) (61.3\,kg) & Y2L & 2016-19 & 170\,kg\,$\cdot$\,y & 0.7\,keV$_{\rm ee}$ & \cite{COSINE-100:2025inn}\\
        \hline
        \multicolumn{7}{|c|}{ Liquid noble-gas scintillators } \\
        \hline
        DarkSide-50 (DM-e\,+\,Mig.) & Ar (19.4\,kg) & LNGS & 2015-17 & 12\,t\,$\cdot$\,d & 4\,e$^-$ (0.06\,keV$_{\rm er}$) & \cite{DarkSide:2022dhx, DarkSide:2022knj} \\    
        LUX (Mig.) & Xe (124\,kg) & SURF & 2013 & 0.01\,t\,$\cdot$\,y & 3.5\,e$^-$ ($\sim 1$\,keV$_{\rm nr}$) & \cite{LUX:2020yym}\\
        LZ (Mig.) & Xe (5.5\,t) & SURF & 2021-22 & 0.91\,t\,$\cdot$\,y & 1.6\,keV$_{\rm er}$ & \cite{LZ:2023poo}\\
        PandaX-II (DM-e) & Xe (117\,kg) & CJPL & 2016-18 & 47\,t\,$\cdot$\,d & 50\,p.e.  (0.08\,keV$_{\rm er}$) & \cite{PandaX-II:2021nsg} \\
        PandaX-4T (Mig.) & Xe (3.1\,t) & CJPL & 2020-21 & 0.55\,t\,$\cdot$\,y & 60\,p.e. (0.07\,keV$_{\rm er}$) & \cite{PandaX:2023xgl}\\
        PandaX-4T (DM-e) & Xe (3.7\,t) & CJPL & 2020-22 & 1.04\,t\,$\cdot$\,y & 4\,e$^-$ (0.04\,keV$_{\rm er}$) & \cite{PandaX:2025rrz, PandaX:2025rrz_erratum}\\
        XENONnT (DM-e) & Xe (5.9\,t) & LNGS & 2021 & $(0.011-3.4)$\,t\,$\cdot$\,d & 10\,p.e. ($\sim 14$\,eV$_{\rm er}$) & \cite{XENON:2024znc}\\
        XENONnT (DM-e) & Xe ($\sim5$\,t) & LNGS & 2021-25 & 7.8\,t\,$\cdot$\,y & $(3-4$)\,e$^-$ (0.04\,keV$_{\rm ee}$) & \cite{XENON:2026qow}\\
        XMASS (Mig.+ Brems.) & Xe (832\,kg) & Kamioka & 2013-19 & $\lesssim 3.6$\,t\,$\cdot$\,y & 1.0\,keV$_{\rm ee}$  & \cite{XMASS:2022tkr} \\
        \hline
        \multicolumn{7}{|c|}{ Other technologies } \\
        \hline
        QROCODILE (DM-e) & W$_{45}$Si$_{55}$ (1.67\,ng) & surface & 2023 & $\lesssim 29$\,ng\,$\cdot$\,d & 0.11\,eV$_{\rm er}$ &  \cite{QROCODILE:2024nqm} \\\hline

    \end{tabular}
    \label{tab:Running_ER}
\end{table}


\subsubsection*{Cryogenic crystal calorimeters}

\begin{figure}[t]
    \centering
    \includegraphics[width=0.49\linewidth]{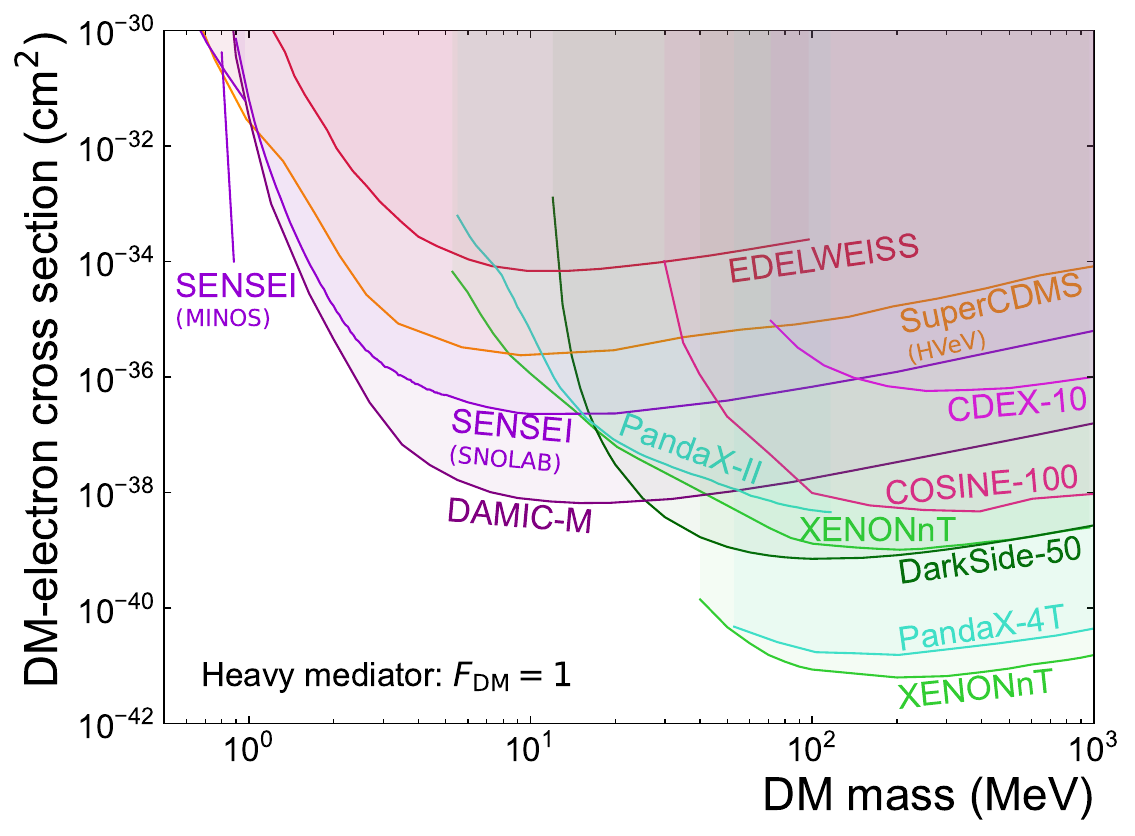}
    \hfill
    \centering
    \includegraphics[width=0.49\linewidth]{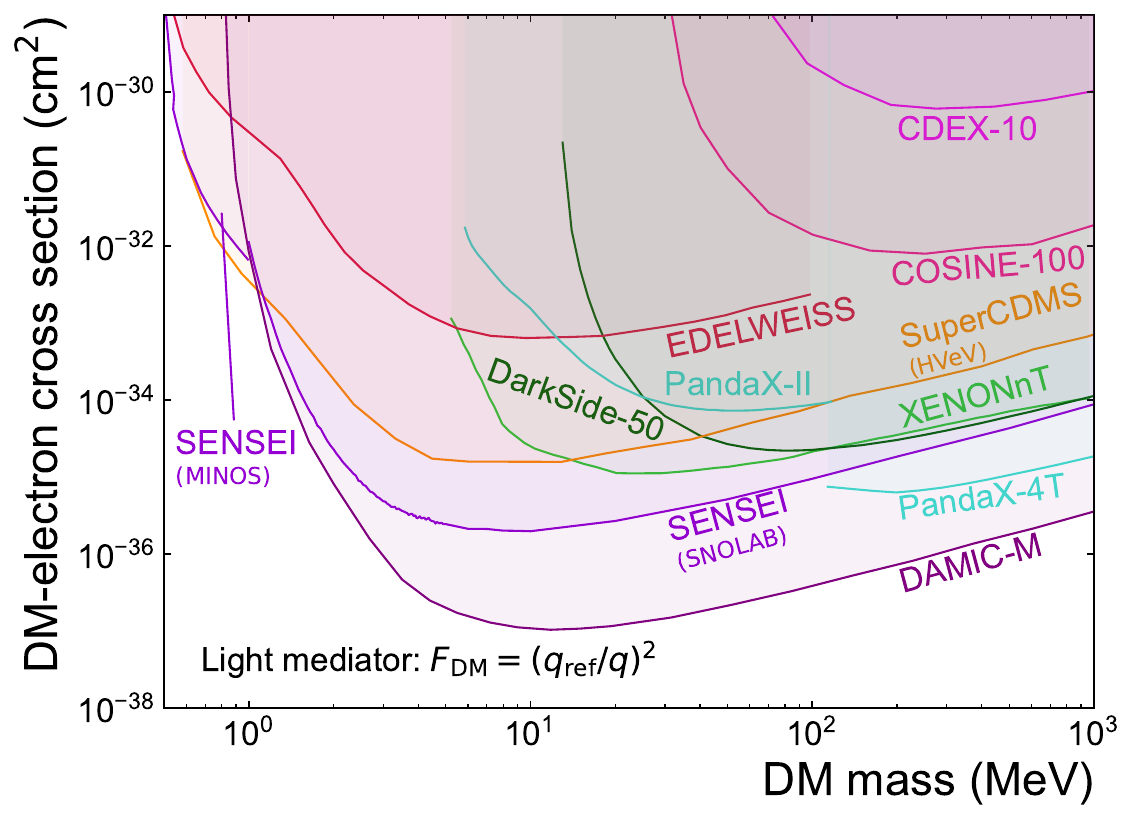}
    \caption{Exclusion region at 90\% C.L. for dark matter-electron scattering for heavy mediator (left panel) and light mediator (right panel). The experiments shown are: 
CDEX-10~\cite{CDEX:2022kcd}, COSINE-100~\cite{COSINE-100:2025inn}, 
DAMIC-M~\cite{DAMIC-M:2023gxo, DAMIC-M:2025luv}, 
DarkSide50~\cite{DarkSide:2022knj}, EDELWEISS~\cite{EDELWEISS:2020fxc},  PandaX-II~\cite{PandaX-II:2021nsg}, PandaX-4T~\cite{PandaX:2022xqx},
SENSEI~\cite{SENSEI:2020dpa,SENSEI:2023zdf}, 
SuperCDMS-HVeV~\cite{SuperCDMS:2025dha},
and XENONnT~\cite{XENON:2024znc,XENON:2026qow}. 
    }
    \label{fig:exp_running_ER}
\end{figure}

The \textbf{EDELWEISS} experiment, formerly operated at LSM, used cryogenic germanium detectors with simultaneous phonon and ionization readout to search for dark matter interactions. 
In a dedicated search for direct dark matter-electron scattering, a 33.4\,g detector was operated at a bias voltage of 78\,V to exploit NTL amplification of the phonon signal, achieving a baseline resolution of 0.53 electron-hole pairs~\cite{EDELWEISS:2020fxc}. 
Using an exposure of 80.72\,g\,$\cdot$\,d, the analysis achieved sensitivity to single electron-hole pairs, corresponding to a reconstructed signal of approximately 3\,eV$_{\rm ee}$ in germanium. 
This provides sensitivity to electronic excitations down to the germanium band-gap energy of approximately 0.7\,eV.
It provided the first germanium-based constraints on direct dark matter-electron scattering for dark matter masses below 1\,MeV (red line in Fig.~\ref{fig:exp_running_ER}).
Constraints were derived for both heavy-mediator and light-mediator interactions, reaching minimum cross sections of approximately $6\times10^{-35}\,\mathrm{cm^2}$ and $6\times10^{-34}\,\mathrm{cm^2}$, respectively, at dark matter masses around $(10-20)$\,MeV. 

In a separate search, a 200\,g germanium detector equipped with an NbSi transition-edge sensor (TES) was operated at 66\,V, achieving a baseline resolution of 4.46\,eV$_{\rm ee}$ and an analysis threshold of 30\,eV$_{\rm ee}$~\cite{EDELWEISS:2022ktt}. 
The threshold corresponds to approximately 690\,eV of NTL-amplified phonon energy and was chosen for the interpretation of spin-independent dark matter-nucleus scattering through the Migdal effect. 
Using 27.9\,d of data, resulting in a nominal exposure of approximately 5.58\,kg\,$\cdot$\,d before event-selection efficiencies, the analysis probes dark matter masses between 32\,MeV and 2\,GeV and excluded previously unexplored parameter space at cross sections around $10^{-29}\,\mathrm{cm^2}$ for masses between 32 and 100\,MeV (red line in Fig.~\ref{fig:exp_running_migdal}). 
Only the electronic contribution to the Migdal signal was considered, avoiding uncertainties associated with the nuclear-recoil quenching factor. 
The sensitivity was limited by a substantial population of heat-only events, a manifestation of the LEE discussed in Sec.~\ref{sssec:instrumental_bck}. 
The absence of associated ionization disfavors ordinary electronic recoils as their origin and points towards non-ionizing energy-release mechanisms, although their microscopic origin remains unresolved. 
These events were partially suppressed in the analysis by requiring an additional ionization signal.
The analysis furthermore explicitly accounts for the attenuation of the dark matter flux in the atmosphere, rock overburden, and detector shielding, resulting in a closed exclusion region at larger cross sections than shown in Fig.~\ref{fig:exp_running_migdal}) and limiting the mass reach to approximately 32\,MeV. 
Although EDELWEISS has concluded its dark matter search program, its cryogenic germanium detector development is being carried forward at LSM within TESSERACT, which builds on this expertise in low-threshold cryogenic Ge detectors~\cite{Billard:2024zvc}.\\

\textbf{SuperCDMS} has performed several searches for electronic-recoil signatures using cryogenic silicon and germanium detectors. 
A series of measurements with gram-scale Si HVeV detectors exploits NTL amplification to resolve individual electron-hole pairs and constrain direct dark matter-electron scattering for both heavy- and light-mediator interactions ~\cite{SuperCDMS:2020ymb,SuperCDMS:2024yiv}. 
These results have been superseded by a search using a 0.93\,g detector operated at a bias voltage of 100\,V and a net exposure of 6.1\,g\,$\cdot$\,d at NEXUS~\cite{SuperCDMS:2025dha}. 
In response to the findings in Ref.~\cite{SuperCDMS:2022zmd} that luminescence from materials surrounding the detector may constitute one source of the LEE, a redesigned detector holder substantially reduced background events associated with luminescence from nearby printed circuit boards, lowering the event rate by at least two orders of magnitude compared with the previous HVeV search. 
The analysis considered observed phonon energies between 85 and 500\,eV, corresponding to the first four electron-hole-pair peaks after NTL amplification (see Sec.~\ref{ssec:tech_bolom}). 
The lower analysis boundary of 85\,eV lies below the first pair peak and enables sensitivity to single electron-hole pairs, corresponding to a reconstructed signal of approximately 3.8\,eV$_{\rm ee}$ in silicon. 
The minimum detectable electronic excitation energy is governed by the silicon band gap of approximately 1.13\,eV. 
The search probes dark matter masses down to 0.5\,MeV and reaches minimum cross sections of approximately $5\times10^{-36}\,\mathrm{cm^2}$ for a heavy mediator and $2\times10^{-35}\,\mathrm{cm^2}$ for a light mediator, at masses of a few tens of MeV (orange lines in Fig.~\ref{fig:exp_running_ER}). 
The remaining LEE background is not sufficiently understood to construct a reliable background model, and the analysis therefore treats all observed events as potential signal. 

In a complementary analysis, SuperCDMS reanalyzed 36.9\,kg\,$\cdot$\,d of CDMSlite data from a 0.6\,kg germanium detector operated in high-voltage mode at Soudan~\cite{SuperCDMS:2023sql}.
With a bias voltage of approximately 70\,V and an analysis threshold of 70\,eV$_{\rm ee}$, the data were used to search for spin-independent dark matter-nucleus scattering through the Migdal effect and bremsstrahlung (orange line in Fig.~\ref{fig:exp_running_migdal}). 
The Migdal search reaches dark matter masses down to 30\,MeV at a cross section of $5.0\times10^{-30}\,\mathrm{cm^2}$, excluding previously unexplored parameter space between approximately 32 and 100\,MeV. 
The less sensitive bremsstrahlung search reaches 220\,MeV at $2.7\times10^{-30}\,\mathrm{cm^2}$, without excluding new parameter space. 
Both analyses account for attenuation of the dark matter flux through the atmosphere and Earth, which sets an upper-cross-section boundary to their exclusion regions.\\

\subsubsection*{Semiconductors}

The \textbf{CDEX} collaboration reanalyzed 205.4\,kg\,$\cdot$\,d of CDEX-10 data with an analysis threshold of 160\,eV$_{\rm ee}$ to constrain direct dark matter-electron scattering, considering a heavy mediator, electric-dipole coupling, and a light mediator~\cite{CDEX:2022kcd}. 
The analysis threshold corresponds to an electron-equivalent energy calibrated using the ionization signal, with a combined detection efficiency of 4.5\% at threshold. 
Despite its higher threshold compared with cryogenic semiconductor detectors, the substantially larger exposure and low radioactive background provided competitive sensitivity at higher dark matter masses, reaching minimum cross sections of approximately $5\times10^{-37}\,\mathrm{cm^2}$ for the heavy mediator and a few $10^{-31}\,\mathrm{cm^2}$ for the light mediator at dark matter masses around 1\,GeV. 
For the heavy-mediator scenario, CDEX-10 provided the strongest constraint among solid-state detectors above 90\,MeV at the time of publication (light pink lines in Fig.~\ref{fig:exp_running_ER})

\begin{figure}[t]
    \centering
    \includegraphics[width=0.65\linewidth]{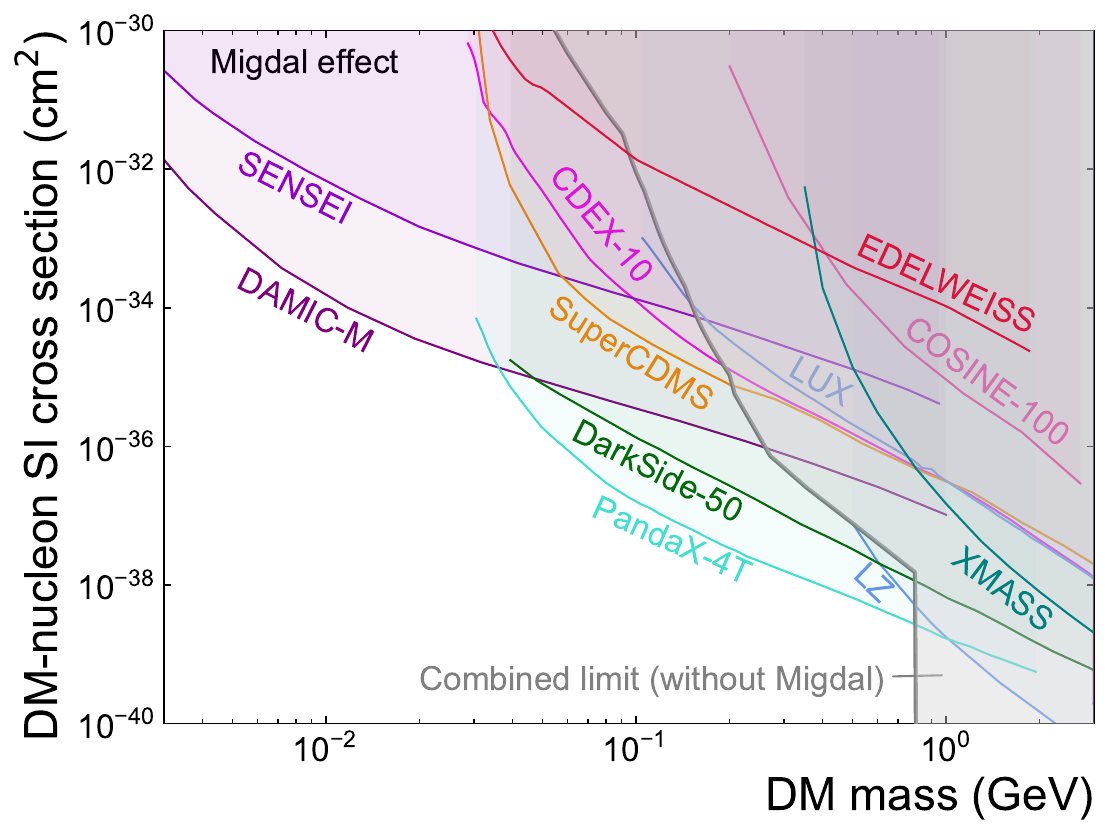}
    \caption{Left: Exclusion region at 90\% C.L. for electron ionisation from nuclear scattering via the Migdal effect. The experiments shown are: CDEX-10~\cite{CDEX:2021cll}, 
COSINE-100~\cite{COSINE-100:2021poy},
DAMIC-M~\cite{DAMIC-M:2025luv},
DarkSide-50~\cite{DarkSide:2022dhx},
EDELWEISS~\cite{EDELWEISS:2022ktt},
LUX~\cite{LUX:2020yym},
LZ~\cite{LZ:2023poo},
PandaX-4T~\cite{PandaX:2023xgl},
SENSEI~\cite{SENSEI:2023zdf},
SuperCDMS~\cite{SuperCDMS:2023sql} and 
XMASS~\cite{XMASS:2022tkr}.
}
    \label{fig:exp_running_migdal}
\end{figure}

CDEX-1B and CDEX-10 data were additionally used to constrain dark matter-nucleus scattering through the Migdal effect, including the attenuation of the dark matter flux in the Earth~\cite{CDEX:2021cll}. 
The analysis modeled the energy deposited in the detector as the sum of the electronic energy from the Migdal process and the quenched nuclear-recoil energy. 
A dedicated Monte Carlo simulation accounted for multiple dark matter-nucleus scatterings in the Earth and the detailed topography of the Jinping Mountain overburden. 
The resulting spin-independent exclusion regions are bounded at both low and high cross sections, extending the sensitivity to dark matter masses down to approximately 30\,MeV and excluding previously unexplored parameter space at large cross sections at the time of publication (light pink line in Fig.~\ref{fig:exp_running_migdal}).
At a dark matter mass of 100\,MeV, including the detailed mountain topography changes the upper exclusion boundary by approximately 45\% compared with a flat-overburden approximation. 
Spin-dependent neutron-only constraints were also derived, exploiting the non-zero nuclear spin of $^{73}$Ge to exclude previously unexplored parameter space at sub-GeV dark matter masses.\\

The \textbf{DAMIC-M} collaboration searches for electronic-recoil signatures using silicon skipper CCDs operated at LSM, achieving sub-electron charge resolution.
The first underground search used two CCDs with a combined target mass of approximately 18\,g and an exposure of 85.23\,g\,$\cdot$\,d, probing direct dark matter-electron scattering down to masses of 0.53\,MeV with a single-electron threshold~\cite{DAMIC-M:2023gxo}.
This threshold corresponds to an effective ionization threshold of approximately 1.1\,eV$_{\rm ee}$, the gap energy of silicon.
The resulting limits reached minimum cross sections of approximately $2\times10^{-37}\,\mathrm{cm^2}$ (heavy mediator) and $3\times10^{-36}\,\mathrm{cm^2}$ (light mediator).

A subsequent analysis of 1.257\,kg\,$\cdot$\,d of selected data achieved a charge resolution of about 0.16\,e$^-$ and searched for spatially correlated clusters containing two to four electrons~\cite{DAMIC-M:2025luv}.
A two-electron charge threshold corresponds approximately to an energy threshold of 4.9\,eV$_{\rm ee}$, which is the sum of the gap energy and the average ionization energy of 3.8\,eV, required to create an electron-hole pair in silicon.
The resulting constraints cover dark matter masses between 1\,MeV and 1\,GeV, reaching minimum cross sections of approximately $8\times10^{-39}\,\mathrm{cm^2}$ and $1\times10^{-37}\,\mathrm{cm^2}$ for heavy and light mediators, respectively (dark purple line in Fig.~\ref{fig:exp_running_ER}).
Notably, these constraints exclude benchmark scenarios in which hidden-sector dark matter is produced through freeze-in or freeze-out in the early Universe.
The same dataset was also used to search for spin-independent dark matter-nucleus scattering through the Migdal effect, yielding leading constraints for dark matter masses between about 1 and 35\,MeV and reaching a cross section of approximately $8\times10^{-35}\,\mathrm{cm^2}$ at 10\,MeV for a heavy mediator~\cite{DAMIC-M:2025luv} (dark purple line in Fig.~\ref{fig:exp_running_migdal}).
At the lowest masses, however, the Migdal constraints may be affected by Earth shielding at large cross sections, an effect that was not fully incorporated in the analysis.
An excess of isolated multi-electron pixels was observed, with indications of a possible readout-related origin, while the previously reported low-energy excess in DAMIC at SNOLAB was not directly investigated in this search.\\

The \textbf{SENSEI} collaboration searches for electronic-recoil signatures using silicon Skipper CCDs with single-electron sensitivity. 
An early measurement with a 1.93\,g detector operated in the MINOS cavern at Fermilab probed direct dark matter-electron scattering down to masses of about 0.5\,MeV (light purple lines in Fig.~\ref{fig:exp_running_ER}), reaching minimum cross sections of approximately $2\times10^{-37}\,\mathrm{cm^2}$ for a heavy mediator and $5\times10^{-36}\,\mathrm{cm^2}$ for a light mediator~\cite{SENSEI:2020dpa}.
The same dataset was also used to constrain dark matter-nucleus scattering through the Migdal effect, extending the sensitivity to masses as low as about 0.6\,MeV for a light mediator, although the collaboration cautions that theoretical predictions become increasingly uncertain at such low masses~\cite{SENSEI:2020dpa}.
These data also revealed a correlation between the rate of single-electron events and higher-energy backgrounds, suggesting that at least part of the LEE population is associated with environmental radiation rather than intrinsic dark current.

A subsequent SENSEI detector at SNOLAB, comprising six 2.19\,g Skipper CCDs with a total active mass of about 13.1\,g, used a net exposure of 100.72\,g\,$\cdot$\,d and charge clusters containing two to ten electrons to obtain stronger constraints above about 1\,MeV~\cite{SENSEI:2023zdf}.
The limits reached minimum cross sections of approximately $2\times10^{-37}\,\mathrm{cm^2}$ and $2\times10^{-36}\,\mathrm{cm^2}$ for the heavy- and light-mediator scenarios, respectively (light purple lines in Fig.~\ref{fig:exp_running_ER}).
The same SNOLAB data were also interpreted in terms of dark matter-nucleus scattering through the Migdal effect for both heavy and light mediators, probing previously unexplored parameter space below approximately 30\,MeV and 40\,MeV, respectively, with cross sections of approximately $5\times10^{-33}\,\mathrm{cm^2}$ and $2\times10^{-31}\,\mathrm{cm^2}$ at 10\,MeV (light purple line in Fig.~\ref{fig:exp_running_migdal}).\\

\subsubsection*{Scintillation crystals}

Beyond its conventional nuclear-recoil searches, \textbf{COSINE-100} has reinterpreted data of its NaI(Tl) crystals in the electronic-recoil channel to probe dark matter interactions at lower masses.
A first dedicated search for dark matter-nucleus scattering through the Migdal effect used an effective exposure of 97.7\,kg\,$\cdot$\,y and a 1\,keV$_{\rm ee}$ analysis threshold, probing spin-independent and spin-dependent proton-only interactions for dark matter masses down to 0.2\,GeV~\cite{COSINE-100:2021poy}.  
The spin-independent search reached a cross section of approximately $1\times10^{-36}\,\mathrm{cm^2}$ at 2\,GeV (dark pink line in Fig.~\ref{fig:exp_running_migdal}).
A subsequent analysis using three years of data from five crystals with a combined mass of 61.3\,kg lowered the analysis threshold to 0.7\,keV$_{\rm ee}$, corresponding to approximately eight detected photoelectrons, and constrained spin-dependent proton scattering including the Migdal effect, superseding the previous results~\cite{COSINE-100:2025kbw}.
The estimated gross exposure was $\lesssim175.3$\,kg\,$\cdot$\,y inferred from the target mass and 2.86 years of data taking.
The resulting limits cover dark matter masses between about 0.1 and 5\,GeV, reaching cross sections of approximately $1\times10^{-33}\,\mathrm{cm^2}$ at 2\,GeV.

More recently, COSINE-100 used an exposure of 172.9\,kg\,$\cdot$\,y with the same 0.7\,keV$_{\rm ee}$ threshold to search directly for dark matter-electron scattering (dark pink line in Fig.~\ref{fig:exp_running_ER}), reaching minimum cross sections of about $5\times10^{-39}\,\mathrm{cm^2}$ at 0.3\,GeV for a heavy mediator and $6.4\times10^{-33}\,\mathrm{cm^2}$ at 0.25\,GeV for a light mediator~\cite{COSINE-100:2025inn}. 
The analysis also estimated the attenuation of the dark matter flux due to Earth shielding at large cross sections, although the resulting upper exclusion boundaries are subject to approximations in the overburden model.\\

\subsubsection*{Liquid noble-gas scintillators}

\textbf{DarkSide-50} has also performed searches for electronic-recoil signatures using its dual-phase liquid-argon target, based on a low-threshold S2-only analysis with a fiducial mass of 19.4\,kg and a net exposure of 12.306\,t\,$\cdot$\,d~\cite{DarkSide:2022dhx,DarkSide:2022knj}.
The analyses considered an ROI from 4 to 170\,extracted electrons, corresponding to approximately $(0.06-21)$\,keV$_{\rm er}$ based on the calibrated ionization response.
For spin-independent dark matter-nucleus scattering, the nuclear-recoil and Migdal-induced electronic-recoil contributions were combined in the signal model, extending the sensitivity down to dark matter masses of about 40\,MeV and reaching a cross section of approximately $6\times10^{-38}\,\mathrm{cm^2}$ at 400\,MeV~\cite{DarkSide:2022dhx} (dark green line in Fig.~\ref{fig:exp_running_migdal}).
As discussed in Sec.~\ref{sssec:exp_running_NR}, the poorly constrained ionization response at very low nuclear-recoil energies was treated using models with and without quenching fluctuations, while below about 0.5\,GeV the sensitivity is dominated by the Migdal contribution and becomes largely independent of this treatment.
DarkSide-50 used the same data set to also search for direct dark matter-electron scattering for both heavy and light mediators (dark green lines in Fig.~\ref{fig:exp_running_ER}), reaching cross sections of approximately $7\times10^{-40}\,\mathrm{cm^2}$ and $2\times10^{-35}\,\mathrm{cm^2}$, respectively, at a dark matter mass of 100\,MeV~\cite{DarkSide:2022knj}.
These are currently the strongest constraints for masses between about 25 and 40\,MeV for the heavy mediator case.\\

The \textbf{LUX-ZEPLIN (LZ)} experiment and its direct predecessor LUX have both searched for dark matter-nucleus scattering through the Migdal effect using their liquid-xenon targets.
The low-threshold S2-only LUX analysis discussed in Sec.~\ref{sssec:exp_running_NR} was additionally interpreted in terms of spin-independent interactions, using an effective exposure of approximately 5\,t\,$\cdot$\,d~\cite{LUX:2020yym}.
A machine-learning classifier based on the S2 pulse shape suppressed backgrounds originating from the TPC electrodes by about a factor of four while retaining approximately 60\% signal efficiency, enabling an analysis threshold of 3.5 detected electrons corresponding to about 1\,keV$_{\rm nr}$.
A direct conversion of the ionization threshold into electron-equivalent energy is not provided and cannot be uniquely inferred. 
However, the relationship between the ionization signal and deposited energy was modeled using NEST, with the electron-recoil charge yield conservatively set to zero below 0.186\,keV$_{\rm ee}$ in the Migdal analysis.
The inclusion of the Migdal effect extended the mass reach down to 0.15\,GeV, with the corresponding cross-section limit improving from about $10^{-33}\,\mathrm{cm^2}$ at 0.15\,GeV to slightly below $10^{-37}\,\mathrm{cm^2}$ at 2\,GeV (blue line in Fig.~\ref{fig:exp_running_migdal}).

The subsequent LZ analysis used a fiducial mass of 5.5\,t and an exposure of 0.91\,t\,$\cdot$\,y, considering electronic-recoil energies between approximately 1 and 15\,keV$_{\rm er}$~\cite{LZ:2023poo}.
The detection efficiency reached 50\% at an electronic-recoil energy of 1.56\,keV${\rm er}$.
Constraints were derived for spin-independent as well as spin-dependent neutron and proton interactions, conservatively including only the electronic-recoil component of the Migdal signal.
The spin-independent search reached a cross section of approximately $2\times10^{-40}\,\mathrm{cm^2}$ at 2\,GeV (blue line in Fig.~\ref{fig:exp_running_migdal}), while the spin-dependent searches reached approximately $3\times10^{-36}\,\mathrm{cm^2}$ and $1\times10^{-34}\,\mathrm{cm^2}$ at 3\,GeV for neutron and proton couplings, respectively.
In particular, the spin-dependent neutron search excluded previously unexplored parameter space between about 1 and 3\,GeV.
However, the requirement of a detectable S1 signal limited the LZ mass reach to approximately 0.5\,GeV, leaving the earlier LUX result relevant at lower masses.
Nuclear-structure uncertainties are not included in the quoted limits and can change the spin-dependent neutron and proton constraints on average by factors of $0.74-2.7$ and $0.81-44$, respectively.\\

The \textbf{PandaX} collaboration has used the low-threshold ionization channel of its liquid-xenon TPCs to search for electronic-recoil signatures of light dark matter (cyan lines in Fig.~\ref{fig:exp_running_ER} and \ref{fig:exp_running_migdal}). 
An initial PandaX-II search for direct dark matter-electron scattering used a fiducial mass of 117\,kg and an exposure of 46.9\,t\,$\cdot$\,d, with an analysis threshold of 50\,p.e. in the S2 signal, corresponding to approximately 0.08\,keV$_{\rm er}$~\cite{PandaX-II:2021nsg}.
This search provided the strongest constraints at the time for a heavy mediator between about 15 and 30\,MeV.
The search was substantially extended in reach with PandaX-4T, which used a fiducial mass of 3.1\,t and an effective exposure of 0.55\,t\,$\cdot$\,y, with an S2-only threshold of 60\,p.e., corresponding to approximately 0.07\,keV$_{\rm er}$ for electronic recoils or 0.8\,keV$_{\rm nr}$ for nuclear recoils, according to the respective ionization-response models~\cite{PandaX:2022xqx}.
The most recent PandaX-4T analysis, using a fiducial mass of approximately 3.7\,t and an effective ionization-only exposure of 1.04\,t\,$\cdot$\,y, lowered the analysis threshold to four detected electrons, corresponding to approximately 0.04\,keV$_{\rm er}$~\cite{PandaX:2025rrz,PandaX:2025rrz_erratum}.
This improved the previous constraints by factors of about 1.5 and 9.3 for heavy and light mediators, respectively, between approximately 50\,MeV and 10\,GeV, reaching cross sections of $1.5\times10^{-41}\,\mathrm{cm^2}$ and $5.6\times10^{-37}\,\mathrm{cm^2}$ at 200\,MeV.

PandaX-4T has additionally searched for spin-independent dark matter-nucleus scattering through Migdal-induced electronic recoils, using the earlier 3.1\,t fiducial target and an effective exposure of 0.55\,t\,$\cdot$\,y~\cite{PandaX:2023xgl}.
The S2-only analysis had an electron-recoil response threshold of approximately 0.07\,keV$_{\rm er}$. 
Considering mediator masses of 0.1 and 1\,MeV as well as the heavy-mediator limit, the analysis extended the sensitivity to dark matter masses between 30\,MeV and 2\,GeV, excluding previously unexplored parameter space for dark-mediator interactions.
At a dark matter mass of 1\,GeV, the corresponding cross-section limits are approximately $5\times10^{-35}\,\mathrm{cm^2}$, $3\times10^{-38}\,\mathrm{cm^2}$, and $2\times10^{-39}\,\mathrm{cm^2}$, respectively.\\

The \textbf{XENONnT} collaboration has performed two complementary ionization-only searches for light dark matter using its liquid-xenon TPC. 
The first analysis exploited single- and few-electron signals with an S2 threshold of 10\,p.e., corresponding to a minimum electronic-recoil energy of 13.7\,eV$_{\rm er}$ in the adopted detector-response model~\cite{XENON:2024znc}.
This energy corresponds to the production of a single detectable quantum, using an average energy of 13.7\,eV per scintillation photon or ionization electron in liquid xenon, rather than a direct conversion of the measured S2 signal.
The analysis used a variable effective exposure between 0.002 and 5.03\,t\,$\cdot$\,d, depending on the reconstructed electron multiplicity and dataset, reflecting increasingly stringent spatial and temporal background vetoes for smaller signals.
It constrained direct dark matter-electron scattering for both heavy and light mediators, assuming standard halo dark matter distribution.
The lowest cross section for the heavy and light mediator case are about $1\times10^{-39}\,\mathrm{cm^2}$ at 200\,MeV and $4\times10^{-35}\,\mathrm{cm^2}$ at 30\,MeV, respectively (light green lines in Fig.~\ref{fig:exp_running_ER}).

A subsequent XENONnT analysis used a fiducial mass of approximately 5\,t and an exposure of 7.83\,t\,$\cdot$\,y, providing sensitivity to electronic recoils between 0.04 and 0.7\,keV$_{\rm ee}$~\cite{XENON:2026qow}.
Here, the electron-equivalent energy scale was obtained from the modeled ionization response to electronic recoils, including the energy-dependent charge yield and the electron extraction and detection efficiencies.
The analysis employed a detailed model of instrumental backgrounds together with machine-learning-based rejection techniques for cathode, delayed-electron, and accidental-electron events.
The resulting data were used to constrain direct dark matter-electron scattering for a heavy mediator, reaching a cross section of $6.3\times10^{-42}\,\mathrm{cm^2}$ at a dark matter mass of 200\,MeV and improving upon previous constraints (light green lines in Fig.~\ref{fig:exp_running_ER}).
An alternative xenon ionization response calculation  was also considered to assess the associated model dependence.\\

\textbf{XMASS-I} extended its search for spin-independent dark matter-nucleon scattering to lower masses using the Migdal effect and bremsstrahlung~\cite{XMASS:2022tkr}. 
The analysis used the full 832\,kg liquid-xenon target and 1590.9\,live days, corresponding to a gross exposure of approximately 3.63\,t\,$\cdot$\,y.
It considered scintillation signals between 1 and 20\,keV$_{\rm ee}$, where the electron-equivalent energy scale was established from gamma-ray calibrations and a model of the energy-dependent scintillation light yield.
The Migdal analysis probed dark matter masses between 0.35 and 4\,GeV and reached an upper limit of $1.4\times10^{-35}\,\mathrm{cm^2}$ at 0.5\,GeV (teal line in Fig.~\ref{fig:exp_running_migdal}), while the bremsstrahlung analysis covered masses between 0.32 and 1\,GeV and constrained the cross section to $1.1\times10^{-33}\,\mathrm{cm^2}$ at 0.5\,GeV.\\

\subsubsection*{Other technologies}

The same \textbf{QROCODILE} dataset as discussed in Sec.~\ref{sssec:exp_running_NR} was also used to constrain direct dark matter-electron scattering for both light- and heavy-mediator scenarios~\cite{QROCODILE:2024nqm}.
Both mediator scenarios probe previously unexplored parameter space below about 500\,keV, reaching cross sections slightly below $10^{-27}\,\mathrm{cm^2}$ and slightly above $10^{-25}\,\mathrm{cm^2}$ for light and heavy mediators, respectively, at 500\,keV.
As discussed in Sec.~\ref{sssec:exp_running_NR}, the collaboration estimated the cross sections at which atmospheric attenuation becomes relevant and truncated the exclusion regions accordingly.


\subsubsection{Modulation}
\label{sssec:exp_running_modulation}

\renewcommand{\arraystretch}{1.2}
\begin{table}[H]
    \centering
    \caption{
    Results of dark matter experiments searching for modulation signals published since 2020 (or before if still world-leading). For each entry, the following information is reported: the experiment name, the signature type (electron or nuclear recoil or model-independent) including if the modulation was annual (AM) or diurnal (DM), the target, the data collection period and accumulated exposure, the energy threshold and references corresponding to the latest results. Masses refer to the target or fiducial volume used for the respective analysis, where available. Exposures are the net exposures, if provided. Else the gross exposure is cited. If neither is provided, it is approximated (indicated with $\sim$ or $\lesssim$) based on the given mass and live time. Thresholds are quoted in the energy scale or experimental observable used by the collaboration, with approximate conversions to comparable energy scales where appropriate. The experiments are grouped according to their technology.
    }
    \footnotesize
    \begin{tabular}{|c|c|c|c|c|c|c|c|}\hline
        {\bf Experiment} & {\bf Signature} & \makecell{\bf Target \\\bf(Mass)} & {\bf Location} &
        \makecell{\bf Data\\\bf Taking} &
        {\bf Exposure} & {\bf Threshold} & {\bf Reference(s)}\\\hline 
        \hline
        \multicolumn{8}{|c|}{ Scintillation crystals } \\
        \hline
        DAMA/LIBRA & MI/AM & NaI(Tl) (250\,kg) & LNGS & 2010-20 & 1\,100\,kg\,$\cdot$\,y  & 1\,keV$_\mathrm{ee}$ & \cite{Bernabei:2020mon}\\
        DAMA/LIBRA & MI/AM & NaI(Tl) (250\,kg) & LNGS & 2021-24 & 540\,kg\,$\cdot$\,y  & 0.5\,keV$_\mathrm{ee}$ & \cite{Bernabei:2026fdk}\\
        ANAIS-112 & NR,ER,MI/AM & NaI(Tl) (112.5\,kg) & LSC & 2017-23 & 630\,kg\,$\cdot$\,y & 1\,keV$_\mathrm{ee}$ & \cite{Amare:2025dfq,ANAIS-112:2025fne}\\
        COSINE-100 & NR,ER,MI/AM & NaI(Tl) (61.3\,kg) & Y2L & 2016-23 & 360\,kg\,$\cdot$\,y & 0.7\,keV$_\mathrm{ee}$ & \cite{COSINE-100:2024nfa,ANAIS-112:2025fne}\\
        \hline
        \multicolumn{8}{|c|}{ Semiconductors } \\
        \hline
        CDEX-1B & NR,MI/AM & Ge (939\,g) & CJPL & 2014-17 & 1\,100 kg\,$\cdot$\,d &  0.2\,keV$_\mathrm{ee}$ &\cite{Li:2021hhx}\\
        DAMIC-M & ER,MI/DM & Si (26.4\,g) & LSM & 2022 & 40\,g\,$\cdot$\,d & $\mathcal{O}$(1\,eV), 0.2\,e$^-$ & \cite{DAMIC-M:2023hgj, DAMIC-M:2025ltz}\\
        SENSEI & ER/DM & Si (4.38\,g) & MINOS & 2024 & $\lesssim 150$\,g\,$\cdot$\,d & $\mathcal{O}$(1\,eV) & \cite{SENSEI:2025qvp}\\
        \hline
        \multicolumn{8}{|c|}{ Liquid Noble Gas scintillators } \\
        \hline
        DarkSide-50 & ER,MI/AM & Ar (19.4\,kg) & LNGS & 2015-18 & $\lesssim$37\,kg\,$\cdot$\,y & 40\,eV$_\mathrm{ee}$ & \cite{DarkSide-50:2023fgf}\\
          XMASS & ER,NR,MI/AM & Xe (832\,kg) & Kamioka & 2013-19 & $\lesssim 3.6$\,t\,$\cdot$\,y & $\sim$\,$(0.5 - 1.0)$\,keV$_{\rm ee}$  & \cite{XMASS:2022tkr}\\
        \hline
    \end{tabular}
    \label{tab:Running_mod}
\end{table}

This section summarizes experiments that are currently running or have been recently decommissioned but have reported results from searches for dark matter time modulations. Table~\ref{tab:Running_mod} lists the target material, the host laboratory for each experiment, the accumulated exposure, data-taking period, energy threshold, and the specific signature investigated, grouped by technology (Sec.~\ref{sec:tech}). 
Regarding potential signatures, they are categorized into annual and diurnal modulation searches (see Sec.~\ref{sec:theory}), which are detailed in the following subsections. Furthermore, these searches can be classified as either model-independent or model-dependent. 
Model-independent analyses look for periodic variations in the event rate without assuming a specific dark matter particle or halo model. 
While such observations may be compatible with a dark matter signal, their interpretation in terms of specific dark matter candidates requires additional assumptions about the particle interactions and the Galactic dark matter distribution.
In Tab.~\ref{tab:Running_mod}, results from experiments without event-by-event discrimination between nuclear and electronic recoils are labeled MI when reporting model-independent modulation searches.
Results interpreted under the assumption of dark matter interactions with nuclei or electrons are instead labeled NR or ER, respectively.
\\

\noindent{}\emph{Annual modulation}

The \textbf{DAMA/LIBRA} experiment using $\sim$250\,kg of NaI(Tl) operated at LNGS from 2003 to 2010 (Phase1) with a 2\,keV$_\mathrm{ee}$ energy threshold and from 2010 to 2024 (Phase2) with a threshold at 1\,keV$_\mathrm{ee}$~\cite{Bernabei:2020mon}, and 0.5\,keV$_\mathrm{ee}$ for the last three years (Phase2-empowered)~\cite{Bernabei:2026fdk}. These data provided consistent and model-independent evidence for a year-period modulation in the detection rate below 6\,keV$_\mathrm{ee}$, with all the key features expected of a signal arising from interactions with dark matter particles distributed in the Galactic halo. Results in the $(1-6)$\,keV$_\mathrm{ee}$ energy region correspond to an exposure of 2.07\,t\,$\cdot$\,y over 11 annual cycles (Phase2\,+\,Phase2-empowered), providing evidence for a modulation with the features of a DM signal with 13.9\,$\sigma$, while in the $(2-6)$\,keV$_\mathrm{ee}$, adding data
from DAMA/NaI and DAMA/LIBRA-phase1, the exposure reaches 3.40\,t\,$\cdot$\,y, and provides a modulated signal with 15.3\,$\sigma$~\cite{Bernabei:2026fdk}. \\

\begin{figure}[h]
\centering
\includegraphics[width=0.9\columnwidth]{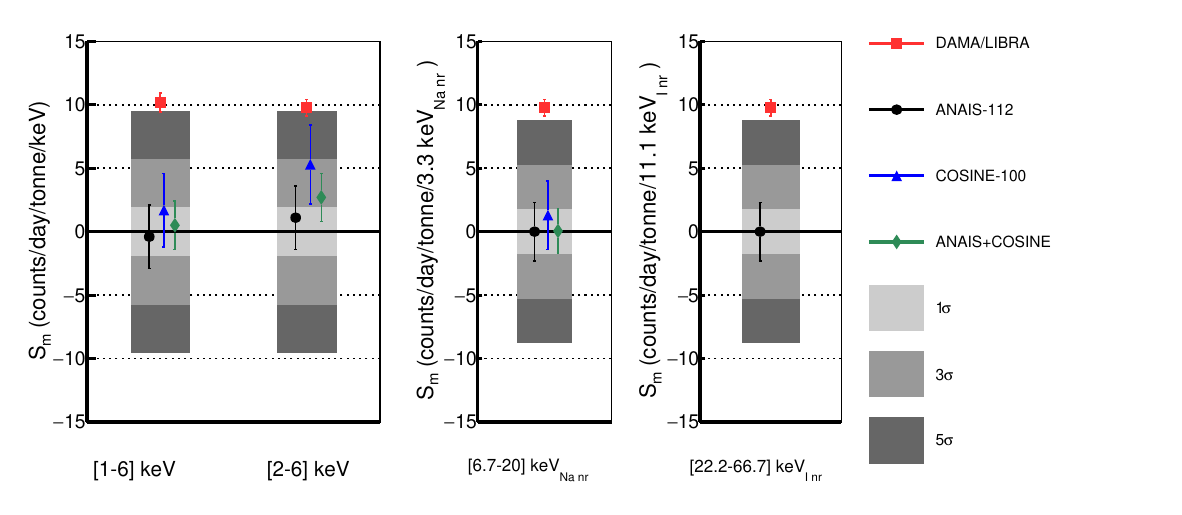}
\caption{Modulation amplitudes derived from NaI(Tl) experiments: DAMA/LIBRA full dataset \cite{Bernabei:2026fdk}, COSINE-100 full dataset (6.4\,years)~\cite{COSINE-100:2024nfa} and 6-year exposure ANAIS-112~\cite{Amare:2025dfq}. ANAIS+COSINE combined modulation result is also shown~\cite{ANAIS-112:2025fne}. Results are shown in the electron equivalent energy regions (left panels) and in the sodium/iodine nuclear recoil scales (middle and right panels). For the nuclear recoil scale results, constant QF$_\text{Na}=0.2$ and QF$_\text{I}=0.06$ for ANAIS crystals have been considered. The gray bands correspond to 1, 3 and 5\,$\sigma$ sensitivity of the 6-year ANAIS-112 result. }
\label{fig:modelindependent_mod_results}
\end{figure}

The \textbf{ANAIS-112} experiment was designed, constructed, and operated with the explicit goal of directly testing the DAMA/LIBRA annual modulation signal. 
It uses the same target material, NaI(Tl), and employs similar analysis techniques, enabling a model-independent comparison while incorporating specific improvements in hardware, software, and analysis procedures.
However, the achieved background level in the ROI is higher than DAMA/LIBRA's~\cite{Martinez:2026thh}.
ANAIS-112 consists of nine NaI(Tl) modules, 12.5\,kg each, with a total target mass of 112.5\,kg installed at the LSC which has been taking data from August 2017 until January 2026. The analysis threshold was set at 1\,keV$_\mathrm{ee}$ in order to guarantee bulk scintillation events efficiency above 20\% while triggering well below 1\,keV$_\mathrm{ee}$ was granted by the detection of 0.87\,keV$_\mathrm{ee}$ events from the decay of $^{22}$Na cosmogenically produced in the crystals, tagged by the detection of the high-energy accompanying photon interacting in a second module~\cite{Amare2019ANAIS112}. 
It has reported results based on six years of data, corresponding to an effective exposure of 625.75\,kg\,$\cdot$\,y, excluding the DAMA/LIBRA annual modulation signal at the 4\,$\sigma$ level (see Fig.~\ref{fig:modelindependent_mod_results}). The experiment is still analyzing its full dataset, which, according to sensitivity estimates, could increase the incompatibility to 5\,$\sigma$, while in parallel, efforts to reduce the analysis threshold are ongoing. \\

Also the \textbf{COSINE-100} experiment (see Sec.~\ref{sssec:exp_running_NR} and \ref{sssec:exp_running_ER}) is specifically built to probe the standing annual modulation signal and follows the same principles as described for ANAIS-112. 
Similarly, the background level in the ROI is higher than in the case of DAMA/LIBRA~\cite{Martinez:2026thh}.
The collaboration has reported annual modulation results based on six years of data (full dataset) amounting to 358\,kg\,$\cdot$\,y, with an analysis threshold reduced to 0.7\,keV$_\mathrm{ee}$ with respect to previous analysis, excluding the DAMA/LIBRA modulation signal at the 3\,$\sigma$ level (see Fig.~\ref{fig:modelindependent_mod_results})~\cite{COSINE-100:2024nfa}. \\

A combined ANAIS-112 and COSINE-100 analysis taking into account both known time-dependent and time-independent backgrounds for each experiment using three years of data has also been published~\cite{ANAIS-112:2025fne}. This publication also includes a simple statistical combination of the results obtained with six years of data from each experiment, also shown in Figure~\ref{fig:modelindependent_mod_results}. 
Even when potential differences in the quenching factor among different detectors, efficiency and resolution are taken into account, the results are very difficult to reconcile with a dark matter interpretation of the DAMA/LIBRA signal without significant tuning~\cite{Busoni:2025tqa}. \\

Experiments using other technologies and targets have searched for an annual modulation in their detection rates.
The comparison with the DAMA/LIBRA result is in those cases dependent on the models chosen for the dark matter particle and halo. For instance, XENON100~\cite{XENON:2017nik} and LUX~\cite{LUX:2018xvj} presented negative results for annual modulation searches in their electron recoil rates.\\

More recently \textbf{CDEX-1B} (see Sec.~\ref{sssec:exp_running_NR} and \ref{sssec:exp_running_ER}) provided results of an annual modulation analysis using p-type point-contact Ge detectors with 939\,g fiducial mass and profiting from a stable data taking of more than four years in two different experimental configurations (with and without an anti-Compton detector, replaced by a passive copper shielding in the second run). The low analysis threshold of about 200\,eV$_\mathrm{ee}$ provides a high sensitivity to low-mass dark matter energy depositions. The total exposure used in the annual modulation analysis is 1\,107.5\,kg\,$\cdot$\,d. No evidence of modulation was found. Interpreted in a model dependent analysis (SI), their data provided upper limits of $1.99\times10^{-41}\,\mathrm{cm^2}$ at 7.9\,GeV mass \cite{Li:2021hhx}.\\

\textbf{DarkSide-50} (see Sec.~\ref{sssec:exp_running_NR} and \ref{sssec:exp_running_ER}) 
was a dual-phase TPC operated at LNGS, featuring an active target mass of $(46.4\pm 0.7)$\,kg active liquid argon target. The annual modulation in the ionization signal was searched through 693.3\,days (live time) of data with the 19.4\,kg of argon (fiducial volume) resulting in an approximate exposure of at most 36.8\,t\,$\cdot$\,y~\cite{DarkSide-50:2023fgf}. The energy threshold was 40\,eV$_\mathrm{ee}$, the lowest achieved in a dark matter annual modulation search. The analysis considers different approaches. On the one hand, maximum likelihood fitting in the energy regions $(0.06-2)$ and $(2-6)$\,keV$_\mathrm{ee}$, without constraining the phase, and in the range $(0.04-20)$\,keV$_\mathrm{ee}$ with fixed phase and bin widths along the energy axis. On the other hand, a periodogram analysis is performed. No significant modulation is found, but the significance of the results was not sufficient to confirm or reject the DAMA/LIBRA claim. \\   

\textbf{XMASS-I} (see Sec.~\ref{sssec:exp_running_NR} and \ref{sssec:exp_running_ER}) searched for WIMP induced annual modulation signatures using 832\,kg of liquid xenon as target with an exposure of 420\,kg\,$\cdot$\,y. After accounting for dead time and event selection efficiencies, the total collected live time was 1\,590.9\,days with the standard threshold of 1\,keV$_\mathrm{ee}$. Additional low threshold (0.5\,keV$_\mathrm{ee}$) data
corresponds to 768.8\,days live time. No modulation was found, enabling to constrain SI cross sections down to $2.3\times10^{-42}\,\mathrm{cm^2}$ for a 8\,GeV WIMP mass \cite{XMASS:2022tkr}. When introducing the Migdal effect and bremsstrahlung in the annual modulation analysis, the upper limits obtained for a 0.5\,GeV mass were $1.4\times10^{-35}\,\mathrm{cm^2}$ and $1.1\times10^{-33}\,\mathrm{cm^2}$, respectively (see also Sec.~\ref{sssec:exp_running_ER}). \\

\noindent{}\emph{Diurnal modulation}

\textbf{SENSEI} (see Sec.~\ref{sssec:exp_running_ER}) presented results on sidereal daily modulation in the signal rate. 
It uses for this search two science-grade Skipper-CCDs with an active mass of 2.19\,g which were operated at Fermilab (MINOS site). 
From the detector masses and acquisition time, a gross exposure of approximately 154\,g\,$\cdot$\,d can be inferred, corresponding to an estimated net exposure of 119\,g\,$\cdot$\,d after accounting for the reported 23\% of pixels excluded by data-quality selections.
The analysis is performed in the single electron event rate on 141 six-hour images, collected over 75\,days. The model-independent analysis of modulation set as 90\% C.L. upper limit for the amplitude of such a modulation of 6.8\,electrons/g/d~\cite{SENSEI:2025qvp}. This analysis enabled a significant improvement in the limits on scattering cross-sections for
DM masses below 1\,MeV with respect to previous results. \\

\textbf{DAMIC-M} has analysed in terms of this daily modulation the same dataset used for the DM search at LSM (see Sec.~\ref{sssec:exp_running_NR} and ~\ref{sssec:exp_running_ER}). The DAMIC-M Low Background Chamber houses two prototype CCD modules, each consisting of four n-type silicon CCDs, 3.3\,g each. For this analysis 84\,days are used, accumulating images corresponding to 1668\,s. After selection cuts, the integrated exposures are 0.139 and 1.257\,kg\,$\cdot$\,d for each module. Periodic signals are searched for in the data, in particular those corresponding to a potential daily modulation. No statistically significant periodic signal is found and limits are set to hidden sector particles interacting with electrons, surpassing previous results in the range $(0.53-1.2)$\,MeV considering both, heavy and light mediator scenarios~\cite{DAMIC-M:2023hgj,DAMIC-M:2025ltz}.


\subsubsection{Directionality}
\label{sssec:exp_running_directionality}

Directional information provides a distinctive signature for dark matter searches, since the motion of the solar system through the Galactic halo is expected to produce an anisotropic flux of dark matter particles in the laboratory frame, as discussed in Sec.~\ref{ssec:modulation}. 
Directional sensitivity can be achieved either by reconstructing the direction of individual nuclear recoils or through an anisotropic detector response that depends on the orientation relative to the incoming dark matter flux. 
Consequently, the experiments shown in Tab.~\ref{tab:Running_dir} and further discussed below probe different dark matter interactions and mass ranges, which are specified separately for each result.
In this section, we discuss experiments that have experimentally demonstrated directional sensitivity, irrespective of whether the directional information was already used to derive the reported dark matter constraints. 
This distinction is important, since some of the directional experiments still derive their strongest limits primarily from the measured event rate, while demonstrating directionality as an additional capability.
Those dark matter search results are discussed in Sec.~\ref{sssec:exp_running_NR} if they were published after 2020.\\

\renewcommand{\arraystretch}{1.2}
\begin{table}[H]
    \centering
    \caption{Summary of direct dark matter experiments investigating directional dark matter signatures, with published results since 2020 (or before if still world-leading), organized by detector technology.
    Provided are the leading results per experiment.
    Masses refer to the target or fiducial volume used for the respective analysis, where available. Exposures are the net exposures, if provided. Else the gross exposure is cited. If neither is provided, it is approximated (indicated with $\sim$ or $\lesssim$) based on the given mass and live time. Thresholds are quoted in the energy scale or experimental observable used by the collaboration, with approximate conversions to comparable energy scales where appropriate.}
    \footnotesize
    \begin{tabular}{|c|c|c|c|c|c|c|}\hline
        \makecell{\bf Experiment\\\bf (Interaction)} & \makecell{\bf Target \\\bf(Mass)} & {\bf Location} &
        \makecell{\bf Data\\\bf Taking} &
        {\bf Exposure} & {\bf Threshold} & {\bf Reference(s)}\\\hline
        \hline
        \multicolumn{7}{|c|}{ Gaseous detectors } \\
        \hline
         DRIFT (SD) & CS$_2$\,+\,CF$_4$\,+\,O$_2$ (140\,g) & BUL & 2014-15 & $\lesssim 7.7$\,kg\,$\cdot$\,d & 700\,NIPs ($\sim$36\,keV$_{\rm nr}$) & \cite{BATTAT201765, Battat_2021} \\
        MIMAC (SD) & i-C$_4$H$_{10}$ + CHF$_3$ ($\sim0.46$\,g) & LSM & n.a. & $\lesssim 0.20$\,kg\,$\cdot$\,d & 0.45\,keV$_{\rm ee}$ & \cite{Beaufort:2023vlb, Ourahou:2026tpf} \\
        NEWAGE (SD) & CF$_4$ (10\,g) & Kamioka & 2013-17 & 4.5\,kg\,$\cdot$\,d & 50\,keV$_{\rm ee}$ & \cite{Yakabe:2020rua} \\
        NEWAGE (SD) & CF$_4$ (10\,g) & Kamioka & 2017-20 & 3.2\,kg\,$\cdot$\,d & 50\,keV$_{\rm ee}$ & \cite{Shimada:2023vky} \\
        \hline
        \multicolumn{7}{|c|}{ Other technologies } \\
        \hline
        QROCODILE (DM-e) & W$_{45}$Si$_{55}$ (1.67\,ng) & surface & 2023 & $\lesssim 29$\,ng\,$\cdot$\,d & 0.11\,eV$_{\rm er}$ &  \cite{QROCODILE:2024nqm} \\\hline

    \end{tabular}
    \label{tab:Running_dir}
\end{table}

\subsubsection*{Gaseous detectors}

\textbf{DRIFT} used a low-pressure negative-ion gas TPC at the Boulby Underground Laboratory to reconstruct the tracks of nuclear recoils and thereby obtain directional information on dark matter interactions. 
The DRIFT-IId detector operated with a mixture of CS$_2$, CF$_4$, and O$_2$, with approximately 140\,g total gas mass containing 34.2\,g of fluorine to test spin-dependent dark matter-proton scattering~\cite{BATTAT201765}.
The negative-ion drift suppresses diffusion, while the different drift velocities of minority ion species enable fiducialization along the drift direction.
Directional sensitivity to nuclear recoils was demonstrated using neutron calibration data, although the dark matter exclusion itself relied on the recoil-energy information rather than the reconstructed recoil directions.
A 54.7-day shielded dark matter search with a gross exposure of approximately 7.66\,kg\,$\cdot$\,d found no events in the fiducial signal region and was interpreted in terms of spin-dependent dark matter-proton scattering on fluorine, reaching a cross section of $2.8\times10^{-37}\,\mathrm{cm^2}$ at 100\,GeV. 
The analysis required ionization signals above 700\,NIPs (number of ion pairs), corresponding to approximately 36\,keV$_{\rm nr}$ for fluorine recoils using the species-dependent ionization response, with the NIPs scale calibrated using $^{55}$Fe X-rays.
A subsequent machine-learning analysis substantially increased the expected signal efficiency at low recoil energies and projected an improvement in sensitivity by about one order of magnitude below 15\,GeV, extending the projected mass reach down to about 9\,GeV~\cite{Battat_2021}.\\

\textbf{MIMAC} searches for directional dark matter signatures using a low-pressure gas micro-TPC at LSM, reconstructing three-dimensional nuclear-recoil tracks with statistical head-tail sensitivity, meaning the start and end of a recoil track can be distinguished on a statistical basis.
The detector consists of two chambers with a combined active volume of about 6\,L, containing approximately 0.46\,g of a 50:50 mixture of isobutane (i-C$_4$H$_{10}$) and CHF$_3$ at 30\,mbar~\cite{Beaufort:2023vlb, Ourahou:2026tpf}.
The gas mixture provides hydrogen and fluorine nuclei for spin-dependent dark matter-proton scattering, although the low-energy analysis is dominated by proton recoils.
The first directional dark matter search used effective live times of 495.1 and 354.2\,days for the two chambers, corresponding to a gross total-gas exposure of approximately 0.195\,kg\,$\cdot$\,d and a hydrogen exposure of about 0.017\,kg$\cdot$d.
With an ionization-energy threshold of 0.45\,keV$_{\rm ee}$, calibrated using X-ray fluorescence lines, the analysis combined three-dimensional track reconstruction and topological background discrimination with a directional ON/OFF analysis comparing the expected dark matter recoil direction with background regions.
No significant directional excess was observed and an upper limit on the spin-dependent dark matter-proton cross section was derived\footnote{The first result from the MIMAC Collaboration~\cite{Ourahou:2026tpf} has not yet been published in a peer-reviewed journal. We therefore do not yet comment on their reach and will update the text if the analysis is published before the publication of this review.}.\\

\textbf{NEWAGE} searches for directional dark matter signatures using a low-pressure CF$_4$ micro-TPC at the Kamioka Observatory, also reconstructing three-dimensional nuclear-recoil tracks with statistical head-tail sensitivity. 
The approximately 10\,g CF$_4$ target contains 8.64\,g of fluorine, whose $^{19}$F nuclei provide sensitivity to spin-dependent dark matter-proton scattering.
The first three-dimensional vector dark matter search with NEWAGE-0.3b' used an exposure of 4.51\,kg\,$\cdot$\,d and reached a spin-dependent proton cross-section limit of $4.3\times10^{-34}\,\mathrm{cm^2}$ at 150\,GeV~\cite{Yakabe:2020rua}.
The analysis applied a lower energy threshold of 50\,keV$_{\rm ee}$, defined on the ionization energy scale calibrated with X-rays, while the corresponding nuclear-recoil energy depends on the fluorine ionization quenching factor.

An upgrade to the NEWAGE-0.3b'' detector introduced a low-alpha-emission $\upmu$-PIC readout, reducing radioactive backgrounds and enabling improved directional analyses~\cite{Ikeda:2021ckk}.
The latest search incorporated the earlier NEWAGE-0.3b'' dataset and additional measurements, reaching an exposure of 3.18\,kg\,$\cdot$\,d~\cite{Shimada:2023vky}.
With improved gamma-ray rejection and head-tail determination, the analysis tested the expected directional anisotropy of dark matter-induced fluorine recoils and found no significant signal.
Using also an analysis threshold of 50\,keV$_{\rm ee}$, primarily determined by the requirements for reliable track reconstruction and background discrimination, it reached a spin-dependent dark matter-proton cross-section limit of $2.57\times10^{-35}\,\mathrm{cm^2}$ at 150\,GeV, currently providing the strongest published constraint from a directional dark matter analysis.\\

\subsubsection*{Other technologies}

\textbf{QROCODILE} investigates directional sensitivity through the anisotropic electronic response of its thin WSi target rather than through event-by-event reconstruction of recoil directions~\cite{QROCODILE:2024nqm} (see also Sec.~\ref{sssec:exp_running_NR}).
The changing orientation of the detector relative to the Galactic dark matter wind over a sidereal day could therefore produce a modulation of the event rate.
However, the first science run recorded only fifteen events, whose arrival times were consistent with an isotropic background at 95\%~C.L.
The reported dark matter constraints consequently rely on the total event rate rather than directional information, while the calculated anisotropic response illustrates the potential for directional dark matter searches with larger exposures.\\


\FloatBarrier

\subsection{Experiments under construction, in commissioning, or running with upcoming results}
\label{ssec:exp_nearfuture}

\begin{figure}[t]
    \centering
    \centering
    \includegraphics[width=0.495\linewidth]{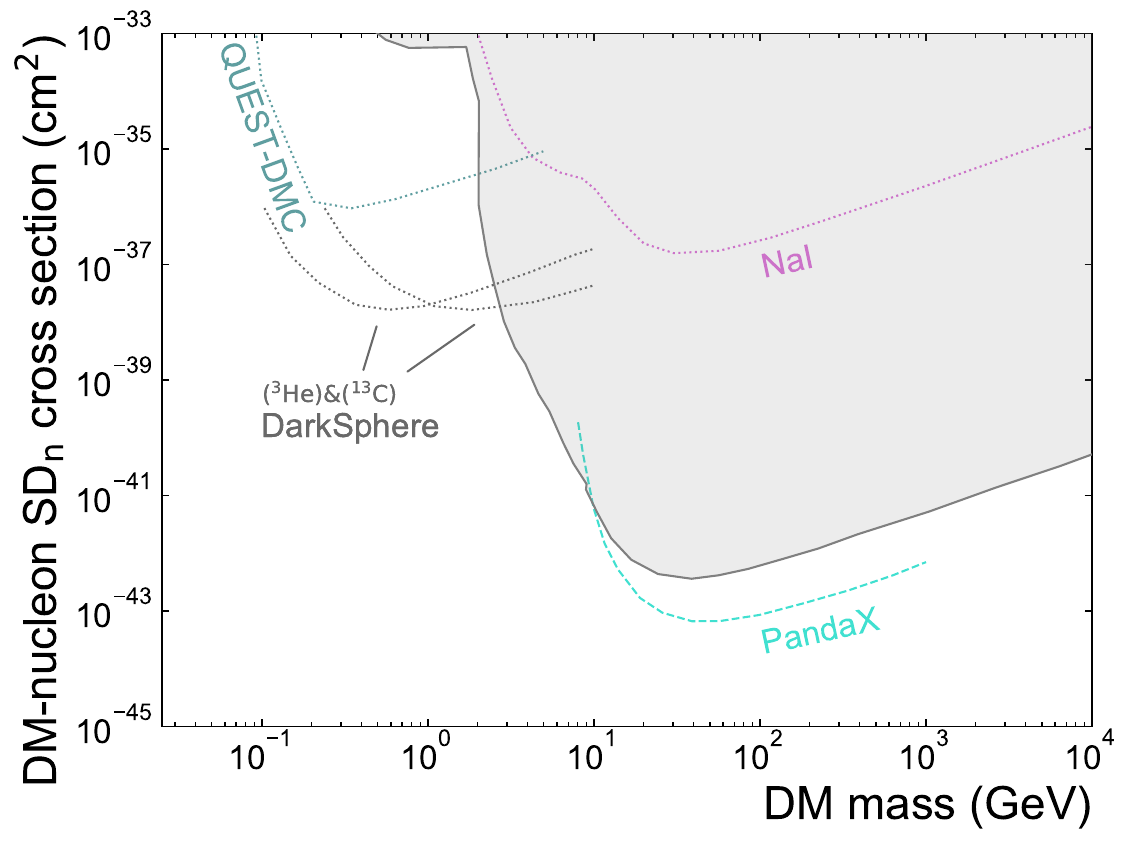}
    \hfill
    \centering
    \includegraphics[width=0.495\linewidth]{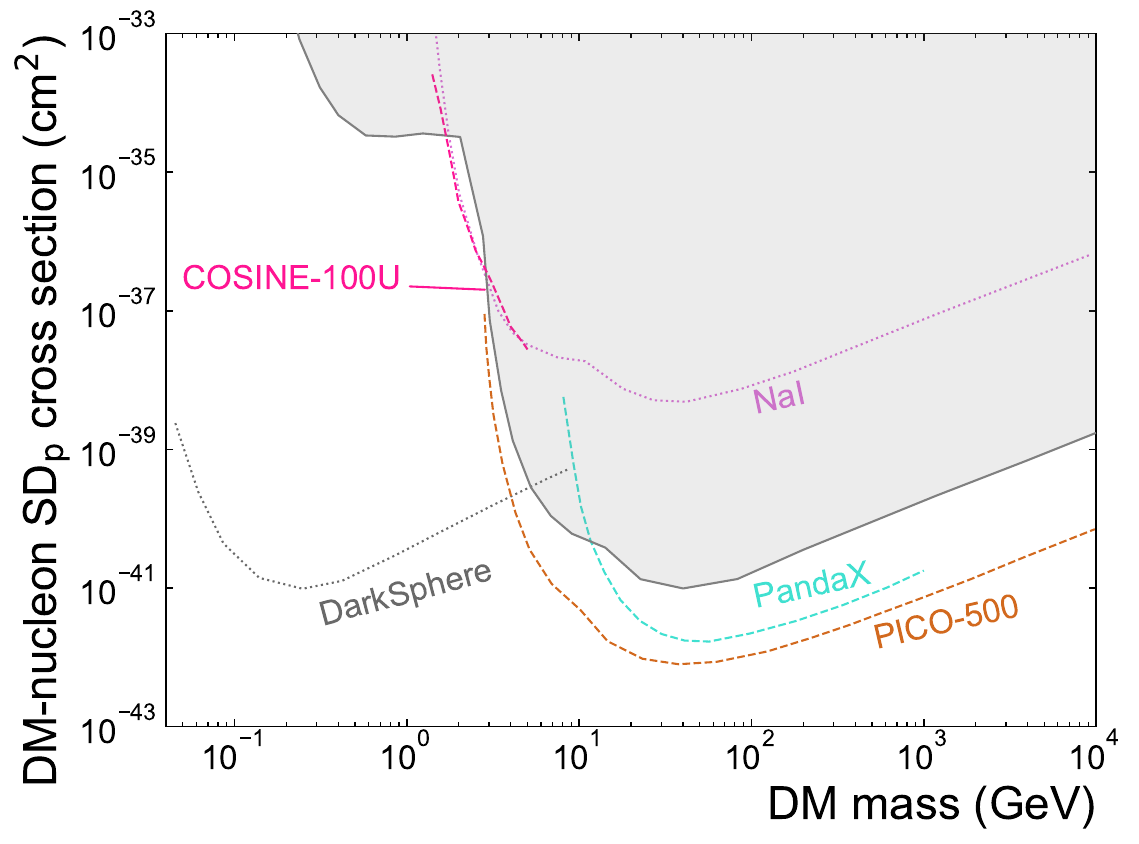}
    \caption{Left: sensitivity projections for dark matter-nucleon spin-dependent (SD) scattering of neutron-only (left panel) and proton-only (right panel). The gray area shows the space already covered by experiments. Dashed lines are upcoming experiments that are (close to) running (Sec.~\ref{ssec:exp_nearfuture}). When a collaboration provides multiple sensitivity projections, the most conservative curve is shown in this figure. Direct comparisons between the projected sensitivities should be made with caution, as they depend on the assumptions adopted by each collaboration, which may range from conservative to optimistic. Dotted lines are experiments in their conceptual design phase (Sec.~\ref{ssec:exp_cd}). The experiments shown are: COSINE-100U~\cite{Lee:2024wzd}, DarkSphere~\cite{NEWS-G:2023qwh}, NaI~\cite{Martinez:2026thh}, PandaX~\cite{PANDA-X:2024dlo}, PICO-500~\cite{Krauss:2024jzg} and QUEST-DMC~\cite{QUEST-DMC:2023nug}.}
    \label{fig:exp_nearfuture_SD}
\end{figure}

This section describes the experiments that are currently under construction, in commissioning, or taking data without published results since 2020. A list of these experiments is reported in Tab.~\ref{tab:UnderConstruction}, which summarizes the hosting laboratory, the target and its dimensions, and their status. 
The following subsections describe each experiment, categorized by technology
(more details about each technology are in Sec.~\ref{sec:tech}).
Figures~\ref{fig:exp_nearfuture_SI}--\ref{fig:exp_nearfuture_ER} summarize the sensitivity projections for these experiments
for spin-independent and spin-dependent dark matter-nucleus scattering searches as well as for dark matter-electron scattering searches.
When collaborations provide multiple sensitivity projections, the most conservative curve is shown in the figures, as it is considered the earliest achievable. Comparisons between different experiments should be made with caution, as the projected sensitivities may not be directly comparable because of differences in the underlying assumptions (optimistic vs. conservative).
The information on the background, energy threshold, and exposure considered in the sensitivity projections are reported in the text when provided by the collaborations.

\renewcommand{\arraystretch}{1.2}
\begin{table}[H]
    \centering
    \caption{List of direct dark matter detection experiments that are currently under construction, in the commissioning phase, or taking data. For each experiment, the hosting laboratory, the target, its size, its status and references are included. If the mass is not available in literature, it is indicated as ``n.a.'' (not available).
    }
    \footnotesize
    \begin{tabular}{|c|c|c|c|c|}\hline
       {\bf Experiment} & {\bf Location} & {\bf Target (size)} & {\bf Status} & {\bf Reference(s)}\\\hline
        \hline
        \multicolumn{5}{|c|}{ Cryogenic crystal calorimeters } \\
        \hline
        BULLKID-DM & LNGS & Si (800\,g) & under construction & \cite{Folcarelli:2026wkv}\\
        COSINUS & LNGS & NaI (278\,g) & commissioning & \cite{Angloher:2025shf,Hughes:2024nub}\\
        SuperCDMS & SNOLAB & Ge (25\,kg) / Si (3.7\,kg) & commissioning & \cite{SuperCDMS:2016wui}\\
        TESSERACT & LSM & Si (n.a.)/Ge (100\,g)/GaAs (n.a.)/Al$_\mathrm{2}$O$_\mathrm{3}$ (n.a.) & under construction & \cite{Billard:2024zvc, McKinsey_TESSERACT_Snowmass,Billard:2023_IN2P3}\\
        \hline
        \multicolumn{5}{|c|}{ Semiconductors } \\
        \hline
        CDEX-50 & CJPL & Ge (50\,kg) & under construction & \cite{CDEX:2023vvc}\\
        DAMIC-M &LSM & Si ($\sim 350$\,g) & under construction & \cite{Lin:2025orh,VilarIDM:2026aaa}\\
        \hline
        \multicolumn{5}{|c|}{ Scintillation crystals } \\
        \hline
        COSINE-100U & Yemilab & NaI(Tl) (99.1\,kg) & data-taking since 2025 & \cite{Lee:2024wzd} \\
        SABRE-North & LNGS & NaI(Tl) ($\sim 45$\,kg) & under construction & \cite{Antonello:2020xhj, Calaprice:2022vte}\\
        SABRE-South & Stawell & NaI(Tl) ($35-50$\,kg) &  under construction & \cite{SABRESouth:2024bpv} \\
        \hline
        \multicolumn{5}{|c|}{ Bubble chambers } \\
        \hline
         PICO-500L & SNOLAB & C$_3$F$_8$ (250\,L) & under construction & \cite{Krauss:2024jzg,Moore:2024que}\\
         SBC & SNOLAB & LAr (10\,L) & under construction & \cite{Alfonso-Pita:2023frp}\\
        \hline
        \multicolumn{5}{|c|}{ Liquid noble-gas scintillators } \\
        \hline
         DarkSide-20k & LNGS & LAr ($\sim$\,50\,t) & under construction & \cite{DarkSide-20k:2017zyg, DarkSide-20k:2024yfq, DarkSide-50:2025lns} \\
         PandaX-20T & CJPL & LXe  ($\sim$\,20\,t) & under construction & \cite{PANDA-X:2024dlo} \\
        \hline
        \multicolumn{5}{|c|}{ Superfluid noble-gas detectors } \\
        \hline
        TESSERACT & LSM & $^4$He ($\sim 1$\,kg) & under construction & \cite{Hertel:2018aal, SPICE:2023tru} \\
        \hline
        \multicolumn{5}{|c|}{ Gaseous detectors } \\
        \hline
        NEWS-G & SNOLAB & Ne + CH$_4$ ($\sim 1\,300$\,L) & data-taking since 2022 & \cite{NEWS-G:2022kon, Coquillat:2024tfa,NEWS-G:2023qwh} \\
        TREX-DM & LSC &  Ar / Ne + iC$_4$H$_{10}$ (20\,L) & commissioning & \cite{Castel:2019ngt, Castel:2025yvh} \\\hline
    \end{tabular}
    \label{tab:UnderConstruction}
\end{table}

\subsubsection*{Cryogenic crystal calorimeters}
\label{sssec:exp_nearfuture_bolo}

\begin{figure}[t]
    \centering
    \centering
    \includegraphics[width=0.49\linewidth]{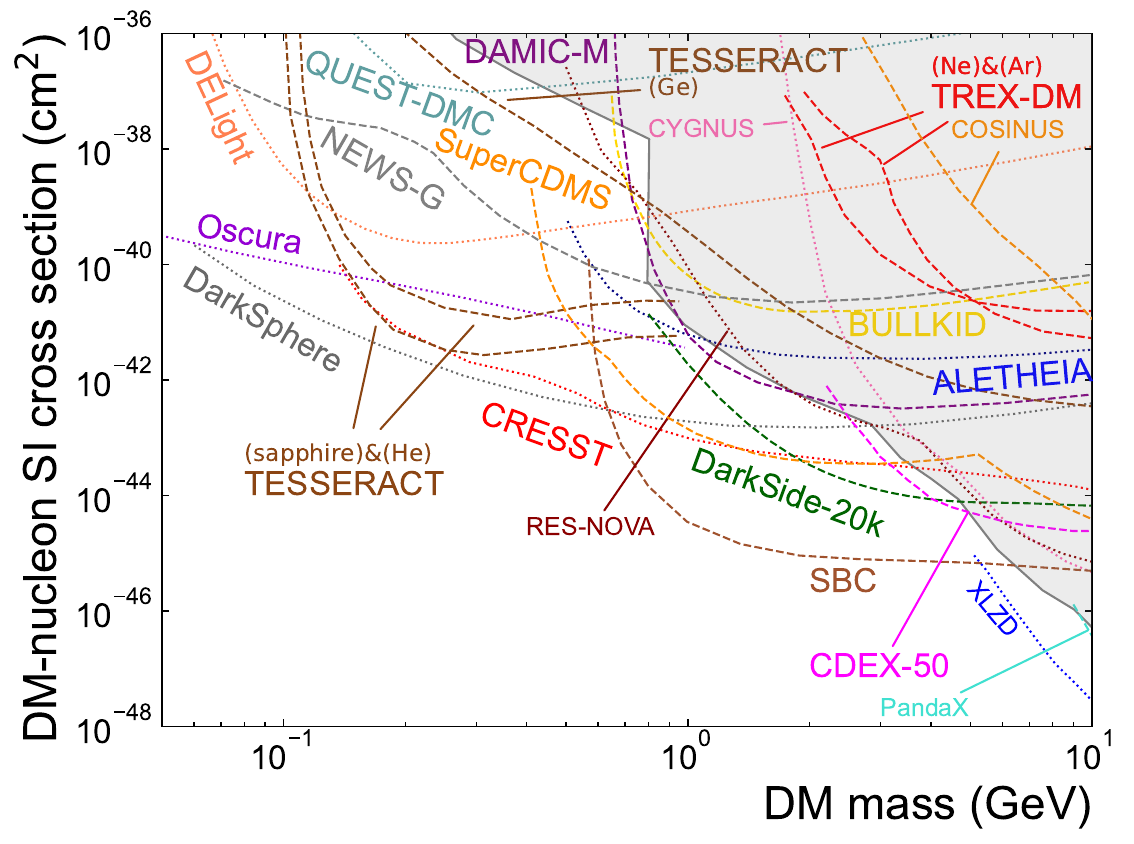}
    \hfill
    \centering
    \includegraphics[width=0.49\linewidth]{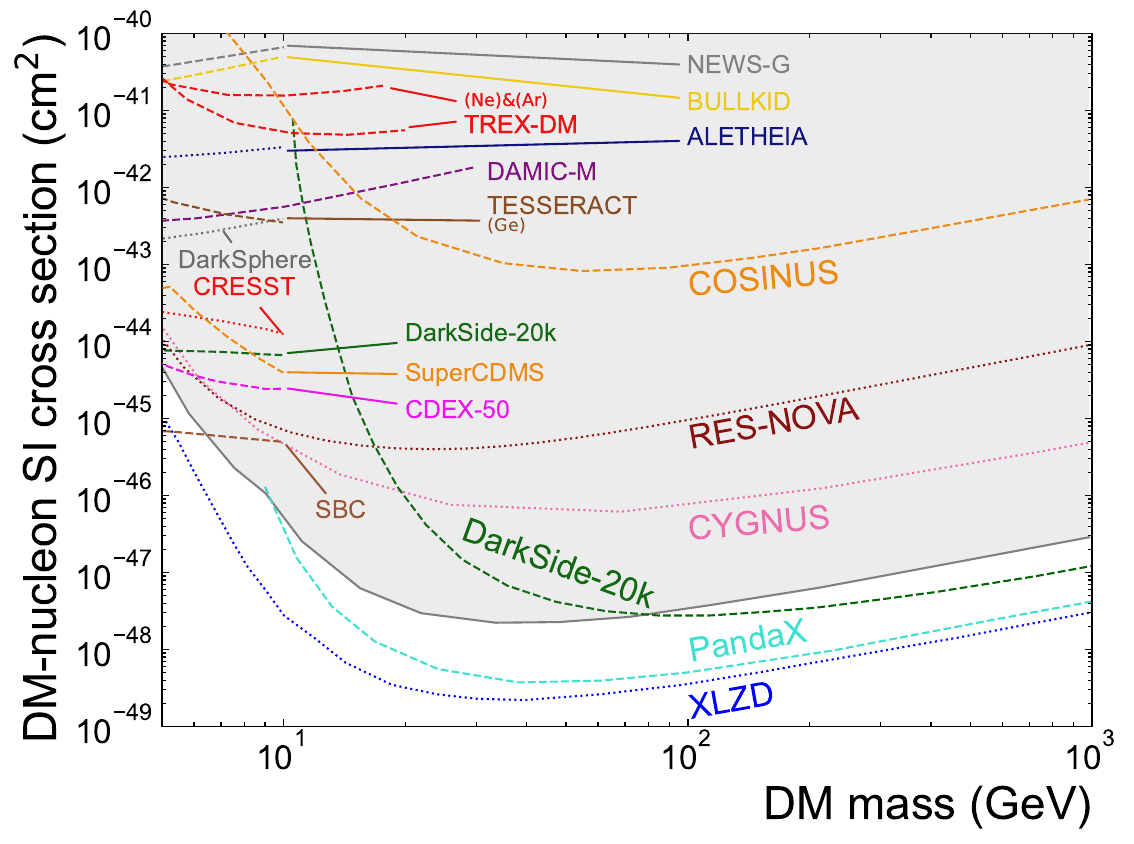}
    \caption{Sensitivity projections for dark matter-nucleon spin-independent (SI) scattering with two dark matter energy ranges below 10\,GeV (left panel) and above 5\,GeV (right panel). The gray area shows the space already covered by experiments. Dashed lines are upcoming experiments that are (close to) running (Sec.~\ref{ssec:exp_nearfuture}). Dotted lines are experiments in their conceptual design phase (Sec.~\ref{ssec:exp_cd}). When a collaboration provides multiple sensitivity projections, the most conservative curve is shown in this figure. Direct comparisons between the projected sensitivities should be made with caution, as they depend on the assumptions adopted by each collaboration, which may range from conservative to optimistic. The following experiments are represented: ALETHEIA~\cite{Liao:2021npo}, BULLKID~\cite{Folcarelli:2026wkv}, CDEX-50~\cite{CDEX:2023vvc}, COSINUS~\cite{Angloher:2025shf}, CRESST~\cite{Angloher:2025fzw}, CYGNUS~\cite{Vahsen:2020pzb}, DAMIC-M~\cite{VilarIDM:2026aaa}, DarkSide-20k~\cite{DarkSide-20k:2017zyg, DarkSide-20k:2024yfq}, DarkSphere~\cite{NEWS-G:2023qwh}, DELight~\cite{Solmaz:2025xiq}, NEWS-G~\cite{NEWS-G:2023qwh}, Oscura~\cite{Oscura:2022vmi}, PandaX-xT~\cite{PANDA-X:2024dlo}, QUEST-DMC~\cite{QUEST-DMC:2023nug}, RES-NOVA~\cite{RES-NOVACollaboration:2025stq}, SBC~\cite{Alfonso-Pita:2023frp}, SuperCDMS~\cite{SuperCDMS:2016wui}, TESSERACT~\cite{McKinsey_TESSERACT_Snowmass, Billard:2023_IN2P3}, TREX-DM~\cite{Castel:2019ngt}, XLZD~\cite{XLZD:2024nsu}.}
    \label{fig:exp_nearfuture_SI}
\end{figure}

The \textbf{BULLKID-DM} experiment aims at detecting nuclear-recoil signatures from dark matter particles with masses between 0.3 and 10\,GeV/c$^2$.
Its chosen target is made of silicon and will be composed of sixteen 5\,mm-thick wafers with a 3-inch diameter, whose surface is partially diced forming single cubic pixels with a mass of 0.34\,g.
Each pixel is read out by a KID (see Sec.~\ref{ssec:tech_bolom}).
The total silicon mass will be 800\,g which will be reduced to 600\,g after fiducialization.
A conservative threshold of 200\,eV$_{\rm{ee}}$ is considered for the sensitivity projections, considering that the BULLKID-DM demonstrator showed a threshold of 160\,eV$_{\rm{ee}}$ when reading out two pixels.
The energy scale was calibrated with a laser and an $^{55}$Fe source.
The sensitivity projections consider a background level of 1\,DRU\footnote{1\,DRU = 1\,count/(keV\,$\cdot$\,kg\,$\cdot$\,d)} and 0.01\,DRU~\cite{Folcarelli:2026wkv}. 
Figure~\ref{fig:exp_nearfuture_SI} (yellow-dashed curve) shows the projected sensitivity for dark matter-nucleon spin-independent scattering of the BULLKID experiment assuming a threshold of 200\,eV, a background of 1\,DRU and with an exposure of 600\,g for one year. 
The strategy to achieve these backgrounds includes a combination of passive shielding and active veto. The outer shielding surrounding the dilution refrigerator will include Roman lead, high-purity copper and polyethylene.
Inside the cryostat, an ultra-pure copper shield will enclose the detector and will be thermalized at base temperature.
In addition, the use of coincidences between pixels will permit to reject events non compatible with dark matter signature and the fiducialization of the detector via the rejection of events occurring on pixels on the boundaries of the wafer.
Last, a cryogenic veto made of scintillating material (such as bismuth germanate or gadolinium aluminum gallium garnet)~\cite{BULLKID:2026aaa} and read out by a light-sensitive KID detector will improve the background discrimination.
The BULLKID collaboration anticipates that these improvements will yield a background reduction between four to six orders of magnitude compared to measurements obtained with an above-ground dilution refrigerator under moderate shielding conditions~\cite{Delicato:2026pbj}.
\\

The \textbf{COSINUS} experiment aims at testing the DAMA/LIBRA claim by using NaI crystals as target. As NaI crystals have a low melting point and are hygroscopic, their surface is not suitable for involved fabrications~\cite{COSINUS:2021onk} including TES depositions. 
This is addressed by depositing the TES sensor on a separate wafer and thermally connecting it to a gold pad deposited on the target with a gold bonding wire.
In order to reduce the cosmogenic neutron background, COSINUS uses an active muon veto system~\cite{Angloher:2024pbw}: the dilution refrigerator is inserted in a cylindrical water tank equipped with 30 PMTs capable of tagging Cherenkov radiation.
The dilution refrigerator is also equipped with an inner shielding made of a copper plug with a mass of 190\,kg.
COSINUS results obtained with a 3.67\,g target are presented in Sec.~\ref{sssec:exp_running_NR}. 
In its first science phase, COSINUS will employ eight detector modules, each of them with a NaI target of 34.8\,g which will allow to confirm or refute DAMA/LIBRA claim in one to one and a half years after collecting an exposure of 100\,kg\,$\cdot$\,d~\cite{Angloher:2025shf}.
Figure~\ref{fig:exp_nearfuture_SI} (orange-dashed curve) shows the projected sensitivity for dark matter-nucleon spin-independent scattering of the COSINUS experiment~\cite{Angloher:2025shf} assuming an energy threshold of 6\,keV and an exposure of 100\,kg\,$\cdot$\,d. The projections consider an electron recoil background of 1\,DRU and an internal $^{40}$K activity of 600\,$\upmu$Bq/kg. The nuclear recoil contribution (0.156\,counts/(kg\,$\cdot$\,y)) was neglected.
The experimental setup is built and commissioned at the LNGS and the first science run is expected to start in late 2026.\\

\textbf{SuperCDMS}~\cite{SuperCDMS:2016wui} is using eighteen germanium and six silicon cryogenic calorimeters corresponding to a total mass of 25\,kg and 3.7\,kg, respectively. 
Each detector is equipped with twelve phonon channels read out by Quasiparticle-trap-assisted Electrothermal-feedback Transition-edge sensors~\cite{Irwin:1995zzb}.
Two detector designs are optimized for operation with low- and high-voltage (see Sec.~\ref{ssec:tech_bolom}). 
Different energy thresholds are expected based on the target material and the voltage applied: 350(175)\,eV$_{\rm{ee}}$ for Ge(Si) in low voltage, and 70(35)\,eV$_{\rm{ee}}$ for Ge(Si) in high voltage.
The shielding consists of three layers, arranged from the inside outward: a neutron shield made of high-density polyethylene, a gamma shield made of lead and an outer neutron shield provided by a water tank.
Figure~\ref{fig:exp_nearfuture_SI} (orange-dashed curve) the projected sensitivity of the SuperCDMS experiment to dark matter–nucleon spin-independent scattering, combining the silicon and germanium targets.
An expected electron recoil background 0.82(1.01)\,DRU is considered for the high(low)-voltage silicon detectors and 0.07(0.06)\,DRU for the high(low)-voltage germanium detectors between 3\,eV and 2\,keV. The contributions for the nuclear recoil backgrounds in the low-voltage detectors are $9\times 10^{-6}$ and $7.9\times 10^{-6}$\,DRU for the germanium and silicon, respectively. The projected sensitivity assumes two years of operation for the high-voltage detectors and five years for the low-voltage detectors.
The SuperCDMS detector payload and cryogenic setup are being commissioned at SNOLAB.\\

\textbf{TESSERACT} is a multi-target experiment aiming to probe dark matter candidates with masses ranging from the meV scale to 1\,GeV through absorption, electron-recoil, and nuclear-recoil signatures. 
Its program combines cryogenic solid-state detectors based on polar crystals and semiconductors, discussed in this section, with a superfluid-helium target discussed separately as part of the ``Superfluid noble-gas detectors'' technologies.

The polar-crystal program considers sapphire and gallium arsenide as target material, whose optical-phonon energies extend up to approximately 100 and 30\,meV, respectively~\cite{Griffin:2018bjn}.
These low-energy excitations provide sensitivity to dark matter through the direct production of optical phonons.
In addition, gallium arsenide exhibits excellent scintillation properties, with a light yield of approximately 125\,photons/keV for a doped crystal~\cite{Vasiukov:2019hwn}, making it suitable for particle identification through the simultaneous readout of phonons and scintillation light. The coincidences between the light and the phonon signal can be also used to reject the LEE background (see Sec.~\ref{sssec:instrumental_bck}).
Figures~\ref{fig:exp_nearfuture_SI} and \ref{fig:exp_nearfuture_ER} (brown-dashed curves) show the projected sensitivity for dark matter-nucleon spin-independent and dark matter-electron scattering of the TESSERACT polar-crystal targets. The assumptions considered are for the present technology: the energy threshold assumed is 3.5\,eV and two-photons for the sapphire and gallium arsenide targets, respectively. A zero-background scenario is considered, together with exposures of 100\,g\,$\cdot$\,y and 1\,kg\,$\cdot$\,y for the sapphire and gallium arsenide targets, respectively.

The semiconductor program leverages a double readout of charge and phonons for particle discrimination. 
Two technologies are optimized to be operated either in low-voltage (LV) or high-voltage (HV) mode (see Sec.~\ref{ssec:tech_bolom}). 
The LV detector will simultaneously read out charge and phonons from four 5\,g germanium crystals~\cite{BleIDM:2026aaa}. 
Figure~\ref{fig:exp_nearfuture_SI} (brown-dashed curves) shows the projected sensitivity for dark matter-nucleon spin-independent scattering of the LV targets. The energy thresholds considered are 1\,eV and 3\,eV$_{\rm{ee}}$ for the phonon and ionization channels, respectively. The exposure considered is 100\,g\,$\cdot$\,y. The background assumes 10\,DRU for electron recoils in addition to the background from Ref.~\cite{EDELWEISS:2016nzl} and a reduction of the LEE background of a 100 factor with respect
to Ref.~\cite{EDELWEISS:2020fxc}. 
For the HV detectors, the charge signal is not readout directly. 
Instead, the electrodes adopt a point-contact geometry, with one phonon sensor located on the point-contact electrode and a second on the opposite electrode. 
The first phase of the HV program will deploy four 40\,g germanium targets~\cite{BleIDM:2026aaa}.
Figure~\ref{fig:exp_nearfuture_ER} (brown-dashed curves) shows the projected sensitivity for dark matter-electron scattering of the HV targets. An exposure of 100\,g\,$\cdot$\,y is assumed with an energy threshold of two photons. The background assumes 10\,DRU for electron recoils in addition to the background from Ref.~\cite{EDELWEISS:2016nzl} and a reduction of the LEE background of a 1\,000 factor with respect to Ref.~\cite{EDELWEISS:2020fxc}. 
The TESSERACT detectors will be operated in two identical setups, each comprising a dilution refrigerator and passive shielding (including lead, polyethylene, and copper thermal screens). 
The radioactive background budget requirement for the TESSERACT experiment is $<10$\,DRU.
TESSERACT is planned to be installed at LSM in 2028, with the science phase starting in 2029.

\subsubsection*{Semiconductors}

The CDEX experiment scaled up its germanium detector array from 10\,kg (in CDEX-10) to 50\,kg~\cite{CDEX:2023vvc}. The new \textbf{CDEX-50} detector array will consist of 50 p-type point-contact germanium detectors with a mass of 1\,kg. 
The experiment is located at CJPL. 
The detectors are hosted in a liquid nitrogen tank with a volume of 1725\,m$^3$. The liquid nitrogen will be used both for cooling the detectors and for shielding environmental radiation. 
The tank will be shared with the CDEX-300 experiment using enriched germanium crystals for the detection of neutrinoless double beta decay.
The expected background level is $\sim$\,0.01\,DRU~\cite{CDEX:2023vvc} between $(2- 2.5)$\,keV$_\mathrm{ee}$.
A threshold of 160\,eV$_\mathrm{ee}$ is expected, which was already demonstrated by the CDEX-10 experiment. 
Figure~\ref{fig:exp_nearfuture_SI} (magenta-dashed curve) shows the projected sensitivity for dark matter-nucleon spin-independent scattering of the CDEX-50 experiment with the reported energy threshold and background, and assuming an exposure of 150\,kg\,$\cdot$\,y.
\\
\pagebreak

\begin{figure}[t]
    \centering
    \centering
    \includegraphics[width=0.495\linewidth]{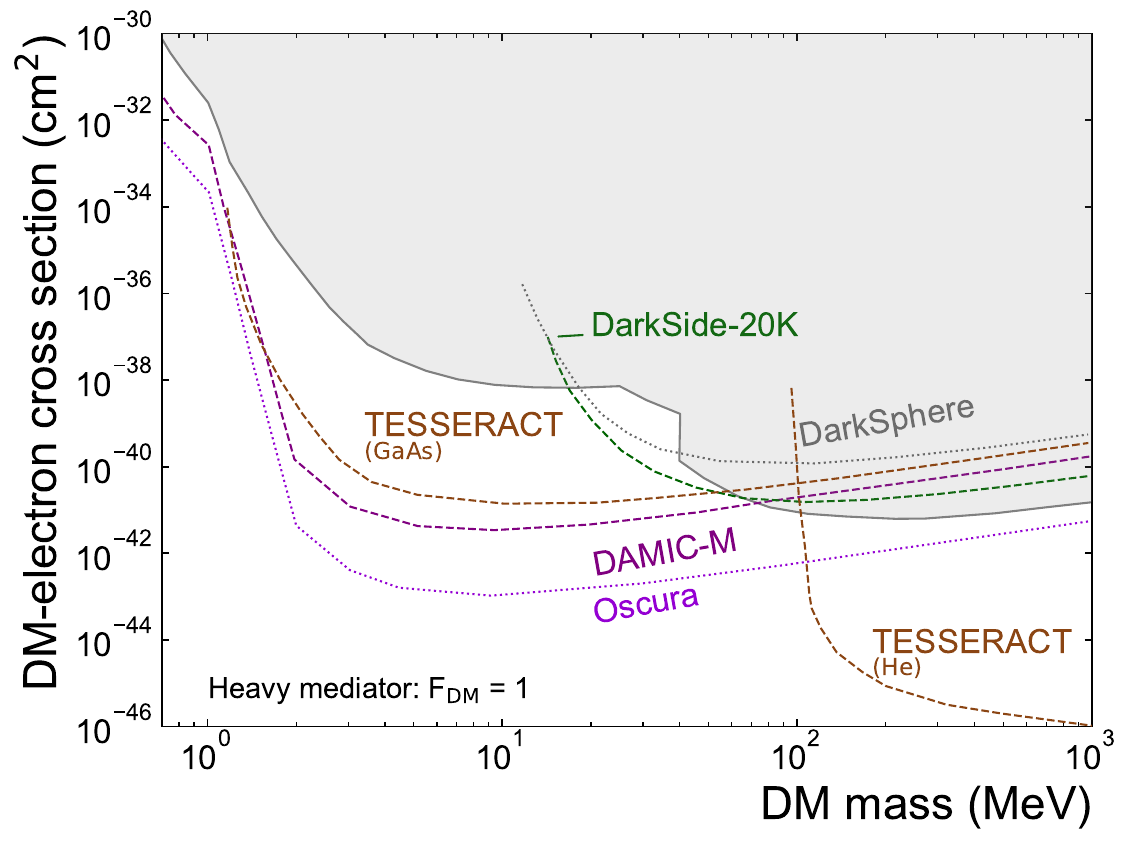}
    \hfill
    \centering
    \includegraphics[width=0.495\linewidth]{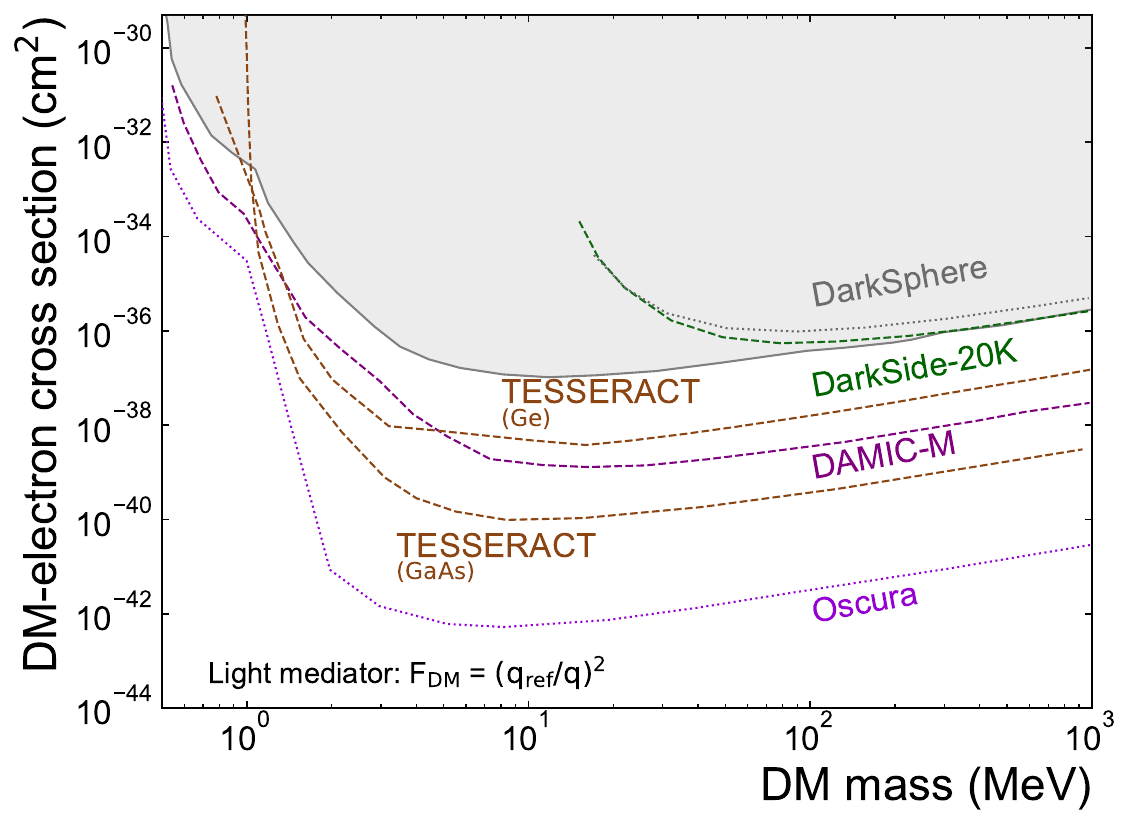}
    \caption{Sensitivity projections for dark matter-electron scattering for heavy mediator (left panel) and light mediator (right panel). Dashed lines are upcoming experiments that are (close to) running (Sec.~\ref{ssec:exp_nearfuture}). Dotted lines are experiments in their conceptual design phase (Sec.~\ref{ssec:exp_cd}). When a collaboration provides multiple sensitivity projections, the most conservative curve is shown in this figure. Direct comparisons between the projected sensitivities should be made with caution, as they depend on the assumptions adopted by each collaboration, which may range from conservative to optimistic. The experiments shown are: DAMIC-M~\cite{VilarIDM:2026aaa}, DarkSide-20k~\cite{DarkSide-20k:2024yfq}, DarkSphere~\cite{NEWS-G:2023qwh}, Oscura~\cite{Oscura:2022vmi}, TESSERACT~\cite{McKinsey_TESSERACT_Snowmass,Billard:2023_IN2P3}.}
    \label{fig:exp_nearfuture_ER}
\end{figure}

\textbf{DAMIC-M} commissioning results are presented in Sec.~\ref{sssec:exp_running_ER}. DAMIC-M is scaling up its detectors payload to twenty-six CCD modules for a total mass of $\sim$\,350\,g, which corresponds to its first phase~\cite{Lin:2025orh}. 
Each CCD module assembles four nine megapixel CCDs. Thanks to the employment of skipper CCDs, the baseline resolution is expected to be $\sim$\,0.1\,eV$_{\rm ee}$.
The CCDs will be hosted in a cryocooler operated at 130\,K.
The cryocooler is made of oxygen free high conductivity copper and it is enclosed in low-background lead and polyethylene movable shield. The detectors are surrounded by an ancient lead enclosure and all the copper in their proximity is electro-formed. 
The exposure to cosmic rays was limited for the copper and the silicon detectors.
In addition, the detectors and surrounding components are handled in a radon-free environment to reduce surface contamination and minimize backgrounds from radon progeny implantation.
The experiment is expected to achieve a background level of less than 1\,DRU thanks to the careful material selection and experimental design.
Figures~\ref{fig:exp_nearfuture_SI} and \ref{fig:exp_nearfuture_ER} (purple-dashed curve) show the projected sensitivities of the experiment~\cite{VilarIDM:2026aaa} to spin-independent scattering and dark matter–electron scattering, respectively. An exposure of 1\,kg\,$\cdot$\,y is considered in both cases. 
DAMIC-M construction and installation are ongoing at LSM as of fall 2026.

\subsubsection*{Scintillation crystals}

\textbf{COSINE-100U} consists of eight NaI(Tl) crystals with a total mass of 99.1\,kg, taking data in Yemilab underground laboratory since September 2025~\cite{Lee:2024wzd}. 
The NaI crystals are immersed in 2\,200\,L of liquid scintillator to suppress backgrounds, which is contained in an acrylic box and a 3-cm-thick copper box. The copper box is surrounded by 20\,cm of lead to shield from gammas and 3-cm-thick plastic scintillator to tag muons.
The detectors used in COSINE-100 (see Sec.~\ref{ssec:exp_running}) were resurfaced and tapered at the ends to 3\,inches to match the diameter of the PMTs, eliminating the quartz light guides, which allows to gain a 50\% in light collection, up to 22\,p.e./keV. All the upgrades are expected to significantly improve the experiment's background and energy threshold, providing increased sensitivity to low-mass dark matter candidates~\cite{Lee:2024wzd}.
Figure~\ref{fig:exp_nearfuture_SD} (pink-dashed curve) shows the projected sensitivity to dark matter–nucleon spin-dependent scattering of the COSINE-100U experiment. The projected sensitivity assumes a background level similar to the one measured in COSINE-100, an energy threshold of 8\,p.e. and one year of operation.
\\

The \textbf{SABRE} (Sodium Iodide with Active Background REjection) collaboration plans to install two experiments, one at LNGS in Italy (Northern hemisphere) and one at Stawell, Australia (Southern hemisphere) to decouple the expected dark matter modulation from site-dependent effects such as those caused by seasonal variations. The goal of SABRE is to develop beyond the state-of-the-art purification of NaI raw material, by using a zone-refinement process. \textbf{SABRE-North} will consist of nine crystals of $\sim$\,5\,kg mass each within a fully passive multilayer shielding system based on copper and polyethylene, optimized to suppress external gamma-ray and neutron backgrounds~\cite{Antonello:2020xhj, Calaprice:2022vte}. \textbf{SABRE-South} crystals will be installed within a shielding combining passive steel and polyethylene with an active liquid scintillator veto and plastic scintillators for muon tagging~\cite{SABRESouth:2024bpv} and two designs are under consideration: one with seven crystals of $\sim$\,7\,kg each for a total mass of $\sim$\,50\,kg, and the other with seven crystals of $\sim$\,5\,kg each for a total of about 35\,kg. Both experiments aim at background levels below $\sim$\,1\,DRU in the $(1-6)$\,keV$_\mathrm{ee}$ energy region. 

\subsubsection*{Bubble chambers}

The \textbf{PICO-500L} experiment employs bubble chamber technology with a C$_3$F$_8$ target, enabling searches for both spin-dependent and spin-independent interacting dark matter.
The pressure vessel was designed to hold 1\,000\,L of gas volume, but methods to produce the fused silica vessel limited the size of the vessel to 250\,L, which corresponds to 420\,kg of C$_3$F$_8$.
The vessel is shielded by a water tank against neutrons~\cite{Krauss:2024jzg}.
Figure~\ref{fig:exp_nearfuture_SD} (brown-dashed curve) shows the projected sensitivity to dark matter–nucleon spin-dependent scattering of the PICO-500L experiment.
Two distinct data taking periods: (1) the first period (corresponding to 80~live days) with a lower threshold (3.3\,keV) and 5 background events expected aims at probing the lower energy region but it will be background limited after one year of operation because of the coherent elastic scattering of neutrinos from the sun; (2) the second period with a higher threshold (8.2\,keV) is expected to collect 600\,days of exposure with 5 background events~\cite{Krauss:2024jzg,Moore:2024que}.
PICO-500L is hosted at SNOLAB in the Cube hall.
The detector system was installed in 2025 and the commissioning is expected to start in late 2026.\\

The \textbf{SBC} (Scintillating Bubble Chamber) experiment~\cite{Alfonso-Pita:2023frp} employs a bubble chamber containing 10\,L of liquid argon as the target medium. In addition to the conventional bubble chamber signal (bubble imaging and acoustic measurement), SBC will exploit the scintillation light produced by interactions with argon to measure the energy deposited by the interacting particle. The combination of these observables is expected to enable energy reconstruction and further enhance background discrimination. 
Figure~\ref{fig:exp_nearfuture_SI} (brown-dashed curve) shows the projected sensitivity to dark matter-nucleon spin-independent scattering for the SBC experiment. The projected sensitivity consider a threshold of 100\,eV$_{\rm{nr}}$, an exposure of 10\,kg\,$\cdot$\,y and zero background. A prototype chamber SBC-LAr10 has already been tested at Fermilab for calibration purposes, providing an opportunity to characterize the detector response and validate its performance ahead of physics searches.
SBC will be deployed at SNOLAB to minimize the muon-induced background rate and a dedicated shield is being designed to minimize the neutron background.

\subsubsection*{Liquid noble-gas scintillators}

The \textbf{DarkSide-20k} detector is composed by three nested detectors: (1) an outer veto with 650\,tonne of atmospheric argon hosted in a ProtoDUNE-like membrane cryostat to tag cosmogenic backgrounds; (2) an inner veto with 32\,tonne of underground argon contained in a stainless steel vessel which is used to tag gammas from neutron capture on the walls of the inner detector; (3) the inner detector made of a dual-phase TPC filled with $\sim$\,50\,tonne of underground argon. 
In the TPC, an electric field of 75\,kV is applied and the signal is readout with $\sim$\,200\,000 silicon photomultipliers. The inner TPC surfaces are covered with reflectors to improve the light collection from the scintillation and ionization signals and are coated with tetraphenyl butadiene to shift the wavelength of argon from UV to visible to maximize its detection efficiency.
The signals from the dual phase TPC will also be fiducialized leading to 34.2\,tonne of fiducial mass, corresponding to an exposure of 17.4\,t\,$\cdot$\,y for one year of data, when considering the effective livetime.
DarkSide-20k reports projected sensitivity both for the dark matter masses below and above 10\,GeV.
Figure~\ref{fig:exp_nearfuture_SI} (green-dashed curve in the right panel) shows the projected sensitivity of dark matter-nucleon spin-independent scattering for energies above 10\,GeV~\cite{DarkSide-20k:2017zyg}. 
An energy threshold of 30\,keV$_{\rm{nr}}$ is considered with an exposure of 100\,t\,$\cdot$\,y.
Figures~\ref{fig:exp_nearfuture_SI} and \ref{fig:exp_nearfuture_ER} (green-dashed curve) show the energy range below 10\,GeV both for dark matter-nucleon spin-independent scattering and dark matter-electron scattering~\cite{DarkSide-20k:2024yfq}.
An energy threshold of 4\,electrons\footnote{4\,electrons correspond to $\sim$\,100\,eV$_\mathrm{ee}$~\cite{DarkSide:2018bpj}} is reported with an exposure of one year of data.
DarkSide-20k is currently in construction in the hall C of the LNGS, DarkSide is expected to start its science phase around 2029.\\

The \textbf{PandaX-xT} experiment is designed to have a dual TPC filled with xenon. The ionization and light signals are readout by PMTs.
The TPC detector is located in a thin-walled ($\sim$\,3\,mm) ultrapure copper inner vessel. The copper vessel has a cylindrical shape with a diameter of 2.5\,m and a height of 2.5\,m~\cite{XiaoIDM:2026aaa}, which is capable to host up to 40\,tonne of liquid xenon. The copper vessel is located into a secondary double-walled vacuum insulated titanium vessel with an outer (inner) diameter of $\sim$\,4\,m (3.5\,m) and an outer (inner) height of $\sim$\,6\,m (5\,m).
The titanium vessel is filled with liquid scintillator which acts as veto detector.
The titanium vessel is nested in a 900\,m$^3$ water tank (13\,m height and 10\,m diameter) read out with PMTs to veto cosmogenic backgrounds.
Figures~\ref{fig:exp_nearfuture_SI} and \ref{fig:exp_nearfuture_SD} (cyan-dashed curve) show the projected sensitivities of the PandaX experiment~\cite{PANDA-X:2024dlo} to dark matter–nucleon spin-independent and spin-dependent scattering, respectively.
The projected sensitivity consider a fiducial volume of 34\,tonne, an exposure of 200\,t\,$\cdot$\,y and a background of 74.8\,event/(t\,$\cdot\,$y) for electron-recoils and 0.02\,event/(t\,$\cdot$\,y) for nuclear recoils in the region of interest between 1 to 10\,keV$_\mathrm{ee}$.
In a first phase which received funding and is referred as \textbf{PandaX-20T}, the TPC will be filled with 20\,tonne of xenon.
This experiment is in construction and is expected to start its science phase in 2028.

\subsubsection*{Superfluid noble-gas detectors}
\label{sssec:exp_nearfuture_superfluid}
As part of its multi-target program, \textbf{TESSERACT} will also operate a superfluid-$^4$He detector~\cite{Hertel:2018aal, SPICE:2023tru}, complementing the solid-state targets described above. 
The superfluid-helium target is contained in a stainless-steel and copper vessel and is read out by silicon detectors instrumented with TES. 
Requiring coincident signals from helium interactions in multiple silicon detectors will provide a powerful means of suppressing the LEE background.
A particular challenge is the superfluid film that would creep along the vessel walls and onto the calorimeters above the liquid target.
To suppress this film flow, a non-oxidized cesium coating~\cite{SPICE:2023tru} is deposited on the rod supporting the detectors to ensure that superfluid helium does not creep on the detectors above the target.
The cesium is evaporated in-situ during the cooldown to prevent its oxidation.
The superfluid-helium program will be integrated into the common TESSERACT experimental infrastructure at LSM, described in Sec.~\ref{sssec:exp_nearfuture_bolo}.
Figures~\ref{fig:exp_nearfuture_SI} and \ref{fig:exp_nearfuture_ER} (brown-dashed curves) show the projected sensitivity for dark matter-nucleon spin-independent and dark matter-electron scattering of the TESSERACT helium target. Assuming
present technology performance, an energy threshold of 18\,eV is reported. A zero-background scenario is considered with an exposure corresponding to 100\,g\,$\cdot$\,y.

\subsubsection*{Gaseous detectors}

\textbf{NEWS-G} (New Experiments With Spheres-Gas) is an experiment searching for sub-GeV dark matter using a spherical proportional counter.
The NEWS-G experiment is operating a 135-cm diameter spherical proportional counter at SNOLAB.
The vessel is made of 99.99\% pure oxygen-free electronic copper with a 500-$\upmu$m-thick inner layer of electro-formed copper.
Despite the presence of the inner layer of electro-formed copper, the dominant background of NEWS-G is radioactive contamination and cosmogenic activation of the copper vessel.
The vessel is shielded by a 25-cm-thick spherical lead shield (including a 3-cm layer of roman lead and a 22-cm layer of modern lead).
The lead shield is contained in a stainless steel shell flushed with nitrogen to minimize the presence of $^{222}$Rn in the air close to the detector, which is then surrounded by an octagonal 40-cm-thick neutron shield made of high density polyethylene.
The NEWS-G collaboration has developed an eleven-anode structure read-out called ACHINOS. Thanks to this approach each anode can tune the volume electric field and provide detector segmentation with independent read-out.
The results of the NEWS-G experiment commissioning at LSM are reported in Sec.~\ref{sssec:exp_running_NR}.
Figure~\ref{fig:exp_nearfuture_SI} (gray-dashed curve) shows the projected sensitivity for dark matter-nucleon spin-independent scattering of the NEWS-G experiment~\cite{NEWS-G:2023qwh}. The projections consider a sphere with a diameter of 140\,cm, a background of 1.7\,DRU,
and exposures of 20\,kg\,$\cdot$\,d. The region of interest is defined between 14\,eV$_\mathrm{ee}$ and 1\,keV$_\mathrm{ee}$.\\

The \textbf{TREX-DM} (TPC for Rare Event eXperiments-Dark Matter) experiment employs a high-pressure gas time-projection chamber to search for low-mass dark matter.
The target consists of a 20\,L of argon-isobutane or neon-isobutane gas mixture at a pressure of 10\,bar. Assuming a single isotope, 20\,L at 10\,bar correspond to 0.30\,kg of argon or 0.16\,kg of neon.
The target is contained in a 6-cm-thick copper vessel and hosted in a shielding composed by a 5-cm-thick copper layer, a 20-cm-thick low-activity lead layer and 40-cm-thick of neutron moderator (water tanks and polyethylene blocks)~\cite{Castel:2018gcp}. In addition the inside of the shielding has a radon-free atmosphere.
A central cathode creates two active volumes.
The readout is performed by micromesh gas amplification structures (Micromegas).
The background model shows levels at $(1-10)$\,DRU in the region of interest~\cite{Castel:2018gcp}.
However, the TREX-DM collaboration is investigating path forwards to reduce the background to 0.1\,DRU.
Using a combined GEM-Micromegas readout and an internal $^{37}$Ar calibration source, TREX-DM has demonstrated an electronics threshold of approximately 20\,eV$_{\rm ee}$ at a gas pressure of about 1\,bar, although the detection efficiency for genuine energy depositions near this threshold remains to be established~\cite{Castel:2025yvh}.
Figure~\ref{fig:exp_nearfuture_SI} (red-dashed curve) shows the projected sensitivities for dark matter-nucleon spin-independent scattering of the TREX-DM experiment. 
The projected sensitivities consider two targets (argon and neon) and assume a background level of 10\,DRU, an exposure of 0.3\,kg\,$\cdot$\,d and an energy threshold of 0.4\,keV$_{\rm{ee}}$.
The detector is currently located in Lab2500 at the LSC, although it was first deployed in 2018 in LSC-hall A.

\FloatBarrier

\subsection{Experiments in a conceptual design phase}
\label{ssec:exp_cd}

Despite the strong performance of current-generation direct detection experiments, a broad range of next-generation concepts is being developed to extend the sensitivity to lower cross sections, lower dark matter masses, and new classes of dark matter interactions. 
The proposed detectors typically either focus on substantially larger target masses or notably lower energy thresholds. 
Several of these concepts explore entirely new target media and detection principles motivated by the challenges posed by the neutrino background and by the motivation to probe dark matter candidates notably beyond the traditional WIMP paradigm.
The experiments discussed in this section, and summarized in Tab.~\ref{tab:Design}, are in a conceptual design or early prototyping phase, and showcase the diverse technological directions currently pursued within the field.
This includes next-generation experiments within already existing experimental programs, if they require significant changes to existing infrastructure or an entirely new set-up.

Just like in Sec.~\ref{ssec:exp_nearfuture}, Fig.~\ref{fig:exp_nearfuture_SI}--\ref{fig:exp_nearfuture_ER} summarize the sensitivity projections for these experiments for spin-independent and spin-dependent dark matter-nucleus scattering searches as well as for dark matter-electron scattering searches.
When collaborations provide multiple projections, the most conservative curve is shown also in this case, as it is considered the earliest achievable. 
We remind the reader to be cautious about comparisons between different experiments, as the underlying assumptions may strongly differ and may be more or less conservative.
The information on the background, energy threshold, and exposure considered in the projections are reported in the text when provided by the collaborations.

\renewcommand{\arraystretch}{1.2}
\begin{table}[H]
    \centering
    \caption{List of direct dark matter detection experiments that are in their conceptual design or prototyping phase. For each experiment, the anticipated target type is provided together with the first planned target size.}
    \footnotesize
    \begin{tabular}{|c|c|c|c|}\hline
       {\bf Experiment / Project} & {\bf Target} &  {\bf Projected target size} & {\bf Reference(s)}\\\hline
        \hline
        \multicolumn{4}{|c|}{ Cryogenic crystal calorimeters } \\
        \hline
        CRESST &  CaWO$_4$  & $\sim 2$\,kg & \cite{Angloher:2025fzw} \\ 
        RES-NOVA & PbWO$_4$ & $\sim 170$\,kg & \cite{RES-NOVACollaboration:2025stq}\\
        \hline
        \multicolumn{4}{|c|}{ Semiconductors } \\
        \hline
        Oscura & Si & 10\,kg & \cite{Oscura:2022vmi}\\
        \hline
        \multicolumn{4}{|c|}{ Scintillation crystals } \\
        \hline
        ADAMO & ZnWO$_4$ & 200\,kg & \cite{Cappella:2013rua, Leoncini:2025haz}\\
        COSINE-200 & NaI(Tl)  & 200\,kg & \cite{COSINE:2020egt}\\
        NaI & NaI(Tl) / NaI & (1-100)\,kg & \cite{Martinez:2026thh}\\
        PICOLON & NaI(Tl) & 250\,kg & \cite{Fushimi:2021mez}\\
        \hline
        \multicolumn{4}{|c|}{ Liquid noble-gas scintillators } \\
        \hline
         ALETHEIA & He & $\sim 10$\,kg & \cite{Liao:2021npo} \\         
         ARGO & Ar & $\sim$\,370\,t & \cite{DarkSide20k:2020ymr}\\
         XLZD & Xe  & $(60-80)$\,t & \cite{XLZD:2024nsu} \\
        \hline
        \multicolumn{4}{|c|}{ Superfluid noble-gas detectors } \\
        \hline
        DELight & $^4$He & $(0.1-1)$\,kg & \cite{vonKrosigk:2022vnf, Solmaz:2025xiq} \\
        QUEST-DMC & $^3$He & $\sim 0.1$\,g & \cite{QUEST-DMC:2023nug} \\
        \hline
        \multicolumn{4}{|c|}{ Gaseous detectors } \\
        \hline
        CYGNO & He:CF$_4$ & $(30\,000 - 100\,000)$\,L & \cite{Amaro:2026mes, DAstolfo:2024cei} \\
        CYGNUS & He:CF$_4$, He:SF$_6$ & $(1\,000 - 1\,000\,000)$\,L & \cite{Vahsen:2020pzb} \\
        NEWS-G: ECuME & Ne:CH$_4$ & $\sim 1\,400$\,L & \cite{Knights:2023fry, NEWS-G:2023qwh} \\
        NEWS-G: DarkSPHERE & He:i-C4H10 & $\sim 14\,000$\,L & \cite{Knights:2023fry, NEWS-G:2023qwh} \\
        \hline
        \multicolumn{4}{|c|}{ Other technologies } \\
        \hline
         NEWSdm & AgBr (incl. H,\,C,\,N,\,O) & 10\,kg & \cite{NEWSdm:2023qyb}\\\hline
    \end{tabular}
    \label{tab:Design}
\end{table}

\subsubsection*{Cryogenic crystal calorimeters}

Future cryogenic calorimeter experiments aim to further reduce recoil-energy thresholds while maintaining excellent energy resolution and particle-discrimination capabilities.
These developments are particularly motivated by searches for sub-GeV dark matter, where sensitivity is driven primarily by threshold performance rather than target mass.
A next-generation upgrade of the \textbf{CRESST} experiment has been proposed to extend the sensitivity of cryogenic crystal detectors to increasingly smaller cross sections and lighter dark matter masses \cite{Angloher:2025fzw}.
It will remain at LNGS and will be based on an upgrade of the existing experimental infrastructure of CRESST.
The default concept foresees an array of $\mathcal{O}(100)$ small (2\,g) and large (24\,g) CaWO$_4$ target crystals with a total target mass of almost 2\,kg.
Building upon the low-threshold phonon and scintillation-light detection techniques demonstrated in previous CRESST generations, the upgraded experiment aims to achieve recoil-energy thresholds around 5\,eV$_{\rm nr}$ (30\,eV$_{\rm nr}$) in the small (large) modules together with improved background suppression.
Specifically, the LEE will be addressed in multiple complementary ways, including further dedicated detector layout adaptations and the exploitation of time and pulse-shape information.
Figure~\ref{fig:exp_nearfuture_SI} (red-dotted line) shows the projected sensitivity to spin-independent dark matter--nucleon scattering.
The projections assume 26 small modules with a total exposure of 15\,kg\,$\cdot$\,d and a threshold of 5\,eV$_{\rm nr}$, together with 70 large modules with an exposure of 500\,kg\,$\cdot$\,d and a threshold of 30.1\,eV$_{\rm nr}$.
The background model is based on the measured spectrum of CRESST-III Detector A, with the LEE component reduced by factors of 10 or 100.
The primary scientific goal of the next-generation CRESST experiment is the exploration of sub-GeV dark matter parameter space, where cryogenic calorimeters remain among the most sensitive direct detection technologies.\\

\textbf{RES-NOVA} is a proposed cryogenic calorimeter experiment primarily designed to detect neutrinos from galactic core-collapse supernovae through CE$\nu$NS, but also offering sensitivity to dark matter interactions~\cite{RES-NOVACollaboration:2025stq}.
While a first dark matter search using a 13\,g PbWO$_4$ prototype has already been reported (see Sec.~\ref{sssec:exp_running_NR}), the following discussion concerns the planned, substantially larger detector demonstrator.
The experiment plans to employ PbWO$_4$ crystals produced using archaeological lead, exploiting its exceptionally low intrinsic radioactivity and the complementary scattering kinematics of heavy Pb and light O nuclei.
The initial detector demonstrator, planned for LNGS, foresees 84 cryogenic modules with a total target mass of approximately 170\,kg, followed by a larger 1.8\,t detector.
Each module combines a phonon-sensitive crystal absorber with a separate cryogenic light detector, allowing discrimination between nuclear and electron recoils.
The demonstrator aims for a nuclear recoil energy threshold of approximately 1\,keV$_{\rm nr}$.
Figure~\ref{fig:exp_nearfuture_SI} (brown-dotted line) shows the projected sensitivity to spin-independent dark matter--nucleon scattering.
The projections assume an exposure of 170\,kg\,$\cdot$\,y, corresponding to one year of operation, an energy resolution of 200\,eV$_{\rm nr}$, and a 50\% trigger efficiency at 1\,keV$_{\rm nr}$, with a hard energy cutoff at 200\,eV$_{\rm nr}$.
The background model includes contributions from environmental radiation, detector materials, and surface contamination.
Two scenarios are considered, with the conservative one assuming no particle discrimination and the optimistic one assuming complete rejection of electron and gamma backgrounds.
In the optimistic scenario, sensitivity projections reach spin-independent dark matter--nucleon cross sections of approximately $1\times10^{-43}$\,cm$^2$ at 2\,GeV and $2\times10^{-46}$\,cm$^2$ at 20\,GeV.
The natural abundance of $^{207}$Pb also provides sensitivity to spin-dependent dark matter-neutron interactions.

\subsubsection*{Semiconductors}

\textbf{Oscura} is a proposed direct dark matter experiment based on silicon skipper charge-coupled devices (CCDs), aiming to substantially improve the sensitivity to sub-GeV dark matter through electron scattering and nuclear scattering accompanied by electronic excitations~\cite{Oscura:2022vmi}. Building upon the skipper-CCD technology demonstrated by SENSEI and DAMIC-M, the experiment plans to deploy approximately 10\,kg of silicon detectors, comprising about 28 billion pixels, with single-electron charge resolution. The detector array will be operated at temperatures between 120 and 140\,K in a liquid nitrogen cryostat and surrounded by lead, copper, and neutron shielding to suppress radioactive backgrounds. The experiment aims to achieve a background level below 0.01\,DRU in the region of interest, corresponding to fewer than one background event between two and ten ionization electrons over the full exposure. Figures~\ref{fig:exp_nearfuture_SI} and \ref{fig:exp_nearfuture_ER} (dark-purple-dotted lines) show the projected sensitivities to spin-independent dark matter--nucleon scattering through the Migdal effect and to dark matter--electron scattering, respectively. The projections assume an exposure of 30\,kg\,$\cdot$\,y, an analysis threshold of two ionization electrons (corresponding approximately to 5\,eV$_{\rm ee}$), and negligible background above this threshold. For the electron scattering projections, a single-electron dark-count rate of $10^{-6}$\,e$^-$/pixel/day is additionally assumed. Oscura aims to probe dark matter--electron scattering for masses down to approximately 0.5\,MeV and also offers sensitivity to bosonic dark matter absorption for masses down to the silicon band gap of approximately 1.1\,eV.

\subsubsection*{Scintillation crystals}

In contrast to NaI-based annual modulation searches, \textbf{ADAMO} pursues a directional dark matter detection approach using ZnWO$_4$ scintillation crystals~\cite{Cappella:2013rua,Leoncini:2025haz}.
Unlike gaseous directional detectors, which aim to reconstruct individual nuclear-recoil tracks, ADAMO exploits the dependence of the scintillation light yield and pulse shape on the direction of the recoil relative to the crystal axes.
As the orientation of the crystal changes with respect to the expected Galactic dark matter wind due to Earth's rotation, this anisotropy could produce a sidereal daily modulation in the detected event rate.
The collaboration has investigated the anisotropic scintillation response using alpha particles and neutron-induced nuclear recoils, providing an experimental basis for this approach.
In particular, anisotropy in the response to oxygen nuclear recoils has been demonstrated with a statistical significance of 5.4$\sigma$ at recoil energies down to approximately 100\,keV$_{\rm nr}$~\cite{Leoncini:2025haz}.
A full-scale detector with approximately 200\,kg of ZnWO$_4$ has been considered in sensitivity projections, although the experiment remains in the R\&D phase.
These projections target the sensitivity to a sidereal modulation signal rather than conventional exclusion limits on the dark matter-nucleon scattering cross section, and are therefore not directly comparable to the sensitivity curves shown in Figs.~\ref{fig:exp_nearfuture_SI} and \ref{fig:exp_nearfuture_SD}.
\\

NaI(Tl)-based dark matter searches continue to pursue annual modulation measurements, motivated in part by the long-standing signal reported by DAMA/LIBRA~\cite{Bernabei:2020mon,Bernabei:2026fdk}.
However, the results from ANAIS-112 and COSINE-100 have already provided important independent tests of this claim and have placed significant constraints on its interpretation as a dark matter signal~\cite{ANAIS-112:2025fne}.
Future NaI(Tl)-based experiments therefore aim to improve radiopurity, detector stability, energy thresholds, and target mass to further strengthen the tests of the DAMA/LIBRA result under comparable experimental conditions.\\

\textbf{COSINE-200} is a proposed successor to COSINE-100 and plans to deploy approximately 200\,kg of NaI(Tl) crystals with substantially reduced internal backgrounds, in particular from $^{210}$Pb and $^{40}$K contamination~\cite{COSINE:2020egt}.
The anticipated energy threshold in 12.5\,kg crystals is 1\,keV$_{\rm ee}$ for an ROI from $(1-6)$\,keV$_{\rm ee}$.
The expected background level in the ROI is below 0.5\,DRU, lower than the DAMA/LIBRA background level.
The experiment is thus designed to improve sensitivity to annual modulation signals and to provide a decisive test of the DAMA/LIBRA parameter space.\\

Beyond testing the DAMA/LIBRA claim, NaI is an interesting target for spin-dependent low-mass dark matter particles because both naturally occurring nuclei, $^{23}$Na and $^{127}$I, have non-zero nuclear spin.
A proposed low-threshold \textbf{NaI} initiative aims to exploit this potential by combining radiopure crystals with substantially reduced energy thresholds.
Reducing the energy threshold down to 100\,eV$_{\rm ee}$ would provide competitive sensitivities for sub-GeV dark matter if combined with radiopure crystals.
Projects such as \textbf{ANAIS+}~\cite{Amaré:2026XQ} and \textbf{ASTAROTH}~\cite{Martinenghi:2025ckb} are exploring the operation of NaI crystals at cryogenic temperatures and the use of SiPMs for the light readout.
Figure~\ref{fig:exp_nearfuture_SD} (light-purple-dotted curve, labeled as NaI) shows the sensitivity prospects for a generic NaI detector achieving a 250\,eV$_{\rm ee}$ threshold for an exposure of 1\,kg$\cdot$y and a background level of 0.2\,DRU in the low-energy region~\cite{Martinez:2026thh}.\\

\textbf{PICOLON} is a staged NaI(Tl)-based dark matter program under development~\cite{Fushimi:2021mez}.
The full-scale experiment envisions a target mass of approximately 250\,kg at Kamioka and emphasizes ultra-high radiopurity crystal development together with improved shielding and background suppression techniques.
The collaboration has demonstrated high radiopurity in a prototype crystal, with activities of $(1.2\pm1.4)\,\upmu$Bq/kg for the $^{232}$Th series and $(13\pm4)\,\upmu$Bq/kg for $^{226}$Ra, together with an upper limit of $5.7\,\upmu$Bq/kg (90\% C.L.) for $^{210}$Pb.
These values are comparable to or below those reported for DAMA/LIBRA crystals, which range from $(2-31)\,\upmu$Bq/kg for the $^{232}$Th series, $(8.7-124)\,\upmu$Bq/kg for $^{226}$Ra, and $(10-30)\,\upmu$Bq/kg for $^{210}$Pb~\cite{Fushimi:2021mez}.
Similar to COSINE-200, its primary scientific objective is a sensitive search for annual modulation signatures in NaI(Tl).

\subsubsection*{Liquid noble-gas scintillators}

Several next-generation liquid noble-gas detectors have been proposed to extend the sensitivity well beyond the reach of currently operating experiments.
All of the proposed experiments are based on time projection chambers (TPCs, see Sec.~\ref{ssec:tech_noble}) using either liquid xenon (LXe), argon (LAr), or helium (LHe) as target.
The LXe and LAr TPCs primarily focus on scaling the target mass to the multi-tonne level while maintaining extremely low radioactive backgrounds to further enhance the sensitivity to dark matter--nucleon scattering in the GeV to TeV mass range, ultimately reaching into the neutrino signal region.
The use of LHe instead supports searches for sub-GeV dark matter particles, given the lighter mass of the target nuclei, with a stronger emphasis on reducing the accessible dark matter mass threshold rather than maximizing sensitivity at higher masses.\\

\textbf{ALETHEIA} is a proposed liquid helium TPC targeting low-mass dark matter through the combined detection of primary scintillation (S1) and ionization signals (S2)~\cite{Liao:2021npo}.
Unlike cryogenic superfluid helium calorimeters, ALETHEIA proposes to operate liquid helium at approximately 4\,K and to exploit the potential for discrimination between electronic and nuclear recoils using scintillation and ionization information.
The development program includes a 30\,g prototype followed by an approximately 10\,kg detector intended to demonstrate the feasibility of scaling the technology towards larger target masses.
The projected spin-independent dark matter--nucleon sensitivity extends over masses of approximately $(0.5-10)$\,GeV, with the strongest sensitivity of order $5\times10^{-42}$\,cm$^2$ near 2\,GeV for an exposure of 1\,kg$\cdot$y.
These projections assume negligible instrumental backgrounds and a nuclear-recoil detection efficiency of 50\% over an energy interval of approximately $(0.5-10)$\,keV$_{\rm nr}$.
The corresponding sensitivity is shown in Fig.~\ref{fig:exp_nearfuture_SI} (blue dotted line).
Larger exposures of 100\,kg$\cdot$y and 1\,t$\cdot$y have also been considered, with the latter intended to approach the neutrino signal region.
However, the projected performance, particularly the low-energy nuclear-recoil response and discrimination capability, remains to be demonstrated experimentally.\\

\textbf{ARGO} is a proposed multi-hundred-tonne liquid argon TPC developed within the DarkSide program~\cite{DarkSide20k:2020ymr}.
Building upon the low-background techniques of DarkSide-50 and DarkSide-20k, the experiment aims to employ underground argon with strongly reduced $^{39}$Ar activity, together with scintillation pulse-shape discrimination and ionization readout to suppress electronic-recoil backgrounds.
The proposed detector has an active target mass of approximately 370\,t, with a fiducial mass of order 300\,t.
An exposure of approximately 3000\,t$\cdot$y over ten years has been considered for searches for spin-independent WIMP--nucleon scattering, with the aim of reaching the neutrino signal region at high dark matter masses.
In addition to its dark matter program, ARGO is expected to be sensitive to astrophysical neutrinos through CE$\upnu$NS.
Studies of supernova-neutrino detection consider an ionization-based nuclear-recoil threshold of approximately 0.5\,keV$_{\rm nr}$, demonstrating the potential of the large target mass for detecting neutrino bursts from Galactic supernovae.
This threshold refers to the dedicated low-energy neutrino analysis and should not be interpreted as the threshold assumed for the conventional WIMP search.
ARGO is not included in the sensitivity projections shown in Fig.~\ref{fig:exp_nearfuture_SI}.\\

\textbf{XLZD} is a planned large-scale liquid xenon observatory combining the expertise and technologies developed by the XENON, LZ, and DARWIN programs~\cite{XLZD:2024nsu}.
The baseline design foresees a dual-phase LXe TPC with an active target mass of 60\,t, potentially expandable to 80\,t, making it one of the largest direct dark matter detectors proposed.
The experiment aims to achieve exceptionally low radioactive backgrounds through material selection, xenon purification, and active veto systems, with particular emphasis on suppressing $^{222}$Rn and krypton contamination.
For a projected fiducial exposure of 200\,t$\cdot$y, assuming a nuclear-recoil energy threshold of 3\,keV$_{\rm nr}$ and radioactive backgrounds suppressed below the irreducible neutrino backgrounds, XLZD expects a 90\% C.L. exclusion sensitivity to spin-independent WIMP--nucleon cross sections of approximately $2\times10^{-49}$\,cm$^2$ at a dark matter mass of 40\,GeV.
The corresponding sensitivity is shown in Fig.~\ref{fig:exp_nearfuture_SI} (blue dotted line).
An extended exposure of up to 1000\,t$\cdot$y is envisioned to further improve its discovery potential within the neutrino signal region, with a projected 3$\sigma$ evidence sensitivity of approximately $3\times10^{-49}$\,cm$^2$ at 40\,GeV.
Beyond dark matter searches, XLZD will also provide sensitivity to rare processes such as neutrinoless double-beta decay of $^{136}$Xe and measurements of solar and supernova neutrinos~\cite{XLZD:2024pdv}.

\subsubsection*{Superfluid noble-gas detectors}

Superfluid helium detectors constitute an emerging class of experiments designed to achieve extremely low recoil-energy thresholds by exploiting the excitations generated upon an energy deposition in the superfluid, combined with highly sensitive quantum-sensing devices (see Sec.~\ref{ssec:tech_superfluid}).
These detectors are particularly well suited for searches for light dark matter particles given the low mass of helium nuclei as scattering partners.
The proposed approaches differ in their detection mechanisms, with superfluid $^4$He experiments exploiting phonons, rotons, scintillation photons, and excimers, while superfluid $^3$He detectors can measure changes in the density of fermionic quasiparticles.\\

\textbf{DELight} is a planned staged superfluid $^4$He calorimetric dark matter search, with a target mass in the range of $(0.1-1)$\,kg in its first phase~\cite{vonKrosigk:2022vnf}.
The experiment will initially employ 56 large-area cryogenic microcalorimeters (LAMCALs) based on MMC technology (see~Sec.~\ref{ssec:tech_bolom}), targeting an intrinsic energy resolution of a few eV.
The calorimeters will be operated at temperatures below 20\,mK, with a film burner employed to suppress superfluid helium film flow onto the sensor wafers positioned above the liquid.
For an initial exposure of 1\,kg$\cdot$d and an anticipated detector threshold of approximately 20\,eV$_{\rm nr}$, the projected sensitivity extends to dark matter masses below 100\,MeV.
The sensitivity projections assume background levels of 1 and 10\,DRU for nuclear and electronic recoils, respectively.
In its second phase, DELight aims to lower the detector threshold to 10\,eV$_{\rm nr}$, increase the exposure to 100\,kg$\cdot$d, and reduce both background levels by a factor of ten.
The corresponding sensitivity projections are shown in Fig.~\ref{fig:exp_nearfuture_SI} (orange dotted line).
The experiment is currently in its prototyping phase, with an operational R\&D helium cell being used to investigate the detector response and validate key design parameters~\cite{Solmaz:2025xiq}.\\

\textbf{QUEST-DMC} is a proposed superfluid $^3$He detector employing superconducting nanomechanical resonators for the calorimetric detection of particle interactions~\cite{QUEST-DMC:2023nug}.
The detector concept foresees five bolometer cells containing approximately 0.03\,g of $^3$He each, operated at temperatures around 100\,$\upmu$K using nuclear demagnetization refrigeration.
The sensitivity projections assume six months of live data taking, corresponding to an exposure of 4.9\,g$\cdot$d, and include simulated radiogenic, cosmic-ray, and solar-neutrino backgrounds.
For conventional and SQUID-based readout, nuclear-recoil energy thresholds of 31 and 0.51\,eV$_{\rm nr}$ are predicted, respectively.
The projections assume operation at the surface with a cosmic-ray veto efficiency of 90\%, while the radiogenic background estimates are based on material radioassay measurements and GEANT4 simulations.
The SQUID-based readout scenario provides sensitivity to dark matter masses down to approximately 25\,MeV, with particularly promising prospects for spin-dependent dark matter-neutron scattering.
The corresponding spin-independent and spin-dependent sensitivity projections are shown in Figs.~\ref{fig:exp_nearfuture_SI} and \ref{fig:exp_nearfuture_SD} (teal dotted lines).

\subsubsection*{Gaseous detectors}

Future gaseous dark matter detectors pursue two complementary approaches: directional detection through the reconstruction of nuclear-recoil tracks and low-threshold searches using spherical proportional counters (see Sec.~\ref{ssec:tech_gas}).
Directional sensitivity would provide a powerful means of distinguishing dark matter signals from backgrounds and could enable confirmation of their Galactic origin.
Spherical proportional counters instead primarily exploit their low energy thresholds and the flexibility of using different target gases to probe light dark matter particles.\\

\textbf{CYGNO} is a proposed directional gaseous TPC employing a He:CF$_4$ mixture of 60:40 at atmospheric pressure, together with optical readout based on scientific CMOS cameras and photodetectors~\cite{Amaro:2026mes,DAstolfo:2024cei}.
The experiment follows a staged development program towards active volumes of approximately $(30\,000-100\,000)$\,L, with the aim of achieving three-dimensional nuclear-recoil track reconstruction at low energies.
The detector technology is being developed and tested at LNGS, with particular emphasis on improving gas amplification, optical readout, and background rejection.
Recent studies have investigated the addition of $(3-10)$\% CH$_4$ to the baseline gas mixture, reporting improved electrical stability and scintillation yield at higher amplification voltages, while introducing hydrogen as an additional scattering target for low-mass dark matter~\cite{Amaro:2026mes}.
CYGNO also serves as a technological pathfinder for larger directional observatories.\\

\textbf{CYGNUS} is an international initiative aiming to establish a network of large-volume directional gaseous detectors, incorporating several of the directional detection efforts discussed in Sec.~\ref{sssec:exp_running_directionality}~\cite{Vahsen:2020pzb}.
The proposed observatory would consist of modular TPCs distributed across multiple underground sites, with different gas mixtures and readout technologies under consideration.
The CYGNUS-1000 reference study considers a total active volume of 1000\,m$^3$, using helium and SF$_6$ at approximately atmospheric pressure.
Depending on the readout technology, the sensitivity projections consider nuclear-recoil thresholds of approximately 6 or 20\,keV$_{\rm nr}$, together with directional information and electronic-recoil background rejection.
The study finds that approximately $(10-20)$ helium recoils above 6\,keV$_{\rm nr}$, or $(3-4)$ above 20\,keV$_{\rm nr}$, could be sufficient to distinguish a 10\,GeV dark matter signal from the solar-neutrino background.
The projected spin-independent sensitivity extends into previously unexplored parameter space below 10\,GeV, as shown in Fig.~\ref{fig:exp_nearfuture_SI} (pink dotted line).
The modular concept also allows for smaller initial detectors, while the ultimate goal is a large-scale observatory capable of probing dark matter interactions in the neutrino signal region.\\

The NEWS-G collaboration is developing two next-generation spherical proportional counters for light dark matter searches, \textbf{ECuME} and \textbf{DarkSPHERE}~\cite{Knights:2023fry,NEWS-G:2023qwh}.
ECuME foresees a 1.4\,m diameter copper vessel with an active volume of approximately 1400\,L, produced by underground electroforming to suppress radioactive backgrounds, particularly those associated with $^{210}$Pb contamination.
Sensitivity projections consider a Ne:CH$_4$ mixture of 90:10 at 2\,bar, an exposure of 200\,kg$\cdot$d, and a background level of 0.3\,DRU.
DarkSPHERE extends this approach to a 3\,m diameter detector with an active volume of approximately 14\,000\,L, proposed for Boulby.
Its baseline configuration employs a He:i-C$_4$H$_{10}$ mixture of 90:10 at 5\,bar, together with an electroformed copper vessel and water shielding to achieve a projected background level below 0.01\,DRU.
For 300 days of operation, the projected spin-independent sensitivity reaches approximately $2\times10^{-41}$\,cm$^2$ at 0.1\,GeV and $2\times10^{-43}$\,cm$^2$ at 1\,GeV.
The hydrogen content of the gas mixture also provides sensitivity to spin-dependent dark matter--proton scattering, while the natural abundance of $^{13}$C allows searches for spin-dependent neutron interactions.
The corresponding sensitivity projections are shown in Figs.~\ref{fig:exp_nearfuture_SI}, \ref{fig:exp_nearfuture_SD}, and \ref{fig:exp_nearfuture_ER} (gray dotted lines).\\

\subsubsection*{Other technologies}

\textbf{NEWSdm} is a proposed nuclear-emulsion-based directional dark matter experiment employing nanometric tracking in solid targets~\cite{NEWSdm:2023qyb}.
The detector concept uses Nano Imaging Trackers (NIT), nuclear emulsions containing AgBr crystals with diameters down to approximately 20\,nm, to preserve the direction of nuclear-recoil tracks.
In addition to Ag and Br, the emulsions contain lighter nuclei, particularly H, C, N, and O, which are the principal recoil targets considered in the sensitivity study.
A proposed 10\,kg detector module would be mounted on an equatorial telescope to compensate for the Earth's rotation and maintain its orientation relative to the expected dark matter arrival direction.
A recent sensitivity study considers cosmic-ray-boosted dark matter, for which the comparatively long recoil tracks allow operation at the surface laboratory of LNGS~\cite{NEWSdm:2023qyb}.
For an exposure of 10\,kg$\cdot$y, the projected directional sensitivity extends to dark matter masses between 1\,keV and 1\,GeV, with scattering cross sections reaching approximately $10^{-30}$\,cm$^2$.
Unlike conventional halo dark matter searches, this projection relies on the acceleration of dark matter particles through scattering with Galactic cosmic rays and their resulting directional flux from the Galactic Center (See Sec.~\ref{ssec:astrophysical}).\\

\textbf{Paleo-detectors} offer a complementary approach to direct dark matter detection by exploiting ancient minerals that have accumulated nuclear-recoil damage tracks over geological timescales~\cite{Theodosopoulos:2026ehn}.
A recent sensitivity study considers mineral samples with an age of 1\,Gyr and two representative readout scenarios: a high-resolution configuration with 10\,mg of material and a spatial resolution of 1\,nm, and a high-exposure configuration with 100\,g of material and a resolution of 15\,nm.
The study investigates several minerals, including gypsum, halite, olivine, and muscovite, accounting for backgrounds from neutrinos, radiogenic neutrons, and radioactive decays.
For dark matter masses between approximately 1 and 10\,GeV, the high-resolution scenario is projected to achieve sensitivities beyond current experimental constraints for a broad range of nonrelativistic effective field theory (NREFT) interactions.
At higher masses, the larger-exposure scenario can provide comparable or improved sensitivity for several NREFT operators.
The approach remains at an early stage of development, with the feasibility of reconstructing sufficiently short recoil tracks and controlling mineral-specific backgrounds representing important experimental challenges.\\

\FloatBarrier
\section{Conclusion and outlook}
\label{sec:conclusion}

Understanding the nature of the dark matter remains an open question at the frontier of particle physics, cosmology and astrophysics. Complementary approaches are required to address this challenge. For about 40 years, direct detection strategy has provided strong constraints on many dark matter candidates and models. 

Numerous candidates have been proposed to explain dark matter within extensions of the Standard Model of particle physics. The microscopic nature of the dark matter particle determines its experimental signatures, resulting in nuclear or electron recoils with distinct spectral shapes. In addition, annual and daily modulations of the signal, as well as directional signatures, are expected because of the variation of the relative velocity between the WIMP and the detector.

The direct search for dark matter has driven the development of advanced particle detection techniques, enabling powerful particle discrimination for background rejection and ultra-high radiopurity in their manufacturing as well as scaling to several tonnes of detector mass.
A multi-target, multi-technique strategy is essential in the direct search for dark matter to disentangle uncertainties in predicted signal rates and spectra. Any potential signal observed with one target should be independently confirmed using different target materials and, where possible, complementary detection techniques, to constrain the properties of the dark matter particle. 

In the absence of a positive detection, dark matter direct-detection experiments face two main sensitivity challenges: probing higher-mass candidates (above a few GeV) down to the neutrino floor requires substantial background reduction and scalability to multi-tonne target masses, whereas extending the reach to lower-mass candidates relies primarily on reducing energy thresholds.

At present, the search for high-mass WIMPs is dominated by liquid xenon experiments, with both xenon and argon-based experiments expected to play a leading role in the future. A major challenge for these experiments is the control and mitigation of internal and accidental backgrounds. 
In the low-mass WIMP regime, a wide range of target materials and detection techniques are being pursued to extend sensitivity to unexplored parameter space regions. Current efforts focus on improving detector performance, reducing the energy threshold, while developing more effective strategies to characterize and mitigate backgrounds, in particular the low-energy excess observed in most low-threshold detectors. Sensitivity can also be enhanced through the selection of target nuclei suited to specific dark matter interaction models, taking advantage of nuclear spin for spin-dependent interactions and favorable recoil kinematics. 
Across direct dark matter detection technologies, improving the calibration of detector responses to both electron recoils and nuclear recoils remains a common challenge. Novel approaches are needed to accommodate the large target masses of tonne-scale experiments and to extend calibration capabilities to the sub-keV and even eV energy scales.
At present, the neutrino fog sets the horizon for the next generation of dark matter detectors. Overcoming this challenge will require novel detection techniques and analysis strategies to distinguish dark matter signals from neutrino-induced backgrounds.

While the nature of dark matter remains elusive, the increasing development of complementary theoretical and experimental approaches offers promising prospects for uncovering its particle nature.
	
	
	\section*{Acknowledgments}
	The authors would like to thank Daniel Baxter, Anna Bertolini, Susana Cebri\'an, Eleanor Fascione, Bradley Kavanagh, Mar\'ia Mart\'inez, Gray Putnam, and Francesco Toschi for their valuable feedback and constructive comments on the text. 
	This work was supported by the Max Planck Society and by MCIN/AEI/10.13039/501100011033 under grants PID2025-170770NB-C22 and PID2022-138357NB-C21.
    
	\section*{Author's contributions}
    All authors contributed equally to the preparation of this review article.

    \bibliographystyle{utphys}
	\bibliography{mybibfile}

\providecommand{\href}[2]{#2}\begingroup\raggedright\begin{thebibliography}{100}

\bibitem{Aghanim:2018eyx}
{\bfseries Planck} Collaboration, N.~Aghanim {\em et~al.}, ``{Planck 2018 results. VI. Cosmological parameters},'' \href{https://dx.doi.org/10.1051/0004-6361/201833910}{{\em Astron. Astrophys.} {\bfseries 641} (2020) A6}, \href{https://arxiv.org/abs/1807.06209}{{\ttfamily arXiv:1807.06209}}.

\bibitem{ParticleDataGroup:2026mpi}
{\bfseries Particle Data Group} Collaboration, F.~Takahashi {\em et~al.}, ``{Review of Particle Physics},'' \href{https://dx.doi.org/10.1142/s0217751x26300115}{{\em Int. J. Mod. Phys. A} {\bfseries 41} no.~22, (2026) 2630011}.

\bibitem{Clowe:2006eq}
D.~Clowe, M.~Bradac, A.~H. Gonzalez, M.~Markevitch, S.~W. Randall, C.~Jones, and D.~Zaritsky, ``{A Direct Empirical Proof of the Existence of Dark Matter},'' \href{https://dx.doi.org/10.1086/508162}{{\em Astrophys. J.} {\bfseries 648} (2006) L109}, \href{https://arxiv.org/abs/astro-ph/0608407}{{\ttfamily arXiv:astro-ph/0608407}}.

\bibitem{BOEHM_2024116503}
C.~Boehm, ``Particle dark matter versus modified gravity,'' \href{https://dx.doi.org/https://doi.org/10.1016/j.nuclphysb.2024.116503}{{\em Nucl. Phys. B} {\bfseries 1003} (2024) 116503}.

\bibitem{Green:2020jor}
A.~M. Green and B.~J. Kavanagh, ``{Primordial black holes as a dark matter candidate},'' \href{https://dx.doi.org/10.1088/1361-6471/abc534}{{\em J. Phys. G} {\bfseries 48} (2021) 043001}, \href{https://arxiv.org/abs/2007.10722}{{\ttfamily arXiv:2007.10722}}.

\bibitem{Feng:2010gw}
J.~L. Feng, ``{Dark Matter Candidates from Particle Physics and Methods of Detection},'' \href{https://dx.doi.org/10.1146/annurev-astro-082708-101659}{{\em Ann. Rev. Astron. Astrophys.} {\bfseries 48} (2010) 495}, \href{https://arxiv.org/abs/1003.0904}{{\ttfamily arXiv:1003.0904}}.

\bibitem{Zurek:2024vma}
K.~M. Zurek, ``{Dark Matter Candidates of a Very Low Mass},'' \href{https://dx.doi.org/10.1146/annurev-nucl-101918-023542}{{\em Ann. Rev. Nucl. Part. Sci.} {\bfseries 74} no.~1, (2024) 287}, \href{https://arxiv.org/abs/2401.03025}{{\ttfamily arXiv:2401.03025}}.

\bibitem{Billard:2022uyg}
J.~Billard, M.~Boulay, {\em et~al.}, ``{Direct detection of dark matter -- APPEC committee report},'' \href{https://dx.doi.org/10.1088/1361-6633/ac5754}{{\em Rept. Prog. Phys.} {\bfseries 85} no.~5, (2022) 056201}, \href{https://arxiv.org/abs/2104.07634}{{\ttfamily arXiv:2104.07634}}.

\bibitem{Conrad:2017pda}
J.~Conrad and O.~Reimer, ``{Indirect dark matter searches in gamma and cosmic rays},'' \href{https://dx.doi.org/10.1038/nphys4049}{{\em Nature Phys.} {\bfseries 13} (2017) 224}, \href{https://arxiv.org/abs/1705.11165}{{\ttfamily arXiv:1705.11165}}.

\bibitem{Boveia:2018yeb}
A.~Boveia and C.~Doglioni, ``{Dark Matter Searches at Colliders},'' \href{https://dx.doi.org/10.1146/annurev-nucl-101918-023501}{{\em Ann. Rev. Nucl. Part. Sci.} {\bfseries 69} (2019) 197}, \href{https://arxiv.org/abs/1810.12238}{{\ttfamily arXiv:1810.12238}}.

\bibitem{DiLuzio:2020wdo}
L.~Di~Luzio, M.~Giannotti, E.~Nardi, and L.~Visinelli, ``{The landscape of QCD axion models},'' \href{https://dx.doi.org/10.1016/j.physrep.2020.06.002}{{\em Phys. Rept.} {\bfseries 870} (2020) 1}, \href{https://arxiv.org/abs/2003.01100}{{\ttfamily arXiv:2003.01100}}.

\bibitem{Irastorza:2018dyq}
I.~G. Irastorza and J.~Redondo, ``{New experimental approaches in the search for axion-like particles},'' \href{https://dx.doi.org/10.1016/j.ppnp.2018.05.003}{{\em Prog. Part. Nucl. Phys.} {\bfseries 102} (2018) 89}, \href{https://arxiv.org/abs/1801.08127}{{\ttfamily arXiv:1801.08127}}.

\bibitem{Ahlen:1987mn}
S.~P. Ahlen, F.~T. Avignone, R.~L. Brodzinski, A.~K. Drukier, G.~Gelmini, and D.~N. Spergel, ``{Limits on Cold Dark Matter Candidates from an Ultralow Background Germanium Spectrometer},'' \href{https://dx.doi.org/10.1016/0370-2693(87)91581-4}{{\em Phys. Lett. B} {\bfseries 195} (1987) 603}.

\bibitem{Bernabei:2020mon}
R.~Bernabei {\em et~al.}, ``{The DAMA project: Achievements, implications and perspectives},'' \href{https://dx.doi.org/10.1016/j.ppnp.2020.103810}{{\em Prog. Part. Nucl. Phys.} {\bfseries 114} (2020) 103810}.

\bibitem{Bernabei:2026fdk}
R.~Bernabei {\em et~al.}, ``{Model-independent results from DAMA/LIBRA{\textendash}phase2-empowered},'' \href{https://dx.doi.org/10.15407/jnpae2026.02.105}{{\em Nucl. Phys. Atom. Energy} {\bfseries 27} no.~2, (2026) 105}.

\bibitem{Essig:2022dfa}
R.~Essig {\em et~al.}, ``{Snowmass2021 Cosmic Frontier: The landscape of low-threshold dark matter direct detection in the next decade},'' in {\em {Snowmass 2021}}.
\newblock 3, 2022.
\newblock \href{https://arxiv.org/abs/2203.08297}{{\ttfamily arXiv:2203.08297}}.

\bibitem{Misiaszek:2023sxe}
M.~Misiaszek and N.~Rossi, ``{Direct Detection of Dark Matter: A Critical Review},'' \href{https://dx.doi.org/10.3390/sym16020201}{{\em Symmetry} {\bfseries 16} no.~2, (2024) 201}, \href{https://arxiv.org/abs/2310.20472}{{\ttfamily arXiv:2310.20472}}.

\bibitem{Baudis:2025yva}
L.~Baudis, ``{Dark matter direct detection: status, results and future plans},'' \href{https://dx.doi.org/10.1088/1742-6596/3162/1/012007}{{\em J. Phys. Conf. Ser.} {\bfseries 3162} no.~1, (2025) 012007}, \href{https://arxiv.org/abs/2512.23039}{{\ttfamily arXiv:2512.23039}}.

\bibitem{Cerdeno:2010jj}
D.~G. Cerdeño and A.~M. Green, {\em Direct detection of WIMPs}, \href{https://dx.doi.org/10.1017/CBO9780511770739.018}{p.~347–369}.
\newblock Cambridge University Press, 2010.
\newblock \href{https://arxiv.org/abs/1002.1912}{{\ttfamily arXiv:1002.1912}}.

\bibitem{Lin:2019uvt}
T.~Lin, ``{Dark matter models and direct detection},'' \href{https://dx.doi.org/10.22323/1.333.0009}{{\em PoS} {\bfseries 333} (2019) 009}, \href{https://arxiv.org/abs/1904.07915}{{\ttfamily arXiv:1904.07915}}.

\bibitem{Bertone:2016nfn}
G.~Bertone and D.~Hooper, ``{History of dark matter},'' \href{https://dx.doi.org/10.1103/RevModPhys.90.045002}{{\em Rev. Mod. Phys.} {\bfseries 90} no.~4, (2018) 045002}, \href{https://arxiv.org/abs/1605.04909}{{\ttfamily arXiv:1605.04909}}.

\bibitem{Bauer:2017qwy}
M.~Bauer and T.~Plehn, \href{https://dx.doi.org/10.1007/978-3-030-16234-4}{{\em {Yet Another Introduction to Dark Matter}: {The Particle Physics Approach}}}, vol.~959 of {\em Lecture Notes in Physics}.
\newblock Springer, 2019.
\newblock \href{https://arxiv.org/abs/1705.01987}{{\ttfamily arXiv:1705.01987}}.

\bibitem{Balazs:2024uyj}
C.~Balazs, T.~Bringmann, F.~Kahlhoefer, and M.~White, ``{A Primer on Dark Matter},'' \href{https://dx.doi.org/10.1016/B978-0-443-21439-4.00070-5}{{\em Astrophysics} {\bfseries 5} (2026) 17}, \href{https://arxiv.org/abs/2411.05062}{{\ttfamily arXiv:2411.05062}}.

\bibitem{Cirelli:2024ssz}
M.~Cirelli, A.~Strumia, and J.~Zupan, ``{Dark Matter},'' \href{https://dx.doi.org/10.21468/SciPostPhysRev.1}{{\em SciPost Phys. Rev.} (6, 2026) 1}, \href{https://arxiv.org/abs/2406.01705}{{\ttfamily arXiv:2406.01705}}.

\bibitem{McCabe:2010zh}
C.~McCabe, ``{The Astrophysical Uncertainties Of Dark Matter Direct Detection Experiments},'' \href{https://dx.doi.org/10.1103/PhysRevD.82.023530}{{\em Phys. Rev. D} {\bfseries 82} (2010) 023530}, \href{https://arxiv.org/abs/1005.0579}{{\ttfamily arXiv:1005.0579}}.

\bibitem{Green:2011bv}
A.~M. Green, ``{Astrophysical uncertainties on direct detection experiments},'' \href{https://dx.doi.org/10.1142/S0217732312300042}{{\em Mod. Phys. Lett. A} {\bfseries 27} (2012) 1230004}, \href{https://arxiv.org/abs/1112.0524}{{\ttfamily arXiv:1112.0524}}.

\bibitem{Green:2017odb}
A.~M. Green, ``{Astrophysical uncertainties on the local dark matter distribution and direct detection experiments},'' \href{https://dx.doi.org/10.1088/1361-6471/aa7819}{{\em J. Phys. G} {\bfseries 44} no.~8, (2017) 084001}, \href{https://arxiv.org/abs/1703.10102}{{\ttfamily arXiv:1703.10102}}.

\bibitem{Read:2014qva}
J.~I. Read, ``{The Local Dark Matter Density},'' \href{https://dx.doi.org/10.1088/0954-3899/41/6/063101}{{\em J. Phys. G} {\bfseries 41} (2014) 063101}, \href{https://arxiv.org/abs/1404.1938}{{\ttfamily arXiv:1404.1938}}.

\bibitem{Baxter:2021pqo}
D.~Baxter {\em et~al.}, ``{Recommended conventions for reporting results from direct dark matter searches},'' \href{https://dx.doi.org/10.1140/epjc/s10052-021-09655-y}{{\em Eur. Phys. J. C} {\bfseries 81} no.~10, (2021) 907}, \href{https://arxiv.org/abs/2105.00599}{{\ttfamily arXiv:2105.00599}}.

\bibitem{Fox:2010bz}
P.~J. Fox, J.~Liu, and N.~Weiner, ``{Integrating Out Astrophysical Uncertainties},'' \href{https://dx.doi.org/10.1103/PhysRevD.83.103514}{{\em Phys. Rev. D} {\bfseries 83} (2011) 103514}, \href{https://arxiv.org/abs/1011.1915}{{\ttfamily arXiv:1011.1915}}.

\bibitem{Frandsen:2011gi}
M.~T. Frandsen, F.~Kahlhoefer, C.~McCabe, S.~Sarkar, and K.~Schmidt-Hoberg, ``{Resolving astrophysical uncertainties in dark matter direct detection},'' \href{https://dx.doi.org/10.1088/1475-7516/2012/01/024}{{\em JCAP} {\bfseries 01} (2012) 024}, \href{https://arxiv.org/abs/1111.0292}{{\ttfamily arXiv:1111.0292}}.

\bibitem{Gondolo:2012rs}
P.~Gondolo and G.~B. Gelmini, ``{Halo independent comparison of direct dark matter detection data},'' \href{https://dx.doi.org/10.1088/1475-7516/2012/12/015}{{\em JCAP} {\bfseries 12} (2012) 015}, \href{https://arxiv.org/abs/1202.6359}{{\ttfamily arXiv:1202.6359}}.

\bibitem{Ferrer:2015bta}
F.~Ferrer, A.~Ibarra, and S.~Wild, ``{A novel approach to derive halo-independent limits on dark matter properties},'' \href{https://dx.doi.org/10.1088/1475-7516/2015/09/052}{{\em JCAP} {\bfseries 09} (2015) 052}, \href{https://arxiv.org/abs/1506.03386}{{\ttfamily arXiv:1506.03386}}.

\bibitem{Feldstein:2014gza}
B.~Feldstein and F.~Kahlhoefer, ``{A new halo-independent approach to dark matter direct detection analysis},'' \href{https://dx.doi.org/10.1088/1475-7516/2014/08/065}{{\em JCAP} {\bfseries 08} (2014) 065}, \href{https://arxiv.org/abs/1403.4606}{{\ttfamily arXiv:1403.4606}}.

\bibitem{Gelmini:2017aqe}
G.~B. Gelmini, J.-H. Huh, and S.~J. Witte, ``{Unified Halo-Independent Formalism From Convex Hulls for Direct Dark Matter Searches},'' \href{https://dx.doi.org/10.1088/1475-7516/2017/12/039}{{\em JCAP} {\bfseries 12} (2017) 039}, \href{https://arxiv.org/abs/1707.07019}{{\ttfamily arXiv:1707.07019}}.

\bibitem{Kahlhoefer:2018knc}
F.~Kahlhoefer, F.~Reindl, K.~Sch{\"a}ffner, K.~Schmidt-Hoberg, and S.~Wild, ``{Model-independent comparison of annual modulation and total rate with direct detection experiments},'' \href{https://dx.doi.org/10.1088/1475-7516/2018/05/074}{{\em JCAP} {\bfseries 05} (2018) 074}, \href{https://arxiv.org/abs/1802.10175}{{\ttfamily arXiv:1802.10175}}.

\bibitem{Bernreuther:2023aqe}
E.~Bernreuther, P.~J. Fox, B.~Lillard, A.-M. Taki, and T.-T. Yu, ``{Extracting halo independent information from dark matter electron scattering data},'' \href{https://dx.doi.org/10.1088/1475-7516/2024/03/047}{{\em JCAP} {\bfseries 03} (2024) 047}, \href{https://arxiv.org/abs/2311.04957}{{\ttfamily arXiv:2311.04957}}.

\bibitem{Abuter:2021yys}
R.~Abuter {\em et~al.}, ``{Improved GRAVITY astrometric accuracy from modeling optical aberrations},'' \href{https://dx.doi.org/10.1051/0004-6361/202040208}{{\em Astron. Astrophys.} {\bfseries 647} (2021) A59}, \href{https://arxiv.org/abs/2101.12098}{{\ttfamily arXiv:2101.12098}}.

\bibitem{Evans:2018bqy}
N.~W. Evans, C.~A.~J. O'Hare, and C.~McCabe, ``{Refinement of the standard halo model for dark matter searches in light of the Gaia Sausage},'' \href{https://dx.doi.org/10.1103/PhysRevD.99.023012}{{\em Phys. Rev. D} {\bfseries 99} no.~2, (2019) 023012}, \href{https://arxiv.org/abs/1810.11468}{{\ttfamily arXiv:1810.11468}}.

\bibitem{OHare:2019qxc}
C.~A.~J. O'Hare, N.~W. Evans, C.~McCabe, G.~Myeong, and V.~Belokurov, ``{Velocity substructure from Gaia and direct searches for dark matter},'' \href{https://dx.doi.org/10.1103/PhysRevD.101.023006}{{\em Phys. Rev. D} {\bfseries 101} no.~2, (2020) 023006}, \href{https://arxiv.org/abs/1909.04684}{{\ttfamily arXiv:1909.04684}}.

\bibitem{Smith-Orlik:2023kyl}
A.~Smith-Orlik {\em et~al.}, ``{The impact of the Large Magellanic Cloud on dark matter direct detection signals},'' \href{https://dx.doi.org/10.1088/1475-7516/2023/10/070}{{\em JCAP} {\bfseries 10} (2023) 070}, \href{https://arxiv.org/abs/2302.04281}{{\ttfamily arXiv:2302.04281}}.

\bibitem{Reynoso-Cordova:2024xqz}
J.~Reynoso-Cordova, N.~Bozorgnia, and M.-C. Piro, ``{The Large Magellanic Cloud: expanding the low-mass parameter space of dark matter direct detection},'' \href{https://dx.doi.org/10.1088/1475-7516/2024/12/037}{{\em JCAP} {\bfseries 12} (2024) 037}, \href{https://arxiv.org/abs/2409.09119}{{\ttfamily arXiv:2409.09119}}.

\bibitem{Bringmann:2018cvk}
T.~Bringmann and M.~Pospelov, ``{Novel direct detection constraints on light dark matter},'' \href{https://dx.doi.org/10.1103/PhysRevLett.122.171801}{{\em Phys. Rev. Lett.} {\bfseries 122} no.~17, (2019) 171801}, \href{https://arxiv.org/abs/1810.10543}{{\ttfamily arXiv:1810.10543}}.

\bibitem{Dent:2019krz}
J.~B. Dent, B.~Dutta, J.~L. Newstead, and I.~M. Shoemaker, ``{Bounds on Cosmic Ray-Boosted Dark Matter in Simplified Models and its Corresponding Neutrino-Floor},'' \href{https://dx.doi.org/10.1103/PhysRevD.101.116007}{{\em Phys. Rev. D} {\bfseries 101} no.~11, (2020) 116007}, \href{https://arxiv.org/abs/1907.03782}{{\ttfamily arXiv:1907.03782}}.

\bibitem{Xia:2021vbz}
C.~Xia, Y.-H. Xu, and Y.-F. Zhou, ``{Production and attenuation of cosmic-ray boosted dark matter},'' \href{https://dx.doi.org/10.1088/1475-7516/2022/02/028}{{\em JCAP} {\bfseries 02} no.~02, (2022) 028}, \href{https://arxiv.org/abs/2111.05559}{{\ttfamily arXiv:2111.05559}}.

\bibitem{An:2017ojc}
H.~An, M.~Pospelov, J.~Pradler, and A.~Ritz, ``{Directly Detecting MeV-scale Dark Matter via Solar Reflection},'' \href{https://dx.doi.org/10.1103/PhysRevLett.120.141801}{{\em Phys. Rev. Lett.} {\bfseries 120} no.~14, (2018) 141801}, \href{https://arxiv.org/abs/1708.03642}{{\ttfamily arXiv:1708.03642}}. [Erratum: Phys.Rev.Lett. 121, 259903 (2018)].

\bibitem{An:2021qdl}
H.~An, H.~Nie, M.~Pospelov, J.~Pradler, and A.~Ritz, ``{Solar reflection of dark matter},'' \href{https://dx.doi.org/10.1103/PhysRevD.104.103026}{{\em Phys. Rev. D} {\bfseries 104} no.~10, (2021) 103026}, \href{https://arxiv.org/abs/2108.10332}{{\ttfamily arXiv:2108.10332}}.

\bibitem{Griffin:2018bjn}
S.~Griffin, S.~Knapen, T.~Lin, and K.~M. Zurek, ``{Directional Detection of Light Dark Matter with Polar Materials},'' \href{https://dx.doi.org/10.1103/PhysRevD.98.115034}{{\em Phys. Rev. D} {\bfseries 98} no.~11, (2018) 115034}, \href{https://arxiv.org/abs/1807.10291}{{\ttfamily arXiv:1807.10291}}.

\bibitem{Geilhufe:2019ndy}
R.~M. Geilhufe, F.~Kahlhoefer, and M.~W. Winkler, ``{Dirac Materials for Sub-MeV Dark Matter Detection: New Targets and Improved Formalism},'' \href{https://dx.doi.org/10.1103/PhysRevD.101.055005}{{\em Phys. Rev. D} {\bfseries 101} no.~5, (2020) 055005}, \href{https://arxiv.org/abs/1910.02091}{{\ttfamily arXiv:1910.02091}}.

\bibitem{Boyd:2022tcn}
C.~Boyd, Y.~Hochberg, Y.~Kahn, E.~D. Kramer, N.~Kurinsky, B.~V. Lehmann, and T.~C. Yu, ``{Directional detection of dark matter with anisotropic response functions},'' \href{https://dx.doi.org/10.1103/PhysRevD.108.015015}{{\em Phys. Rev. D} {\bfseries 108} no.~1, (2023) 015015}, \href{https://arxiv.org/abs/2212.04505}{{\ttfamily arXiv:2212.04505}}.

\bibitem{Stratman:2024sng}
C.~Stratman and T.~Lin, ``{Daily modulation of low-energy nuclear recoils from sub-GeV dark matter},'' \href{https://dx.doi.org/10.1103/PhysRevD.111.043030}{{\em Phys. Rev. D} {\bfseries 111} no.~4, (2025) 043030}, \href{https://arxiv.org/abs/2411.03433}{{\ttfamily arXiv:2411.03433}}.

\bibitem{Vahsen:2020pzb}
S.~E. Vahsen {\em et~al.}, ``{CYGNUS: Feasibility of a nuclear recoil observatory with directional sensitivity to dark matter and neutrinos},'' \href{https://arxiv.org/abs/2008.12587}{{\ttfamily arXiv:2008.12587}}.

\bibitem{Mayet:2016zxu}
F.~Mayet {\em et~al.}, ``{A review of the discovery reach of directional Dark Matter detection},'' \href{https://dx.doi.org/10.1016/j.physrep.2016.02.007}{{\em Phys. Rept.} {\bfseries 627} (2016) 1}, \href{https://arxiv.org/abs/1602.03781}{{\ttfamily arXiv:1602.03781}}.

\bibitem{Vahsen:2021gnb}
S.~E. Vahsen, C.~A.~J. O'Hare, and D.~Loomba, ``{Directional Recoil Detection},'' \href{https://dx.doi.org/10.1146/annurev-nucl-020821-035016}{{\em Ann. Rev. Nucl. Part. Sci.} {\bfseries 71} (2021) 189}, \href{https://arxiv.org/abs/2102.04596}{{\ttfamily arXiv:2102.04596}}.

\bibitem{Kavanagh:2016pyr}
B.~J. Kavanagh, R.~Catena, and C.~Kouvaris, ``{Signatures of Earth-scattering in the direct detection of Dark Matter},'' \href{https://dx.doi.org/10.1088/1475-7516/2017/01/012}{{\em JCAP} {\bfseries 01} (2017) 012}, \href{https://arxiv.org/abs/1611.05453}{{\ttfamily arXiv:1611.05453}}.

\bibitem{Bertou:2025adb}
X.~Bertou, A.~Desai, T.~Emken, R.~Essig, T.~Volansky, and T.-T. Yu, ``{Earth-scattering induced modulation in low-threshold dark matter experiments},'' \href{https://dx.doi.org/10.1007/JHEP11(2025)042}{{\em JHEP} {\bfseries 11} (2025) 042}, \href{https://arxiv.org/abs/2507.00344}{{\ttfamily arXiv:2507.00344}}.

\bibitem{Emken:2021vmf}
T.~Emken, J.~Frerick, S.~Heeba, and F.~Kahlhoefer, ``{Electron recoils from terrestrial upscattering of inelastic dark matter},'' \href{https://dx.doi.org/10.1103/PhysRevD.105.055023}{{\em Phys. Rev. D} {\bfseries 105} no.~5, (2022) 055023}, \href{https://arxiv.org/abs/2112.06930}{{\ttfamily arXiv:2112.06930}}.

\bibitem{Kouvaris:2015xga}
C.~Kouvaris and N.~G. Nielsen, ``{Daily modulation and gravitational focusing in direct dark matter search experiments},'' \href{https://dx.doi.org/10.1103/PhysRevD.92.075016}{{\em Phys. Rev. D} {\bfseries 92} no.~7, (2015) 075016}, \href{https://arxiv.org/abs/1505.02615}{{\ttfamily arXiv:1505.02615}}.

\bibitem{Bishara:2017pfq}
F.~Bishara, J.~Brod, B.~Grinstein, and J.~Zupan, ``{From quarks to nucleons in dark matter direct detection},'' \href{https://dx.doi.org/10.1007/JHEP11(2017)059}{{\em JHEP} {\bfseries 11} (2017) 059}, \href{https://arxiv.org/abs/1707.06998}{{\ttfamily arXiv:1707.06998}}.

\bibitem{Lewin:1995rx}
J.~D. Lewin and P.~F. Smith, ``{Review of mathematics, numerical factors, and corrections for dark matter experiments based on elastic nuclear recoil},'' \href{https://dx.doi.org/10.1016/S0927-6505(96)00047-3}{{\em Astropart. Phys.} {\bfseries 6} (1996) 87}.

\bibitem{Hoferichter:2018acd}
M.~Hoferichter, P.~Klos, J.~Men{\'e}ndez, and A.~Schwenk, ``{Nuclear structure factors for general spin-independent WIMP-nucleus scattering},'' \href{https://dx.doi.org/10.1103/PhysRevD.99.055031}{{\em Phys. Rev. D} {\bfseries 99} no.~5, (2019) 055031}, \href{https://arxiv.org/abs/1812.05617}{{\ttfamily arXiv:1812.05617}}.

\bibitem{Feng:2011vu}
J.~L. Feng, J.~Kumar, D.~Marfatia, and D.~Sanford, ``{Isospin-Violating Dark Matter},'' \href{https://dx.doi.org/10.1016/j.physletb.2011.07.083}{{\em Phys. Lett. B} {\bfseries 703} (2011) 124}, \href{https://arxiv.org/abs/1102.4331}{{\ttfamily arXiv:1102.4331}}.

\bibitem{Fitzpatrick:2012ix}
A.~L. Fitzpatrick, W.~Haxton, E.~Katz, N.~Lubbers, and Y.~Xu, ``{The Effective Field Theory of Dark Matter Direct Detection},'' \href{https://dx.doi.org/10.1088/1475-7516/2013/02/004}{{\em JCAP} {\bfseries 02} (2013) 004}, \href{https://arxiv.org/abs/1203.3542}{{\ttfamily arXiv:1203.3542}}.

\bibitem{Anand:2013yka}
N.~Anand, A.~L. Fitzpatrick, and W.~C. Haxton, ``{Weakly interacting massive particle-nucleus elastic scattering response},'' \href{https://dx.doi.org/10.1103/PhysRevC.89.065501}{{\em Phys. Rev. C} {\bfseries 89} no.~6, (2014) 065501}, \href{https://arxiv.org/abs/1308.6288}{{\ttfamily arXiv:1308.6288}}.

\bibitem{Klos:2013rwa}
P.~Klos, J.~Men{\'e}ndez, D.~Gazit, and A.~Schwenk, ``{Large-scale nuclear structure calculations for spin-dependent WIMP scattering with chiral effective field theory currents},'' \href{https://dx.doi.org/10.1103/PhysRevD.88.083516}{{\em Phys. Rev. D} {\bfseries 88} no.~8, (2013) 083516}, \href{https://arxiv.org/abs/1304.7684}{{\ttfamily arXiv:1304.7684}}. [Erratum: Phys.Rev.D 89, 029901 (2014)].

\bibitem{GAMBIT:2021rlp}
{\bfseries GAMBIT} Collaboration, P.~Athron {\em et~al.}, ``{Thermal WIMPs and the scale of new physics: global fits of Dirac dark matter effective field theories},'' \href{https://dx.doi.org/10.1140/epjc/s10052-021-09712-6}{{\em Eur. Phys. J. C} {\bfseries 81} no.~11, (2021) 992}, \href{https://arxiv.org/abs/2106.02056}{{\ttfamily arXiv:2106.02056}}.

\bibitem{Tucker-Smith:2001myb}
D.~Tucker-Smith and N.~Weiner, ``{Inelastic dark matter},'' \href{https://dx.doi.org/10.1103/PhysRevD.64.043502}{{\em Phys. Rev. D} {\bfseries 64} (2001) 043502}, \href{https://arxiv.org/abs/hep-ph/0101138}{{\ttfamily arXiv:hep-ph/0101138}}.

\bibitem{Izaguirre:2015zva}
E.~Izaguirre, G.~Krnjaic, and B.~Shuve, ``{Discovering Inelastic Thermal-Relic Dark Matter at Colliders},'' \href{https://dx.doi.org/10.1103/PhysRevD.93.063523}{{\em Phys. Rev. D} {\bfseries 93} no.~6, (2016) 063523}, \href{https://arxiv.org/abs/1508.03050}{{\ttfamily arXiv:1508.03050}}.

\bibitem{Graham:2010ca}
P.~W. Graham, R.~Harnik, S.~Rajendran, and P.~Saraswat, ``{Exothermic Dark Matter},'' \href{https://dx.doi.org/10.1103/PhysRevD.82.063512}{{\em Phys. Rev. D} {\bfseries 82} (2010) 063512}, \href{https://arxiv.org/abs/1004.0937}{{\ttfamily arXiv:1004.0937}}.

\bibitem{McCabe:2015eia}
C.~McCabe, ``{Prospects for dark matter detection with inelastic transitions of xenon},'' \href{https://dx.doi.org/10.1088/1475-7516/2016/05/033}{{\em JCAP} {\bfseries 05} (2016) 033}, \href{https://arxiv.org/abs/1512.00460}{{\ttfamily arXiv:1512.00460}}.

\bibitem{Dolan:2017xbu}
M.~J. Dolan, F.~Kahlhoefer, and C.~McCabe, ``{Directly detecting sub-GeV dark matter with electrons from nuclear scattering},'' \href{https://dx.doi.org/10.1103/PhysRevLett.121.101801}{{\em Phys. Rev. Lett.} {\bfseries 121} no.~10, (2018) 101801}, \href{https://arxiv.org/abs/1711.09906}{{\ttfamily arXiv:1711.09906}}.

\bibitem{Ibe:2017yqa}
M.~Ibe, W.~Nakano, Y.~Shoji, and K.~Suzuki, ``{Migdal Effect in Dark Matter Direct Detection Experiments},'' \href{https://dx.doi.org/10.1007/JHEP03(2018)194}{{\em JHEP} {\bfseries 03} (2018) 194}, \href{https://arxiv.org/abs/1707.07258}{{\ttfamily arXiv:1707.07258}}.

\bibitem{MIGDAL:2022yip}
{\bfseries MIGDAL} Collaboration, H.~M. Ara{\'u}jo {\em et~al.}, ``{The MIGDAL experiment: Measuring a rare atomic process to aid the search for dark matter},'' \href{https://dx.doi.org/10.1016/j.astropartphys.2023.102853}{{\em Astropart. Phys.} {\bfseries 151} (2023) 102853}, \href{https://arxiv.org/abs/2207.08284}{{\ttfamily arXiv:2207.08284}}.

\bibitem{Xu:2023wev}
J.~Xu {\em et~al.}, ``{Search for the Migdal effect in liquid xenon with keV-level nuclear recoils},'' \href{https://dx.doi.org/10.1103/PhysRevD.109.L051101}{{\em Phys. Rev. D} {\bfseries 109} no.~5, (2024) L051101}, \href{https://arxiv.org/abs/2307.12952}{{\ttfamily arXiv:2307.12952}}.

\bibitem{Yi:2026fmf}
D.~Yi {\em et~al.}, ``{Direct observation of the Migdal effect induced by neutron bombardment},'' \href{https://dx.doi.org/10.1038/s41586-025-09918-8}{{\em Nature} {\bfseries 649} no.~8097, (2026) 580}.

\bibitem{Kouvaris:2016afs}
C.~Kouvaris and J.~Pradler, ``{Probing sub-GeV Dark Matter with conventional detectors},'' \href{https://dx.doi.org/10.1103/PhysRevLett.118.031803}{{\em Phys. Rev. Lett.} {\bfseries 118} no.~3, (2017) 031803}, \href{https://arxiv.org/abs/1607.01789}{{\ttfamily arXiv:1607.01789}}.

\bibitem{Essig:2011nj}
R.~Essig, J.~Mardon, and T.~Volansky, ``{Direct Detection of Sub-GeV Dark Matter},'' \href{https://dx.doi.org/10.1103/PhysRevD.85.076007}{{\em Phys. Rev. D} {\bfseries 85} (2012) 076007}, \href{https://arxiv.org/abs/1108.5383}{{\ttfamily arXiv:1108.5383}}.

\bibitem{Bloch:2020uzh}
I.~M. Bloch, A.~Caputo, R.~Essig, D.~Redigolo, M.~Sholapurkar, and T.~Volansky, ``{Exploring new physics with O(keV) electron recoils in direct detection experiments},'' \href{https://dx.doi.org/10.1007/JHEP01(2021)178}{{\em JHEP} {\bfseries 01} (2021) 178}, \href{https://arxiv.org/abs/2006.14521}{{\ttfamily arXiv:2006.14521}}.

\bibitem{Krnjaic:2024bdd}
G.~Krnjaic, D.~Rocha, and T.~Trickle, ``{The non-relativistic effective field theory of dark matter-electron interactions},'' \href{https://dx.doi.org/10.1007/JHEP03(2025)165}{{\em JHEP} {\bfseries 03} (2025) 165}, \href{https://arxiv.org/abs/2407.14598}{{\ttfamily arXiv:2407.14598}}.

\bibitem{Knapen:2021bwg}
S.~Knapen, J.~Kozaczuk, and T.~Lin, ``{python package for dark matter scattering in dielectric targets},'' \href{https://dx.doi.org/10.1103/PhysRevD.105.015014}{{\em Phys. Rev. D} {\bfseries 105} no.~1, (2022) 015014}, \href{https://arxiv.org/abs/2104.12786}{{\ttfamily arXiv:2104.12786}}.

\bibitem{Trickle:2022fwt}
T.~Trickle, ``{Extended calculation of electronic excitations for direct detection of dark matter},'' \href{https://dx.doi.org/10.1103/PhysRevD.107.035035}{{\em Phys. Rev. D} {\bfseries 107} no.~3, (2023) 035035}, \href{https://arxiv.org/abs/2210.14917}{{\ttfamily arXiv:2210.14917}}.

\bibitem{Hochberg:2021pkt}
Y.~Hochberg {\em et~al.}, ``{Determining Dark-Matter{\textendash}Electron Scattering Rates from the Dielectric Function},'' \href{https://dx.doi.org/10.1103/PhysRevLett.127.151802}{{\em Phys. Rev. Lett.} {\bfseries 127} no.~15, (2021) 151802}, \href{https://arxiv.org/abs/2101.08263}{{\ttfamily arXiv:2101.08263}}.

\bibitem{Knapen:2021run}
S.~Knapen, J.~Kozaczuk, and T.~Lin, ``{Dark matter-electron scattering in dielectrics},'' \href{https://dx.doi.org/10.1103/PhysRevD.104.015031}{{\em Phys. Rev. D} {\bfseries 104} no.~1, (2021) 015031}, \href{https://arxiv.org/abs/2101.08275}{{\ttfamily arXiv:2101.08275}}.

\bibitem{Hochberg:2025rjs}
Y.~Hochberg, M.~Khalaf, A.~Lenoci, and R.~Ovadia, ``{Determining (All) Dark Matter-Electron Scattering Rates From Material Properties},'' \href{https://arxiv.org/abs/2510.25835}{{\ttfamily arXiv:2510.25835}}.

\bibitem{Trickle:2019nya}
T.~Trickle, Z.~Zhang, K.~M. Zurek, K.~Inzani, and S.~M. Griffin, ``{Multi-Channel Direct Detection of Light Dark Matter: Theoretical Framework},'' \href{https://dx.doi.org/10.1007/JHEP03(2020)036}{{\em JHEP} {\bfseries 03} (2020) 036}, \href{https://arxiv.org/abs/1910.08092}{{\ttfamily arXiv:1910.08092}}.

\bibitem{Kahn:2021ttr}
Y.~Kahn and T.~Lin, ``{Searches for light dark matter using condensed matter systems},'' \href{https://dx.doi.org/10.1088/1361-6633/ac5f63}{{\em Rept. Prog. Phys.} {\bfseries 85} no.~6, (2022) 066901}, \href{https://arxiv.org/abs/2108.03239}{{\ttfamily arXiv:2108.03239}}.

\bibitem{Campbell-Deem:2022fqm}
B.~Campbell-Deem, S.~Knapen, T.~Lin, and E.~Villarama, ``{Dark matter direct detection from the single phonon to the nuclear recoil regime},'' \href{https://dx.doi.org/10.1103/PhysRevD.106.036019}{{\em Phys. Rev. D} {\bfseries 106} no.~3, (2022) 036019}, \href{https://arxiv.org/abs/2205.02250}{{\ttfamily arXiv:2205.02250}}.

\bibitem{Kozaczuk:2020uzb}
J.~Kozaczuk and T.~Lin, ``{Plasmon production from dark matter scattering},'' \href{https://dx.doi.org/10.1103/PhysRevD.101.123012}{{\em Phys. Rev. D} {\bfseries 101} no.~12, (2020) 123012}, \href{https://arxiv.org/abs/2003.12077}{{\ttfamily arXiv:2003.12077}}.

\bibitem{Kurinsky:2020dpb}
N.~Kurinsky, D.~Baxter, Y.~Kahn, and G.~Krnjaic, ``{Dark matter interpretation of excesses in multiple direct detection experiments},'' \href{https://dx.doi.org/10.1103/PhysRevD.102.015017}{{\em Phys. Rev. D} {\bfseries 102} no.~1, (2020) 015017}, \href{https://arxiv.org/abs/2002.06937}{{\ttfamily arXiv:2002.06937}}.

\bibitem{Trickle:2019ovy}
T.~Trickle, Z.~Zhang, and K.~M. Zurek, ``{Detecting Light Dark Matter with Magnons},'' \href{https://dx.doi.org/10.1103/PhysRevLett.124.201801}{{\em Phys. Rev. Lett.} {\bfseries 124} no.~20, (2020) 201801}, \href{https://arxiv.org/abs/1905.13744}{{\ttfamily arXiv:1905.13744}}.

\bibitem{Caputo:2019cyg}
A.~Caputo, A.~Esposito, and A.~D. Polosa, ``{Sub-MeV Dark Matter and the Goldstone Modes of Superfluid Helium},'' \href{https://dx.doi.org/10.1103/PhysRevD.100.116007}{{\em Phys. Rev. D} {\bfseries 100} no.~11, (2019) 116007}, \href{https://arxiv.org/abs/1907.10635}{{\ttfamily arXiv:1907.10635}}.

\bibitem{Acanfora:2019con}
F.~Acanfora, A.~Esposito, and A.~D. Polosa, ``{Sub-GeV Dark Matter in Superfluid He-4: an Effective Theory Approach},'' \href{https://dx.doi.org/10.1140/epjc/s10052-019-7057-0}{{\em Eur. Phys. J. C} {\bfseries 79} no.~7, (2019) 549}, \href{https://arxiv.org/abs/1902.02361}{{\ttfamily arXiv:1902.02361}}.

\bibitem{Knapen:2016cue}
S.~Knapen, T.~Lin, and K.~M. Zurek, ``{Light Dark Matter in Superfluid Helium: Detection with Multi-excitation Production},'' \href{https://dx.doi.org/10.1103/PhysRevD.95.056019}{{\em Phys. Rev. D} {\bfseries 95} no.~5, (2017) 056019}, \href{https://arxiv.org/abs/1611.06228}{{\ttfamily arXiv:1611.06228}}.

\bibitem{Knapen:2017ekk}
S.~Knapen, T.~Lin, M.~Pyle, and K.~M. Zurek, ``{Detection of Light Dark Matter With Optical Phonons in Polar Materials},'' \href{https://dx.doi.org/10.1016/j.physletb.2018.08.064}{{\em Phys. Lett. B} {\bfseries 785} (2018) 386}, \href{https://arxiv.org/abs/1712.06598}{{\ttfamily arXiv:1712.06598}}.

\bibitem{Pospelov:2008jk}
M.~Pospelov, A.~Ritz, and M.~B. Voloshin, ``{Bosonic super-WIMPs as keV-scale dark matter},'' \href{https://dx.doi.org/10.1103/PhysRevD.78.115012}{{\em Phys. Rev. D} {\bfseries 78} (2008) 115012}, \href{https://arxiv.org/abs/0807.3279}{{\ttfamily arXiv:0807.3279}}.

\bibitem{Hochberg:2016sqx}
Y.~Hochberg, T.~Lin, and K.~M. Zurek, ``{Absorption of light dark matter in semiconductors},'' \href{https://dx.doi.org/10.1103/PhysRevD.95.023013}{{\em Phys. Rev. D} {\bfseries 95} no.~2, (2017) 023013}, \href{https://arxiv.org/abs/1608.01994}{{\ttfamily arXiv:1608.01994}}.

\bibitem{Mitridate:2021ctr}
A.~Mitridate, T.~Trickle, Z.~Zhang, and K.~M. Zurek, ``{Dark matter absorption via electronic excitations},'' \href{https://dx.doi.org/10.1007/JHEP09(2021)123}{{\em JHEP} {\bfseries 09} (2021) 123}, \href{https://arxiv.org/abs/2106.12586}{{\ttfamily arXiv:2106.12586}}.

\bibitem{An:2014twa}
H.~An, M.~Pospelov, J.~Pradler, and A.~Ritz, ``{Direct Detection Constraints on Dark Photon Dark Matter},'' \href{https://dx.doi.org/10.1016/j.physletb.2015.06.018}{{\em Phys. Lett. B} {\bfseries 747} (2015) 331}, \href{https://arxiv.org/abs/1412.8378}{{\ttfamily arXiv:1412.8378}}.

\bibitem{Mitridate:2023izi}
A.~Mitridate, K.~Pardo, T.~Trickle, and K.~M. Zurek, ``{Effective field theory for dark matter absorption on single phonons},'' \href{https://dx.doi.org/10.1103/PhysRevD.109.015010}{{\em Phys. Rev. D} {\bfseries 109} no.~1, (2024) 015010}, \href{https://arxiv.org/abs/2308.06314}{{\ttfamily arXiv:2308.06314}}.

\bibitem{Bloch:2016sjj}
I.~M. Bloch, R.~Essig, K.~Tobioka, T.~Volansky, and T.-T. Yu, ``{Searching for Dark Absorption with Direct Detection Experiments},'' \href{https://dx.doi.org/10.1007/JHEP06(2017)087}{{\em JHEP} {\bfseries 06} (2017) 087}, \href{https://arxiv.org/abs/1608.02123}{{\ttfamily arXiv:1608.02123}}.

\bibitem{Hochberg:2021zrf}
Y.~Hochberg, B.~von Krosigk, E.~Kuflik, and T.~C. Yu, ``{Impact of Dark Compton Scattering on Direct Dark Matter Absorption Searches},'' \href{https://dx.doi.org/10.1103/PhysRevLett.128.191801}{{\em Phys. Rev. Lett.} {\bfseries 128} no.~19, (2022) 191801}, \href{https://arxiv.org/abs/2109.08168}{{\ttfamily arXiv:2109.08168}}.

\bibitem{Dror:2019onn}
J.~A. Dror, G.~Elor, and R.~Mcgehee, ``{Directly Detecting Signals from Absorption of Fermionic Dark Matter},'' \href{https://dx.doi.org/10.1103/PhysRevLett.124.181301}{{\em Phys. Rev. Lett.} {\bfseries 124} no.~18, (2020) 18}, \href{https://arxiv.org/abs/1905.12635}{{\ttfamily arXiv:1905.12635}}.

\bibitem{Dror:2020czw}
J.~A. Dror, G.~Elor, R.~McGehee, and T.-T. Yu, ``{Absorption of sub-MeV fermionic dark matter by electron targets},'' \href{https://dx.doi.org/10.1103/PhysRevD.103.035001}{{\em Phys. Rev. D} {\bfseries 103} no.~3, (2021) 035001}, \href{https://arxiv.org/abs/2011.01940}{{\ttfamily arXiv:2011.01940}}. [Erratum: Phys.Rev.D 105, 119903 (2022)].

\bibitem{Ge:2022ius}
S.-F. Ge, X.-G. He, X.-D. Ma, and J.~Sheng, ``{Revisiting the fermionic dark matter absorption on electron target},'' \href{https://dx.doi.org/10.1007/JHEP05(2022)191}{{\em JHEP} {\bfseries 05} (2022) 191}, \href{https://arxiv.org/abs/2201.11497}{{\ttfamily arXiv:2201.11497}}.

\bibitem{Ferreira:2022egk}
R.~Z. Ferreira, M.~C.~D. Marsh, and E.~M{\"u}ller, ``{Do Direct Detection Experiments Constrain Axionlike Particles Coupled to Electrons?},'' \href{https://dx.doi.org/10.1103/PhysRevLett.128.221302}{{\em Phys. Rev. Lett.} {\bfseries 128} no.~22, (2022) 221302}, \href{https://arxiv.org/abs/2202.08858}{{\ttfamily arXiv:2202.08858}}.

\bibitem{Perez:2016tcq}
K.~Perez, K.~C.~Y. Ng, J.~F. Beacom, C.~Hersh, S.~Horiuchi, and R.~Krivonos, ``{Almost closing the {\ensuremath{\nu}}MSM sterile neutrino dark matter window with NuSTAR},'' \href{https://dx.doi.org/10.1103/PhysRevD.95.123002}{{\em Phys. Rev. D} {\bfseries 95} no.~12, (2017) 123002}, \href{https://arxiv.org/abs/1609.00667}{{\ttfamily arXiv:1609.00667}}.

\bibitem{Raffelt:1996wa}
G.~G. Raffelt, {\em {Stars as laboratories for fundamental physics}: {The astrophysics of neutrinos, axions, and other weakly interacting particles}}.
\newblock University of Chicago Press, 5, 1996.

\bibitem{Arisaka:2012pb}
K.~Arisaka {\em et~al.}, ``{Expected Sensitivity to Galactic/Solar Axions and Bosonic Super-WIMPs based on the Axio-electric Effect in Liquid Xenon Dark Matter Detectors},'' \href{https://dx.doi.org/10.1016/j.astropartphys.2012.12.009}{{\em Astropart. Phys.} {\bfseries 44} (2013) 59}, \href{https://arxiv.org/abs/1209.3810}{{\ttfamily arXiv:1209.3810}}.

\bibitem{Athron:2020maw}
P.~Athron {\em et~al.}, ``{Global fits of axion-like particles to XENON1T and astrophysical data},'' \href{https://dx.doi.org/10.1007/JHEP05(2021)159}{{\em JHEP} {\bfseries 05} (2021) 159}, \href{https://arxiv.org/abs/2007.05517}{{\ttfamily arXiv:2007.05517}}.

\bibitem{Kavanagh:2017cru}
B.~J. Kavanagh, ``{Earth scattering of superheavy dark matter: Updated constraints from detectors old and new},'' \href{https://dx.doi.org/10.1103/PhysRevD.97.123013}{{\em Phys. Rev. D} {\bfseries 97} no.~12, (2018) 123013}, \href{https://arxiv.org/abs/1712.04901}{{\ttfamily arXiv:1712.04901}}.

\bibitem{Emken:2018run}
T.~Emken and C.~Kouvaris, ``{How blind are underground and surface detectors to strongly interacting Dark Matter?},'' \href{https://dx.doi.org/10.1103/PhysRevD.97.115047}{{\em Phys. Rev. D} {\bfseries 97} no.~11, (2018) 115047}, \href{https://arxiv.org/abs/1802.04764}{{\ttfamily arXiv:1802.04764}}.

\bibitem{Emken:2019tni}
T.~Emken, R.~Essig, C.~Kouvaris, and M.~Sholapurkar, ``{Direct Detection of Strongly Interacting Sub-GeV Dark Matter via Electron Recoils},'' \href{https://dx.doi.org/10.1088/1475-7516/2019/09/070}{{\em JCAP} {\bfseries 09} (2019) 070}, \href{https://arxiv.org/abs/1905.06348}{{\ttfamily arXiv:1905.06348}}.

\bibitem{Du:2024afd}
P.~Du, R.~Essig, B.~J. Rauscher, and H.~Xu, ``{Direct Constraints on Strongly Interacting Dark Matter from the James Webb Space Telescope},'' \href{https://dx.doi.org/10.1103/s2q8-rzb3}{{\em Phys. Rev. Lett.} {\bfseries 135} no.~5, (2025) 051002}, \href{https://arxiv.org/abs/2412.13131}{{\ttfamily arXiv:2412.13131}}.

\bibitem{Mack:2007xj}
G.~D. Mack, J.~F. Beacom, and G.~Bertone, ``{Towards Closing the Window on Strongly Interacting Dark Matter: Far-Reaching Constraints from Earth's Heat Flow},'' \href{https://dx.doi.org/10.1103/PhysRevD.76.043523}{{\em Phys. Rev. D} {\bfseries 76} (2007) 043523}, \href{https://arxiv.org/abs/0705.4298}{{\ttfamily arXiv:0705.4298}}.

\bibitem{Xu:2018efh}
W.~L. Xu, C.~Dvorkin, and A.~Chael, ``{Probing sub-GeV Dark Matter-Baryon Scattering with Cosmological Observables},'' \href{https://dx.doi.org/10.1103/PhysRevD.97.103530}{{\em Phys. Rev. D} {\bfseries 97} no.~10, (2018) 103530}, \href{https://arxiv.org/abs/1802.06788}{{\ttfamily arXiv:1802.06788}}.

\bibitem{Boddy:2018wzy}
K.~K. Boddy {\em et~al.}, ``{Critical assessment of CMB limits on dark matter-baryon scattering: New treatment of the relative bulk velocity},'' \href{https://dx.doi.org/10.1103/PhysRevD.98.123506}{{\em Phys. Rev. D} {\bfseries 98} no.~12, (2018) 123506}, \href{https://arxiv.org/abs/1808.00001}{{\ttfamily arXiv:1808.00001}}.

\bibitem{Nadler:2019zrb}
E.~O. Nadler, V.~Gluscevic, K.~K. Boddy, and R.~H. Wechsler, ``{Constraints on Dark Matter Microphysics from the Milky Way Satellite Population},'' \href{https://dx.doi.org/10.3847/2041-8213/ab1eb2}{{\em Astrophys. J. Lett.} {\bfseries 878} no.~2, (2019) 32}, \href{https://arxiv.org/abs/1904.10000}{{\ttfamily arXiv:1904.10000}}. [Erratum: Astrophys.J.Lett. 897, L46 (2020), Erratum: Astrophys.J. 897, L46 (2020)].

\bibitem{Conrad:2014nna}
J.~Conrad, ``{Statistical Issues in Astrophysical Searches for Particle Dark Matter},'' \href{https://dx.doi.org/10.1016/j.astropartphys.2014.09.003}{{\em Astropart. Phys.} {\bfseries 62} (2015) 165}, \href{https://arxiv.org/abs/1407.6617}{{\ttfamily arXiv:1407.6617}}.

\bibitem{ParticleDataGroup:2024cfk}
{\bfseries Particle Data Group} Collaboration, S.~Navas {\em et~al.}, ``{Review of particle physics},'' \href{https://dx.doi.org/10.1103/PhysRevD.110.030001}{{\em Phys. Rev. D} {\bfseries 110} no.~3, (2024) 030001}.

\bibitem{Feldman:1997qc}
G.~J. Feldman and R.~D. Cousins, ``{A Unified approach to the classical statistical analysis of small signals},'' \href{https://dx.doi.org/10.1103/PhysRevD.57.3873}{{\em Phys. Rev. D} {\bfseries 57} (1998) 3873}, \href{https://arxiv.org/abs/physics/9711021}{{\ttfamily arXiv:physics/9711021}}.

\bibitem{Barlow:1990vc}
R.~J. Barlow, ``{Extended maximum likelihood},'' \href{https://dx.doi.org/10.1016/0168-9002(90)91334-8}{{\em Nucl. Instrum. Meth. A} {\bfseries 297} (1990) 496}.

\bibitem{Rolke:2004mj}
W.~A. Rolke, A.~M. Lopez, and J.~Conrad, ``{Limits and confidence intervals in the presence of nuisance parameters},'' \href{https://dx.doi.org/10.1016/j.nima.2005.05.068}{{\em Nucl. Instrum. Meth. A} {\bfseries 551} (2005) 493}, \href{https://arxiv.org/abs/physics/0403059}{{\ttfamily arXiv:physics/0403059}}.

\bibitem{Cowan:2010js}
G.~Cowan, K.~Cranmer, E.~Gross, and O.~Vitells, ``{Asymptotic formulae for likelihood-based tests of new physics},'' \href{https://dx.doi.org/10.1140/epjc/s10052-011-1554-0}{{\em Eur. Phys. J. C} {\bfseries 71} (2011) 1554}, \href{https://arxiv.org/abs/1007.1727}{{\ttfamily arXiv:1007.1727}}. [Erratum: Eur.Phys.J.C 73, 2501 (2013)].

\bibitem{Cowan:2011an}
G.~Cowan, K.~Cranmer, E.~Gross, and O.~Vitells, ``{Power-Constrained Limits},'' \href{https://arxiv.org/abs/1105.3166}{{\ttfamily arXiv:1105.3166}}.

\bibitem{Yellin:2002xd}
S.~Yellin, ``{Finding an upper limit in the presence of unknown background},'' \href{https://dx.doi.org/10.1103/PhysRevD.66.032005}{{\em Phys. Rev. D} {\bfseries 66} (2002) 032005}, \href{https://arxiv.org/abs/physics/0203002}{{\ttfamily arXiv:physics/0203002}}.

\bibitem{DEramo:2016gos}
F.~D'Eramo, B.~J. Kavanagh, and P.~Panci, ``{You can hide but you have to run: direct detection with vector mediators},'' \href{https://dx.doi.org/10.1007/JHEP08(2016)111}{{\em JHEP} {\bfseries 08} (2016) 111}, \href{https://arxiv.org/abs/1605.04917}{{\ttfamily arXiv:1605.04917}}.

\bibitem{Bishara:2018vix}
F.~Bishara, J.~Brod, B.~Grinstein, and J.~Zupan, ``{Renormalization Group Effects in Dark Matter Interactions},'' \href{https://dx.doi.org/10.1007/JHEP03(2020)089}{{\em JHEP} {\bfseries 03} (2020) 089}, \href{https://arxiv.org/abs/1809.03506}{{\ttfamily arXiv:1809.03506}}.

\bibitem{Belyaev:2018pqr}
A.~Belyaev, E.~Bertuzzo, C.~Caniu~Barros, O.~Eboli, G.~Grilli Di~Cortona, F.~Iocco, and A.~Pukhov, ``{Interplay of the LHC and non-LHC Dark Matter searches in the Effective Field Theory approach},'' \href{https://dx.doi.org/10.1103/PhysRevD.99.015006}{{\em Phys. Rev. D} {\bfseries 99} no.~1, (2019) 015006}, \href{https://arxiv.org/abs/1807.03817}{{\ttfamily arXiv:1807.03817}}.

\bibitem{Bondarenko:2019vrb}
K.~Bondarenko, A.~Boyarsky, T.~Bringmann, M.~Hufnagel, K.~Schmidt-Hoberg, and A.~Sokolenko, ``{Direct detection and complementary constraints for sub-GeV dark matter},'' \href{https://dx.doi.org/10.1007/JHEP03(2020)118}{{\em JHEP} {\bfseries 03} (2020) 118}, \href{https://arxiv.org/abs/1909.08632}{{\ttfamily arXiv:1909.08632}}.

\bibitem{IceCube:2021xzo}
{\bfseries IceCube} Collaboration, R.~Abbasi {\em et~al.}, ``{Search for GeV-scale dark matter annihilation in the Sun with IceCube DeepCore},'' \href{https://dx.doi.org/10.1103/PhysRevD.105.062004}{{\em Phys. Rev. D} {\bfseries 105} no.~6, (2022) 062004}, \href{https://arxiv.org/abs/2111.09970}{{\ttfamily arXiv:2111.09970}}.

\bibitem{Giuliani:2026syi}
A.~Giuliani, ``{The Quest for Neutrinoless Double Beta Decay: Progress and Prospects},'' \href{https://dx.doi.org/10.53941/pac.2026.100006}{{\em Phys. Cosmos} {\bfseries 1} no.~1, (2026) 6}, \href{https://arxiv.org/abs/2604.13696}{{\ttfamily arXiv:2604.13696}}.

\bibitem{Irwin:1995zzb}
K.~D. Irwin, S.~W. Nam, B.~Cabrera, B.~Chugg, and B.~A. Young, ``{A quasiparticle-trap-assisted transition-edge sensor for phonon-mediated particle detection},'' \href{https://dx.doi.org/10.1063/1.1146105}{{\em Rev. Sci. Instrum.} {\bfseries 66} (1995) 5322}.

\bibitem{Pyle:2015pya}
M.~Pyle, E.~Feliciano-Figueroa, and B.~Sadoulet, ``{Optimized Designs for Very Low Temperature Massive Calorimeters},'' \href{https://arxiv.org/abs/1503.01200}{{\ttfamily arXiv:1503.01200}}.

\bibitem{Krantz:2023vpe}
M.~Krantz, F.~Toschi, B.~Maier, G.~Heine, C.~Enss, and S.~Kempf, ``{Magnetic microcalorimeter with paramagnetic temperature sensors and integrated dc-SQUID readout for high-resolution x-ray emission spectroscopy},'' \href{https://dx.doi.org/10.1063/5.0180903}{{\em Appl. Phys. Lett.} {\bfseries 124} no.~3, (2024) 032601}, \href{https://arxiv.org/abs/2310.08698}{{\ttfamily arXiv:2310.08698}}.

\bibitem{Angloher:2025fzw}
G.~Angloher {\em et~al.}, ``{The CRESST experiment towards the next generation of sub GeV direct dark matter detection},'' \href{https://dx.doi.org/10.1038/s42005-025-02476-5}{{\em Commun. Phys.} {\bfseries 9} no.~1, (2026) 163}, \href{https://arxiv.org/abs/2505.01183}{{\ttfamily arXiv:2505.01183}}.

\bibitem{Billard:2024zvc}
J.~Billard, J.~Gascon, S.~Marnieros, and S.~Scorza, ``{Transition Edge Sensors with Sub-eV Resolution And Cryogenic Targets (TESSERACT) at the underground laboratory of Modane (LSM)},'' \href{https://dx.doi.org/10.1016/j.nuclphysb.2024.116465}{{\em Nucl. Phys. B} {\bfseries 1003} (2024) 116465}.

\bibitem{Angloher:2025shf}
{\bfseries COSINUS} Collaboration, G.~Angloher {\em et~al.}, ``{COSINUS: a model-independent challenge to the DAMA/LIBRA dark matter claim using cryogenic NaI detectors},'' \href{https://dx.doi.org/10.1038/s42005-026-02620-9}{{\em Commun. Phys.} {\bfseries 9} (7, 2026) 213}, \href{https://arxiv.org/abs/2507.02429}{{\ttfamily arXiv:2507.02429}}.

\bibitem{Neganov:1985khw}
B.~S. Neganov and V.~N. Trofimov, ``{Colorimetric method measuring ionizing radiation},'' {\em Otkryt. Izobret.} {\bfseries 146} (1985) 215.

\bibitem{Luke:1988xcw}
P.~N. Luke, ``{Voltage-assisted calorimetric ionization detector},'' \href{https://dx.doi.org/10.1063/1.341976}{{\em J. Appl. Phys.} {\bfseries 64} no.~12, (1988) 6858}.

\bibitem{SuperCDMS:2016wui}
{\bfseries SuperCDMS} Collaboration, R.~Agnese {\em et~al.}, ``{Projected Sensitivity of the SuperCDMS SNOLAB experiment},'' \href{https://dx.doi.org/10.1103/PhysRevD.95.082002}{{\em Phys. Rev. D} {\bfseries 95} no.~8, (2017) 082002}, \href{https://arxiv.org/abs/1610.00006}{{\ttfamily arXiv:1610.00006}}.

\bibitem{EDELWEISS:2017lvq}
{\bfseries EDELWEISS} Collaboration, E.~Armengaud {\em et~al.}, ``{Performance of the EDELWEISS-III experiment for direct dark matter searches},'' \href{https://dx.doi.org/10.1088/1748-0221/12/08/P08010}{{\em JINST} {\bfseries 12} no.~08, (2017) P08010}, \href{https://arxiv.org/abs/1706.01070}{{\ttfamily arXiv:1706.01070}}.

\bibitem{Brink:2006rs}
P.~L. Brink {\em et~al.}, ``{First test runs of a dark-matter detector with interleaved ionization electrodes and phonon sensors for surface-event rejection},'' \href{https://dx.doi.org/10.1016/j.nima.2005.12.026}{{\em Nucl. Instrum. Meth. A} {\bfseries 559} (2006) 414}.

\bibitem{Broniatowski:2009qu}
A.~Broniatowski {\em et~al.}, ``{A new high-background-rejection dark matter Ge cryogenic detector},'' \href{https://dx.doi.org/10.1016/j.physletb.2009.10.036}{{\em Phys. Lett. B} {\bfseries 681} (2009) 305}, \href{https://arxiv.org/abs/0905.0753}{{\ttfamily arXiv:0905.0753}}.

\bibitem{CUORE:2021mvw}
{\bfseries CUORE} Collaboration, D.~Q. Adams {\em et~al.}, ``{Search for Majorana neutrinos exploiting millikelvin cryogenics with CUORE},'' \href{https://dx.doi.org/10.1038/s41586-022-04497-4}{{\em Nature} {\bfseries 604} no.~7904, (2022) 53}, \href{https://arxiv.org/abs/2104.06906}{{\ttfamily arXiv:2104.06906}}.

\bibitem{Baxter:2025odk}
D.~Baxter, R.~Essig, Y.~Hochberg, M.~Kaznacheeva, B.~von Krosigk, F.~Reindl, R.~K. Romani, and F.~Wagner, ``{Low-Energy Backgrounds in Solid-State Phonon and Charge Detectors},'' \href{https://dx.doi.org/10.1146/annurev-nucl-121423-100849}{{\em Ann. Rev. Nucl. Part. Sci.} {\bfseries 75} no.~1, (2025) 301}, \href{https://arxiv.org/abs/2503.08859}{{\ttfamily arXiv:2503.08859}}.

\bibitem{Fuss:2022fxe}
P.~Adari {\em et~al.}, ``{EXCESS workshop: Descriptions of rising low-energy spectra},'' \href{https://dx.doi.org/10.21468/SciPostPhysProc.9.001}{{\em SciPost Phys. Proc.} {\bfseries 9} (2022) 001}, \href{https://arxiv.org/abs/2202.05097}{{\ttfamily arXiv:2202.05097}}.

\bibitem{Knoll:2010xta}
G.~Knoll, {\em {Radiation Detection and Measurement (4th ed.)}}.
\newblock John Wiley, Hoboken, NJ, 2010.

\bibitem{Pehl:1968}
R.~Pehl, F.~Goulding, D.~Landis, and M.~Lenzlinger, ``Accurate determination of the ionization energy in semiconductor detectors,'' \href{https://dx.doi.org/https://doi.org/10.1016/0029-554X(68)90342-X}{{\em Nucl. Instrum. Meth. A} {\bfseries 59} no.~1, (1968) 45}.

\bibitem{Ramanathan:2020fwm}
K.~Ramanathan and N.~Kurinsky, ``{Ionization yield in silicon for eV-scale electron-recoil processes},'' \href{https://dx.doi.org/10.1103/PhysRevD.102.063026}{{\em Phys. Rev. D} {\bfseries 102} no.~6, (2020) 063026}, \href{https://arxiv.org/abs/2004.10709}{{\ttfamily arXiv:2004.10709}}.

\bibitem{Varshni:1967}
Y.~Varshni, ``Temperature dependence of the energy gap in semiconductors,'' \href{https://dx.doi.org/https://doi.org/10.1016/0031-8914(67)90062-6}{{\em Physica} {\bfseries 34} no.~1, (1967) 149}.

\bibitem{Bludau:1974}
W.~Bludau, A.~Onton, and W.~Heinke, ``Temperature dependence of the band gap of silicon,'' \href{https://dx.doi.org/10.1063/1.1663501}{{\em J. Appl. Phys.} {\bfseries 45} no.~4, (04, 1974) 1846}.

\bibitem{Smith:2010}
G.~E. Smith, ``{Nobel Lecture: The invention and early history of the CCD},'' \href{https://dx.doi.org/10.1103/RevModPhys.82.2307}{{\em Rev. Mod. Phys.} {\bfseries 82} (2010) 2307}.

\bibitem{DAMIC:2016lrs}
{\bfseries DAMIC} Collaboration, A.~Aguilar-Arevalo {\em et~al.}, ``{Search for low-mass WIMPs in a 0.6 kg day exposure of the DAMIC experiment at SNOLAB},'' \href{https://dx.doi.org/10.1103/PhysRevD.94.082006}{{\em Phys. Rev. D} {\bfseries 94} no.~8, (2016) 082006}, \href{https://arxiv.org/abs/1607.07410}{{\ttfamily arXiv:1607.07410}}.

\bibitem{Tiffenberg:2017aac}
{\bfseries SENSEI} Collaboration, J.~Tiffenberg, M.~Sofo-Haro, A.~Drlica-Wagner, R.~Essig, Y.~Guardincerri, S.~Holland, T.~Volansky, and T.-T. Yu, ``{Single-electron and single-photon sensitivity with a silicon Skipper CCD},'' \href{https://dx.doi.org/10.1103/PhysRevLett.119.131802}{{\em Phys. Rev. Lett.} {\bfseries 119} no.~13, (2017) 131802}, \href{https://arxiv.org/abs/1706.00028}{{\ttfamily arXiv:1706.00028}}.

\bibitem{Essig:2015cda}
R.~Essig, M.~Fernandez-Serra, J.~Mardon, A.~Soto, T.~Volansky, and T.-T. Yu, ``{Direct Detection of sub-GeV Dark Matter with Semiconductor Targets},'' \href{https://dx.doi.org/10.1007/JHEP05(2016)046}{{\em JHEP} {\bfseries 05} (2016) 046}, \href{https://arxiv.org/abs/1509.01598}{{\ttfamily arXiv:1509.01598}}.

\bibitem{SENSEI:2020dpa}
{\bfseries SENSEI} Collaboration, L.~Barak {\em et~al.}, ``{SENSEI: Direct-Detection Results on sub-GeV Dark Matter from a New Skipper-CCD},'' \href{https://dx.doi.org/10.1103/PhysRevLett.125.171802}{{\em Phys. Rev. Lett.} {\bfseries 125} no.~17, (2020) 171802}, \href{https://arxiv.org/abs/2004.11378}{{\ttfamily arXiv:2004.11378}}.

\bibitem{DAMIC:2020cut}
{\bfseries DAMIC} Collaboration, A.~Aguilar-Arevalo {\em et~al.}, ``{Results on low-mass weakly interacting massive particles from a 11 kg-day target exposure of DAMIC at SNOLAB},'' \href{https://dx.doi.org/10.1103/PhysRevLett.125.241803}{{\em Phys. Rev. Lett.} {\bfseries 125} (2020) 241803}, \href{https://arxiv.org/abs/2007.15622}{{\ttfamily arXiv:2007.15622}}.

\bibitem{DAMIC-M:2024ooa}
{\bfseries DAMIC-M} Collaboration, I.~Arnquist {\em et~al.}, ``{The DAMIC-M Low Background Chamber},'' \href{https://dx.doi.org/10.1088/1748-0221/19/11/T11010}{{\em JINST} {\bfseries 19} no.~11, (2024) T11010}, \href{https://arxiv.org/abs/2407.17872}{{\ttfamily arXiv:2407.17872}}.

\bibitem{Avalos:2021fxm}
N.~{\'A}valos {\em et~al.}, ``{Skipper CCDs for the search of a daily modulation of Dark Matter signal in the DMSQUARE experiment},'' \href{https://dx.doi.org/10.1088/1742-6596/2156/1/012074}{{\em J. Phys. Conf. Ser.} {\bfseries 2156} (2021) 012074}.

\bibitem{Oscura:2022vmi}
{\bfseries Oscura} Collaboration, A.~Aguilar-Arevalo {\em et~al.}, ``{The Oscura Experiment},'' \href{https://arxiv.org/abs/2202.10518}{{\ttfamily arXiv:2202.10518}}.

\bibitem{Pfann:1966}
W.~G. Pfann, {\em Zone Melting}.
\newblock John Wiley \& Sons, New York, 2~ed., 1966.

\bibitem{Sumathi:2023}
R.~R. Sumathi, A.~Gybin, K.-P. Gradwohl, P.~C. Palleti, M.~Pietsch, K.~Irmscher, N.~Dropka, and U.~Juda, ``Development of large-diameter and very high purity ge crystal growth technology for devices,'' \href{https://dx.doi.org/https://doi.org/10.1002/crat.202200286}{{\em Cryst. Res. Technol.} {\bfseries 58} no.~5, (2023) 2200286}.

\bibitem{Abrosimov:2019qeb}
N.~Abrosimov {\em et~al.}, ``{Technology development of high purity germanium crystals for radiation detectors},'' \href{https://dx.doi.org/10.1016/j.jcrysgro.2019.125396}{{\em J. Cryst. Growth} {\bfseries 532} (2020) 125396}.

\bibitem{Wei:2021drf}
W.~Z. Wei {\em et~al.}, ``{Development of planar P-type point contact germanium detectors for low-mass dark matter searches},'' \href{https://dx.doi.org/10.1140/epjc/s10052-022-10162-x}{{\em Eur. Phys. J. C} {\bfseries 82} no.~3, (2022) 203}, \href{https://arxiv.org/abs/2105.02109}{{\ttfamily arXiv:2105.02109}}.

\bibitem{Yang:2016crf}
L.~T. Yang {\em et~al.}, ``{Bulk and Surface Event Identification in p-type Germanium Detectors},'' \href{https://dx.doi.org/10.1016/j.nima.2017.12.078}{{\em Nucl. Instrum. Meth. A} {\bfseries 886} (2018) 13}, \href{https://arxiv.org/abs/1611.03357}{{\ttfamily arXiv:1611.03357}}.

\bibitem{CDEX:2023vvc}
{\bfseries CDEX} Collaboration, X.~P. Geng {\em et~al.}, ``{Projected WIMP sensitivity of the CDEX-50 dark matter experiment},'' \href{https://dx.doi.org/10.1088/1475-7516/2024/07/009}{{\em JCAP} {\bfseries 07} (2024) 009}, \href{https://arxiv.org/abs/2309.01843}{{\ttfamily arXiv:2309.01843}}.

\bibitem{Crookes1903}
W.~Crookes, ``The emanations of radium,'' {\em Proceedings of the Royal Society of London} {\bfseries 71} (1903) 405.

\bibitem{Martinez:2026thh}
M.~Martinez, R.~H. Maruyama, and M.~L. Sarsa, ``{Dark Matter Searches with Sodium Iodide Detectors},'' \href{https://arxiv.org/abs/2607.24448}{{\ttfamily arXiv:2607.24448}}.

\bibitem{Bernabei1999DAMA_NaI}
R.~Bernabei, P.~Belli, {\em et~al.}, ``{Performances of the $\sim$100 kg NaI(Tl) set-up of the DAMA experiment at Gran Sasso},'' \href{https://dx.doi.org/10.1007/BF03035868}{{\em Nuovo Cim. A} {\bfseries 112} no.~6, (1999) 545}.

\bibitem{Bernabei2008DAMA_LIBRA}
R.~Bernabei, P.~Belli, {\em et~al.}, ``{The DAMA/LIBRA apparatus},'' \href{https://dx.doi.org/10.1016/j.nima.2008.04.082}{{\em Nucl. Instrum. Meth. A} {\bfseries 592} no.~3, (2008) 297}.

\bibitem{Lee2006KIMS}
H.~S. Lee, H.~Bhang, {\em et~al.}, ``{First limit on WIMP cross section with low background CsI(Tl) crystal detector},'' \href{https://dx.doi.org/10.1016/j.physletb.2005.12.035}{{\em Phys. Lett. B.} {\bfseries 633} (2006) 201}.

\bibitem{Amare2019ANAIS112}
J.~Amar{\'e}, S.~Cebri{\'a}n, {\em et~al.}, ``{Performance of ANAIS-112 experiment after the first year of data taking},'' \href{https://dx.doi.org/10.1140/epjc/s10052-019-6697-4}{{\em Eur. Phys. J. C} {\bfseries 79} (2019) 228}.

\bibitem{Adhikari:2017esn}
G.~Adhikari {\em et~al.}, ``{Initial Performance of the COSINE-100 Experiment},'' \href{https://dx.doi.org/10.1140/epjc/s10052-018-5590-x}{{\em Eur. Phys. J. C} {\bfseries 78} no.~2, (2018) 107}, \href{https://arxiv.org/abs/1710.05299}{{\ttfamily arXiv:1710.05299}}.

\bibitem{Lee:2024wzd}
D.~Lee {\em et~al.}, ``{Upgrading the COSINE-100 experiment for enhanced sensitivity to low-mass dark matter detection},'' \href{https://dx.doi.org/10.1038/s42005-025-02067-4}{{\em Commun. Phys.} {\bfseries 8} no.~1, (2025) 135}, \href{https://arxiv.org/abs/2409.15748}{{\ttfamily arXiv:2409.15748}}.

\bibitem{Antonello2019SABRE}
M.~Antonello, E.~Barberio, {\em et~al.}, ``{The SABRE project and the SABRE Proof-of-Principle},'' \href{https://dx.doi.org/10.1140/epjc/s10052-019-6860-y}{{\em Eur. Phys. J. C} {\bfseries 79} (2019) 363}.

\bibitem{COSINE:2020egt}
{\bfseries COSINE} Collaboration, B.~J. Park {\em et~al.}, ``{Development of ultra-pure NaI(Tl) detectors for the COSINE-200 experiment},'' \href{https://dx.doi.org/10.1140/epjc/s10052-020-8386-8}{{\em Eur. Phys. J. C} {\bfseries 80} no.~9, (2020) 814}, \href{https://arxiv.org/abs/2004.06287}{{\ttfamily arXiv:2004.06287}}.

\bibitem{Fushimi:2021mez}
K.~Fushimi {\em et~al.}, ``{Development of highly radiopure NaI(Tl) scintillator for PICOLON dark matter search project},'' \href{https://dx.doi.org/10.1093/ptep/ptab020}{{\em PTEP} {\bfseries 2021} no.~4, (2021) 043F01}, \href{https://arxiv.org/abs/2101.00759}{{\ttfamily arXiv:2101.00759}}.

\bibitem{Martinenghi:2025ckb}
E.~Martinenghi {\em et~al.}, ``{ASTAROTH: a novel detector for dark matter direct detection using cryogenic SiPMs},'' \href{https://dx.doi.org/10.1088/1748-0221/20/12/C12019}{{\em JINST} {\bfseries 20} no.~12, (2025) C12019}, \href{https://arxiv.org/abs/2510.24418}{{\ttfamily arXiv:2510.24418}}.

\bibitem{Amaré:2026XQ}
J.~Amaré, J.~Apilluelo, {\em et~al.}, ``{Advanced NaI detectors for Dark Matter search and other applications: the ANAIS+ project},'' \href{https://dx.doi.org/10.22323/1.511.0047}{{\em PoS} {\bfseries TAUP2025} (2026) 047}.

\bibitem{Glaser1952}
D.~A. Glaser, ``Some effects of ionizing radiation on the formation of bubbles in liquids,'' \href{https://dx.doi.org/10.1103/PhysRev.87.665}{{\em Phys. Rev.} {\bfseries 87} (1952) 665}.

\bibitem{Seitz}
F.~Seitz, ``On the theory of the bubble chamber,'' \href{https://dx.doi.org/http://dx.doi.org/10.1063/1.1724333}{{\em Phys. Fluids. (1958-1988)} {\bfseries 1} (1958) 2}.

\bibitem{Behnke:2010xt}
E.~Behnke {\em et~al.}, ``{Improved Limits on Spin-Dependent WIMP-Proton Interactions from a Two Liter CF$_3$I Bubble Chamber},'' \href{https://dx.doi.org/10.1103/PhysRevLett.106.021303}{{\em Phys. Rev. Lett.} {\bfseries 106} (2011) 021303}, \href{https://arxiv.org/abs/1008.3518}{{\ttfamily arXiv:1008.3518}}.

\bibitem{Behnke2012COUPP}
{\bfseries COUPP} Collaboration, E.~Behnke {\em et~al.}, ``First dark matter search results from a 4-kg cf3i bubble chamber operated in a deep underground site,'' \href{https://dx.doi.org/10.1103/PhysRevD.86.052001}{{\em Phys. Rev. D} {\bfseries 86} (2012) 052001}.

\bibitem{PICASSO:2012ngj}
{\bfseries PICASSO} Collaboration, S.~Archambault {\em et~al.}, ``{Constraints on Low-Mass WIMP Interactions on $^{19}F$ from PICASSO},'' \href{https://dx.doi.org/10.1016/j.physletb.2012.03.078}{{\em Phys. Lett. B} {\bfseries 711} (2012) 153}, \href{https://arxiv.org/abs/1202.1240}{{\ttfamily arXiv:1202.1240}}.

\bibitem{SIMPLE:2014pun}
{\bfseries SIMPLE} Collaboration, M.~Felizardo {\em et~al.}, ``{The SIMPLE Phase II Dark Matter Search},'' \href{https://dx.doi.org/10.1103/PhysRevD.89.072013}{{\em Phys. Rev. D} {\bfseries 89} no.~7, (2014) 072013}, \href{https://arxiv.org/abs/1404.4309}{{\ttfamily arXiv:1404.4309}}.

\bibitem{PICO:2019vsc}
{\bfseries PICO} Collaboration, C.~Amole {\em et~al.}, ``{Dark Matter Search Results from the Complete Exposure of the PICO-60 C$_3$F$_8$ Bubble Chamber},'' \href{https://dx.doi.org/10.1103/PhysRevD.100.022001}{{\em Phys. Rev. D} {\bfseries 100} no.~2, (2019) 022001}, \href{https://arxiv.org/abs/1902.04031}{{\ttfamily arXiv:1902.04031}}.

\bibitem{Giroux:2021vpy}
G.~Giroux, ``{Search for Dark Matter with the PICO-500 Experiment},'' \href{https://dx.doi.org/10.1088/1742-6596/2156/1/012068}{{\em J. Phys. Conf. Ser.} {\bfseries 2156} (2021) 012068}.

\bibitem{Alfonso-Pita:2022akn}
E.~Alfonso-Pita {\em et~al.}, ``{Snowmass 2021 Scintillating Bubble Chambers: Liquid-noble Bubble Chambers for Dark Matter and CE$\nu$NS Detection},'' in {\em {Snowmass 2021}}.
\newblock 7, 2022.
\newblock \href{https://arxiv.org/abs/2207.12400}{{\ttfamily arXiv:2207.12400}}.

\bibitem{Alfonso-Pita:2023frp}
E.~Alfonso-Pita {\em et~al.}, ``{Scintillating Bubble Chambers for Rare Event Searches},'' \href{https://dx.doi.org/10.3390/universe9080346}{{\em Universe} {\bfseries 9} no.~8, (2023) 346}.

\bibitem{Das:2025coa}
{\bfseries InDEx} Collaboration, S.~Das, M.~Das, V.~Kumar, S.~Ali, N.~Biswas, S.~Sahoo, and N.~Chaddha, ``{Dark matter direct search result from InDEx run2 at JUSL},'' \href{https://dx.doi.org/10.1103/vpb7-kdrz}{{\em Phys. Rev. D} {\bfseries 112} no.~4, (2025) 042003}, \href{https://arxiv.org/abs/2510.06921}{{\ttfamily arXiv:2510.06921}}.

\bibitem{Lippincott:2011zr}
{\bfseries MiniCLEAN} Collaboration, W.~H. Lippincott, K.~J. Coakley, D.~Gastler, E.~Kearns, D.~N. McKinsey, and J.~A. Nikkel, ``{Scintillation yield and time dependence from electronic and nuclear recoils in liquid neon},'' \href{https://dx.doi.org/10.1103/PhysRevC.86.015807}{{\em Phys. Rev. C} {\bfseries 86} (2012) 015807}, \href{https://arxiv.org/abs/1111.3260}{{\ttfamily arXiv:1111.3260}}.

\bibitem{DEAP:2025igy}
{\bfseries DEAP} Collaboration, P.~Adhikari {\em et~al.}, ``{Position Reconstruction in the DEAP-3600 Dark Matter Search Experiment},'' \href{https://dx.doi.org/10.1088/1748-0221/20/07/P07012}{{\em JINST} {\bfseries 20} (3, 2025) P07012}, \href{https://arxiv.org/abs/2503.10383}{{\ttfamily arXiv:2503.10383}}.

\bibitem{Abe:2013tc}
K.~Abe {\em et~al.}, ``{XMASS detector},'' \href{https://dx.doi.org/10.1016/j.nima.2013.03.059}{{\em Nucl. Instrum. Meth. A} {\bfseries 716} (2013) 78}, \href{https://arxiv.org/abs/1301.2815}{{\ttfamily arXiv:1301.2815}}.

\bibitem{Boulay:2006mb}
M.~Boulay and A.~Hime, ``{Technique for direct detection of weakly interacting massive particles using scintillation time discrimination in liquid argon},''
\href{https://dx.doi.org/10.1016/j.astropartphys.2005.12.009}{{\em Astropart. Phys.} {\bfseries 25} (2006) 179}.

\bibitem{Cichon:2022gbu}
D.~Cichon, G.~Eurin, F.~J{\"o}rg, T.~{Marrod\'an Undagoitia}, and N.~Rupp, ``{Scintillation decay-time constants for alpha particles and electrons in liquid xenon},'' {\em Rev. Sci. Instrum.} {\bfseries 93} no.~11, (2022) 113302, \href{https://arxiv.org/abs/2201.12483}{{\ttfamily arXiv:2201.12483}}.

\bibitem{DEAP:2019yzn}
{\bfseries DEAP} Collaboration, R.~Ajaj {\em et~al.}, ``{Search for dark matter with a 231-day exposure of liquid argon using DEAP-3600 at SNOLAB},'' \href{https://dx.doi.org/10.1103/PhysRevD.100.022004}{{\em Phys. Rev. D} {\bfseries 100} no.~2, (2019) 022004}, \href{https://arxiv.org/abs/1902.04048}{{\ttfamily arXiv:1902.04048}}.

\bibitem{XMASS:2022tkr}
{\bfseries XMASS} Collaboration, K.~Abe {\em et~al.}, ``{Direct dark matter searches with the full data set of XMASS-I},'' \href{https://dx.doi.org/10.1103/PhysRevD.108.083022}{{\em Phys. Rev. D} {\bfseries 108} no.~8, (2023) 083022}, \href{https://arxiv.org/abs/2211.06204}{{\ttfamily arXiv:2211.06204}}.

\bibitem{Lansiart:1976}
A.~Lansiart, A.~Seigneur, J.-L. Moretti, and J.-P. Morucci, ``Development research on a highly luminous condensed xenon scintillator,'' {\em Nucl. Instrum. Meth.} {\bfseries 135} no.~1, (1976) 47.

\bibitem{DarkSide-20k:2024usz}
{\bfseries DarkSide-20k} Collaboration, F.~Acerbi {\em et~al.}, ``{Quality assurance and quality control of the $26~\text {m}^2$ SiPM production for the DarkSide-20k dark matter experiment},'' \href{https://dx.doi.org/10.1140/epjc/s10052-025-14196-9}{{\em Eur. Phys. J. C} {\bfseries 85} no.~5, (2025) 534}, \href{https://arxiv.org/abs/2412.18867}{{\ttfamily arXiv:2412.18867}}.

\bibitem{Baudis:2018pdv}
L.~Baudis, M.~Galloway, A.~Kish, C.~Marentini, and J.~Wulf, ``{Characterisation of Silicon Photomultipliers for Liquid Xenon Detectors},'' \href{https://dx.doi.org/10.1088/1748-0221/13/10/P10022}{{\em JINST} {\bfseries 13} no.~10, (2018) P10022}, \href{https://arxiv.org/abs/1808.06827}{{\ttfamily arXiv:1808.06827}}.

\bibitem{Xu:2015xx}
J.~Xu {\em et~al.}, ``A study of the trace 39ar content in argon from deep underground sources,'' \href{https://dx.doi.org/https://doi.org/10.1016/j.astropartphys.2015.01.002}{{\em Astrop. Phys.} {\bfseries 66} (2015) 53}.

\bibitem{XLZD:2024nsu}
{\bfseries XLZD} Collaboration, J.~Aalbers {\em et~al.}, ``{The XLZD Design Book: towards the next-generation liquid xenon observatory for dark matter and neutrino physics},'' \href{https://dx.doi.org/10.1140/epjc/s10052-025-14810-w}{{\em Eur. Phys. J. C} {\bfseries 85} no.~10, (2025) 1192}, \href{https://arxiv.org/abs/2410.17137}{{\ttfamily arXiv:2410.17137}}.

\bibitem{Kopec:2023uii}
A.~Kopec, ``{Design Challenges for a Future Liquid Xenon Observatory},'' \href{https://dx.doi.org/10.31526/jais.2024.480}{{\em JAIS} {\bfseries 2024} (2024) 480}, \href{https://arxiv.org/abs/2310.00722}{{\ttfamily arXiv:2310.00722}}.

\bibitem{LZ:2024zvo}
{\bfseries LZ} Collaboration, J.~Aalbers {\em et~al.}, ``{Dark Matter Search Results from 4.2{\,}{\,}Tonne-Years of Exposure of the LUX-ZEPLIN (LZ) Experiment},'' \href{https://dx.doi.org/10.1103/4dyc-z8zf}{{\em Phys. Rev. Lett.} {\bfseries 135} no.~1, (2025) 011802}, \href{https://arxiv.org/abs/2410.17036}{{\ttfamily arXiv:2410.17036}}.

\bibitem{PandaX:2024qfu}
{\bfseries PandaX} Collaboration, Z.~Bo {\em et~al.}, ``{Dark Matter Search Results from 1.54{\,}{\,}Tonne{\textperiodcentered}Year Exposure of PandaX-4T},'' \href{https://dx.doi.org/10.1103/PhysRevLett.134.011805}{{\em Phys. Rev. Lett.} {\bfseries 134} no.~1, (2025) 011805}, \href{https://arxiv.org/abs/2408.00664}{{\ttfamily arXiv:2408.00664}}.

\bibitem{XENON:2025vwd}
{\bfseries XENON} Collaboration, E.~Aprile {\em et~al.}, ``{WIMP Dark Matter Search Using a 3.1 Tonne-Year Exposure of the XENONnT Experiment},'' \href{https://dx.doi.org/10.1103/msw4-t342}{{\em Phys. Rev. Lett.} {\bfseries 135} no.~22, (2025) 221003}, \href{https://arxiv.org/abs/2502.18005}{{\ttfamily arXiv:2502.18005}}.

\bibitem{DarkSide:2018ppu}
{\bfseries DarkSide} Collaboration, P.~Agnes {\em et~al.}, ``{Constraints on Sub-GeV Dark-Matter{\textendash}Electron Scattering from the DarkSide-50 Experiment},'' \href{https://dx.doi.org/10.1103/PhysRevLett.121.111303}{{\em Phys. Rev. Lett.} {\bfseries 121} no.~11, (2018) 111303}, \href{https://arxiv.org/abs/1802.06998}{{\ttfamily arXiv:1802.06998}}.

\bibitem{DarkSide:2018bpj}
{\bfseries DarkSide} Collaboration, P.~Agnes {\em et~al.}, ``{Low-Mass Dark Matter Search with the DarkSide-50 Experiment},'' \href{https://dx.doi.org/10.1103/PhysRevLett.121.081307}{{\em Phys. Rev. Lett.} {\bfseries 121} no.~8, (2018) 081307}, \href{https://arxiv.org/abs/1802.06994}{{\ttfamily arXiv:1802.06994}}.

\bibitem{XENON:2026qow}
{\bfseries XENON} Collaboration, E.~Aprile {\em et~al.}, ``{Light Dark Matter Search with 7.8~Tonne-Year of Ionization-Only Data in XENONnT},'' \href{https://dx.doi.org/10.1103/2lrq-f6bk}{{\em Phys. Rev. Lett.} {\bfseries 137} no.~5, (2026) 051003}, \href{https://arxiv.org/abs/2601.11296}{{\ttfamily arXiv:2601.11296}}.

\bibitem{XENON:2024ijk}
{\bfseries XENON} Collaboration, E.~Aprile {\em et~al.}, ``{First Indication of Solar B8 Neutrinos via Coherent Elastic Neutrino-Nucleus Scattering with XENONnT},'' \href{https://dx.doi.org/10.1103/PhysRevLett.133.191002}{{\em Phys. Rev. Lett.} {\bfseries 133} no.~19, (2024) 191002}, \href{https://arxiv.org/abs/2408.02877}{{\ttfamily arXiv:2408.02877}}.

\bibitem{LUX:2019npm}
{\bfseries LUX} Collaboration, D.~S. Akerib {\em et~al.}, ``{Extending light WIMP searches to single scintillation photons in LUX},'' \href{https://dx.doi.org/10.1103/PhysRevD.101.042001}{{\em Phys. Rev. D} {\bfseries 101} no.~4, (2020) 042001}, \href{https://arxiv.org/abs/1907.06272}{{\ttfamily arXiv:1907.06272}}.

\bibitem{LUX:2020yym}
{\bfseries LUX} Collaboration, D.~S. Akerib {\em et~al.}, ``{Improving sensitivity to low-mass dark matter in LUX using a novel electrode background mitigation technique},'' \href{https://dx.doi.org/10.1103/PhysRevD.104.012011}{{\em Phys. Rev. D} {\bfseries 104} no.~1, (2021) 012011}, \href{https://arxiv.org/abs/2011.09602}{{\ttfamily arXiv:2011.09602}}.

\bibitem{Cocco:2011zz}
A.~G. Cocco {\em et~al.}, ``{First physics results from WARP 2.3 litre prototype},'' \href{https://dx.doi.org/10.1016/j.nuclphysbps.2011.03.093}{{\em Nucl. Phys. B Proc. Suppl.} {\bfseries 221} (2011) 53}.

\bibitem{DarkSide-20k:2017zyg}
{\bfseries DarkSide-20k} Collaboration, C.~E. Aalseth {\em et~al.}, ``{DarkSide-20k: A 20 tonne two-phase LAr TPC for direct dark matter detection at LNGS},'' \href{https://dx.doi.org/10.1140/epjp/i2018-11973-4}{{\em Eur. Phys. J. Plus} {\bfseries 133} (2018) 131}, \href{https://arxiv.org/abs/1707.08145}{{\ttfamily arXiv:1707.08145}}.

\bibitem{Aprile_2011}
{\bfseries XENON} Collaboration, E.~Aprile {\em et~al.}, ``{Design and performance of the XENON10 dark matter experiment},'' \href{https://dx.doi.org/10.1016/j.astropartphys.2011.01.006}{{\em Astrop. Phys.} {\bfseries 34} no.~9, (Apr., 2011) 679}.

\bibitem{Ivaniouchenkov:2000}
I.~Ivaniouchenkov {\em et~al.}, \href{https://dx.doi.org/10.1109/NSSMIC.2000.949135}{``{ZEPLIN-I}: a single phase liquid {Xe} detector for dark matter search,''} in {\em 2000 IEEE Nuclear Science Symposium Conference Record}, vol.~1.
\newblock 2000.

\bibitem{PandaX-4T:2021bab}
{\bfseries PandaX-4T} Collaboration, Y.~Meng {\em et~al.}, ``{Dark Matter Search Results from the PandaX-4T Commissioning Run},'' \href{https://dx.doi.org/10.1103/PhysRevLett.127.261802}{{\em Phys. Rev. Lett.} {\bfseries 127} no.~26, (2021) 261802}, \href{https://arxiv.org/abs/2107.13438}{{\ttfamily arXiv:2107.13438}}.

\bibitem{Akerib:2020lzi}
D.~S. Akerib {\em et~al.}, ``The {LUX-ZEPLIN} ({LZ}) experiment,'' \href{https://dx.doi.org/10.1016/j.nima.2019.163047}{{\em NIM A} {\bfseries 953} (2020) 163047}.

\bibitem{XENON:2024}
{\bfseries XENON} Collaboration, E.~Aprile {\em et~al.}, ``The xenonnt dark matter experiment,'' \href{https://dx.doi.org/10.1140/epjc/s10052-024-12982-5}{{\em Eur. Phys. J. C} {\bfseries 84} (2024) 784}.

\bibitem{Lanou:1987}
R.~E. Lanou, H.~J. Maris, and G.~M. Seidel, ``Detection of solar neutrinos in superfluid helium,'' \href{https://dx.doi.org/10.1103/PhysRevLett.58.2498}{{\em Phys. Rev. Lett.} {\bfseries 58} (Jun, 1987) 2498--2501}. \url{https://link.aps.org/doi/10.1103/PhysRevLett.58.2498}.

\bibitem{Zhang_2016}
G.~Zhang, X.~Xiang, F.~Yang, L.~Liu, T.~Tang, Y.~Shi, and X.~Wang, ``First principles investigation of helium physisorption on an {\ensuremath{\alpha}}-al2o3(0001) surface,'' \href{https://dx.doi.org/10.1039/c6cp01517d}{{\em Physical Chemistry Chemical Physics} {\bfseries 18} no.~23, (2016) 15711--15718}, \href{https://arxiv.org/abs/https://pubs.rsc.org/cp/article-pdf/18/23/15711/4771284/c6cp01517d.pdf}{{\ttfamily https://pubs.rsc.org/cp/article-pdf/18/23/15711/4771284/c6cp01517d.pdf}}. \url{https://doi.org/10.1039/c6cp01517d}.

\bibitem{DELight:2024bgv}
{\bfseries DELight} Collaboration, F.~Toschi {\em et~al.}, ``{Signal partitioning in superfluid He4: A Monte~Carlo approach},'' \href{https://dx.doi.org/10.1103/PhysRevD.111.032013}{{\em Phys. Rev. D} {\bfseries 111} no.~3, (2025) 032013}, \href{https://arxiv.org/abs/2410.13684}{{\ttfamily arXiv:2410.13684 [hep-ex]}}.

\bibitem{vonKrosigk:2022vnf}
B.~von Krosigk {\em et~al.}, ``{DELight: A Direct search Experiment for Light dark matter with superfluid helium},'' \href{https://dx.doi.org/10.21468/SciPostPhysProc.12.016}{{\em SciPost Phys. Proc.} {\bfseries 12} (2023) 016}, \href{https://arxiv.org/abs/2209.10950}{{\ttfamily arXiv:2209.10950}}.

\bibitem{Hertel:2018aal}
S.~A. Hertel, A.~Biekert, J.~Lin, V.~Velan, and D.~N. McKinsey, ``{Direct detection of sub-GeV dark matter using a superfluid $^4$He target},'' \href{https://dx.doi.org/10.1103/PhysRevD.100.092007}{{\em Phys. Rev. D} {\bfseries 100} no.~9, (2019) 092007}, \href{https://arxiv.org/abs/1810.06283}{{\ttfamily arXiv:1810.06283}}.

\bibitem{Pickett1988}
G.~R. Pickett, ``{Superfluid $^{3}$He as a Possible Medium for Detecting Dark Matter},'' in {\em Proceedings of the Second European Workshop on Neutrinos and Dark Matter Detectors}, L.~Gonzalez-Mestres and D.~Perret-Gallix, eds., p.~377.
\newblock Editions Fronti{\`e}res, 1988.

\bibitem{Mayet:2000aa}
F.~Mayet, D.~Santos, G.~Perrin, Y.~M. Bunkov, and H.~Godfrin, ``{Design optimization of MACHe3, a project of superfluid He-3 detector for direct dark matter search},'' \href{https://dx.doi.org/10.1016/S0168-9002(00)00540-4}{{\em Nucl. Instrum. Meth. A} {\bfseries 455} (2000) 554}, \href{https://arxiv.org/abs/astro-ph/0002303}{{\ttfamily arXiv:astro-ph/0002303}}.

\bibitem{Winkelmann:2006pw}
C.~B. Winkelmann, J.~Elbs, Y.~M. Bunkov, E.~Collin, H.~Godfrin, and M.~Krusius, ``{Bolometric calibration of a superfluid He-3 detector for dark matter search: Direct measurement of the scintillated energy fraction for neutron, electron and muon events},'' \href{https://dx.doi.org/10.1016/j.nima.2007.01.180}{{\em Nucl. Instrum. Meth. A} {\bfseries 574} (2007) 264}, \href{https://arxiv.org/abs/physics/0611273}{{\ttfamily arXiv:physics/0611273}}.

\bibitem{QUEST-DMC:2023nug}
{\bfseries QUEST-DMC} Collaboration, S.~Autti {\em et~al.}, ``{QUEST-DMC superfluid $^3$He detector for sub-GeV dark matter},'' \href{https://dx.doi.org/10.1140/epjc/s10052-024-12410-8}{{\em Eur. Phys. J. C} {\bfseries 84} no.~3, (2024) 248}, \href{https://arxiv.org/abs/2310.11304}{{\ttfamily arXiv:2310.11304}}.

\bibitem{Nikolopoulos:2025kws}
K.~Nikolopoulos and P.~Knights, ``{Searches for Light Dark Matter with Spherical Proportional Counters},'' \href{https://dx.doi.org/10.31526/JAIS.2025.548}{{\em JAIS} {\bfseries 2025} (2025) 548}, \href{https://arxiv.org/abs/2502.11870}{{\ttfamily arXiv:2502.11870}}.

\bibitem{NEWS-G:2022kon}
{\bfseries NEWS-G} Collaboration, L.~Balogh {\em et~al.}, ``{The NEWS-G detector at SNOLAB},'' \href{https://dx.doi.org/10.1088/1748-0221/18/02/T02005}{{\em JINST} {\bfseries 18} no.~02, (2023) T02005}, \href{https://arxiv.org/abs/2205.15433}{{\ttfamily arXiv:2205.15433}}.

\bibitem{Castel:2025yvh}
J.~F. Castel {\em et~al.}, ``{Low-energy threshold demonstration for dark matter searches in TREX-DM with an $^{37}$Ar source produced at CNA HiSPANoS},'' \href{https://dx.doi.org/10.1088/1475-7516/2026/04/072}{{\em JCAP} {\bfseries 04} (2026) 072}, \href{https://arxiv.org/abs/2510.05877}{{\ttfamily arXiv:2510.05877}}.

\bibitem{Castel:2019ngt}
J.~Castel {\em et~al.}, ``{The TREX-DM experiment at the Canfranc Underground Laboratory},'' \href{https://dx.doi.org/10.1088/1742-6596/1468/1/012063}{{\em J. Phys. Conf. Ser.} {\bfseries 1468} no.~1, (2020) 012063}, \href{https://arxiv.org/abs/1910.13957}{{\ttfamily arXiv:1910.13957}}.

\bibitem{DAstolfo:2024cei}
{\bfseries CYGNO} Collaboration, M.~D'Astolfo, ``{The CYGNO experiment},'' \href{https://dx.doi.org/10.22323/1.476.0763}{{\em PoS} {\bfseries ICHEP2024} (2025) 763}.

\bibitem{Ikeda:2021ckk}
T.~Ikeda {\em et~al.}, ``{Direction-sensitive dark matter search with the low-background gaseous detector NEWAGE-0.3b{\textquotedblright}},'' \href{https://dx.doi.org/10.1093/ptep/ptab053}{{\em PTEP} {\bfseries 2021} no.~6, (2021) 063F01}, \href{https://arxiv.org/abs/2101.09921}{{\ttfamily arXiv:2101.09921}}.

\bibitem{Beaufort:2023vlb}
C.~Beaufort, O.~Guillaudin, D.~Santos, N.~Sauzet, E.~Mobio, R.~Babut, and C.~Tao, ``{Directional detection of keV proton and carbon recoils with MIMAC},'' \href{https://dx.doi.org/10.1088/1748-0221/19/05/P05052}{{\em JINST} {\bfseries 19} no.~05, (2024) P05052}, \href{https://arxiv.org/abs/2312.12842}{{\ttfamily arXiv:2312.12842}}.

\bibitem{Coskuner:2021qxo}
A.~Coskuner, T.~Trickle, Z.~Zhang, and K.~M. Zurek, ``{Directional detectability of dark matter with single phonon excitations: Target comparison},'' \href{https://dx.doi.org/10.1103/PhysRevD.105.015010}{{\em Phys. Rev. D} {\bfseries 105} no.~1, (2022) 015010}, \href{https://arxiv.org/abs/2102.09567}{{\ttfamily arXiv:2102.09567}}.

\bibitem{Baiocco:2025omh}
D.~Baiocco, D.~Marian, G.~Marino, P.~Panci, M.~Polini, and A.~Tredicucci, ``{Direct dark matter searches with metal halide perovskites},'' \href{https://dx.doi.org/10.1103/qzlg-3dp3}{{\em Phys. Rev. D} {\bfseries 112} no.~9, (2025) 092005}, \href{https://arxiv.org/abs/2502.01732}{{\ttfamily arXiv:2502.01732}}.

\bibitem{Hochberg:2017wce}
Y.~Hochberg {\em et~al.}, ``{Detection of sub-MeV Dark Matter with Three-Dimensional Dirac Materials},'' \href{https://dx.doi.org/10.1103/PhysRevD.97.015004}{{\em Phys. Rev. D} {\bfseries 97} no.~1, (2018) 015004}, \href{https://arxiv.org/abs/1708.08929}{{\ttfamily arXiv:1708.08929}}.

\bibitem{Coskuner:2019odd}
A.~Coskuner, A.~Mitridate, A.~Olivares, and K.~M. Zurek, ``{Directional Dark Matter Detection in Anisotropic Dirac Materials},'' \href{https://dx.doi.org/10.1103/PhysRevD.103.016006}{{\em Phys. Rev. D} {\bfseries 103} no.~1, (2021) 016006}, \href{https://arxiv.org/abs/1909.09170}{{\ttfamily arXiv:1909.09170}}.

\bibitem{Hochberg:2016ntt}
Y.~Hochberg {\em et~al.}, ``{Directional detection of dark matter with two-dimensional targets},'' \href{https://dx.doi.org/10.1016/j.physletb.2017.06.051}{{\em Phys. Lett. B} {\bfseries 772} (2017) 239}, \href{https://arxiv.org/abs/1606.08849}{{\ttfamily arXiv:1606.08849}}.

\bibitem{Catena:2023qkj}
R.~Catena, T.~Emken, M.~Matas, N.~A. Spaldin, and E.~Urdshals, ``{Direct searches for general dark matter-electron interactions with graphene detectors: Part I. Electronic structure calculations},'' \href{https://dx.doi.org/10.1103/PhysRevResearch.5.043257}{{\em Phys. Rev. Res.} {\bfseries 5} no.~4, (2023) 043257}, \href{https://arxiv.org/abs/2303.15497}{{\ttfamily arXiv:2303.15497}}.

\bibitem{Catena:2023awl}
R.~Catena, T.~Emken, M.~Matas, N.~A. Spaldin, and E.~Urdshals, ``{Direct searches for general dark matter-electron interactions with graphene detectors: Part II. Sensitivity studies},'' \href{https://dx.doi.org/10.1103/PhysRevResearch.5.043258}{{\em Phys. Rev. Res.} {\bfseries 5} no.~4, (2023) 043258}, \href{https://arxiv.org/abs/2303.15509}{{\ttfamily arXiv:2303.15509}}.

\bibitem{Cavoto:2019flp}
G.~Cavoto, M.~G. Betti, C.~Mariani, F.~Pandolfi, A.~D. Polosa, I.~Rago, and A.~Ruocco, ``{Carbon nanotubes as anisotropic target for dark matter},'' \href{https://dx.doi.org/10.1088/1742-6596/1468/1/012232}{{\em J. Phys. Conf. Ser.} {\bfseries 1468} no.~1, (2020) 012232}, \href{https://arxiv.org/abs/1911.01122}{{\ttfamily arXiv:1911.01122}}.

\bibitem{Hochberg:2019cyy}
Y.~Hochberg, I.~Charaev, S.-W. Nam, V.~Verma, M.~Colangelo, and K.~K. Berggren, ``{Detecting Sub-GeV Dark Matter with Superconducting Nanowires},'' \href{https://dx.doi.org/10.1103/PhysRevLett.123.151802}{{\em Phys. Rev. Lett.} {\bfseries 123} no.~15, (2019) 151802}, \href{https://arxiv.org/abs/1903.05101}{{\ttfamily arXiv:1903.05101}}.

\bibitem{QROCODILE:2024nqm}
{\bfseries QROCODILE} Collaboration, L.~Baudis {\em et~al.}, ``{First Sub-MeV Dark Matter Search with the QROCODILE Experiment Using Superconducting Nanowire Single-Photon Detectors},'' \href{https://dx.doi.org/10.1103/4hb6-f6jl}{{\em Phys. Rev. Lett.} {\bfseries 135} no.~8, (2025) 081002}, \href{https://arxiv.org/abs/2412.16279}{{\ttfamily arXiv:2412.16279}}.

\bibitem{Hochberg:2026mdl}
Y.~Hochberg, M.~Khalaf, N.~Kurinsky, A.~Lenoci, and R.~Ovadia, ``{Probing Dark Matter-Electron Interactions with Superconducting Qubits},'' \href{https://arxiv.org/abs/2601.02474}{{\ttfamily arXiv:2601.02474}}.

\bibitem{Schwemmbauer:2025evp}
C.~Schwemmbauer {\em et~al.}, ``{First direct search for light dark matter interactions in a transition-edge sensor},'' \href{https://dx.doi.org/10.1103/zytz-n25v}{{\em Phys. Rev. D} {\bfseries 114} no.~1, (2026) 012006}, \href{https://arxiv.org/abs/2506.18982}{{\ttfamily arXiv:2506.18982}}.

\bibitem{Gao:2024irf}
J.~Gao, Y.~Hochberg, B.~V. Lehmann, S.~W. Nam, P.~Szypryt, M.~R. Vissers, and T.~Xu, ``{Detecting Light Dark Matter with Kinetic Inductance Detectors},'' \href{https://arxiv.org/abs/2403.19739}{{\ttfamily arXiv:2403.19739}}.

\bibitem{Blanco:2019lrf}
C.~Blanco, J.~I. Collar, Y.~Kahn, and B.~Lillard, ``{Dark Matter-Electron Scattering from Aromatic Organic Targets},'' \href{https://dx.doi.org/10.1103/PhysRevD.101.056001}{{\em Phys. Rev. D} {\bfseries 101} no.~5, (2020) 056001}, \href{https://arxiv.org/abs/1912.02822}{{\ttfamily arXiv:1912.02822}}.

\bibitem{Blanco:2021hlm}
C.~Blanco, Y.~Kahn, B.~Lillard, and S.~D. McDermott, ``{Dark Matter Daily Modulation With Anisotropic Organic Crystals},'' \href{https://dx.doi.org/10.1103/PhysRevD.104.036011}{{\em Phys. Rev. D} {\bfseries 104} (2021) 036011}, \href{https://arxiv.org/abs/2103.08601}{{\ttfamily arXiv:2103.08601}}.

\bibitem{Blanco:2022cel}
C.~Blanco, R.~Essig, M.~Fernandez-Serra, H.~Ramani, and O.~Slone, ``{Dark matter direct detection with quantum dots},'' \href{https://dx.doi.org/10.1103/PhysRevD.107.095035}{{\em Phys. Rev. D} {\bfseries 107} no.~9, (2023) 095035}, \href{https://arxiv.org/abs/2208.05967}{{\ttfamily arXiv:2208.05967}}.

\bibitem{Bunting:2017net}
P.~C. Bunting, G.~Gratta, T.~Melia, and S.~Rajendran, ``{Magnetic Bubble Chambers and Sub-GeV Dark Matter Direct Detection},'' \href{https://dx.doi.org/10.1103/PhysRevD.95.095001}{{\em Phys. Rev. D} {\bfseries 95} no.~9, (2017) 095001}, \href{https://arxiv.org/abs/1701.06566}{{\ttfamily arXiv:1701.06566}}.

\bibitem{Alves:2026sgo}
J.~R. Alves, M.~Lindner, F.~S. Queiroz, and M.~S. Vasconcelos, ``{Search for Axions and Dark Photons Using Single Molecule Magnets},'' \href{https://arxiv.org/abs/2601.01043}{{\ttfamily arXiv:2601.01043}}.

\bibitem{Rajendran:2017ynw}
S.~Rajendran, N.~Zobrist, A.~O. Sushkov, R.~Walsworth, and M.~Lukin, ``{A method for directional detection of dark matter using spectroscopy of crystal defects},'' \href{https://dx.doi.org/10.1103/PhysRevD.96.035009}{{\em Phys. Rev. D} {\bfseries 96} no.~3, (2017) 035009}, \href{https://arxiv.org/abs/1705.09760}{{\ttfamily arXiv:1705.09760}}.

\bibitem{Budnik:2017sbu}
R.~Budnik, O.~Chesnovsky, O.~Slone, and T.~Volansky, ``{Direct Detection of Light Dark Matter and Solar Neutrinos via Color Center Production in Crystals},'' \href{https://dx.doi.org/10.1016/j.physletb.2018.04.063}{{\em Phys. Lett. B} {\bfseries 782} (2018) 242}, \href{https://arxiv.org/abs/1705.03016}{{\ttfamily arXiv:1705.03016}}.

\bibitem{Araujo:2025rhr}
G.~R. Araujo {\em et~al.}, ``{Nuclear recoil detection with color centers in bulk lithium fluoride},'' \href{https://dx.doi.org/10.1140/epjc/s10052-026-15920-9}{{\em Eur. Phys. J. C} {\bfseries 86} no.~7, (2026) 887}, \href{https://arxiv.org/abs/2503.20732}{{\ttfamily arXiv:2503.20732}}.

\bibitem{Baum:2018tfw}
S.~Baum, A.~K. Drukier, K.~Freese, M.~G{\'o}rski, and P.~Stengel, ``{Searching for Dark Matter with Paleo-Detectors},'' \href{https://dx.doi.org/10.1016/j.physletb.2020.135325}{{\em Phys. Lett. B} {\bfseries 803} (2020) 135325}, \href{https://arxiv.org/abs/1806.05991}{{\ttfamily arXiv:1806.05991}}.

\bibitem{Baum:2023cct}
S.~Baum {\em et~al.}, ``{Mineral detection of neutrinos and dark matter. A whitepaper},'' \href{https://dx.doi.org/10.1016/j.dark.2023.101245}{{\em Phys. Dark Univ.} {\bfseries 41} (2023) 101245}, \href{https://arxiv.org/abs/2301.07118}{{\ttfamily arXiv:2301.07118}}.

\bibitem{Fung:2025cub}
A.~Fung, T.~Lucas, L.~Balogh, M.~Leybourne, and A.~C. Vincent, ``{Refining the sensitivity of new physics searches with ancient minerals},'' \href{https://dx.doi.org/10.1103/clvr-mhx5}{{\em Phys. Rev. D} {\bfseries 112} no.~4, (2025) 043040}, \href{https://arxiv.org/abs/2504.08885}{{\ttfamily arXiv:2504.08885}}.

\bibitem{Theodosopoulos:2026ehn}
D.~P. Theodosopoulos, K.~Freese, C.~Kelso, and P.~Stengel, ``{Projected sensitivity of paleodetectors to dark matter effective interactions with nuclei},'' \href{https://dx.doi.org/10.1103/gnbr-ck58}{{\em Phys. Rev. D} {\bfseries 113} no.~12, (2026) 123044}, \href{https://arxiv.org/abs/2603.13629}{{\ttfamily arXiv:2603.13629}}.

\bibitem{Drukier:2012hj}
A.~Drukier, K.~Freese, A.~Lopez, D.~Spergel, C.~Cantor, G.~Church, and T.~Sano, ``{New Dark Matter Detectors using DNA or RNA for Nanometer Tracking},'' \href{https://arxiv.org/abs/1206.6809}{{\ttfamily arXiv:1206.6809}}.

\bibitem{OHare:2021cgj}
C.~A.~J. O'Hare {\em et~al.}, ``{Particle detection and tracking with DNA},'' \href{https://dx.doi.org/10.1140/epjc/s10052-022-10264-6}{{\em Eur. Phys. J. C} {\bfseries 82} no.~4, (2022) 306}, \href{https://arxiv.org/abs/2105.11949}{{\ttfamily arXiv:2105.11949}}.

\bibitem{Cook:2024cgm}
C.~Cook, C.~Blanco, and J.~Smirnov, ``{Deep learning optimal molecular scintillators for dark matter direct detection},'' \href{https://dx.doi.org/10.1103/89gh-lwcd}{{\em Phys. Rev. D} {\bfseries 112} no.~8, (2025) 083005}, \href{https://arxiv.org/abs/2501.00091}{{\ttfamily arXiv:2501.00091}}.

\bibitem{Geilhufe:2018gry}
R.~M. Geilhufe, B.~Olsthoorn, A.~Ferella, T.~Koski, F.~Kahlhoefer, J.~Conrad, and A.~V. Balatsky, ``{Materials Informatics for Dark Matter Detection},'' \href{https://dx.doi.org/10.1002/pssr.201800293}{{\em Phys. Status Solidi RRL} {\bfseries 12} no.~11, (2018) 1800293}, \href{https://arxiv.org/abs/1806.06040}{{\ttfamily arXiv:1806.06040}}.

\bibitem{Griffin:2025wew}
S.~M. Griffin {\em et~al.}, ``{First High-Throughput Evaluation of Dark Matter Detector Materials},'' \href{https://dx.doi.org/10.1103/bb4g-7939}{{\em Phys. Rev. Lett.} {\bfseries 136} no.~19, (2026) 191801}, \href{https://arxiv.org/abs/2506.19905}{{\ttfamily arXiv:2506.19905}}.

\bibitem{Lindhard:1961}
J.~Lindhard and M.~Scharff, ``Energy dissipation by ions in the kev region,'' \href{https://dx.doi.org/10.1103/PhysRev.124.128}{{\em Phys. Rev.} {\bfseries 124} (Oct, 1961) 128}.

\bibitem{Lindhard:1963}
J.~Lindhard, M.~Scharff, and H.~E. Schi{\o}tt, ``Range concepts and heavy ion ranges (notes on atomic collisions, ii),'' {\em Mat. Fys. Medd. Dan. Vid. Selsk.} {\bfseries 33} no.~14, (1963) 1.

\bibitem{Lindhard:1963b}
J.~Lindhard, V.~Nielsen, M.~Scharff, and P.~V. Thomsen, ``Integral equations governing radiation effects (notes on atomic collisions, iii),'' {\em Mat. Fys. Medd. Dan. Vid. Selsk.} {\bfseries 33} no.~10, (1963) 1.

\bibitem{Xu:2023rzm}
J.~Xu, P.~S. Barbeau, and Z.~Hong, ``{Detection and Calibration of Low-Energy Nuclear Recoils for Dark Matter and Neutrino Scattering Experiments},'' \href{https://dx.doi.org/10.1146/annurev-nucl-111722-025122}{{\em Ann. Rev. Nucl. Part. Sci.} {\bfseries 73} no.~1, (2023) 95}, \href{https://arxiv.org/abs/2311.02257}{{\ttfamily arXiv:2311.02257}}.

\bibitem{Angloher:2025fnl}
G.~Angloher {\em et~al.}, ``{Quenching factors for Na recoils as a function of Tl dopant concentrations in NaI(Tl) crystals},'' \href{https://dx.doi.org/10.1140/epjc/s10052-026-15738-5}{{\em Eur. Phys. J. C} {\bfseries 86} no.~5, (2026) 485}, \href{https://arxiv.org/abs/2512.05980}{{\ttfamily arXiv:2512.05980}}.

\bibitem{Szydagis:2022ikv}
M.~Szydagis {\em et~al.}, ``{A review of NEST models for liquid xenon and an exhaustive comparison with other approaches},'' \href{https://dx.doi.org/10.3389/fdest.2024.1480975}{{\em Front. Detect. Sci. Tech.} {\bfseries 2} (2024) 1480975}, \href{https://arxiv.org/abs/2211.10726}{{\ttfamily arXiv:2211.10726}}.

\bibitem{DAMIC-M:2022xtp}
{\bfseries DAMIC-M} Collaboration, D.~Norcini {\em et~al.}, ``{Precision measurement of Compton scattering in silicon with a skipper CCD for dark matter detection},'' \href{https://dx.doi.org/10.1103/PhysRevD.106.092001}{{\em Phys. Rev. D} {\bfseries 106} no.~9, (2022) 092001}, \href{https://arxiv.org/abs/2207.00809}{{\ttfamily arXiv:2207.00809}}.

\bibitem{SuperCDMS:2025rgn}
{\bfseries SuperCDMS} Collaboration, M.~F. Albakry {\em et~al.}, ``{Low-energy calibration of SuperCDMS HVeV cryogenic silicon calorimeters using Compton steps},'' \href{https://dx.doi.org/10.1103/jj7w-gkgg}{{\em Phys. Rev. D} {\bfseries 112} no.~9, (2025) 092014}, \href{https://arxiv.org/abs/2508.02402}{{\ttfamily arXiv:2508.02402}}.

\bibitem{DarkSide-20k:2026xdv}
{\bfseries DarkSide-20k} Collaboration, L.~Pandola {\em et~al.}, ``{Characterization of argon recoils at the keV scale with ReD and ReD+},'' \href{https://dx.doi.org/10.1088/1748-0221/21/04/C04050}{{\em JINST} {\bfseries 21} no.~04, (2026) C04050}, \href{https://arxiv.org/abs/2602.14538}{{\ttfamily arXiv:2602.14538}}.

\bibitem{Cintas:2024pdu}
D.~Cintas {\em et~al.}, ``{Measurement of the sodium and iodine scintillation quenching factors across multiple NaI(Tl) detectors to identify systematics},'' \href{https://dx.doi.org/10.1103/PhysRevC.110.014613}{{\em Phys. Rev. C} {\bfseries 110} no.~1, (2024) 014613}, \href{https://arxiv.org/abs/2402.12480}{{\ttfamily arXiv:2402.12480}}.

\bibitem{SuperCDMS:2023geu}
{\bfseries SuperCDMS} Collaboration, M.~F. Albakry {\em et~al.}, ``{First Measurement of the Nuclear-Recoil Ionization Yield in Silicon at 100~eV},'' \href{https://dx.doi.org/10.1103/PhysRevLett.131.091801}{{\em Phys. Rev. Lett.} {\bfseries 131} no.~9, (2023) 091801}, \href{https://arxiv.org/abs/2303.02196}{{\ttfamily arXiv:2303.02196}}.

\bibitem{LUX:2016ezw}
{\bfseries LUX} Collaboration, D.~S. Akerib {\em et~al.}, ``{Low-energy (0.7-74 keV) nuclear recoil calibration of the LUX dark matter experiment using D-D neutron scattering kinematics},'' \href{https://arxiv.org/abs/1608.05381}{{\ttfamily arXiv:1608.05381}}.

\bibitem{XENON:2024kbh}
{\bfseries XENON} Collaboration, E.~Aprile {\em et~al.}, ``{Low-energy nuclear recoil calibration of XENONnT with a Y88Be photoneutron source},'' \href{https://dx.doi.org/10.1103/rsl9-j1ky}{{\em Phys. Rev. D} {\bfseries 113} no.~11, (2026) 112017}, \href{https://arxiv.org/abs/2412.10451}{{\ttfamily arXiv:2412.10451}}.

\bibitem{Wagner:2022ckd}
{\bfseries CRAB} Collaboration, V.~Wagner, ``{Accurate Calibration of Nuclear Recoils at the 100 eV Scale Using Neutron Capture},'' \href{https://dx.doi.org/10.1007/s10909-022-02816-7}{{\em J. Low Temp. Phys.} {\bfseries 209} no.~5-6, (2022) 1055}.

\bibitem{Abele:2025vrj}
H.~Abele {\em et~al.}, ``{The CRAB facility at the TU Wien TRIGA reactor: status and related physics program},'' \href{https://dx.doi.org/10.1140/epjc/s10052-025-14809-3}{{\em Eur. Phys. J. C} {\bfseries 85} no.~10, (2025) 1188}, \href{https://arxiv.org/abs/2505.15227}{{\ttfamily arXiv:2505.15227}}.

\bibitem{CRAB:2022rcm}
{\bfseries CRAB, NUCLEUS} Collaboration, H.~Abele {\em et~al.}, ``{Observation of a Nuclear Recoil Peak at the 100~eV Scale Induced by Neutron Capture},'' \href{https://dx.doi.org/10.1103/PhysRevLett.130.211802}{{\em Phys. Rev. Lett.} {\bfseries 130} no.~21, (2023) 211802}, \href{https://arxiv.org/abs/2211.03631}{{\ttfamily arXiv:2211.03631}}.

\bibitem{CRESST:2023cxk}
{\bfseries CRESST} Collaboration, G.~Angloher {\em et~al.}, ``{Observation of a low energy nuclear recoil peak in the neutron calibration data of the CRESST-III experiment},'' \href{https://dx.doi.org/10.1103/PhysRevD.108.022005}{{\em Phys. Rev. D} {\bfseries 108} no.~2, (2023) 022005}, \href{https://arxiv.org/abs/2303.15315}{{\ttfamily arXiv:2303.15315}}.

\bibitem{CRAB:2025ydy}
{\bfseries CRAB} Collaboration, H.~Abele {\em et~al.}, ``{Observation of two nuclear recoil peaks induced by neutron capture on Al2O3},'' \href{https://dx.doi.org/10.1103/nt33-5z88}{{\em Phys. Rev. C} {\bfseries 114} no.~1, (2026) 014621}, \href{https://arxiv.org/abs/2512.06507}{{\ttfamily arXiv:2512.06507}}.

\bibitem{Formaggio:2004}
J.~A. Formaggio and C.~Martoff, ``{Backgrounds to sensitive experiments underground},'' \href{https://dx.doi.org/10.1146/annurev.ns.45.120195.002551}{{\em Ann. Rev. Nucl. Part. Sci.} {\bfseries 54} (2004) 361}.

\bibitem{Wu:2013cno}
Y.-C. Wu {\em et~al.}, ``{Measurement of Cosmic Ray Flux in China JinPing underground Laboratory},'' \href{https://dx.doi.org/10.1088/1674-1137/37/8/086001}{{\em Chin. Phys. C} {\bfseries 37} no.~8, (2013) 086001}, \href{https://arxiv.org/abs/1305.0899}{{\ttfamily arXiv:1305.0899}}.

\bibitem{Mei:2005gm}
D.~Mei and A.~Hime, ``{Muon-induced background study for underground laboratories},'' \href{https://dx.doi.org/10.1103/PhysRevD.73.053004}{{\em Phys. Rev. D} {\bfseries 73} (2006) 053004}, \href{https://arxiv.org/abs/astro-ph/0512125}{{\ttfamily arXiv:astro-ph/0512125}}. [Erratum: Phys.Rev.D 109, 019901 (2024)].

\bibitem{LZ:2026kit}
{\bfseries LZ} Collaboration, D.~S. Akerib {\em et~al.}, ``{Measurement of the Muon Flux at the Sanford Underground Research Facility with the LUX-ZEPLIN Dark Matter Detector},'' \href{https://arxiv.org/abs/2602.16799}{{\ttfamily arXiv:2602.16799}}.

\bibitem{Fu:2026ubt}
G.~Fu {\em et~al.}, ``{Measurement of the cosmic muon flux at the Stawell Underground Physics Laboratory},'' \href{https://dx.doi.org/10.1016/j.astropartphys.2026.103240}{{\em Astropart. Phys.} {\bfseries 179} (2026) 103240}, \href{https://arxiv.org/abs/2603.11981}{{\ttfamily arXiv:2603.11981}}.

\bibitem{Trzaska:2019kuk}
W.~H. Trzaska {\em et~al.}, ``{Cosmic-ray muon flux at Canfranc Underground Laboratory},'' \href{https://dx.doi.org/10.1140/epjc/s10052-019-7239-9}{{\em Eur. Phys. J. C} {\bfseries 79} no.~8, (2019) 721}, \href{https://arxiv.org/abs/1902.00868}{{\ttfamily arXiv:1902.00868}}.

\bibitem{Sharan:2021tnj}
M.~K. Sharan, R.~N. Singaraju, T.~Sinha, T.~Ghosh, and V.~N. Jha, ``{Measurement of cosmic-ray muon flux in the underground laboratory at UCIL, India, using plastic scintillators and SiPM},'' \href{https://dx.doi.org/10.1016/j.nima.2021.165083}{{\em Nucl. Instrum. Meth. A} {\bfseries 994} (2021) 165083}.

\bibitem{Garrison:2014nla}
L.~Garrison, \href{https://dx.doi.org/10.2172/1151758}{{\em {Measurement of Neutron and Muon Fluxes 100{\textasciitilde}m Underground with the SciBath Detector}}}.
\newblock PhD thesis, Indiana U., Bloomington (main), 2014.

\bibitem{JNE:2020bwn}
{\bfseries JNE} Collaboration, Z.~Guo {\em et~al.}, ``{Muon flux measurement at China Jinping Underground Laboratory},'' \href{https://dx.doi.org/10.1088/1674-1137/abccae}{{\em Chin. Phys. C} {\bfseries 45} no.~2, (2021) 025001}, \href{https://arxiv.org/abs/2007.15925}{{\ttfamily arXiv:2007.15925}}.

\bibitem{LZ:2025igz}
{\bfseries LZ} Collaboration, D.~S. Akerib {\em et~al.}, ``{Searches for Light Dark Matter and Evidence of Coherent Elastic Neutrino-Nucleus Scattering of Solar Neutrinos with the LUX-ZEPLIN (LZ) Experiment},'' \href{https://dx.doi.org/10.1103/jvqf-njpj}{{\em Phys. Rev. Lett.} {\bfseries 137} no.~9, (2026) 091806}, \href{https://arxiv.org/abs/2512.08065}{{\ttfamily arXiv:2512.08065}}.

\bibitem{XENON:2026smt}
{\bfseries XENON} Collaboration, E.~Aprile {\em et~al.}, ``{Probing the Solar B8 Neutrino Fog with XENONnT},'' \href{https://dx.doi.org/10.1103/xvpn-w3q9}{{\em Phys. Rev. Lett.} {\bfseries 137} no.~9, (2026) 091807}, \href{https://arxiv.org/abs/2604.06002}{{\ttfamily arXiv:2604.06002}}.

\bibitem{PandaX:2024muv}
{\bfseries PandaX} Collaboration, Z.~Bo {\em et~al.}, ``{First Indication of Solar B8 Neutrinos through Coherent Elastic Neutrino-Nucleus Scattering in PandaX-4T},'' \href{https://dx.doi.org/10.1103/PhysRevLett.133.191001}{{\em Phys. Rev. Lett.} {\bfseries 133} no.~19, (2024) 191001}, \href{https://arxiv.org/abs/2407.10892}{{\ttfamily arXiv:2407.10892}}.

\bibitem{OHare:2021utq}
C.~A.~J. O'Hare, ``{New Definition of the Neutrino Floor for Direct Dark Matter Searches},'' \href{https://dx.doi.org/10.1103/PhysRevLett.127.251802}{{\em Phys. Rev. Lett.} {\bfseries 127} no.~25, (2021) 251802}, \href{https://arxiv.org/abs/2109.03116}{{\ttfamily arXiv:2109.03116}}.

\bibitem{PandaX:2026qdp}
{\bfseries PandaX} Collaboration, P.~Chen {\em et~al.}, ``{Measurement of solar $pp$ neutrino flux with the new PandaX-4T data},'' \href{https://arxiv.org/abs/2607.02405}{{\ttfamily arXiv:2607.02405}}.

\bibitem{XENON:2026rbz}
{\bfseries XENON} Collaboration, E.~Aprile {\em et~al.}, ``{First Measurement of Solar Neutrinos through Elastic Neutrino-Electron Scattering at the keV Scale},'' \href{https://arxiv.org/abs/2608.29450}{{\ttfamily arXiv:2608.29450}}.

\bibitem{SuperCDMS:2024yiv}
{\bfseries SuperCDMS} Collaboration, M.~F. Albakry {\em et~al.}, ``{Light dark matter constraints from SuperCDMS HVeV detectors operated underground with an anticoincidence event selection},'' \href{https://dx.doi.org/10.1103/PhysRevD.111.012006}{{\em Phys. Rev. D} {\bfseries 111} no.~1, (2025) 012006}, \href{https://arxiv.org/abs/2407.08085}{{\ttfamily arXiv:2407.08085}}.

\bibitem{COSINUS:2021bdj}
{\bfseries COSINUS} Collaboration, G.~Angloher {\em et~al.}, ``{Simulation-based design study for the passive shielding of the COSINUS dark matter experiment},'' \href{https://dx.doi.org/10.1140/epjc/s10052-022-10184-5}{{\em Eur. Phys. J. C} {\bfseries 82} no.~3, (2022) 248}, \href{https://arxiv.org/abs/2106.07390}{{\ttfamily arXiv:2106.07390}}.

\bibitem{Angloher:2024pbw}
G.~Angloher {\em et~al.}, ``{Water Cherenkov muon veto for the COSINUS experiment: design and simulation optimization},'' \href{https://dx.doi.org/10.1140/epjc/s10052-024-12923-2}{{\em Eur. Phys. J. C} {\bfseries 84} no.~5, (2024) 551}, \href{https://arxiv.org/abs/2406.12870}{{\ttfamily arXiv:2406.12870}}.

\bibitem{XENON1T:2014eqx}
{\bfseries XENON1T} Collaboration, E.~Aprile {\em et~al.}, ``{Conceptual design and simulation of a water Cherenkov muon veto for the XENON1T experiment},'' \href{https://dx.doi.org/10.1088/1748-0221/9/11/P11006}{{\em JINST} {\bfseries 9} (2014) P11006}, \href{https://arxiv.org/abs/1406.2374}{{\ttfamily arXiv:1406.2374}}.

\bibitem{Heusser:1995wd}
G.~Heusser, ``{Low-radioactivity background techniques},'' \href{https://dx.doi.org/10.1146/annurev.ns.45.120195.002551}{{\em Ann. Rev. Nucl. Part. Sci.} {\bfseries 45} (1995) 543}.

\bibitem{Laubenstein:2020rbe}
M.~Laubenstein and I.~Lawson, ``{Low Background Radiation Detection Techniques and Mitigation of Radioactive Backgrounds},'' \href{https://dx.doi.org/10.3389/fphy.2020.577734}{{\em Front. in Phys.} {\bfseries 8} (2020) 506}.

\bibitem{XENON:2021mrg}
{\bfseries XENON} Collaboration, E.~Aprile {\em et~al.}, ``{Material radiopurity control in the XENONnT experiment},'' \href{https://dx.doi.org/10.1140/epjc/s10052-022-10345-6}{{\em Eur. Phys. J. C} {\bfseries 82} no.~7, (2022) 599}, \href{https://arxiv.org/abs/2112.05629}{{\ttfamily arXiv:2112.05629}}.

\bibitem{radiopurity}
``Radiopurity.org,'' \url{https://www.radiopurity.org}.
\newblock Accessed: 2026-08-01.

\bibitem{Amare:2017roa}
J.~Amare {\em et~al.}, ``{Cosmogenic production of tritium in dark matter detectors},'' \href{https://dx.doi.org/10.1016/j.astropartphys.2017.11.004}{{\em Astropart. Phys.} {\bfseries 97} (2018) 96}, \href{https://arxiv.org/abs/1706.05818}{{\ttfamily arXiv:1706.05818}}.

\bibitem{DarkSide-20k:2023fbh}
{\bfseries DarkSide-20k} Collaboration, A.~Elersich {\em et~al.}, ``{Study of cosmogenic activation above ground for the DarkSide-20k experiment},'' \href{https://dx.doi.org/10.1016/j.astropartphys.2023.102878}{{\em Astropart. Phys.} {\bfseries 152} (2023) 102878}, \href{https://arxiv.org/abs/2301.12970}{{\ttfamily arXiv:2301.12970}}.

\bibitem{Westerdale:2017kml}
S.~Westerdale and P.~D. Meyers, ``{Radiogenic Neutron Yield Calculations for Low-Background Experiments},'' \href{https://dx.doi.org/10.1016/j.nima.2017.09.007}{{\em Nucl. Instrum. Meth. A} {\bfseries 875} (2017) 57}, \href{https://arxiv.org/abs/1702.02465}{{\ttfamily arXiv:1702.02465}}.

\bibitem{Cano-Ott:2024elo}
D.~Cano-Ott {\em et~al.}, ``{Review of neutron yield from ({\ensuremath{\alpha}}, n) reactions: data, methods, and prospects},'' \href{https://dx.doi.org/10.1088/1361-6471/adeffa}{{\em J. Phys. G} {\bfseries 53} no.~2, (2026) 023001}, \href{https://arxiv.org/abs/2405.07952}{{\ttfamily arXiv:2405.07952}}.

\bibitem{Veselska_2024}
O.~Veselska {\em et~al.}, ``Exploring the potential use of silver-exchanged zeolites for adsorption of radon traces in low background experiments,'' \href{https://dx.doi.org/10.1093/ptep/ptad160}{{\em PTEP} {\bfseries 2024} no.~2, (2024) 023C01}.

\bibitem{Pelczar:2021}
K.~Pelczar, G.~Zuzel, M.~Wójcik, {\em et~al.}, ``{An online radon monitor for low-background detector assembly facilities},'' \href{https://dx.doi.org/10.1140/epjc/s10052-020-08793-z}{{\em EPJC} {\bfseries 81} (2021) 86}.

\bibitem{Acosta-Kane:2007you}
D.~Acosta-Kane {\em et~al.}, ``{Discovery of underground argon with low level of radioactive 39Ar and possible applications to WIMP dark matter detectors},'' \href{https://dx.doi.org/10.1016/j.nima.2007.12.032}{{\em Nucl. Instrum. Meth. A} {\bfseries 587} (2008) 46}, \href{https://arxiv.org/abs/0712.0381}{{\ttfamily arXiv:0712.0381}}.

\bibitem{LVD:2019zlh}
{\bfseries LVD} Collaboration, N.~Y. Agafonova {\em et~al.}, ``{Characterization of the varying flux of atmospheric muons measured with the Large Volume Detector for 24 years},'' \href{https://dx.doi.org/10.1103/PhysRevD.100.062002}{{\em Phys. Rev. D} {\bfseries 100} no.~6, (2019) 062002}, \href{https://arxiv.org/abs/1909.04579}{{\ttfamily arXiv:1909.04579}}.

\bibitem{Bernabei:2000ew}
R.~Bernabei {\em et~al.}, ``{On the investigation of possible systematics in WIMP annual modulation search},'' \href{https://dx.doi.org/10.1007/s100520000540}{{\em Eur. Phys. J. C} {\bfseries 18} (2000) 283}.

\bibitem{Messina:2020pnt}
A.~Messina, M.~Nardecchia, and S.~Piacentini, ``{Annual modulations from secular variations: not relaxing DAMA?},'' \href{https://dx.doi.org/10.1088/1475-7516/2020/04/037}{{\em JCAP} {\bfseries 04} (2020) 037}, \href{https://arxiv.org/abs/2003.03340}{{\ttfamily arXiv:2003.03340}}.

\bibitem{Buttazzo:2020bto}
D.~Buttazzo, P.~Panci, N.~Rossi, and A.~Strumia, ``{Annual modulations from secular variations: relaxing DAMA?},'' \href{https://dx.doi.org/10.1007/JHEP04(2020)137}{{\em JHEP} {\bfseries 04} (2020) 137}, \href{https://arxiv.org/abs/2002.00459}{{\ttfamily arXiv:2002.00459}}.

\bibitem{DarkSide-50:2025umf}
{\bfseries DarkSide-50} Collaboration, P.~Agnes {\em et~al.}, ``{Characterization of spurious-electron signals in the double-phase argon TPC of the DarkSide-50 experiment},'' \href{https://dx.doi.org/10.1140/epjc/s10052-026-16193-y}{{\em Eur. Phys. J. C} {\bfseries 86} no.~9, (2026) 1099}, \href{https://arxiv.org/abs/2507.23003}{{\ttfamily arXiv:2507.23003}}.

\bibitem{XENON:2021qze}
{\bfseries XENON} Collaboration, E.~Aprile {\em et~al.}, ``{Emission of single and few electrons in XENON1T and limits on light dark matter},'' \href{https://dx.doi.org/10.1103/PhysRevD.106.022001}{{\em Phys. Rev. D} {\bfseries 106} no.~2, (2022) 022001}, \href{https://arxiv.org/abs/2112.12116}{{\ttfamily arXiv:2112.12116}}. [Erratum: Phys.Rev.D 110, 109903 (2024)].

\bibitem{LUX-ZEPLIN:2025uaf}
{\bfseries LUX-ZEPLIN} Collaboration, D.~S. Akerib {\em et~al.}, ``{Study of few-electron backgrounds in the LUX-ZEPLIN detector},'' \href{https://dx.doi.org/10.1103/p2vl-rx74}{{\em Phys. Rev. D} {\bfseries 113} no.~7, (2026) 072018}, \href{https://arxiv.org/abs/2510.06500}{{\ttfamily arXiv:2510.06500}}.

\bibitem{COSINUS:2023kqd}
{\bfseries COSINUS} Collaboration, G.~Angloher {\em et~al.}, ``{Deep-underground dark matter search with a COSINUS detector prototype},'' \href{https://dx.doi.org/10.1103/PhysRevD.110.043010}{{\em Phys. Rev. D} {\bfseries 110} no.~4, (2024) 043010}, \href{https://arxiv.org/abs/2307.11139}{{\ttfamily arXiv:2307.11139}}.

\bibitem{CRESST:2019jnq}
{\bfseries CRESST} Collaboration, A.~H. Abdelhameed {\em et~al.}, ``{First results from the CRESST-III low-mass dark matter program},'' \href{https://dx.doi.org/10.1103/PhysRevD.100.102002}{{\em Phys. Rev. D} {\bfseries 100} no.~10, (2019) 102002}, \href{https://arxiv.org/abs/1904.00498}{{\ttfamily arXiv:1904.00498}}.

\bibitem{CRESST:2020tlq}
{\bfseries CRESST} Collaboration, A.~H. Abdelhameed {\em et~al.}, ``{Cryogenic characterization of a $\hbox {LiAlO}_{2}$ crystal and new results on spin-dependent dark matter interactions with ordinary matter},'' \href{https://dx.doi.org/101140/epjc/s10052-020-8329-4}{{\em Eur. Phys. J. C} {\bfseries 80} no.~9, (2020) 834}, \href{https://arxiv.org/abs/2005.02692}{{\ttfamily arXiv:2005.02692}}.

\bibitem{CRESST:2022jig}
{\bfseries CRESST} Collaboration, G.~Angloher {\em et~al.}, ``{Probing spin-dependent dark matter interactions with $^6$Li: CRESST Collaboration},'' \href{https://dx.doi.org/10.1140/epjc/s10052-022-10140-3}{{\em Eur. Phys. J. C} {\bfseries 82} no.~3, (2022) 207}, \href{https://arxiv.org/abs/2201.03863}{{\ttfamily arXiv:2201.03863}}.

\bibitem{CRESST:2022dtl}
{\bfseries CRESST} Collaboration, G.~Angloher {\em et~al.}, ``{Testing spin-dependent dark matter interactions with lithium aluminate targets in CRESST-III},'' \href{https://dx.doi.org/10.1103/PhysRevD.106.092008}{{\em Phys. Rev. D} {\bfseries 106} no.~9, (2022) 092008}, \href{https://arxiv.org/abs/2207.07640}{{\ttfamily arXiv:2207.07640}}.

\bibitem{CRESST:2022lqw}
{\bfseries CRESST} Collaboration, G.~Angloher {\em et~al.}, ``{Results on sub-GeV dark matter from a 10~eV threshold CRESST-III silicon detector},'' \href{https://dx.doi.org/10.1103/PhysRevD.107.122003}{{\em Phys. Rev. D} {\bfseries 107} no.~12, (2023) 122003}, \href{https://arxiv.org/abs/2212.12513}{{\ttfamily arXiv:2212.12513}}.

\bibitem{CRESST:2024cpr}
{\bfseries CRESST} Collaboration, G.~Angloher {\em et~al.}, ``{First observation of single photons in a CRESST detector and new dark matter exclusion limits},'' \href{https://dx.doi.org/10.1103/PhysRevD.110.083038}{{\em Phys. Rev. D} {\bfseries 110} no.~8, (2024) 083038}, \href{https://arxiv.org/abs/2405.06527}{{\ttfamily arXiv:2405.06527}}.

\bibitem{Alloni:2026vak}
D.~Alloni {\em et~al.}, ``{Probing dark matter interactions with a RES-NOVA prototype cryogenic detector},'' \href{https://dx.doi.org/10.1140/epjc/s10052-026-16215-9}{{\em Eur. Phys. J. C} {\bfseries 86} no.~9, (2026) 1117}, \href{https://arxiv.org/abs/2601.16251}{{\ttfamily arXiv:2601.16251}}.

\bibitem{SuperCDMS:2017nns}
{\bfseries SuperCDMS} Collaboration, R.~Agnese {\em et~al.}, ``{Low-mass dark matter search with CDMSlite},'' \href{https://dx.doi.org/10.1103/PhysRevD.97.022002}{{\em Phys. Rev. D} {\bfseries 97} no.~2, (2018) 022002}, \href{https://arxiv.org/abs/1707.01632}{{\ttfamily arXiv:1707.01632}}.

\bibitem{SuperCDMS:2020aus}
{\bfseries SuperCDMS} Collaboration, I.~Alkhatib {\em et~al.}, ``{Light Dark Matter Search with a High-Resolution Athermal Phonon Detector Operated Above Ground},'' \href{https://dx.doi.org/10.1103/PhysRevLett.127.061801}{{\em Phys. Rev. Lett.} {\bfseries 127} (2021) 061801}, \href{https://arxiv.org/abs/2007.14289}{{\ttfamily arXiv:2007.14289}}.

\bibitem{SuperCDMS:2022zmd}
{\bfseries SuperCDMS} Collaboration, M.~F. Albakry {\em et~al.}, ``{Investigating the sources of low-energy events in a SuperCDMS-HVeV detector},'' \href{https://dx.doi.org/10.1103/PhysRevD.105.112006}{{\em Phys. Rev. D} {\bfseries 105} no.~11, (2022) 112006}, \href{https://arxiv.org/abs/2204.08038}{{\ttfamily arXiv:2204.08038}}.

\bibitem{TESSERACT:2025tfw}
{\bfseries TESSERACT} Collaboration, T.~K. Bui {\em et~al.}, ``{First Limits on Light Dark Matter Interactions in a Low Threshold Two-Channel Athermal Phonon Detector from the TESSERACT Collaboration},'' \href{https://dx.doi.org/10.1103/hsrl-crvf}{{\em Phys. Rev. Lett.} {\bfseries 135} no.~16, (2025) 161002}, \href{https://arxiv.org/abs/2503.03683}{{\ttfamily arXiv:2503.03683}}.

\bibitem{CDEX:2021cll}
{\bfseries CDEX} Collaboration, Z.~Z. Liu {\em et~al.}, ``{Studies of the Earth shielding effect to direct dark matter searches at the China Jinping Underground Laboratory},'' \href{https://dx.doi.org/10.1103/PhysRevD.105.052005}{{\em Phys. Rev. D} {\bfseries 105} no.~5, (2022) 052005}, \href{https://arxiv.org/abs/2111.11243}{{\ttfamily arXiv:2111.11243}}.

\bibitem{CDEX:2024rfi}
{\bfseries CDEX} Collaboration, Q.-Y. Nie {\em et~al.}, ``{Constraining dark matter models with a light mediator from the CDEX-10 experiment at China Jinping Underground Laboratory*},'' \href{https://dx.doi.org/10.1088/1674-1137/ada914}{{\em Chin. Phys. C} {\bfseries 49} no.~4, (2025) 043002}, \href{https://arxiv.org/abs/2411.07543}{{\ttfamily arXiv:2411.07543}}.

\bibitem{COSINE-100:2025kbw}
{\bfseries COSINE-100} Collaboration, G.~H. Yu {\em et~al.}, ``{Limits on WIMP Dark Matter with NaI(Tl) Crystals in Three Years of COSINE-100 Data},'' \href{https://dx.doi.org/10.1103/77p2-rnbw}{{\em Phys. Rev. Lett.} {\bfseries 135} no.~23, (2025) 231001}, \href{https://arxiv.org/abs/2501.13665}{{\ttfamily arXiv:2501.13665}}.

\bibitem{DarkSide-50:2022qzh}
{\bfseries DarkSide-50} Collaboration, P.~Agnes {\em et~al.}, ``{Search for low-mass dark matter WIMPs with 12~ton-day exposure of DarkSide-50},'' \href{https://dx.doi.org/10.1103/PhysRevD.107.063001}{{\em Phys. Rev. D} {\bfseries 107} no.~6, (2023) 063001}, \href{https://arxiv.org/abs/2207.11966}{{\ttfamily arXiv:2207.11966}}.

\bibitem{DarkSide-50:2025lns}
{\bfseries DarkSide-20k} Collaboration, F.~Acerbi {\em et~al.}, ``{Sensitivity to low-mass WIMPs with an improved liquid argon ionization response model within the DarkSide program},'' \href{https://dx.doi.org/10.1103/x3vt-q676}{{\em Phys. Rev. D} {\bfseries 113} no.~12, (2026) 123045}, \href{https://arxiv.org/abs/2511.13629}{{\ttfamily arXiv:2511.13629}}.

\bibitem{PandaX:2025rrz}
{\bfseries PandaX} Collaboration, M.~Zhang {\em et~al.}, ``{Search for Light Dark Matter with 259 Days of Data in PandaX-4T},'' \href{https://dx.doi.org/10.1103/rtnh-jn8s}{{\em Phys. Rev. Lett.} {\bfseries 135} no.~21, (2025) 211001}, \href{https://arxiv.org/abs/2507.11930}{{\ttfamily arXiv:2507.11930}}. [Erratum: Phys.Rev.Lett. 136, 069901 (2026)].

\bibitem{PandaX:2025rrz_erratum}
{\bfseries PandaX} Collaboration, M.~Zhang {\em et~al.}, ``Erratum: Search for light dark matter with 259 days of data in pandax-4t [phys. rev. lett. 135, 211001 (2025)],'' \href{https://dx.doi.org/10.1103/dq19-jkqk}{{\em Phys. Rev. Lett.} {\bfseries 136} (Feb, 2026) 069901}.

\bibitem{NEWS-G:2024jms}
{\bfseries NEWS-G} Collaboration, M.~M. Arora {\em et~al.}, ``{Search for Light Dark Matter with NEWS-G at the Laboratoire Souterrain de Modane Using a Methane Target},'' \href{https://dx.doi.org/10.1103/PhysRevLett.134.141002}{{\em Phys. Rev. Lett.} {\bfseries 134} no.~14, (2025) 141002}, \href{https://arxiv.org/abs/2407.12769}{{\ttfamily arXiv:2407.12769}}.

\bibitem{GRAPES-3:2024yym}
{\bfseries CRESST} Collaboration, G.~Angloher {\em et~al.}, ``{DoubleTES detectors to investigate the CRESST low energy background: results from above-ground prototypes},'' \href{https://dx.doi.org/10.1140/epjc/s10052-024-13282-8}{{\em Eur. Phys. J. C} {\bfseries 84} no.~10, (2024) 1001}, \href{https://arxiv.org/abs/2404.02607}{{\ttfamily arXiv:2404.02607}}. [Erratum: Eur.Phys.J.C 84, 1227 (2024)].

\bibitem{CDMS:2013juh}
{\bfseries CDMS} Collaboration, R.~Agnese {\em et~al.}, ``{Silicon Detector Dark Matter Results from the Final Exposure of CDMS II},'' \href{https://dx.doi.org/10.1103/PhysRevLett.111.251301}{{\em Phys. Rev. Lett.} {\bfseries 111} no.~25, (2013) 251301}, \href{https://arxiv.org/abs/1304.4279}{{\ttfamily arXiv:1304.4279}}.

\bibitem{SENSEI:2023rcc}
{\bfseries SENSEI, DAMIC-M, DAMIC} Collaboration, A.~Aguilar-Arevalo {\em et~al.}, ``{Confirmation of the spectral excess in DAMIC at SNOLAB with skipper CCDs},'' \href{https://dx.doi.org/10.1103/PhysRevD.109.062007}{{\em Phys. Rev. D} {\bfseries 109} no.~6, (2024) 062007}, \href{https://arxiv.org/abs/2306.01717}{{\ttfamily arXiv:2306.01717}}.

\bibitem{COSINE-100:2021xqn}
{\bfseries COSINE-100} Collaboration, G.~Adhikari {\em et~al.}, ``{Strong constraints from COSINE-100 on the DAMA dark matter results using the same sodium iodide target},'' \href{https://dx.doi.org/10.1126/sciadv.abk2699}{{\em Sci. Adv.} {\bfseries 7} no.~46, (2021) abk2699}, \href{https://arxiv.org/abs/2104.03537}{{\ttfamily arXiv:2104.03537}}.

\bibitem{Kumar:2025ofs}
V.~Kumar, S.~Ali, M.~Das, N.~Biswas, S.~Das, S.~Sahoo, N.~Chaddha, J.~Basu, and V.~N. Jha, ``{First result from tetrafluoroethane (C$_2$H$_2$F$_4$) superheated emulsion detector for dark matter search at JUSL},'' \href{https://arxiv.org/abs/2501.02833}{{\ttfamily arXiv:2501.02833}}.

\bibitem{DEAP:2026orr}
{\bfseries DEAP} Collaboration, P.~Adhikari {\em et~al.}, ``{Dark Matter Search with the DEAP-3600 Detector using the Profile Likelihood Ratio Method},'' \href{https://arxiv.org/abs/2603.13965}{{\ttfamily arXiv:2603.13965}}.

\bibitem{LZ:2022lsv}
{\bfseries LZ} Collaboration, J.~Aalbers {\em et~al.}, ``{First Dark Matter Search Results from the LUX-ZEPLIN (LZ) Experiment},'' \href{https://dx.doi.org/10.1103/PhysRevLett.131.041002}{{\em Phys. Rev. Lett.} {\bfseries 131} no.~4, (2023) 041002}, \href{https://arxiv.org/abs/2207.03764}{{\ttfamily arXiv:2207.03764}}.

\bibitem{LZ:2026axp}
{\bfseries LZ} Collaboration, D.~S. Akerib {\em et~al.}, ``{Search for dark matter particle interactions in an extended nuclear recoil energy window with the LUX-ZEPLIN (LZ) experiment},'' \href{https://arxiv.org/abs/2609.02823}{{\ttfamily arXiv:2609.02823}}.

\bibitem{XENON:2017lvq}
{\bfseries XENON} Collaboration, E.~Aprile {\em et~al.}, ``{The XENON1T Dark Matter Experiment},'' \href{https://dx.doi.org/10.1140/epjc/s10052-017-5326-3}{{\em Eur. Phys. J. C} {\bfseries 77} no.~12, (2017) 881}, \href{https://arxiv.org/abs/1708.07051}{{\ttfamily arXiv:1708.07051}}.

\bibitem{XENON:2023cxc}
{\bfseries XENON} Collaboration, E.~Aprile {\em et~al.}, ``{First Dark Matter Search with Nuclear Recoils from the XENONnT Experiment},'' \href{https://dx.doi.org/10.1103/PhysRevLett.131.041003}{{\em Phys. Rev. Lett.} {\bfseries 131} no.~4, (2023) 041003}, \href{https://arxiv.org/abs/2303.14729}{{\ttfamily arXiv:2303.14729}}.

\bibitem{XENON:2018voc}
{\bfseries XENON} Collaboration, E.~Aprile {\em et~al.}, ``{Dark Matter Search Results from a One Ton-Year Exposure of XENON1T},'' \href{https://dx.doi.org/10.1103/PhysRevLett.121.111302}{{\em Phys. Rev. Lett.} {\bfseries 121} no.~11, (2018) 111302}, \href{https://arxiv.org/abs/1805.12562}{{\ttfamily arXiv:1805.12562}}.

\bibitem{EDELWEISS:2020fxc}
{\bfseries EDELWEISS} Collaboration, Q.~Arnaud {\em et~al.}, ``{First germanium-based constraints on sub-MeV Dark Matter with the EDELWEISS experiment},'' \href{https://dx.doi.org/10.1103/PhysRevLett.125.141301}{{\em Phys. Rev. Lett.} {\bfseries 125} no.~14, (2020) 141301}, \href{https://arxiv.org/abs/2003.01046}{{\ttfamily arXiv:2003.01046}}.

\bibitem{EDELWEISS:2022ktt}
{\bfseries EDELWEISS} Collaboration, E.~Armengaud {\em et~al.}, ``{Search for sub-GeV dark matter via the Migdal effect with an EDELWEISS germanium detector with NbSi transition-edge sensors},'' \href{https://dx.doi.org/10.1103/PhysRevD.106.062004}{{\em Phys. Rev. D} {\bfseries 106} no.~6, (2022) 062004}, \href{https://arxiv.org/abs/2203.03993}{{\ttfamily arXiv:2203.03993}}.

\bibitem{SuperCDMS:2025dha}
{\bfseries SuperCDMS} Collaboration, M.~F. Albakry {\em et~al.}, ``{Search for low-mass electron-recoil dark matter using a single-charge sensitive SuperCDMS-HVeV detector},'' \href{https://dx.doi.org/10.1103/5lnp-6mng}{{\em Phys. Rev. D} {\bfseries 113} no.~3, (2026) 032001}, \href{https://arxiv.org/abs/2509.03608}{{\ttfamily arXiv:2509.03608}}.

\bibitem{SuperCDMS:2023sql}
{\bfseries SuperCDMS} Collaboration, M.~F. Albakry {\em et~al.}, ``{Search for low-mass dark matter via bremsstrahlung radiation and the Migdal effect in SuperCDMS},'' \href{https://dx.doi.org/10.1103/PhysRevD.107.112013}{{\em Phys. Rev. D} {\bfseries 107} no.~11, (2023) 112013}, \href{https://arxiv.org/abs/2302.09115}{{\ttfamily arXiv:2302.09115}}.

\bibitem{CDEX:2022kcd}
{\bfseries CDEX} Collaboration, Z.~Y. Zhang {\em et~al.}, ``{Constraints on Sub-GeV Dark Matter{\textendash}Electron Scattering from the CDEX-10 Experiment},'' \href{https://dx.doi.org/10.1103/PhysRevLett.129.221301}{{\em Phys. Rev. Lett.} {\bfseries 129} no.~22, (2022) 221301}, \href{https://arxiv.org/abs/2206.04128}{{\ttfamily arXiv:2206.04128}}.

\bibitem{DAMIC-M:2023gxo}
{\bfseries DAMIC-M} Collaboration, I.~Arnquist {\em et~al.}, ``{First Constraints from DAMIC-M on Sub-GeV Dark-Matter Particles Interacting with Electrons},'' \href{https://dx.doi.org/10.1103/PhysRevLett.130.171003}{{\em Phys. Rev. Lett.} {\bfseries 130} no.~17, (2023) 171003}, \href{https://arxiv.org/abs/2302.02372}{{\ttfamily arXiv:2302.02372}}.

\bibitem{DAMIC-M:2025luv}
{\bfseries DAMIC-M} Collaboration, K.~Aggarwal {\em et~al.}, ``{Probing Benchmark Models of Hidden-Sector Dark Matter with DAMIC-M},'' \href{https://dx.doi.org/10.1103/2tcc-bqck}{{\em Phys. Rev. Lett.} {\bfseries 135} no.~7, (2025) 071002}, \href{https://arxiv.org/abs/2503.14617}{{\ttfamily arXiv:2503.14617}}.

\bibitem{SENSEI:2023zdf}
{\bfseries SENSEI} Collaboration, P.~Adari {\em et~al.}, ``{First Direct-Detection Results on Sub-GeV Dark Matter Using the SENSEI Detector at SNOLAB},'' \href{https://dx.doi.org/10.1103/PhysRevLett.134.011804}{{\em Phys. Rev. Lett.} {\bfseries 134} no.~1, (2025) 011804}, \href{https://arxiv.org/abs/2312.13342}{{\ttfamily arXiv:2312.13342}}.

\bibitem{COSINE-100:2025inn}
{\bfseries COSINE-100} Collaboration, L.~E. Fran{\c{c}}a {\em et~al.}, ``{Constraints on sub-GeV dark matter scattering on electrons with COSINE-100},'' \href{https://dx.doi.org/10.1103/kszv-g7tl}{{\em Phys. Rev. D} {\bfseries 113} no.~7, (2026) 072012}, \href{https://arxiv.org/abs/2510.01927}{{\ttfamily arXiv:2510.01927}}.

\bibitem{DarkSide:2022dhx}
{\bfseries DarkSide} Collaboration, P.~Agnes {\em et~al.}, ``{Search for Dark-Matter{\textendash}Nucleon Interactions via Migdal Effect with DarkSide-50},'' \href{https://dx.doi.org/10.1103/PhysRevLett.130.101001}{{\em Phys. Rev. Lett.} {\bfseries 130} no.~10, (2023) 101001}, \href{https://arxiv.org/abs/2207.11967}{{\ttfamily arXiv:2207.11967}}.

\bibitem{DarkSide:2022knj}
{\bfseries DarkSide} Collaboration, P.~Agnes {\em et~al.}, ``{Search for Dark Matter Particle Interactions with Electron Final States with DarkSide-50},'' \href{https://dx.doi.org/10.1103/PhysRevLett.130.101002}{{\em Phys. Rev. Lett.} {\bfseries 130} no.~10, (2023) 101002}, \href{https://arxiv.org/abs/2207.11968}{{\ttfamily arXiv:2207.11968}}.

\bibitem{LZ:2023poo}
{\bfseries LZ} Collaboration, J.~Aalbers {\em et~al.}, ``{Search for new physics in low-energy electron recoils from the first LZ exposure},'' \href{https://dx.doi.org/10.1103/PhysRevD.108.072006}{{\em Phys. Rev. D} {\bfseries 108} no.~7, (2023) 072006}, \href{https://arxiv.org/abs/2307.15753}{{\ttfamily arXiv:2307.15753}}.

\bibitem{PandaX-II:2021nsg}
{\bfseries PandaX-II} Collaboration, C.~Cheng {\em et~al.}, ``{Search for Light Dark Matter-Electron Scatterings in the PandaX-II Experiment},'' \href{https://dx.doi.org/10.1103/PhysRevLett.126.211803}{{\em Phys. Rev. Lett.} {\bfseries 126} no.~21, (2021) 211803}, \href{https://arxiv.org/abs/2101.07479}{{\ttfamily arXiv:2101.07479}}.

\bibitem{PandaX:2023xgl}
{\bfseries PandaX} Collaboration, D.~Huang {\em et~al.}, ``{Search for Dark-Matter{\textendash}Nucleon Interactions with a Dark Mediator in PandaX-4T},'' \href{https://dx.doi.org/10.1103/PhysRevLett.131.191002}{{\em Phys. Rev. Lett.} {\bfseries 131} no.~19, (2023) 191002}, \href{https://arxiv.org/abs/2308.01540}{{\ttfamily arXiv:2308.01540}}.

\bibitem{XENON:2024znc}
{\bfseries XENON} Collaboration, E.~Aprile {\em et~al.}, ``{Search for Light Dark Matter in Low-Energy Ionization Signals from XENONnT},'' \href{https://dx.doi.org/10.1103/PhysRevLett.134.161004}{{\em Phys. Rev. Lett.} {\bfseries 134} no.~16, (2025) 161004}, \href{https://arxiv.org/abs/2411.15289}{{\ttfamily arXiv:2411.15289}}.

\bibitem{SuperCDMS:2020ymb}
{\bfseries SuperCDMS} Collaboration, D.~W. Amaral {\em et~al.}, ``{Constraints on low-mass, relic dark matter candidates from a surface-operated SuperCDMS single-charge sensitive detector},'' \href{https://dx.doi.org/10.1103/PhysRevD.102.091101}{{\em Phys. Rev. D} {\bfseries 102} no.~9, (2020) 091101}, \href{https://arxiv.org/abs/2005.14067}{{\ttfamily arXiv:2005.14067}}.

\bibitem{PandaX:2022xqx}
{\bfseries PandaX} Collaboration, S.~Li {\em et~al.}, ``{Search for Light Dark Matter with Ionization Signals in the PandaX-4T Experiment},'' \href{https://dx.doi.org/10.1103/PhysRevLett.130.261001}{{\em Phys. Rev. Lett.} {\bfseries 130} no.~26, (2023) 261001}, \href{https://arxiv.org/abs/2212.10067}{{\ttfamily arXiv:2212.10067}}.

\bibitem{COSINE-100:2021poy}
{\bfseries COSINE-100} Collaboration, G.~Adhikari {\em et~al.}, ``{Searching for low-mass dark matter via the Migdal effect in COSINE-100},'' \href{https://dx.doi.org/10.1103/PhysRevD.105.042006}{{\em Phys. Rev. D} {\bfseries 105} no.~4, (2022) 042006}, \href{https://arxiv.org/abs/2110.05806}{{\ttfamily arXiv:2110.05806}}.

\bibitem{Amare:2025dfq}
J.~Amar{\'e} {\em et~al.}, ``{Towards a Robust Model-Independent Test of the DAMA/LIBRA Dark Matter Signal: ANAIS-112 Results with Six Years of Data},'' \href{https://dx.doi.org/10.1103/ntnl-zrn9}{{\em Phys. Rev. Lett.} {\bfseries 135} no.~5, (2025) 051001}, \href{https://arxiv.org/abs/2502.01542}{{\ttfamily arXiv:2502.01542}}.

\bibitem{ANAIS-112:2025fne}
{\bfseries ANAIS-112, COSINE-100} Collaboration, N.~Carlin {\em et~al.}, ``{Combined Annual Modulation Dark Matter Search with COSINE-100 and ANAIS-112},'' \href{https://dx.doi.org/10.1103/9j7w-qp1c}{{\em Phys. Rev. Lett.} {\bfseries 135} no.~12, (2025) 121002}, \href{https://arxiv.org/abs/2503.19559}{{\ttfamily arXiv:2503.19559}}.

\bibitem{COSINE-100:2024nfa}
{\bfseries COSINE-100} Collaboration, N.~Carlin {\em et~al.}, ``{COSINE-100 full dataset challenges the annual modulation signal of DAMA/LIBRA},'' \href{https://dx.doi.org/10.1126/sciadv.adv6503}{{\em Sci. Adv.} {\bfseries 11} no.~36, (2025) adv6503}, \href{https://arxiv.org/abs/2409.13226}{{\ttfamily arXiv:2409.13226}}.

\bibitem{Li:2021hhx}
{\bfseries CDEX} Collaboration, H.-B. Li, ``{New Limits on WIMP Dark Matter from Annual Modulation Analysis of the CDEX Experiment at the China Jinping Underground Laboratory},'' \href{https://dx.doi.org/10.22323/1.390.0643}{{\em PoS} {\bfseries ICHEP2020} (2021) 643}.

\bibitem{DAMIC-M:2023hgj}
{\bfseries DAMIC-M} Collaboration, I.~Arnquist {\em et~al.}, ``{Search for Daily Modulation of MeV Dark Matter Signals with DAMIC-M},'' \href{https://dx.doi.org/10.1103/PhysRevLett.132.101006}{{\em Phys. Rev. Lett.} {\bfseries 132} no.~10, (2024) 101006}, \href{https://arxiv.org/abs/2307.07251}{{\ttfamily arXiv:2307.07251}}.

\bibitem{DAMIC-M:2025ltz}
{\bfseries DAMIC-M} Collaboration, K.~Aggarwal {\em et~al.}, ``{Daily modulation constraints on light dark matter with DAMIC-M},'' \href{https://dx.doi.org/10.1103/c9fm-ymf2}{{\em Phys. Rev. D} {\bfseries 114} no.~3, (2026) 032010}, \href{https://arxiv.org/abs/2511.13962}{{\ttfamily arXiv:2511.13962}}.

\bibitem{SENSEI:2025qvp}
{\bfseries SENSEI} Collaboration, I.~M. Bloch {\em et~al.}, ``{SENSEI: A Search for Diurnal Modulation in Sub-GeV Dark Matter Scattering},'' \href{https://dx.doi.org/10.1103/1gpp-3tqd}{{\em Phys. Rev. Lett.} {\bfseries 136} no.~21, (2026) 211001}, \href{https://arxiv.org/abs/2510.20889}{{\ttfamily arXiv:2510.20889}}.

\bibitem{DarkSide-50:2023fgf}
{\bfseries DarkSide-50} Collaboration, P.~Agnes {\em et~al.}, ``{Search for dark matter annual modulation with DarkSide-50},'' \href{https://dx.doi.org/10.1103/PhysRevD.110.102006}{{\em Phys. Rev. D} {\bfseries 110} no.~10, (2024) 102006}, \href{https://arxiv.org/abs/2307.07249}{{\ttfamily arXiv:2307.07249}}.

\bibitem{Busoni:2025tqa}
G.~Busoni, J.~M. Cornell, W.~Handley, F.~Kahlhoefer, A.~Kvellestad, M.~Pitts, L.~Street, A.~C. Vincent, and M.~White, ``{DAMA/LIBRA and dark matter: decisive tension or contrived cancellation},'' \href{https://dx.doi.org/10.1088/1475-7516/2026/08/047}{{\em JCAP} {\bfseries 08} (2026) 047}, \href{https://arxiv.org/abs/2510.05216}{{\ttfamily arXiv:2510.05216}}.

\bibitem{XENON:2017nik}
{\bfseries XENON} Collaboration, E.~Aprile {\em et~al.}, ``{Search for Electronic Recoil Event Rate Modulation with 4 Years of XENON100 Data},'' \href{https://dx.doi.org/10.1103/PhysRevLett.118.101101}{{\em Phys. Rev. Lett.} {\bfseries 118} no.~10, (2017) 101101}, \href{https://arxiv.org/abs/1701.00769}{{\ttfamily arXiv:1701.00769}}.

\bibitem{LUX:2018xvj}
{\bfseries LUX} Collaboration, D.~S. Akerib {\em et~al.}, ``{Search for annual and diurnal rate modulations in the LUX experiment},'' \href{https://dx.doi.org/10.1103/PhysRevD.98.062005}{{\em Phys. Rev. D} {\bfseries 98} no.~6, (2018) 062005}, \href{https://arxiv.org/abs/1807.07113}{{\ttfamily arXiv:1807.07113}}.

\bibitem{BATTAT201765}
J.~Battat, A.~Ezeribe, {\em et~al.}, ``Low threshold results and limits from the drift directional dark matter detector,'' \href{https://dx.doi.org/https://doi.org/10.1016/j.astropartphys.2017.03.007}{{\em Astrop. Phys.} {\bfseries 91} (2017) 65}.

\bibitem{Battat_2021}
{\bfseries DRIFT} Collaboration, J.~Battat {\em et~al.}, ``{Improved sensitivity of the DRIFT-IId directional dark matter experiment using machine learning},'' \href{https://dx.doi.org/10.1088/1475-7516/2021/07/014}{{\em JCAP} {\bfseries 2021} no.~07, (Jul, 2021) 014}.

\bibitem{Ourahou:2026tpf}
I.~Ourahou, D.~Santos, O.~Guillaudin, J.-F. Muraz, N.~Sauzet, and C.~Tao, ``{First Directional Dark Matter Limits from the MIMAC {\ensuremath{\mu}}-TPC Detector},'' \href{https://arxiv.org/abs/2607.23527}{{\ttfamily arXiv:2607.23527}}.

\bibitem{Yakabe:2020rua}
R.~Yakabe {\em et~al.}, ``{First limits from a 3D-vector directional dark matter search with the NEWAGE-0.3b{\textquoteright} detector},'' \href{https://dx.doi.org/10.1093/ptep/ptaa147}{{\em PTEP} {\bfseries 2020} no.~11, (2020) 113F01}, \href{https://arxiv.org/abs/2005.05157}{{\ttfamily arXiv:2005.05157}}.

\bibitem{Shimada:2023vky}
T.~Shimada {\em et~al.}, ``{Direction-sensitive dark matter search with 3D-vector-type tracking in NEWAGE},'' \href{https://dx.doi.org/10.1093/ptep/ptad120}{{\em PTEP} {\bfseries 2023} no.~10, (2023) 103F01}, \href{https://arxiv.org/abs/2301.04779}{{\ttfamily arXiv:2301.04779}}.

\bibitem{Folcarelli:2026wkv}
M.~Folcarelli {\em et~al.}, ``{Projected sensitivity to light WIMP-like particles of the BULLKID-DM experiment},'' \href{https://arxiv.org/abs/2601.13766}{{\ttfamily arXiv:2601.13766}}.

\bibitem{Hughes:2024nub}
{\bfseries COSINUS} Collaboration, M.~Hughes, ``{The COSINUS Underground Cryogenic Facility},'' \href{https://dx.doi.org/10.22323/1.441.0096}{{\em PoS} {\bfseries TAUP2023} (2024) 096}.

\bibitem{McKinsey_TESSERACT_Snowmass}
{\bfseries TESSERACT} Collaboration, D.~N. McKinsey {\em et~al.}, ``{The TESSERACT Dark Matter Project},'' SNOWMASS Letter of Intent (LOI).

\bibitem{Billard:2023_IN2P3}
J.~Billar {\em et~al.}, ``{Transition Edge Sensors with Sub-eV Resolution And Cryogenic Targets (TESSERACT) FRANCE},'' IN2P3 Scientific Council (October session on direct dark matter search), 2023.
\newblock \url{https://atrium.in2p3.fr/nuxeo/nxfile/default/72936158-0c83-4c23-8696-7582148b4097/blobholder:0/4-Doc-TESSERACT-Billard-Scorza.pdf}.

\bibitem{Lin:2025orh}
H.~Lin {\em et~al.}, ``{First Production of Skipper-CCD Modules for the DAMIC-M Experiment},'' \href{https://arxiv.org/abs/2509.06943}{{\ttfamily arXiv:2509.06943}}.

\bibitem{VilarIDM:2026aaa}
R.~Vilar, ``{The DAMIC-M experiment: an Skipper CCD detector at LSM},'' 2026.
\newblock Identification of Dark Matter 2026.

\bibitem{Antonello:2020xhj}
M.~Antonello {\em et~al.}, ``{Characterization of SABRE crystal NaI-33 with direct underground counting},'' \href{https://dx.doi.org/10.1140/epjc/s10052-021-09098-5}{{\em Eur. Phys. J. C} {\bfseries 81} no.~4, (2021) 299}, \href{https://arxiv.org/abs/2012.02610}{{\ttfamily arXiv:2012.02610}}.

\bibitem{Calaprice:2022vte}
F.~Calaprice {\em et~al.}, ``{Performance of the SABRE detector module in a purely passive shielding},'' \href{https://dx.doi.org/10.1140/epjc/s10052-022-11108-z}{{\em Eur. Phys. J. C} {\bfseries 82} no.~12, (2022) 1158}, \href{https://arxiv.org/abs/2205.13876}{{\ttfamily arXiv:2205.13876}}.

\bibitem{SABRESouth:2024bpv}
{\bfseries SABRE South} Collaboration, E.~Barberio {\em et~al.}, ``{The SABRE South technical design report executive summary},'' \href{https://dx.doi.org/10.1088/1748-0221/20/04/T04001}{{\em JINST} {\bfseries 20} no.~04, (2025) T04001}, \href{https://arxiv.org/abs/2411.13889}{{\ttfamily arXiv:2411.13889}}.

\bibitem{Krauss:2024jzg}
C.~Krauss and S.~Viel, ``{Direct detection searches for dark matter particles using superheated bubble chambers and cryogenic liquid argon detectors},'' \href{https://dx.doi.org/10.1139/cjp-2024-0131}{{\em Can. J. Phys.} {\bfseries 103} no.~8, (2025) 704}.

\bibitem{Moore:2024que}
C.~Moore, {\em {Dark Matter Search with the PICO-40L Bubble Chamber}}.
\newblock PhD thesis, Queen's U., Kingston, Queen's U., Kingston, 2024.

\bibitem{DarkSide-20k:2024yfq}
{\bfseries DarkSide-20k} Collaboration, F.~Acerbi {\em et~al.}, ``{DarkSide-20k sensitivity to light dark matter particles},'' \href{https://dx.doi.org/10.1038/s42005-024-01896-z}{{\em Commun. Phys.} {\bfseries 7} no.~1, (2024) 422}, \href{https://arxiv.org/abs/2407.05813}{{\ttfamily arXiv:2407.05813}}.

\bibitem{PANDA-X:2024dlo}
{\bfseries PANDA-X, PandaX} Collaboration, A.~Abdukerim {\em et~al.}, ``{PandaX-xT{\textemdash}A deep underground multi-ten-tonne liquid xenon observatory},'' \href{https://dx.doi.org/10.1007/s11433-024-2539-y}{{\em Sci. China Phys. Mech. Astron.} {\bfseries 68} no.~2, (2025) 221011}, \href{https://arxiv.org/abs/2402.03596}{{\ttfamily arXiv:2402.03596}}.

\bibitem{SPICE:2023tru}
{\bfseries SPICE, HeRALD} Collaboration, R.~Anthony-Petersen {\em et~al.}, ``{Demonstration of the HeRALD superfluid helium detector concept},'' \href{https://dx.doi.org/10.1103/PhysRevD.110.072006}{{\em Phys. Rev. D} {\bfseries 110} no.~7, (2024) 072006}, \href{https://arxiv.org/abs/2307.11877}{{\ttfamily arXiv:2307.11877}}.

\bibitem{Coquillat:2024tfa}
{\bfseries NEWS-G} Collaboration, J.-M. Coquillat, ``{Latest results from the NEWS-G dark matter experiment},'' \href{https://dx.doi.org/10.22323/1.441.0042}{{\em PoS} {\bfseries TAUP2023} (2024) 042}.

\bibitem{NEWS-G:2023qwh}
{\bfseries NEWS-G} Collaboration, L.~Balogh {\em et~al.}, ``{Exploring light dark matter with the DarkSPHERE spherical proportional counter electroformed underground at the Boulby Underground Laboratory},'' \href{https://dx.doi.org/10.1103/PhysRevD.108.112006}{{\em Phys. Rev. D} {\bfseries 108} no.~11, (2023) 112006}, \href{https://arxiv.org/abs/2301.05183}{{\ttfamily arXiv:2301.05183}}.

\bibitem{BULLKID:2026aaa}
T.~Lari {\em et~al.}, ``{Developing the Cryogenic Veto for BULLKID-DM Experiment},'' \href{https://dx.doi.org/10.1109/TASC.2026.3665727}{{\em IEEE Trans. Appl. Supercond.} {\bfseries 36} no.~6, (2026) 1}.

\bibitem{Delicato:2026pbj}
D.~Delicato {\em et~al.}, ``{Background of the BULLKID detector array operated with moderate shield on surface},'' \href{https://arxiv.org/abs/2605.02468}{{\ttfamily arXiv:2605.02468}}.

\bibitem{COSINUS:2021onk}
{\bfseries COSINUS} Collaboration, G.~Angloher {\em et~al.}, ``{First measurements of remoTES cryogenic calorimeters: Easy-to-fabricate particle detectors for a wide choice of target materials},'' \href{https://dx.doi.org/10.1016/j.nima.2022.167532}{{\em Nucl. Instrum. Meth. A} {\bfseries 1045} (2023) 167532}, \href{https://arxiv.org/abs/2111.00349}{{\ttfamily arXiv:2111.00349}}.

\bibitem{Liao:2021npo}
J.~Liao, Y.~Gao, Z.~Liang, Z.~Peng, Z.~Ouyang, L.~Zhang, L.~Zhang, and J.~Zhou, ``{A low-mass dark matter project, ALETHEIA: A Liquid hElium Time projection cHambEr In dArk matter},'' \href{https://arxiv.org/abs/2103.02161}{{\ttfamily arXiv:2103.02161}}.

\bibitem{Solmaz:2025xiq}
{\bfseries DELight} Collaboration, M.~Solmaz, ``{The Direct Search Experiment for Light Dark Matter (DELight): Overview and Perspectives},'' \href{https://dx.doi.org/10.22323/1.511.0052}{{\em PoS(TAUP2025)} (12, 2025) 0052}, \href{https://arxiv.org/abs/2512.15278}{{\ttfamily arXiv:2512.15278}}.

\bibitem{RES-NOVACollaboration:2025stq}
{\bfseries RES-NOVA} Collaboration, D.~Alloni {\em et~al.}, ``{New dark matter direct search based on archaeological Pb},'' \href{https://dx.doi.org/10.1103/wcvd-rk1f}{{\em Phys. Rev. D} {\bfseries 111} no.~10, (2025) 103050}, \href{https://arxiv.org/abs/2501.12409}{{\ttfamily arXiv:2501.12409}}.

\bibitem{Vasiukov:2019hwn}
S.~Vasiukov, F.~Chiossi, C.~Braggio, G.~Carugno, F.~Moretti, E.~Bourret, and S.~Derenzo, ``{GaAs as a bright cryogenic scintillator for the direct dark matter detection},'' \href{https://arxiv.org/abs/1904.09362}{{\ttfamily arXiv:1904.09362}}.

\bibitem{BleIDM:2026aaa}
J.~Blé, ``{The semiconductor technology for the Tesseract direct dark matter search experiment at LSM},'' 2026.
\newblock Identification of Dark Matter 2026.

\bibitem{EDELWEISS:2016nzl}
{\bfseries EDELWEISS} Collaboration, L.~Hehn {\em et~al.}, ``{Improved EDELWEISS-III sensitivity for low-mass WIMPs using a profile likelihood approach},'' \href{https://dx.doi.org/10.1140/epjc/s10052-016-4388-y}{{\em Eur. Phys. J. C} {\bfseries 76} no.~10, (2016) 548}, \href{https://arxiv.org/abs/1607.03367}{{\ttfamily arXiv:1607.03367}}.

\bibitem{XiaoIDM:2026aaa}
X.~Xiao, ``{PandaX-20T: the next generation xenon detector for dark matter and neutrino searches},'' 2026.
\newblock Identification of Dark Matter 2026.

\bibitem{Castel:2018gcp}
J.~Castel {\em et~al.}, ``{Background assessment for the TREX Dark Matter experiment},'' \href{https://dx.doi.org/10.1140/epjc/s10052-019-7282-6}{{\em Eur. Phys. J. C} {\bfseries 79} no.~9, (2019) 782}, \href{https://arxiv.org/abs/1812.04519}{{\ttfamily arXiv:1812.04519}}.

\bibitem{Cappella:2013rua}
F.~Cappella {\em et~al.}, ``{On the potentiality of the $ZnWO_{4}$ anisotropic detectors to measure the directionality of Dark Matter},'' \href{https://dx.doi.org/10.1140/epjc/s10052-013-2276-2}{{\em Eur. Phys. J. C} {\bfseries 73} no.~1, (2013) 2276}.

\bibitem{Leoncini:2025haz}
A.~Leoncini {\em et~al.}, ``{Study of Dark Matter with directionality approach using ZnWO4 crystal scintillators},'' \href{https://dx.doi.org/10.1051/epjconf/202531910003}{{\em EPJ Web Conf.} {\bfseries 319} (2025) 10003}.

\bibitem{DarkSide20k:2020ymr}
{\bfseries DarkSide 20k} Collaboration, P.~Agnes {\em et~al.}, ``{Sensitivity of future liquid argon dark matter search experiments to core-collapse supernova neutrinos},'' \href{https://dx.doi.org/10.1088/1475-7516/2021/03/043}{{\em JCAP} {\bfseries 03} (2021) 043}, \href{https://arxiv.org/abs/2011.07819}{{\ttfamily arXiv:2011.07819}}.

\bibitem{Amaro:2026mes}
F.~D. Amaro {\em et~al.}, ``The cygno experiment: a gaseous tpc with optical readout for rare events searches,'' \href{https://dx.doi.org/10.1088/1748-0221/21/06/C06006}{{\em JINST} {\bfseries 21} no.~06, (Jun, 2026) C06006}, \href{https://arxiv.org/abs/2603.22595}{{\ttfamily 2603.22595}}.

\bibitem{Knights:2023fry}
{\bfseries NEWS-G} Collaboration, P.~Knights, ``{Status and future prospects of the NEWS-G experiment},'' \href{https://dx.doi.org/10.21468/SciPostPhysProc.12.024}{{\em SciPost Phys. Proc.} {\bfseries 12} (2023) 024}.

\bibitem{NEWSdm:2023qyb}
{\bfseries NEWSdm} Collaboration, N.~Y. Agafonova {\em et~al.}, ``{Directional sensitivity of the NEWSdm experiment to cosmic ray boosted dark matter},'' \href{https://dx.doi.org/10.1088/1475-7516/2023/07/067}{{\em JCAP} {\bfseries 07} (2023) 067}, \href{https://arxiv.org/abs/2305.00112}{{\ttfamily arXiv:2305.00112}}.

\bibitem{XLZD:2024pdv}
{\bfseries XLZD} Collaboration, J.~Aalbers {\em et~al.}, ``{Neutrinoless double beta decay sensitivity of the XLZD rare event observatory},'' \href{https://dx.doi.org/10.1088/1361-6471/adb900}{{\em J. Phys. G} {\bfseries 52} no.~4, (2025) 045102}, \href{https://arxiv.org/abs/2410.19016}{{\ttfamily arXiv:2410.19016}}.

\end{thebibliography}\endgroup

	\newpage
	\appendix
	\renewcommand*{\thesection}{\Alph{section}}

\end{document}